\documentclass[10pt,twoside,letterpaper]{book}
\usepackage[english,german]{babel}
\usepackage{afterpage,caption,framed,graphicx,marginnote,pdfpages,subfig}
\usepackage{multirow,float}
\usepackage{amsmath,amsfonts,amssymb}
\usepackage[left=25mm,textwidth=110mm,marginparsep=7mm,marginparwidth=65mm]{geometry}
\usepackage[nottoc]{tocbibind}
\usepackage[numbers,sort]{natbib}

\newcommand{\mychapter}[2]{
    \setcounter{chapter}{#1}
    \setcounter{section}{0}
    \chapter*{#2}
    \addcontentsline{toc}{chapter}{#2}}

\newcommand\blfootnote[1]{
    \begingroup
    \renewcommand\thefootnote{}\footnote{#1}
    \addtocounter{footnote}{-1}
    \endgroup}

\begin{document}
        \newgeometry{}

    \pagenumbering{gobble}
    \includepdf{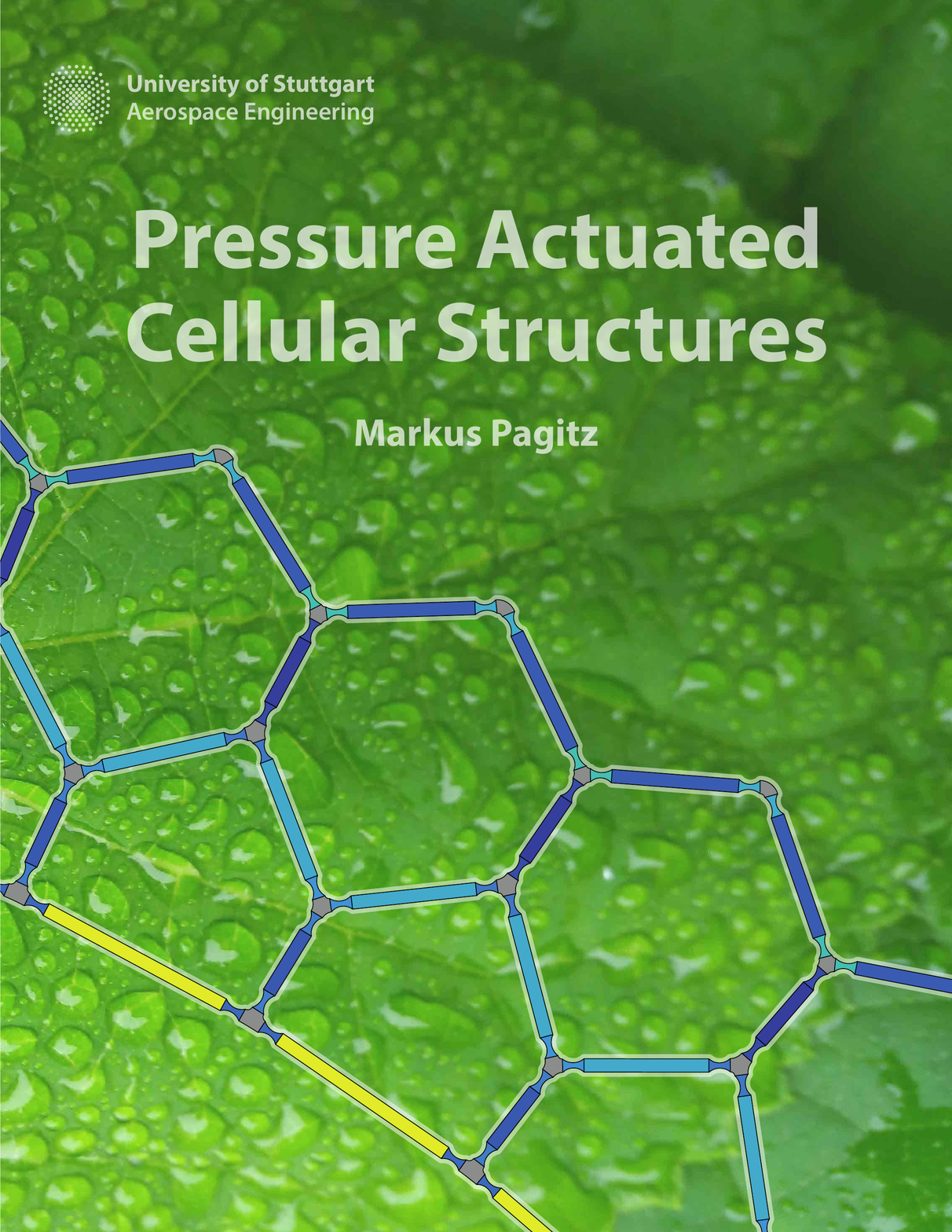}


    \selectlanguage{english}

    \cleardoublepage

    \begin{center}
        The reasonable man adapts himself to the world;\\
        the unreasonable one persists in trying to adapt the world to himself.\\
        Therefore all progress depends on the unreasonable man.\\
        \vspace{3mm}
        \textbf{George Bernard Shaw}
    \end{center}

    \vspace{137mm}\hspace{100mm}
    \includegraphics[width=\marginparwidth]{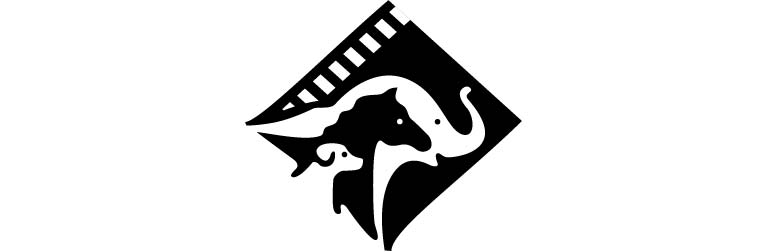}

    \vspace{2mm}\hspace{100mm}
    \begin{minipage}{\marginparwidth}
        \begin{center}
            No students were used\\ or harmed in the\\ making of this thesis.
        \end{center}
    \end{minipage}


    \restoregeometry

    \frontmatter
    \cleardoublepage
    \mychapter{0}{Preface}
    \marginnote{
    \begin{center}
        \includegraphics[width=\marginparwidth]{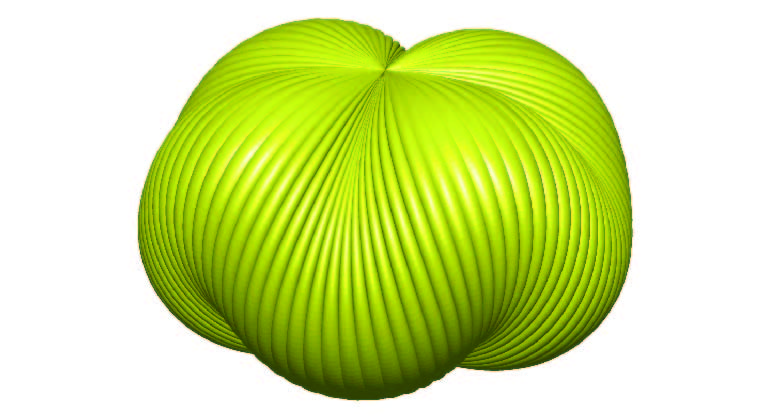}
        \small superpressure balloons\\
        \vspace{14mm}
        \includegraphics[width=\marginparwidth]{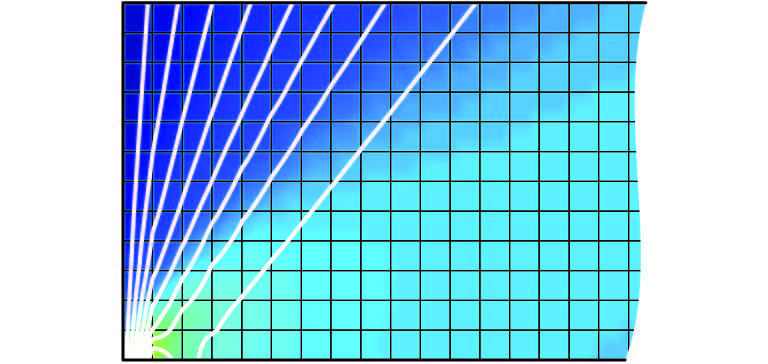}
        \small wrinkling of membranes\\
        \vspace{5mm}
        \includegraphics[width=\marginparwidth]{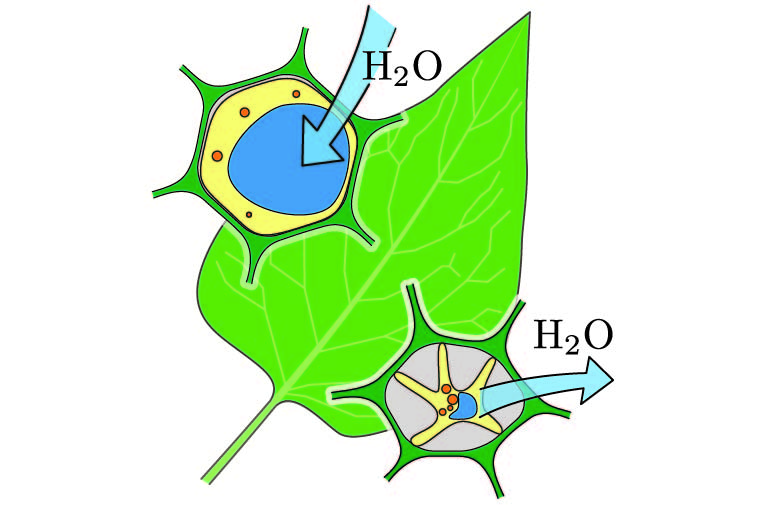}
        \small shape changing structures
    \end{center}}

    \noindent During my studies for a PhD at the University of Cambridge I have worked on the development of NASA's superpressure balloons. These balloons can partially replace satellites and serve as platforms for long-term high altitude experiments. The development of this technology started in the 1990s and was severely hindered for more than a decade by stability problems. My main contribution to this project was a redesigned cutting pattern that increased stability by up to 300~\% without increasing the overall weight and membrane, tendon stresses \cite{Pagitz2007-1, Pagitz2007-2, Pagitz2007-3, Pagitz2010-1, Pagitz2011-1}. A NASA balloon with a diameter of 80~m that is based on this design made a world record flight above Antarctica in 2009. I received a research fellowship at St John's College, Cambridge for my work in this area. Google currently investigates the use of these balloons to establish a global, low cost access to the internet.\\

    After my PhD I have been on the lookout for a new problem. I experimented, among others, with tensegrity structures and tension field theory. During that time I have developed a multi-grid approach for the simulation of wrinkled membranes \cite{Pagitz2010-2}. This approach is capable of matching the analytical tension field of a rectangular membrane for pure shear deformations in the whole domain. Another result from that period is an algorithm for the form finding of tensegrity structures that is fast, robust and possesses an asymptotic quadratic convergence rate \cite{Pagitz2009-1}.\\

    Subsequent to my time as a Research Fellow I have worked at TU Delft as an Assistant Professor. I was still on the lookout for a new problem when I started to investigate the nastic movement of plants. What was supposed to be a short stint turned out to be a problem that kept me occupied for more than seven years. During that time I have found that it is possible to create shape changing structures by combining independently pressurized rows of specifically tailored prismatic cells \cite{Pagitz2012-1, Pagitz2012-2, Pagitz2012-3, Pagitz2013-1, Pagitz2014-1, Pagitz2017-1, Pagitz2017-2}. A key advantage of these pressure actuated cellular structures is their complete fusion between actuators and mechanisms. Furthermore, they are lightweight, strong and capable of considerable shape changes. These properties make them particularly attractive for the aerospace industry that requires advanced high lift devices. At this point I moved from TU Delft to the German Aerospace Agency in Braunschweig to build a number of prototype structures and to investigate various manufacturing techniques. However, I started to miss the rigor and relevance of an academic environment. Therefore, I am indebted to Professor Ricken for giving me the opportunity to write my postdoctoral thesis about pressure actuated cellular structures at the University of Stuttgart.


    \newgeometry{}

    \cleardoublepage
    \mychapter{0}{Abstract}

    A large number of important developments were initiated at the beginning of the last century by a series of manned flights with controlled airplanes. The underpowered vehicles that were used during these early days were mostly constructed from wood and textiles. Despite these humble beginnings, airplanes with a range of more than 40,000 km, a maximum speed of more than 7,000 km/h and a flight ceiling above the von K\'arm\'an line at 100 km were constructed within the following century.\\

    In view of these advancements, it is remarkable that the development approach for airplanes remained largely unchanged. Passenger airplanes still consist of a cylindrical fuselage, wings, high lift devices, engines and control surfaces that are, as far as possible, independently developed and manufactured. The performance improvements that were achieved so far are thus mostly driven by the steady enhancements of individual components. With regard to airplane structures, these enhancements are mostly based on the availability of new materials and the development of computer aided simulation and manufacturing techniques.\\

    The same can be said about the individual components. For example, high lift devices consist of several subcomponents that are, as far as possible, independently developed and manufactured. Aerodynamically shaped rigid bodies are usually connected to wings at discrete points via simple mechanisms. Their positions are controlled by linear or rotary hydraulic or electric actuators. This design methodology inevitably leads to discontinuities in an airplanes skin and thus increases the overall noise emission, fuel consumption and radar signature. Furthermore, current high lift devices do not contribute significantly to a wings stiffness and load carrying capacity.\\

    Potential improvements that can be achieved by a decoupled optimization of individual components or subcomponents are largely exploited. This is reflected by the latest generation of passenger airplanes that achieved only a slight reduction of noise emissions and fuel consumptions while cruise speeds remained largely unchanged. On the other hand, fuel costs, passenger volumes and population densities around airports increase steadily. This puts additional pressure on the airplane industry as current noise and pollutant emissions are critically viewed by a large part of the general public.\\

    It is likely that the improvements needed to counter these trends can only be achieved by an integral consideration of different airplane components. A fusion between high lift devices and wings seems to be particularly promising in this regard. The use of shape changing, gapless high lift devices that are tightly integrated into the wings would not only reduce the noise emissions during takeoff and landing but additionally enable the use of advanced technologies that improve the laminarity of the airflow. These technologies have the potential to increase the lift to drag ratio of wings and thus to significantly reduce the weight and fuel consumption of airplanes.\\

    In contrast, the noise emissions and fuel consumption of military airplanes are not a main concern. More important are their flight speeds and stealth properties as they determine their survivability. A large part of the radar reflections of modern fighter jets and bombers can be traced back to discontinuities in their skins at component boundaries. As a consequence, stealth properties can only be significantly improved by a fusion of neighbouring components. For example, the use of gapless high lift devices would be beneficial as it eliminates many of the gaps that can be found in current airplane skins.\\

    The realization of gapless high lift devices requires skins that are stiff enough to carry the aerodynamic loads and soft enough to undergo the required shape changes. It is usually only possible to satisfy these conflicting demands with the help of complex mechanisms that tightly support the skins. This is one of the reasons why previously developed gapless high lift devices that are based on a separate consideration of actuators, skins and mechanisms are too heavy.\\

    The development of a novel concept for gapless high lift devices that fuses actuators and mechanisms into a single structure is thus indispensable. The nastic movement of plants that is based on the pressure driven shape changes of a large number of cells with individually tailored geometries is subsequently used as a source of inspiration. A main advantage of these ``pressure actuated cellular structures'' is that they can be manufactured from a single piece of material that can range from plastics over composites and metals to ceramics. The required cell side thicknesses of these structures are relatively small as the pressure induced tension forces eliminate compression forces due to external loads. Another advantage of pressure actuated cellular structures is that their stiffness and shape can be independently varied.\\

    The design and manufacturing of structures that consist of a large number of three-dimensional cells with individually tailored geometries and pressures is difficult. It is therefore advantageous that cell geometries and pressure supplies can be greatly simplified for prismatic high lift devices. The pressure actuated cellular structures that are considered in this thesis are based on individually pressurized rows of prismatic cells that are connected to each other. These structures can be efficiently manufactured from fiber reinforced composites with the help of modern manufacturing techniques.\\

    This postdoctoral thesis starts by reviewing the historic development of airplane structures, actuators, high lift devices and skins. However, the main purpose of this work is the development of a novel, plant inspired concept for shape changing, gapless high lift devices. This is done in a first step by investigating the working principles of a wide range of plant movements. The principles that form the basis of nastic movements are subsequently abstracted to create manageable engineering models. A sophisticated algorithm for the simulation and optimization of these models is introduced in the remainder of this thesis. It is demonstrated with the help of several examples that pressure actuated cellular structures can be designed to undergo surprisingly large shape changes while being lightweight, energy efficient and capable of carrying large loads.


    \selectlanguage{german}

    \cleardoublepage
    \mychapter{0}{Zusammenfassung}
        Die rasante Entwicklung der Luft- und Raumfahrt wurde zu Beginn des letzten Jahrhunderts durch eine Reihe von kurzen, bemannten Fl\"ugen eingeleitet. Die daf\"ur verwendeten, untermotorisierten Flugger\"ate, welche vorwiegend aus Holz und Textilien gefertigt wurden, waren nur sehr schwer zu kontrollieren. Trotz dieser bescheidenen Anf\"ange wurden innerhalb eines Jahrhunderts Flugzeuge mit einer Reichweite von mehr als 40,000 km, einer Maximalgeschwindigkeit von mehr als 7,000 km/h und einer Flugh\"ohe von mehr als 100 km gebaut.\\

        In Anbetracht dieser gewaltigen Fortschritte ist es bemerkenswert, dass sich der Ansatz f\"ur die Gesamtentwicklung vor allem von Passagierflugzeugen nur unwesentlich ver\"andert hat. Moderne Passagierflugzeuge bestehen nach wie vor aus zylindrischen R\"umpfen, Tragfl\"achen, Hochauftriebssystemen, Triebwerke und Leitwerke, welche so weit wie m\"oglich unabh\"angig voneinander entwickelt und hergestellt werden. Die bisher erzielten Leistungssteigerungen gehen daher vorwiegend auf die stetigen Verbesserungen einzelner Komponenten zur\"uck. Bez\"uglich der Flugzeugstrukturen wurde dies vor allem durch den Einsatz neuer Metalllegierungen, Faserverbundwerkstoffen und der Entwicklung von rechnergest\"utzten Simulationswerkzeugen und Fertigungsverfahren erreicht.\\

        Dasselbe kann auch \"uber die einzelnen Komponenten gesagt werden. Zum Beispiel werden die Hochauftriebssysteme moderner Flugzeuge, wie auch die Flugzeuge selbst, aus Subkomponenten hergestellt welche so weit wie m\"oglich unabh\"angig voneinander entwickelt werden. Aerodynamisch geformte Starrk\"orper werden \"ublicherweise \"uber relativ einfache Mechanismen an einzelnen Punkten mit den Tragfl\"achen verbunden und mithilfe hydraulisch oder elektrisch betriebener Aktuatoren ein- und ausgefahren. Diese Art der Konstruktion f\"uhrt unweigerlich zu Diskontinuit\"aten in der Flugzeughaut welche die L\"armemissionen, den Treibstoffverbrauch und die Radarsignatur vergr\"o\ss ern. Desweiteren tragen diese Art von Hochauftriebssystemen nur unwesentlich zur Lastabtragung und Steifigkeit der Tragfl\"achen bei.\\

        Die m\"oglichen Verbesserungen, welche durch eine entkoppelte Optimierung einzelner Komponenten erreicht werden k\"onnen, sind weitgehend ausgesch\"opft. Dies spiegelt sich vor allem in den stagnierenden Reisegeschwindigkeiten, moderaten L\"armreduktionen und Treibstoffeinsparungen der letzten Flugzeuggeneration wieder. Auf der anderen Seite steht ein zunehmendes Fluggastaufkommen, eine Verdichtung der Bev\"olkerung in der N\"ahe von Flugh\"afen und steigende Treibstoffkosten. Des Weiteren sinkt die Akzeptanz der Bev\"olkerung gegen\"uber L\"arm- und Schadstoffemissionen, was eine betr\"achtliche Verbesserung von Flugzeugen erforderlich macht.\\

        Derart weitgehende Verbesserungen sind wahrscheinlich nur durch eine integrale Betrachtung verschiedener Flugzeugkomponenten m\"oglich. Am vielversprechendsten erscheint im Moment die Fusion von Hochauftriebssystemen und Tragfl\"achen zu sein. Der Einsatz von formvariablen Hochauftriebssystemen welche formschl\"ussig mit den Tragfl\"achen verbunden sind w\"urde nicht nur die L\"armemissionen besonders w\"ahrend der Start- und Landephasen reduzieren, sondern auch den Einsatz neuartiger Technologien erm\"oglichen, welche die Laminarit\"at der Fl\"ugelumstr\"omung verbessern und dadurch sowohl den Treibstoffverbrauch als auch die L\"armemissionen senken.\\

        Eine \"ahnliche Entwicklung hat auch im milit\"arischen Flugzeugbau stattgefunden, wo die \"Uber- lebensf\"ahigkeit von Flugzeugen zunehmend von ihrer Radartarnung abh\"angt. Auch hier f\"uhrten die bisherigen Entwicklungen dazu das weitere, signifikante Fortschritte nicht mehr nur durch die Optimierung einzelner Komponenten erzielt werden k\"onnen. Ein gro\ss er Teil der Radarreflektionen von modernen Kampfflugzeugen haben ihren Ursprung an den Spalten und Kanten der Flugzeugh\"ullen. Die Entwicklung von spaltfreien Hochauftriebssystemen ist daher auch hier von zentraler Bedeutung.\\

        Die Realisierung formvariabler, spaltfreier Hochauftriebssysteme erfordert Flugzeugh\"aute die steif genug sind, um die aerodynamischen Lasten ohne gro\ss e Verformungen zu tragen und elastisch genug sind um die ben\"otigten Form\"anderungen zu erm\"oglichen. Diese gegens\"atzlichen Anforderungen k\"onnen in der Regel nur mithilfe komplexer Mechanismen erf\"ullt werden welche die Flugzeugh\"aute an zahlreichen Stellen unterst\"utzen beziehungsweise versteifen. Dies ist einer der Hauptgr\"unde, warum bisherige Hochauftriebssysteme welche auf einer Trennung zwischen Flugzeugh\"auten, Mechanismen und Aktuatoren basieren in der Regel deutlich zu schwer sind.\\

        Die Entwicklung eines neuen Konzepts f\"ur spaltfreie Hochauftriebssysteme, welches das Gesamtgewicht durch eine Fusion von Mechanismen und Aktuatoren signifikant reduziert, ist daher unumg\"anglich. Als Vorbild hierf\"ur dient im Folgenden die nastische Bewegung von Pflanzen. Diese basiert auf den Gestalts\"anderungen einer gro\ss en Anzahl von komplex geformten, dreidimensionalen Zellen, welche durch relative Zelldruck\"anderungen hervorgerufen werden. Ein Hauptvorteil dieser ``Druck Aktuierten Zellul\"aren Strukturen'' ist das sie in einem St\"uck und je nach Einsatzzweck aus verschiedenen Materialien gefertigt werden k\"onnen. Des Weiteren werden Instabilit\"aten durch die hohen Zelldr\"ucke weitgehend verhindert, was den Einsatz d\"unner und dadurch leichter Zellw\"ande erm\"oglicht. Weitere Vorteile liegen in einer ver\"anderlichen Steifigkeit welche von den Zelldr\"ucken abh\"angt und der m\"oglichen Zirkulation und Temperierung des Druckmediums um Vereisungen an einer Hochauftriebsfl\"ache vorzubeugen.\\

        Die Herstellung von dreidimensionalen Zellstrukturen, deren Zelldr\"ucke einzeln geregelt werden ist schwierig. Es ist daher von gro\ss em Vorteil das die Zellgeometrien sowie deren Bedruckung f\"ur den Einsatz in prismatischen Hochauftriebssystemen betr\"achtlich vereinfacht werden k\"onnen. Dies erm\"oglicht die Nutzung von miteinander verbundenen prismatischen Zellen, welche unter anderem mit modernen Webverfahren aus Faserverbundwerkstoffen in nahezu beliebigen L\"angen hergestellt werden k\"onnen.\\

        Diese Habilitationsschrift gibt zuerst einen Einblick in die geschichtliche Entwicklung von Flugzeugstrukturen und Hochauftriebssystemen. Ein besonderer Schwerpunkt wird hierbei auf die verwendeten Mechanismen und Aktuatoren gelegt. Das Hauptziel dieser Arbeit liegt jedoch in der Entwicklung eines neuartigen Konzepts f\"ur form\"andernde, spaltfreie Hochauftriebssysteme, welches auf dem Funktionsprinzip von nastischen Pflanzen basiert. Daf\"ur werden in einem ersten Schritt die verschiedenen Pflanzenbewegungen und deren zugrunde liegenden Mechanismen genauer untersucht. Die Funktionsprinzipien welche die Grundlage f\"ur nastische Pflanzenbewegungen bilden werden daraufhin abstrahiert und realisierbare Ingenieurmodelle entwickelt. Besonderes Augenmerk wird hierbei auf einen effizienten Ansatz f\"ur die Simulation und Optimierung von ``Druck Aktuierten Zellul\"aren Strukturen" gelegt. Mithilfe zahlreicher Beispiele wird gezeigt das diese Strukturen eine Reihe von Vorteilen besitzen, die sie f\"ur einen m\"oglichen Einsatz in zuk\"unftigen Hochauftriebssystemen pr\"adestinieren.


    \selectlanguage{english}

    \cleardoublepage
    \tableofcontents
    \cleardoublepage

    \restoregeometry 
        \mainmatter
    \chapter{Motivation}
        The first powered and controlled flights over very short distances in a heavier than air machine were most likely undertaken in 1901 by the German Gustave Wei\ss kopf. Since then, airplanes were developed with a range of more than 40,000~km, a maximum speed of more than 7,000~km/h and a ceiling above the von K\'arm\'an line at 100~km that represents, by definition, the boundary between the Earth's atmosphere and outer space. Today, flying is one of the safest modes of transport where civil airplanes such as the Airbus A380 carry routinely more than 850 passengers over distances of up to 15,000~km at a travel speed of 900~km/h.\\

        More than a century of research and development led to airplane structures that optimally combine existing materials, structural concepts and manufacturing techniques. The topology of a Boeing 777 wing that made its maiden flight in 1994 was optimized by Aage et al \cite{Aage2017-1} in 2017. They discretized the wing with more than one billion voxels and achieved a potential weight reduction of 2-5\% compared to the flight proven wing. However, it is currently not possible to economically manufacture the complex geometry of such an optimized wing. The small potential weight savings that were found in this study are remarkable and a testimony of the maturity of modern airplane structures. As a consequence, significant performance improvements can only be achieved by using new technologies or an integral consideration of existing actuators, manufacturing techniques, materials, mechanisms and structures. Furthermore, solely reducing the weight of an airplane is not necessarily optimal. More important optimality criterions are, for example, the fuel consumption, noise emissions, maintenance requirements and stealth properties.\\

        New technologies such as carbon fiber reinforced composites were widely introduced in the latest generation of civil airplanes to further reduce their overall weight \cite{Slayton2016-1}. As it appears today, one of the largest potential performance improvements in upcoming civil and military airplane generations might be achieved by a seamless integration of high lift devices into the wing structures. This is highlighted by the development history of fighter jets that started towards the end of World War II with propeller machines that were upgraded with simple jet engines \cite{Lorell1998-1} as illustrated in Figure~\ref{pic:Figure_1_1}. Aerodynamic improvements during the last seven decades were mostly driven by the rapid development of computers that allowed the use of unstable designs. The relaxation of current design constraints is continued by the shift towards unmanned airplanes \cite{Ehrhard2010-1} with advanced stealth properties \cite{Grant2010-1} and electronics. On the other hand, changes in the structural design were not as radical. For example, control surfaces are still rigid bodies that require gaps in an airplanes skin and thus create a source of radar reflections \cite{Jenn2005-1}. This is an increasingly big problem since these gaps limit an airplanes stealth properties and thus its survivability \cite{Barrett2017-1}. Hence, the development of technologies that minimize their impact is vital. One of the most promising approaches in this direction is the use of shape changing structures that avoid these gaps altogether.\\

        \marginnote{
        \begin{center}
            \includegraphics[width=\marginparwidth]{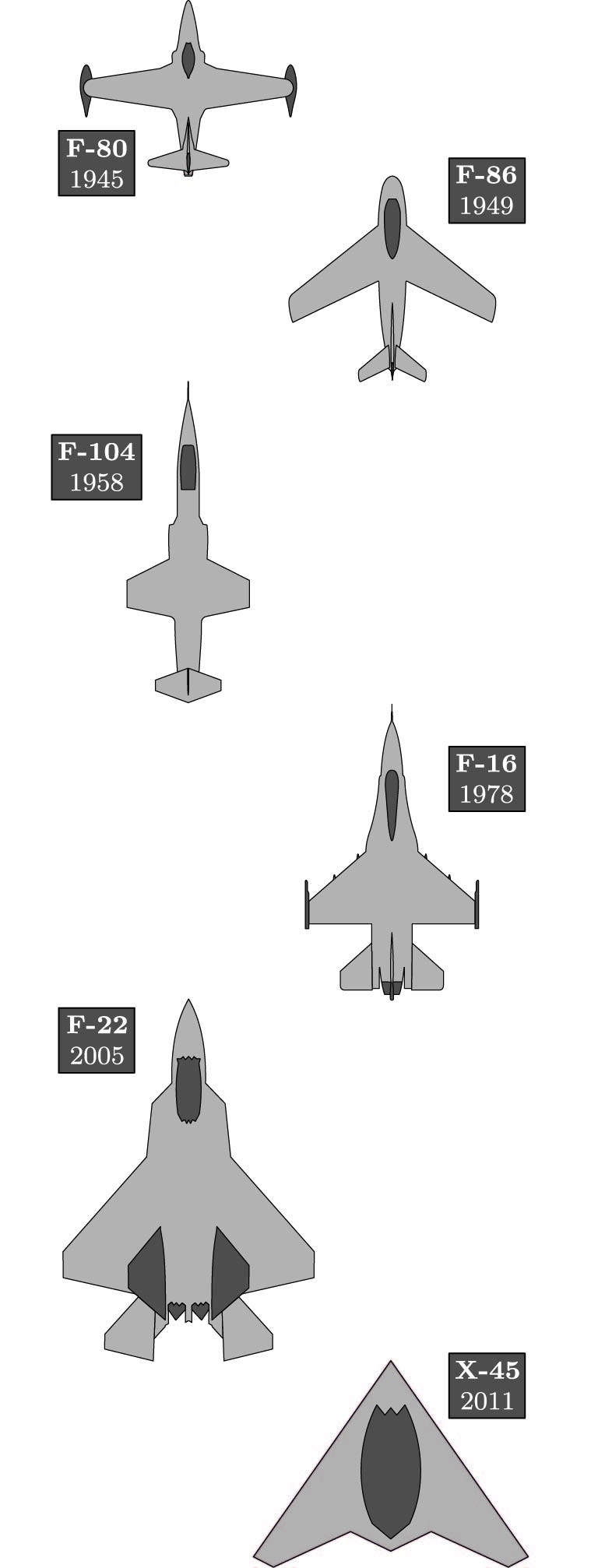}
            \captionof{figure}{Typical shapes of fighter jets during the last seven decades. Sixth generation fighter jets will most likely be unmanned.}
            \label{pic:Figure_1_1}
        \end{center}}[-82mm]

        In contrast, the radar signature of civilian airplanes is unimportant as long as they can be securely detected at all times. However, the weight and aerodynamic conditions of both civilian and military airplanes differ significantly during takeoff, cruise and landing. As a consequence, their wing shapes are usually designed for average flight conditions during the cruise phase and additional high lift devices are deployed during low flight speeds to increase their lift. However, this approach is disadvantageous as the gaps between the wings and high lift devices increase noise emissions particularly during takeoff and landing. This is a large problem for most airport operators due to the increasing number of flights and population densities. Furthermore, the joints between wings and high lift devices increase the turbulence of the flow during cruise and thus the fuel consumption. This was demonstrated by the flight testing of a Gulfstream III in 2014 that was retrofitted with shape changing, gapless trailing edges \cite{Kota2016-2}. These flights indicated potential fuel savings of up to 10\% and noise reductions during landings of up to 40\%. The civilian airplane industry is highly competitive and economically driven. Nonetheless it is relatively conservative due to the increased passenger risk that comes with the introduction of new, safety critical technologies. On the other hand, the benefits of using shape changing structures in unmanned aerial vehicles is significantly higher and the risk for the loss of lives on the operator side is minimal. Hence it is likely that gapless high lift devices will be first used in military airplanes.\\

        Many applications both within and outside of the aerospace industry would benefit from structures that can continuously change their shapes such that they optimize one or more objectives in a varying environment. These kind of structures are often referred to as being active, adaptive or even smart. Buzzwords like that are subsequently avoided since they are misleading and not clearly defined. Instead they are referred to as shape changing structures. Many authors have an illusive perception of potential performance improvements that come with the use of shape changing structures. High expectations are usually unrealistic since a shape changing structure is never lighter than an optimal rigid structure. This holds even if the structural loads are reduced by the shape changing capabilities. For example, safety considerations require an airplane to perform a wide range of flight maneuvers with deployed and retracted high lift devices \cite{Rudolph1996-1}. These failsafe requirements prohibit the utilization of potential load reductions during nominal operations. Hence, the benefits of using a shape changing structure must clearly outweigh its additional complexity, weight and cost. This is reflected in nature by the fact that only a small number of plants are capable of nastic movements. These plants are predominantly found in biotopes with challenging environmental conditions where the exploitation of an additional food source is vital.\\

        The development of new manufacturing techniques particularly in the field of three-dimensional weaving and the exponential growth of computing power enables the fast and inexpensive simulation, optimization and construction of advanced shape changing structures. The availability of these tools led to a point where an engineers competence is often the limiting factor. Competence can be rarely replaced by funding or a large workforce so that progress often depends on individuals. The aim of this postdoctoral thesis is the development of plant inspired shape changing structures. Unlike animals and humans, plants do not possess a central control system. Yet they are capable of fast reversible movements. Their relative simplicity and extremely tight integration of actuators, mechanisms and structures serves in the following as a source of inspiration for the design of future airplane structures.\\


    \newpage

    \sectionmark{Preview}
    \begin{framed}
        \noindent \textbf{Preview}\\

        \noindent The remainder of this thesis can be split into two parts. \textit{Chapters~2-5} review the historic development of conventional airplane components that range from structures \textit{(Chapter~2)}, actuators \textit{(Chapter~3)} and high lift devices \textit{(Chapter~4)} to flexible airplane skins \textit{(Chapter~5)}. The purpose of these chapters is to highlight the developments that led to the current state of the art which can be considered, at least from some perspectives, to be a dead end. The first part concludes by reviewing a wide range of plant movements \textit{(Chapter~6)} that serve in the following as an inspiration for the development of novel, gapless high lift devices. Readers that are familiar with these topics can gladly skip them.\\

        The second part starts by reviewing plant inspired structures that are capable of large shape changes \textit{(Chapter~7)}. This review is followed by the introduction of ``Pressure Actuated Cellular Structures", a novel concept \textit{(Chapter~8)} on which the remainder of this thesis is based. Their simulation and optimization is far from trivial and extensively described in \textit{Chapters~9-11}. A self-contained geometric model and a corresponding set of irreducible variables for the simulation, optimization and fabrication of pressure actuated cellular structures is introduced in \textit{Chapter~9}. \textit{Chapter~10} shows how the geometric model can be translated into a mechanical model for given material parameters. \textit{Chapter~11} combines the geometric and mechanical models for the efficient simulation and optimization of pressure actuated cellular structures. Various examples are used to demonstrate the potential of the presented approach. \textit{Chapter~12} concludes this thesis.
    \end{framed} 
        \cleardoublepage
    \chapter{Airplane Structures}
        Architecture is, most likely, the oldest art of designing and building structures. Vitruvius \cite{Rowland2001-1} wrote in the 1st century BC that a good structure should be durable, suitable for the purpose and aesthetically pleasing. Durability and utility is easy to define but it is considerably harder to find optimality criterions for beauty. Engineers are thus fortunate that the design of structural components in aerospace, civil and mechanical engineering is usually driven by the intended function. For example, an airplane is designed for given flight profiles and cargo properties such that its life-cycle costs are minimal. This is done by taking aerodynamic, control, mass, propulsion and structural considerations into account. These aspects are coupled to each other so that any small changes can significantly affect the whole design. Nonetheless, these dependencies can be relaxed by the use of advanced structural concepts and materials that minimize the overall weight and installation space.


        \section{Structural Concepts}
            \subsection{Spaceframe}
                The symbiosis between structural concepts and the availability of large amounts of inexpensive materials is particularly highlighted by bridges. Nearly all major bridges before the 19th century were made out of stone, a material that can sustain large compression but only small tension forces. The Romans \cite{Denison2012-1} took this into account by using arch like structures that carry the loads solely via compression forces. For example, the Alc\'antara bridge in Spain is one of the oldest arch bridges that is still fully functional. It was build in 106~AD and has a length of 194~m and a height of 71~m. Other examples include the relatively simple suspension bridges that were used in South America. They were made from ropes that can carry large tension but only small compression forces. As a consequence, their structural concept is based on an inverted arch.\\

                Large amounts of relatively inexpensive and durable materials that can carry tension as well as compression forces became available during the Industrial Revolution ($\sim$1760-1830) \cite{Ashton1924-1}. Furthermore, new manufacturing techniques and joining methods enabled the use of advanced structural concepts. As a consequence, truss bridges that combine tension and compression elements started to be widely used in the 19th century. For example, the first truss bridges in the United States used wood for the compression and iron for the tension elements whereas later bridges were fully made out of iron.\\

                A similar concept was used for the first airplanes. Their fabric covered wings were made out of wooden spars and ribs that defined the cross sectional shape. However, these wings could not sustain their flight loads so that they were supported by external wood struts and steel wires. Fuselages were build in a similar manner by using truss like structures where compression and tension forces were carried by struts and wires, respectively. For example, the Bl\'eriot XI that made its maiden flight in 1909 is based on this design principle, Figure~\ref{pic:Figure_2_1}. It was successfully used in many competitions and races and was the first airplane that crossed the English Channel.

                \marginnote{
                \begin{center}
                    \includegraphics[width=\marginparwidth]{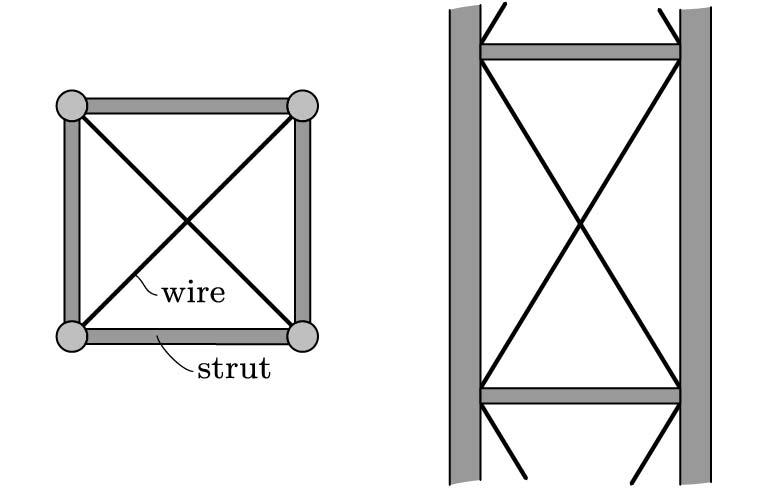}
                    \small(a)\vspace{10mm}
                    \includegraphics[width=\marginparwidth]{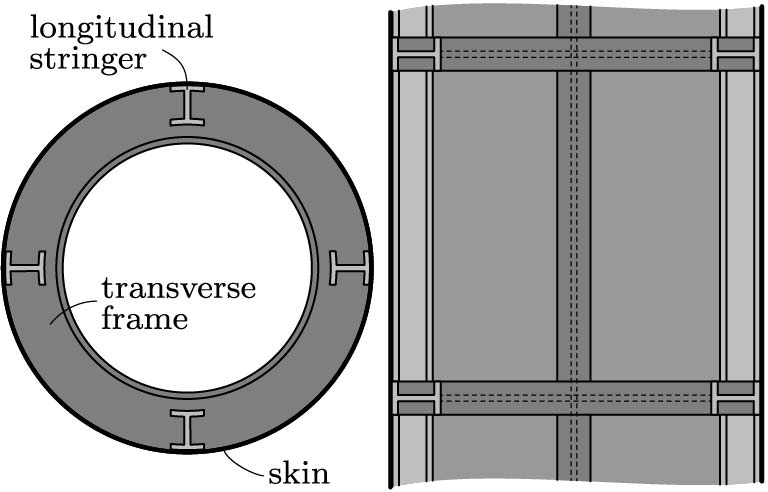}
                    \small(b)\vspace{10mm}
                    \includegraphics[width=\marginparwidth]{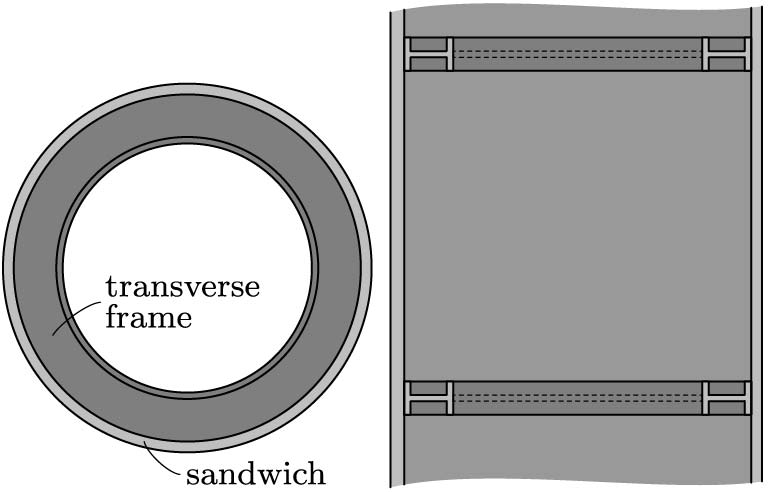}
                    \small(c)
                    \captionof{figure}{Cross sectional and top view of different construction principles for fuselages. (a) Spaceframe, (b) semi-monocoque and (c) monocoque construction.}
                    \label{pic:Figure_2_1}
                \end{center}}[-90mm]
                \vspace{-3mm}


            \subsection{Semi-Monocoque}
                The aerodynamic properties of fuselages that are based on truss like structures were improved by covering them with fabrics. It was soon realized that their tension bearing elements are redundant if the fabrics are sufficiently stiff and strong. As a consequence, the use of fabrics did not only improve the aerodynamics but additionally reduced the overall weight and complexity. The corresponding construction principle where a thin, load carrying skin is supported by transverse frames and longitudinal stringers is known as a semi-monocoque design.\\

                The fabric skins were subsequently replaced by plywood sheets with a relatively large out-of-plane bending stiffness. This further improved the aerodynamic properties of the fuselages and reduced the weight and complexity of the underlying support structures by eliminating at least some of the required compression elements. The Deperdussin Monocoque that made its maiden flight in 1912 was based on such a design. It raised the air speed record for aircraft\footnote{The speed record for cars was 228~km/h at that time. However, cars were soon eclipsed by airplanes.} to 204~km/h and was the fastest airplane for several years.\\

                Biplanes were widely used during the First World War. The truss like structures between the upper and lower wings increased their stiffness and reduced the overall weight. However, the use of two wings did not double the lift and the additional support structures increased the drag. This was problematic as improved engines led to higher airspeeds so that aerodynamic requirements became increasingly dominant. As a consequence, monoplanes without external reinforcements were often used instead. This led to considerably higher wing forces that required plywood instead of fabric covered wings.\\

                Semi-monocoque designs are widely used in modern airplanes. Their efficiency stems from the relatively thin skin that is locally supported against out-of-plane deformations such that it can carry most of the airplane loads. Furthermore, the construction and maintenance of semi-monocoque designs is relatively simple due to their clear separation between skins and support structures.


            \subsection{Monocoque}
                The skins of semi-monocoque designs can sustain large longitudinal compression forces as long as they are sufficiently supported against out-of-plane deformations. As a consequence, support structures are superfluous if compression forces can be avoided or if the bending stiffness of skins and therefore their resistance against buckling deformations is sufficiently large. Structures that consist of skins with a large bending stiffness that are at most supported by transverse frames are known as monocoque designs.


                \subsubsection{Pressure Supported Structures}
                    Compressive skin forces can be avoided by the application of sufficiently large prestresses. This approach was already used in the first manned flight of an aircraft that dates back to 1783 when the Montgolfier brothers pressurized their paper based envelope with the help of hot air \cite{Pagitz2007-2}. To this day, pressure supported structures can be commonly found in aerospace and civil engineering. For example, Atlas rockets \cite{Dawson2004-1} can only carry their self weight if all tanks are sufficiently pressurized. Other rockets such as the Falcon~9 can carry their self weight during manufacturing, storage and transport but require pressurized tanks during flights.\\

                    Hundreds of pressurized radar domes were used by the United States Air Force during the 1950s \cite{Lutes1971-1}. They consisted of an air supported membrane and airlocks that minimized pressure losses during entries and exits as illustrated in Figure~\ref{pic:Figure_2_2}. Current applications of this concept include the Expandable Activity Module of the International Space Station \cite{Seedhouse2014-1} and sport stadiums that span large unobstructed spaces. Air supported buildings are based on a single, pressurized and inhabited space that usually requires a continuous operation of fans and an emergency power supply. In contrast, air inflated buildings are based on a large number of small, self contained and pressurized elements that are connected to each other. Depending on their use, these elements \cite{Krauel2013-1} may or may not be independently pressurized. Basic structural elements for air inflated buildings that are based on straight and curved prismatic tubes were developed by Sumovski \cite{Sumovski1893-1} in 1893. One of the largest air inflated hangars that is based on these elements was constructed by Airbus in 2013. It has a 23~m high ceiling and covers an area of nearly 6,000~m$^2$.\\

                    \marginnote{
                    \begin{center}
                        \includegraphics[width=\marginparwidth]{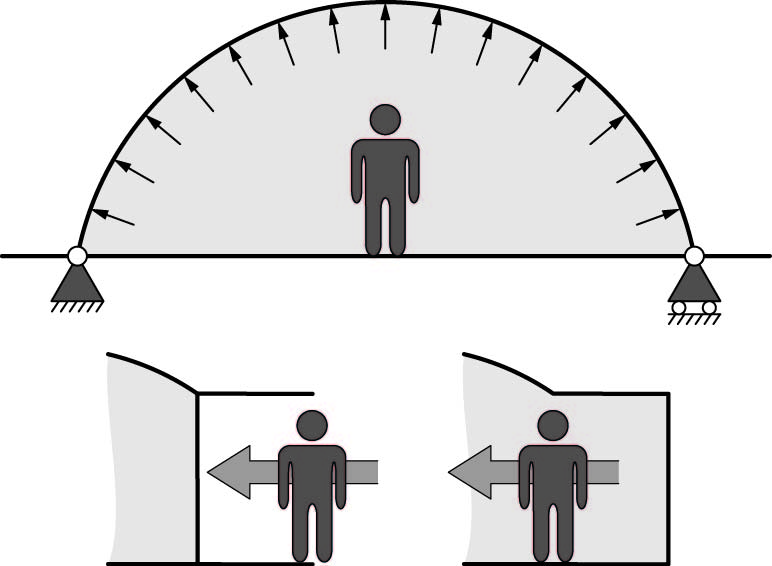}
                        \small(a)\vspace{10mm}
                        \includegraphics[width=\marginparwidth]{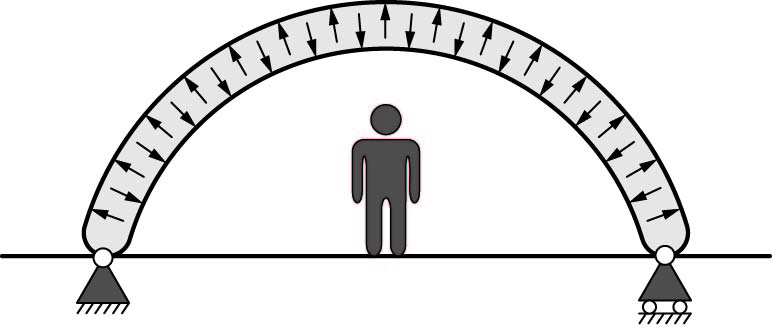}
                        \small(b)\vspace{10mm}
                        \includegraphics[width=\marginparwidth]{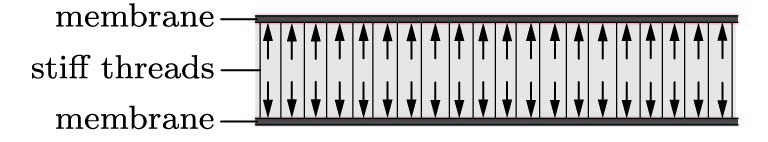}
                        \small(c)\vspace{10mm}
                        \includegraphics[width=\marginparwidth]{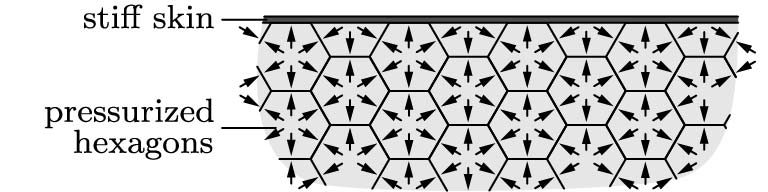}
                        \small(d)
                        \captionof{figure}{Pressure supported structures. (a) Membrane with an airlock that minimizes pressure losses. (b) Tubular arch, (c) Airmat and (d) honeycomb like structure.}
                        \label{pic:Figure_2_2}
                    \end{center}}[-28mm]

                    Early work on pressure stiffened wings and fuselages \cite{Goddard1934-1} was conducted by Robert Goddard in 1934. However, it took another two decades until the Inflatoplane from the Goodyear Tire \& Rubber Company flew for the first time in 1956. The fully inflatable airplane was constructed from pressurized plate and shell like elements that are known as Airmats \cite{Goodyear1961-1}. They consist of two parallel, airtight membranes that are connected by a large number of stiff threads. These threads preserve the membranes overall shape and carry most of its pressure loads. The Inflatoplane could be tightly packed and inflated by a bottle of compressed air or a manual pump. It was supposed to be dropped behind enemy lines to rescue grounded pilots. However, the work on the Inflatoplane was discontinued due to its limited military use. Nonetheless, Airmats are still widely used in applications that range from mattresses to Zorbing.\\

                    The current development of gun launched and airdropped drone swarms \cite{Sanders2017-1} created a renewed interest in inflatable wings. They might be used in the near future in conjunction with expendable inflation systems \cite{Norris2009-1} to increase the packing density of drones. Cellular\footnote{The word ``cell'' was coined by Robert Hooke \cite{Inwood2005-1} in his book micrographia \cite{Hooke1665-1} that was published in 1665. He noticed the similarities between the structure of cork \cite{Gibson1981-1} and the cells that are inhabited by Christian monks in monasteries.} wings that are based on a honeycomb like arrangement of separately pressurized tubes were proposed by Chutter \cite{Chutter1967-1} in 1967. The hexagonal grid of a honeycomb divides a surface into regions of equal area with the least total perimeter \cite{Hales2001-1}. Hence it leads to regular, prismatic structures that can effectively carry internal cell pressures with a minimum amount of material. Modern manufacturing techniques such as the three-dimensional weaving of high strength fibers enables the monolithic construction of pressurized honeycomb wings with tailored cell geometries \cite{Chen2008-1}. Furthermore, an automated integration of subcellular threads that support cell sides that are exposed to differential pressures seems to be possible.


                \subsubsection{Sandwich Structures}
                    Monocoque skins are not supported in the longitudinal direction so that they are prone to buckling if subjected to compressive forces. As a consequence, they require an increased bending stiffness that limits their out-of-plane deformations. The efficiency of monocoque designs thus depends on the tradeoff between the additional weight that comes with an increased stiffness and the savings that can be made due to the use of simpler support structures. Hence, it is essential to design monocoque skins such that their in-plane stiffness is preserved while their out-of-plane stiffness is maximized for a given weight constraint.\\

                    \marginnote{
                    \begin{center}
                        \includegraphics[width=\marginparwidth]{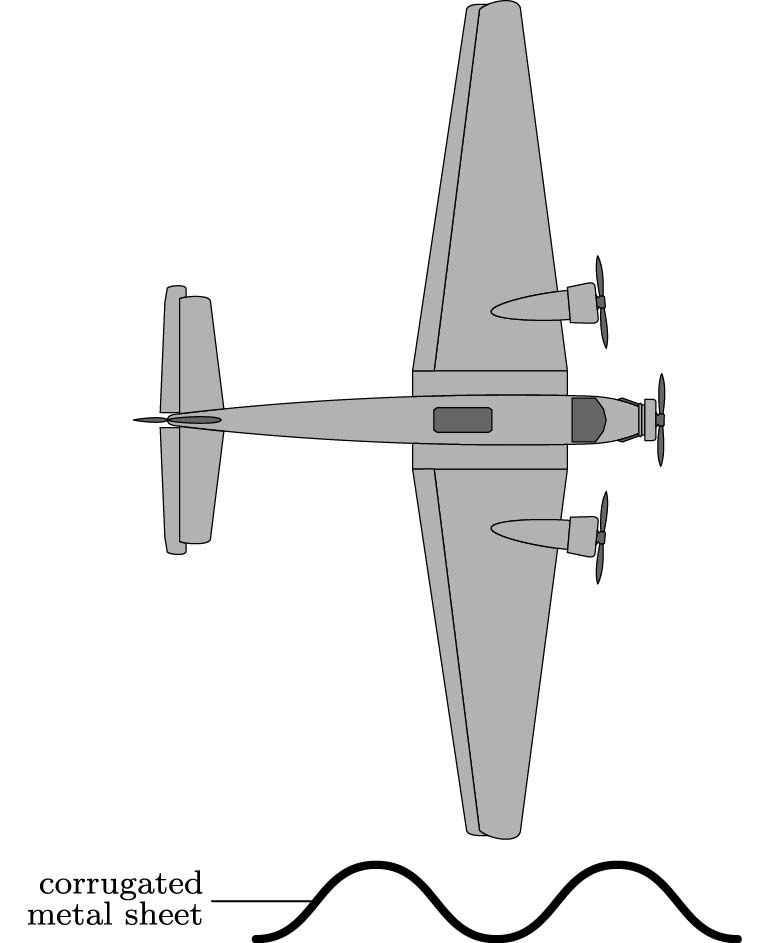}
                        \small(a)\vspace{7mm}
                        \includegraphics[width=\marginparwidth]{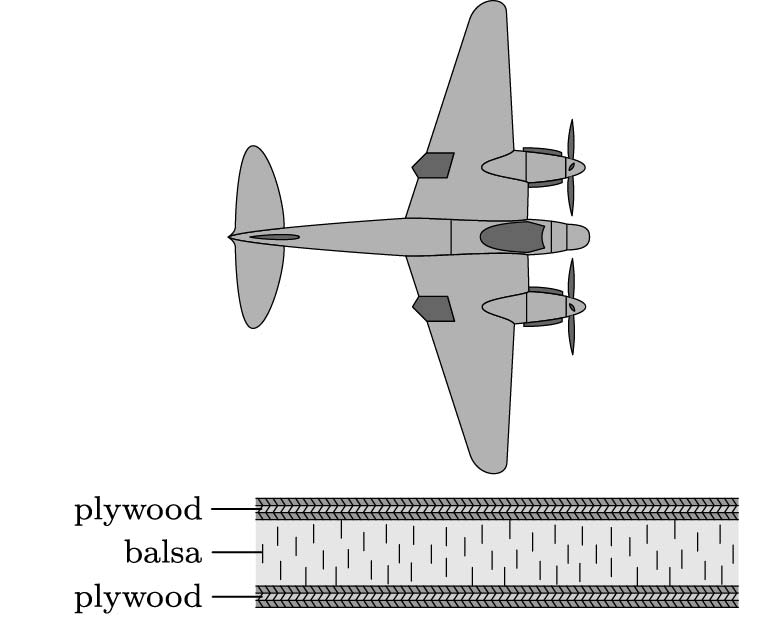}
                        \small(b)\vspace{7mm}
                        \includegraphics[width=\marginparwidth]{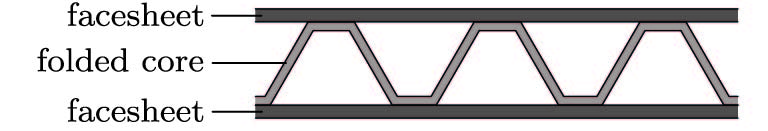}
                        \small(c)\vspace{0mm}
                        \captionof{figure}{(a) Corrugated skin of a Ju 52. (b) Honeycomb sandwich of a mosquito airplane and (c) folded sandwich core.}
                        \label{pic:Figure_2_3}
                    \end{center}}[-29mm]

                    Lightweight skins with a large in- and out-of-plane stiffness can be created by covering a relatively thick honeycomb core with a stiff face plate on each side. The lightweight core increases the distance between both plates and provides closely spaced supports against out-of-plane deformations \cite{Allen1969-1}. The application of these kind of sandwich structures to airplanes \cite{Junkers1915-1} was patented by Junkers in 1915. However, the connection between face sheets and cell cores turned out to be a major problem so that the Junkers J.I. \cite{Kay2004-1} used corrugated metal sheets instead, Figure~\ref{pic:Figure_2_3}. The de Havilland Airplane Company circumvented these problems by gluing sheets of plywood to a lightweight balsa core, a naturally occurring material with a honeycomb like structure \cite{Borrega2015-1}. Their Mosquito airplane that was almost entirely made out of wood was one of the fastest operational airplanes at its introduction in 1941.\\

                    The remarkable performance of the Mosquito can be attributed to its superior structural concept. However, the balsa cores susceptibility to moisture led to weight increases, bonding problems and a degradation of mechanical properties in humid environments. This hindered the widespread use of the Mosquito during the Second World War particularly in the Pacific theater. Water ingress and condensation is still a big problem of modern honeycomb sandwich structures \cite{Shafizadeh1999-1}. This is a reason why their use is commonly restricted to secondary airplane structures such as interior designs, fairings and control surfaces \cite{Herrmann2005-1}. As a consequence, current developments focus on folded cores that ensure a sufficient ventilation through the introduction of drainage channels \cite{Heimbs2010-1}.


            \section{Materials}
                \subsection{Wood}
                    Wood is a widely available and inexpensive material that was used by humans at all times. It possesses anisotropic material properties and can carry relatively large tension and compression forces in the direction of its fibers. Selected lumbers reach the specific strength and modulus of aluminium alloys that were available in the first half of the twentieth century. It is thus not surprising that early airplane structures were made out of wood. However, this changed in the second half of the twentieth century where the development of new alloys incrementally doubled the yield strength of aluminium \cite{Campbell2006-1}. The specific strength and modulus of various materials are summarized in Figure~\ref{pic:Figure_2_4}.\\

                    The lamination of several layers of veneer with glue was proposed by Samuel Bentham in 1797. The mechanical properties of this composite material that is known as plywood can be tailored by varying the fiber direction of each layer. Furthermore, plywood defects such as veneer knots are better distributed and locally restricted to single layers. This minimizes potential stress concentrations and thus improves the average material properties. The inexpensive production of veneers became possible in the mid nineteenth century with the invention of the rotary lathe by Immanuel Nobel\footnote{Immanuel Nobel was an inventor who mainly worked in the defense industry. The Nobel Prize was endowed by his son Alfred who invented the dynamite.}. Material properties of veneers are limited by existing tree species so that the development potential of wood based materials was exhausted after the invention of the plywood. In contrast, the discovery of increasingly strong metal alloys continues to this day.

                    \marginnote{
                    \begin{center}
                        \includegraphics[width=\marginparwidth]{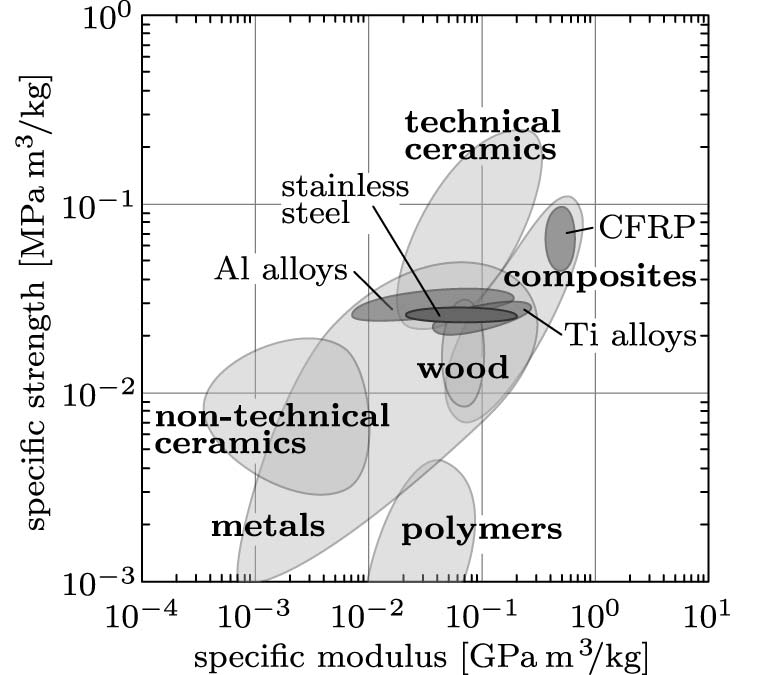}
                        \captionof{figure}{Specific strength and modulus of various materials (data from \cite{Ashby2016-1}).}
                        \label{pic:Figure_2_4}
                    \end{center}}[-57mm]
                    \vspace{-3mm}


                \subsection{Metals}
                    The transition from wood to metal started in 1915 with the maiden flight of the Junkers J~1 that was made out of steel. However, steel was soon replaced by aluminium due to its greater specific strength. The all aluminium DC-3 that made its maiden flight in 1935 is one of the most significant airplanes of all times. Many of the more than 16,000 DC-3's that were build until 1945 are still in service today. In contrast, the era of wooden airplanes ended in 1950 with the production of the last Mosquito. The current use of structural components that are made out of wood is limited to small airplanes.\\

                    The relatively fast transition from wood to metal is interesting since it was only partially driven by performance considerations \cite{Jakab1999-1}. Instead, a general inclination towards change and innovation might have been the main reason. This is underpinned by the performance of the Mosquito that could compete with comparable all metal airplanes of its time. Furthermore, wooden airplanes were cheap and easy to manufacture and their relatively short service life was not a limiting factor during war times. Hence it remains an open question if the rapid focus on all metal airplanes was, at least in the European theater, advantageous.\\

                    The first titanium alloys \cite{Peters2003-1} were developed at the end of the 1940s. Their mechanical properties and corrosion resistance were found to be superior to those of aluminium so that they have been used in airplanes ever since. However, the complex and energy intensive production of titanium makes it relatively expensive. Furthermore, titanium alloys are difficult to machine \cite{Lopez2000-1} so that their use is often restricted to structural components that can not be made out of aluminium. An extreme example is the Lockheed SR-71 that raised the current air speed record for aircraft to 3,530~km/h in 1976. Its sustained high speed flights led to an aerodynamic heating so that 93\% of its structure needed to be made out of titanium \cite{Merlin2009-1}. The performance improvements of military airplanes and the growing demand for lightweight commercial airplanes steadily increases the use of titanium as shown in Figure~\ref{pic:Figure_2_5}. This trend will most likely continue due to the development of advanced manufacturing techniques such as rapid prototyping.

                    \marginnote{
                    \begin{center}
                        \includegraphics[width=\marginparwidth]{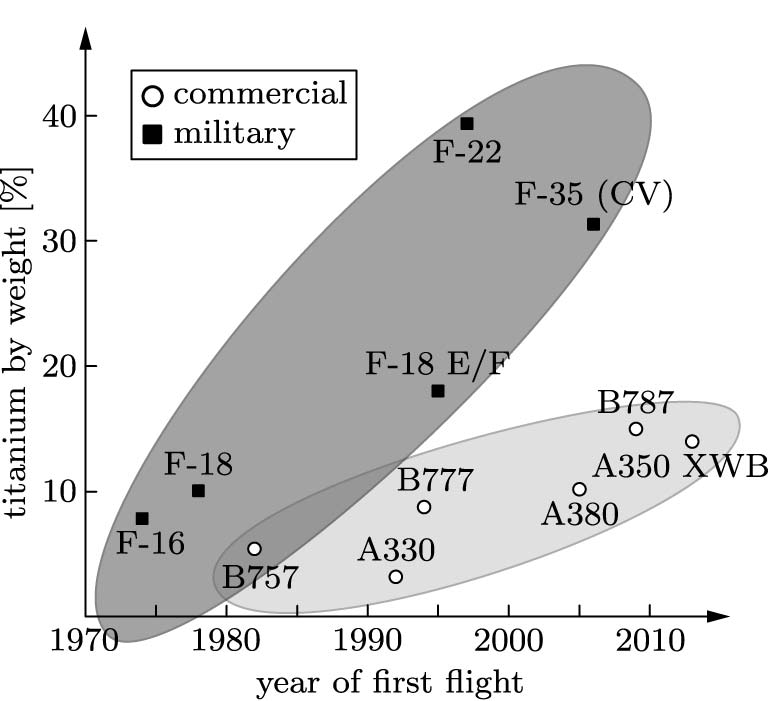}
                        \small(a)\vspace{1mm}
                        \includegraphics[width=\marginparwidth]{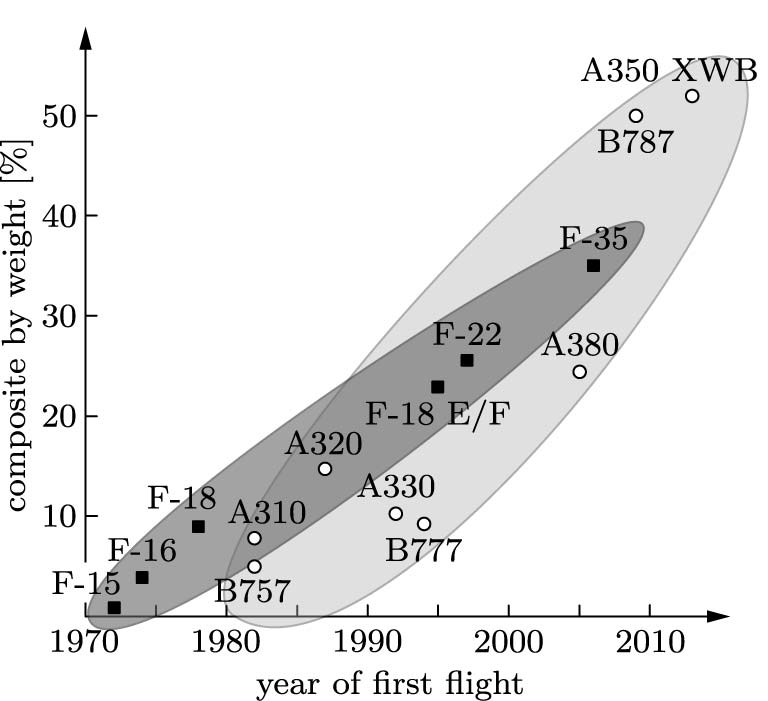} \small(b)\vspace{0mm}
                        \captionof{figure}{Percentage of (a) titanium and (b) composite structural weight of various commercial and military airplanes.}
                        \label{pic:Figure_2_5}
                    \end{center}}[-19mm]


                \subsection{Composites}
                    The use of aluminium in airplane structures diminished after the development of titanium alloys. This trend accelerated after the availability of high strength composite materials. Today, carbon fiber reinforced plastics are the most commonly used materials in commercial and military airplanes.\\

                    Carbon fibers with a yield strength that vastly exceeds those of aluminium and titan were developed in the 1960s. These fibers are embedded in a polymer matrix to create structural components with tailored, anisotropic properties similar to the plywood that was used in the early days of aviation. Modern passenger airplanes from Airbus or Boeing are about 50\% composite by weight. For example, the Boeing 787 airframe consists of 10\% steel, 15\% titanium, 20\% aluminium and 50\% composite. This material mix reduces the total weight by about 20\% compared to an all aluminium design. However, the transition from aluminium to composite is considerably slower than the previous transition from wood to aluminium. This is even more striking since the performance differences between aluminium and carbon fiber reinforced plastics are considerably larger than the differences between wood and aluminium.\\

                    Nonetheless, the trend towards composite airplanes will continue due to the development of increasingly strong fibers and the maturity of design, manufacturing and maintenance tools. The relatively slow progress in the airplane industry today might be attributed to the small number of competitors that have survived. Mergers and withdrawals led to a duopoly in the large jet airliner market that exists since the 1990s. Other aspects are the relatively high regulatory requirements and the increasing complexity of new airplanes.


    \newpage

    \sectionmark{Summary}
    \begin{framed}
        \noindent \textbf{Summary}\\

        \noindent Structures have been build by humans since the beginning of times. The rapid development of airplane structures in the first half of the last century is thus not particularly surprising. Early fuselages were based on truss like structures that consisted of wood struts and steel wires. Soon after, these fuselages were covered by fabrics to improve their aerodynamic properties. The presence of a stiff fabric enabled the removal of the steel wires which led to the semi-monocoque designs that are still used today. The remaining compression struts can be removed if the out-of-plane stiffness of the airplane skin is sufficiently large (monocoque design) or if the skin is prestressed such that compression forces are avoided altogether. The latter concept is commonly used in rockets and lighter than air aircraft that are among the most weight sensitive structures build by man.\\

        The first airplanes were mostly made from wood, a naturally growing material whose yield strength and shapes are limited by existing tree species. It was thus often processed into plywood that consists of several layers of veneers. Unlike wood, metals are far less susceptible to degradation. Furthermore, their development potential was much larger at that time. For example, the yield strength of aluminium alloys was doubled during the second half of the last century. It is thus not surprising that wood was soon replaced by aluminium whereas the latter is currently replaced, although at a much slower pace, by titanium. Metal alloys are slowly reaching their full potential so that further improvements are relatively small. Structural parts that are not subjected to high temperatures are increasingly manufactured from composites such as glass and carbon fiber reinforced plastics. The specific strength of these fibers is far greater than that of metal and their full potential is not reached yet.\\

        \noindent \textbf{Conclusion}\\

        \noindent Some of the most efficient and lightweight structures avoid compression forces with the help of pressure gradients by introducing a sufficiently large prestress into the material. A combination of this concept with modern materials such as carbon fiber reinforced plastics is thus an excellent starting point for the development of shape changing structures.
    \end{framed} 
        \cleardoublepage
    \chapter{Actuators}
        Humans usually approach complex tasks by splitting them into subtasks that are easier to solve. This is certainly true for airplanes that consist of a number of components that are, as far as possible, independently developed and manufactured. For example, commonly used high lift devices consist of aerodynamically shaped rigid bodies and mechanisms that convert the simple input movements of one or more actuators into the desired output movements. Actuators themselves are subcomponents that convert various kinds of energy into linear and rotary motions. In reverse conclusion, they can often be used as generators for the reciprocal task of converting motions into energy. Energy forms such as electric, magnetic, pressure and thermal are commonly used in engineering. These energy forms can be converted into each other although this inevitably leads to energy losses. Hence, the optimal actuator design depends, among others, on the available energy form, the required actuation stress, strain and speed. Achievable stresses, strains and frequencies of various actuation principles are summarized in Figure~\ref{pic:Figure_3_1}.\\

        \marginnote{
        \begin{center}
            \includegraphics[width=\marginparwidth]{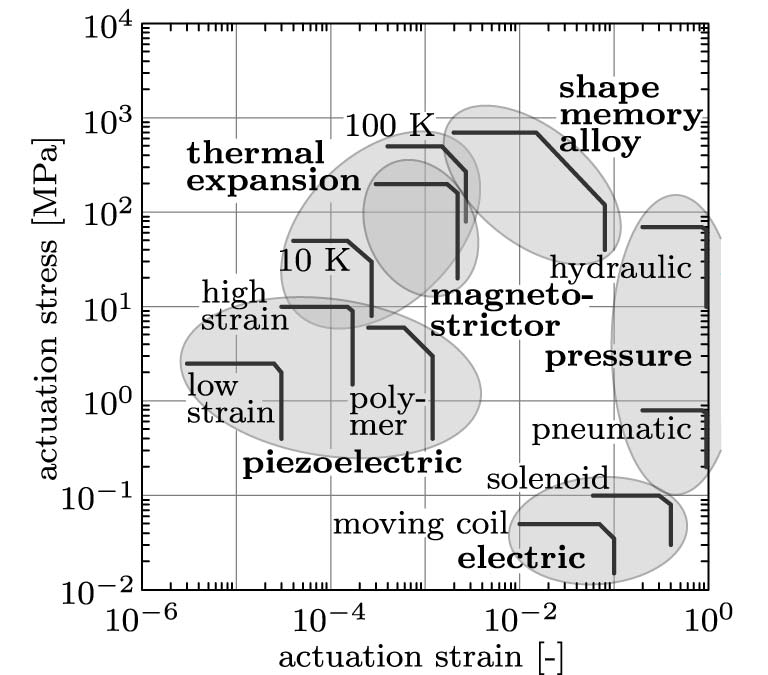}
            \small(a)\vspace{0mm}
            \includegraphics[width=\marginparwidth]{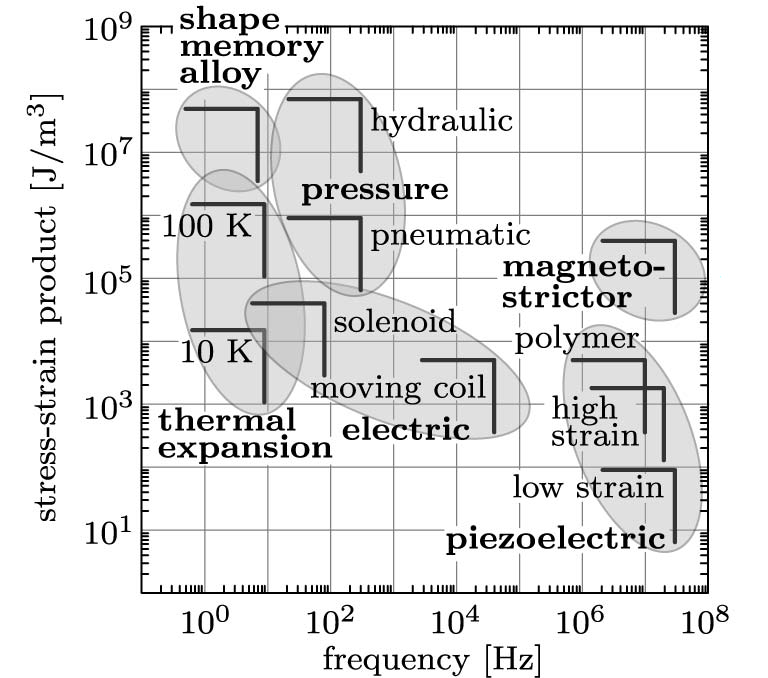}
            \small(b)
            \captionof{figure}{(a) Actuation stress versus strain and (b) stress-strain product versus frequency of various actuation principles (data from \cite{Huber1997-1}).}
            \label{pic:Figure_3_1}
        \end{center}}[-91mm]

        Actuators can be solid state materials or mechanisms themselves. Solid sate actuators are usually driven by variations in their crystal or molecular structure so that they need to be described on a continuum level. In contrast, mechanical actuators are driven by the translation or rotation of rigid bodies or the deformation of compliant mechanisms and are thus described on a structural level. Mechanical actuators can generate large actuation strains or, in terms of motors, unbounded rotations. In contrast, solid state actuators can generate only small strains but are usually capable of operating at higher frequencies. Mechanical actuators are widely used in engineering whereas solid state actuators are predominantly found in nature. A rare exception is the rotation motor that is used for the propulsion of bacteria \cite{Berg1973-1}.\\

        A brief overview of different mechanical and solid state actuators is subsequently provided. The focus for the mechanical actuators is on the main types that are commonly used in heavy machinery such as airplanes. In contrast, currently available solid state actuators often lack the potential or maturity to replace mechanical actuators in these applications. This is taken into account by discussing a large number of different actuation principles, their potential, limitations and current developments.


        \section{Mechanical Actuators}
            \subsection{Electromechanical}
                Electromechanical actuators are mechanisms that convert electrical energy into a rotary or linear motion. There exists a large number of different motor designs \cite{Hughes2013-1} that are widely used in engineering. All of them have in common that they are based on the interaction between winding currents and an electromagnetic field that is generated by permanent magnets or current carrying coils.\\

                Linear actuators usually consist of an electric motor that is connected to a lead screw and a nut that is threaded into it. The nut is prevented from rotating with the lead screw so that a linear motion is created by driving it along the threads \cite{Mare2017-1}. These kind of actuators are often combined with a brake system so that static loads can be carried by unpowered actuators. Solenoids and moving coils \cite{Boldea1997-1} are other types of linear actuators. Solenoids consist of an electromagnetic coil, a spring, and a ferromagnetic cylinder, rod. The coil is wound around the cylinder in which the rod can freely move. The electromagnetic field of the current carrying coil forces the rod towards its center and the spring returns it to its initial position after switch off. Moving coil actuators consist of a permanent magnet and an electromagnetic coil that is loosely wound around it. The electromagnetic field of the current carrying coil interacts with the magnetic field of the magnet and generates a force along its axis. Moving coils possess, unlike solenoids, a linear force displacement relationship where the force is proportional to the current.\\

                Electric motors can be found in applications that range from endoscopes \cite{Yeung2016-1} to large ships \cite{Parker1998-1}. Solenoids are used, for example, in pneumatic or hydraulic control valves whereas moving coils are used in loud speakers that require large accelerations and a linear force displacement relationship. There is currently a strong trend towards the electrification of chemically powered machines. For example, trucks \cite{Tate2009-1} and airplanes \cite{Rosero2007-1} use various kind of actuators that range from hydraulic cylinders to combustion engines. Each kind of actuator requires a different energy form so that the weight and complexity of these machines can be often decreased and their energy efficiency increased by minimizing the number of energy conversions. Hence it is not surprising that, wherever possible, hydraulic and pneumatic cylinders are replaced by lead screw driven actuators.\\

                Most of the global electricity is generated by electric motors that convert motion into energy. On the other hand, electric motors consume about 45\% of the electric energy \cite{Waide2011-1}. This dual use magnifies the impact of new technologies that improve their energy efficiency. Current research in this field focuses particularly on the development of advanced magnetic materials \cite{Gutfleisch2011-1} and optimized motor designs \cite{Duan2013-1}.


            \subsection{Pressure}
                Pressure actuators are mechanisms that convert a fluid flow into a rotary or linear motion. Pneumatic actuators use a gas whereas hydraulic actuators use a liquid for the power transmission. There exists a large number of different motor designs \cite{Parr2011-1} where gear-, vane- and radial piston motors are among the most common. Their working principles are illustrated in Figure~\ref{pic:Figure_3_2}.\\

                \marginnote{
                \begin{center}
                    \includegraphics[width=\marginparwidth]{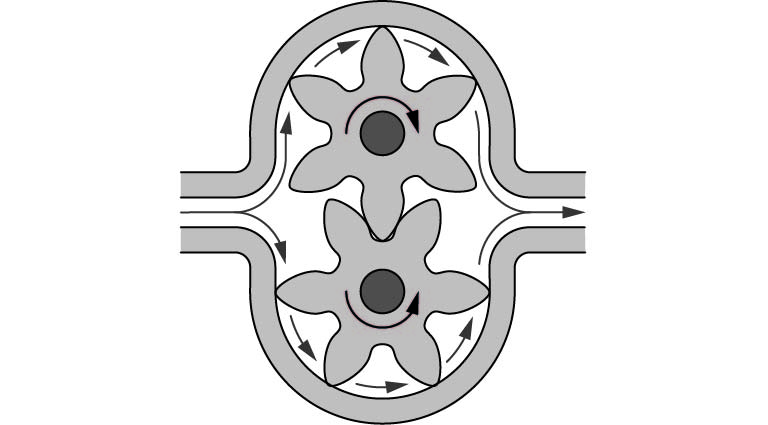}
                    \small(a)\vspace{7mm}
                    \includegraphics[width=\marginparwidth]{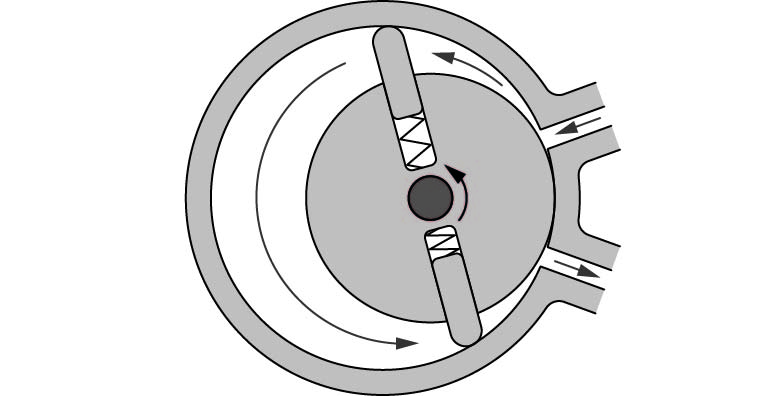}
                    \small(b)\vspace{7mm}
                    \includegraphics[width=\marginparwidth]{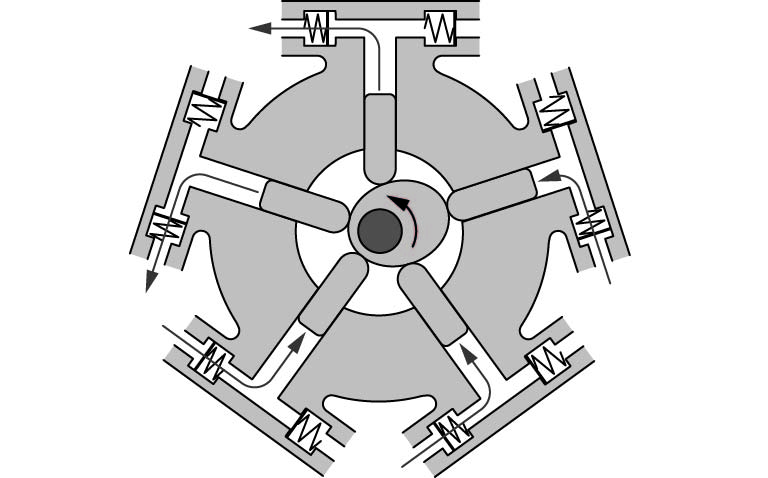}
                    \small(c)
                    \captionof{figure}{Different pressure based motors. (a) Gear-, (b) vane- and (c) radial piston motor.}
                    \label{pic:Figure_3_2}
                \end{center}}[-57mm]

                Pneumatic and hydraulic cylinders are linear actuators that are based on a rigid body mechanism where a single degree of freedom is provided by a prismatic joint between a piston rod and a cylinder. Pneumatic artificial muscles and bellows actuators are the compliant counterparts of pneumatic and hydraulic cylinders. Despite their name, both kind of actuators can be hydraulically and pneumatically operated \cite{Tiwari2012-1}. Pneumatic muscles can generate large tension whereas bellows actuators can generate significant compression forces. The working principle behind both kind of actuators is based on the anisotropic stiffness of their cylindrical structure. Pneumatic muscles are relatively soft in radial and stiff in axial direction whereas bellows actuators are relatively stiff in radial and soft in axial direction. The anisotropy of pneumatic muscles is usually introduced on a material level by embedding axially oriented braided cords into an elastic cylinder \cite{Daerden2002-1}. In contrast, the anisotropy of bellows actuators is often introduced on a geometric level by using corrugated cylinders or on a material level by embedding equally spaced belt ribs around the circumference of an elastic cylinder. Pneumatic muscles and bellows actuators generate forces and displacements that are in the range of cylinder based actuators \cite{Plettenburg2005-1}. Unlike double acting cylinders they can perform work only in one direction so that they are often used in antagonistic pairs \cite{Cerda1995-1}. Furthermore, they possess a nonlinear force-displacement relationship \cite{Chou1994-1} that depends on their pressurization, deformation and temperature. The latter is particularly difficult to describe since it fluctuates during operations and thus increases the control complexity \cite{Thanh2006-1}.\\

                Pneumatic and hydraulic cylinders are relatively simple and inexpensive. They generate large actuation stresses and strains and can operate under challenging conditions. Pneumatic cylinders can be used in hazardous environments with temperatures between -40 to $300^\circ$C as they do not contain any flammable liquids. On the other hand, hydraulic cylinders have a very high power to weight ratio and their incompressible fluid allows them to steadily hold varying loads without a fluid supply. Due to their outstanding properties, pneumatic and hydraulic actuators are widely used. However, they often require additional parts such as fluid reservoirs, pumps, heat exchangers and valves which increases their overall complexity and weight. In contrast, pneumatic artificial muscles are rarely used since they have no clear advantage over their rigid body counterparts. Bellows actuators can be found in the suspension systems of some high end cars since they enable a smoother, adjustable ride than conventional steel springs \cite{Cao2011-1}. Furthermore, they are used as lightweight lifting bags in recovery and rescue operations.\\

                Particularly hydraulic actuators have not reached their full potential yet. Current research focuses on low friction seals \cite{Nikas2010-1} and energy efficient fluids \cite{Totten2012-1} that allow the use of increasingly large operating pressures. Advances in manufacturing techniques \cite{Yap2015-1} enabled a tight integration of actuators, fluid lines, valves and control electronics \cite{Semini2016-1} which led to considerable weight savings. The current development of electro hydraulic actuators is mostly driven by the ongoing trend towards the electrification of airplanes. These actuators are a combination of hydraulic and electromechanical \cite{Bossche2006-1} or solid state actuators \cite{Chaudhuri2011-1}. The main advantage of electrohydraulic actuators is their use of a localized hydraulic system that is electrically powered and electronically controlled. This approach leads to significant weight savings \cite{Bossche2006-1} while it avoids some of the safety critical drawbacks of electromechanical actuators \cite{Balaban2009-1}. Nonetheless, purely pressure based actuators are still used in many heavy machines and it remains to be seen in how far they will be replaced by electromechanical actuators.


        \section{Solid State Actuators}
            Sarcomeres are the smallest building blocks of muscles. They consist of three protein filaments that slide past one another at a length scale of about 2~$\mu$m. Muscles consist of a large number of serial and parallel arranged sarcomeres so that their possible actuation strains and forces depend on their length and thickness. The underlying mechanism of single sarcomeres shares many similarities with mechanical actuators. However, they function at a different length scale and are only used in large numbers. Hence, a muscle is commonly considered to be a solid state actuator.\\

            Even within a single material class, solid state actuators can be based on a wide range of actuation principles. For example, polymer based actuators can utilize various energy forms that range from electric and magnetic to thermal. Generic terms such as electroactive polymers \cite{Carpi2011-1} or polymer artificial muscles \cite{Mirfakhrai2007-1} are thus not helpful. As a consequence, various solid state actuators are subsequently discussed on the basis of their actuation principle.


            \subsection{Piezoelectric and Piezomagnetic}
                Piezoelectric \textit{(piezomagnetic)} materials generate an electric \textit{(magnetic)} field in response to a deformation. Conversely, a deformation occurs when an electric \textit{(magnetic)} field is applied. Piezoelectricity was discovered in 1880 by the brothers Jacques and Pierre Curie\footnote{Pierre Curie married Maria Sklodowska (Marie Curie) in 1895.} \cite{Curie1882-1} and piezomagnetism in 1960 by Borovik-Romanov \cite{Borovik1960-1}. Piezoelectricity \textit{(piezomagneticity)} is a first order effect since deformations are proportional to the applied electric \textit{(magnetic)} fields as illustrated in Figure~\ref{pic:Figure_3_3}. The subsequent focus is on piezoelectricity since piezomagnetic materials are rarely used.\\

                \marginnote{
                \begin{center}
                    \includegraphics[width=\marginparwidth]{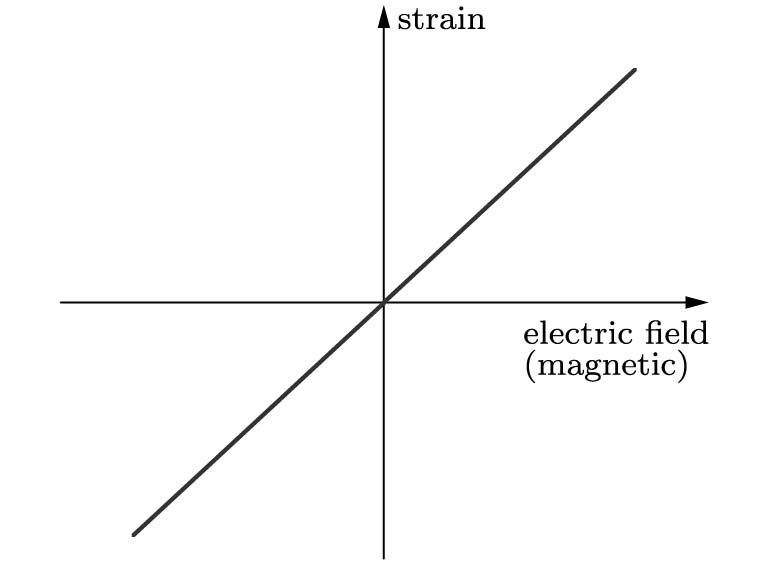}
                    \small(a)\vspace{8mm}
                    \includegraphics[width=\marginparwidth]{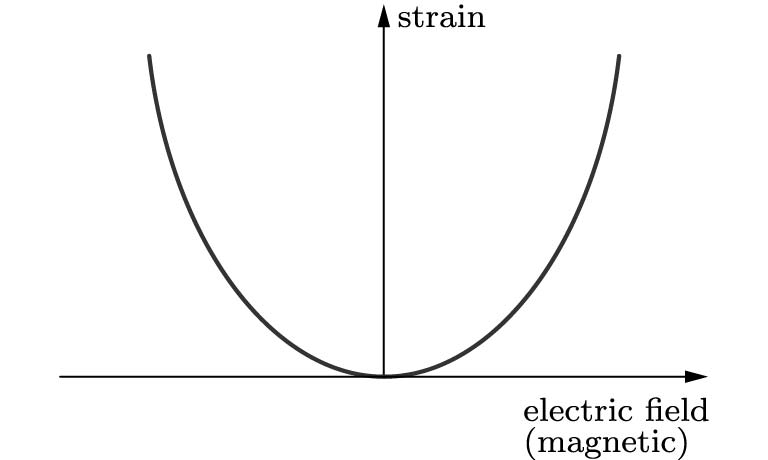}
                    \small(b)
                    \captionof{figure}{(a) First order effect of piezoelectric \textit{(piezomagnetic)} materials. The actuation strain depends on the polarity of the electric \textit{(magnetic)} field. (b) Second order effect of electrostrictive \textit{(magnetostrictive)} materials.}
                    \label{pic:Figure_3_3}
                \end{center}}[-30mm]

                Piezoelectricity of inorganic materials is understood as the interaction between the mechanical and the electric state of crystalline materials with no inversion symmetry. Out of the 32 crystal classes there are 20 that exhibit direct piezoelectricity \cite{Voigt1966-1}. Some organic materials such as the amorphous vinylidene cyanide \cite{Miyata1980-1} and the semi-crystalline polyvinylidene fluoride \cite{Kawai1969-1} are piezoelectric due to their molecular structure and arrangement. The properties of piezoelectric materials are described by strain and voltage coefficients. Strain coefficients are the ratios between the actuation strains and the applied electric fields whereas voltage coefficients are the ratios between the open circuit electric fields and the applied stresses. Large strain coefficients are required for actuators whereas large voltage coefficients are advantageous for sensors. Piezoelectric actuators are energy efficient and capable of high frequencies but limited to small strains and stresses. This makes them the ideal choice for applications such as vibration suppressors, precision positioners and ultrasonic motors \cite{Uchino1997-1}.\\

                The most common piezoelectric materials are the inorganic lead zirconate titanate that is toxic and brittle and the organic polyvinylidene fluoride that is biocompatible and flexible. Lead zirconate titanate has relatively large strain and small voltage coefficients whereas polyvinylidene fluoride has small strain and large voltage coefficients. The advantages of organic and inorganic piezoelectric materials are often combined in composite materials \cite{Venkatragavaraj2001-1}. Current research focuses on the development of lead free inorganic materials \cite{Jo2012-1} with a reduced toxicity and piezoelectric polymer composites for microfabricated devices \cite{Ramadan2014-1}.


            \subsection{Electrostrictive and Magnetostrictive}
                All non-conducting or dielectric \textit{(ferromagnetic)} materials undergo a deformation in response to an electric \textit{(magnetic)} field. These second order effects are independent of the polarity and proportional to the square of the electric \textit{(magnetic)} field. Magnetostriction was first described by Joule \cite{Joule1847-1} in 1842 and electrostriction by Volpicelli \cite{Korteweg1880-1} in 1862.\\

                Electrostriction of dielectric materials is due to the slight displacement of ions in a crystal lattice. These displacements accumulate throughout the material and lead to strains in the direction of the electric field. Some engineered ceramics that are known as relaxor ferroelectrics exhibit a remarkably large electrostriction. However, their underlying mechanism is not fully understood \cite{Cowley2011-1}. In contrast, magnetostriction occurs only in ferromagnetic materials. These materials possess uniformly polarized regions that align themselves to a magnetic field. Deformations from these rotations accumulate throughout the material and lead to strains in the direction of the magnetic field. Electrostrictive \textit{(magnetostrictive)} actuators are energy efficient and capable of high frequencies. They can generate larger stresses and strains than piezoelectric actuators. However, electrostrictive \textit{(magnetostrictive)} actuators are nonlinear and temperature sensitive. Nonetheless, particularly magnetostrictive actuators are widely used in vibration suppressors, sonar transducers and ultra sonic measuring devices \cite{Olabi2006-1}.\\

                Commonly used electrostrictive materials comprise the inorganic lead magnesium niobate \cite{Smolenskii1958-1} and some organic copolymers of polyvinylidene fluoride \cite{Xu2001-1}. Similar to piezoelectric materials, current research focuses on the development of inorganic materials that are lead free \cite{Zhang2009-1}. Furthermore, the material properties of polyvinylidene fluorides are tailored for various applications by using fillers and different processing procedures \cite{Martins2014-1}. The highest known magnetostriction among alloys is exhibited by terbium iron dysprosium (Terfenol-D) \cite{Wohlfarth1980-1}. Despite its huge magnetostriction, the use of Terfenol-D is often limited by its susceptibility to fracture and its small tensile strength. Potential alternatives are iron gallium alloys that exhibit a moderate magnetostriction at very low magnetic fields. These alloys, that possess a large tensile strength and a small hysteresis, can be welded \cite{Atulasimha2011-1}.


            \subsection{Electrostatic and Magnetostatic}
                The electrostatic \textit{(magnetostatic)} force between two charged \textit{(magnetized)} plates is proportional to the square of the electric \textit{(magnetic)} field and inversely proportional to the square of the distance between the plates and the relative permittivity \textit{(permeability)} of the medium in-between. This relationship was discovered by Cavendish \cite{Maxwell1879-1} in the early 1770s for the elactrostatic and by Amp\`ere \cite{Ampere1823-1} in 1820 for the magnetostatic force.\\

                An electrostatic actuator that consists of two parallel plates that are separated from each other is illustrated in Figure~\ref{pic:Figure_3_4}. The lower plate is fixed and the upper plate is connected to a spring. It can be seen that the nonlinear relationship between the electrostatic force and the gap between both plates leads to two potential equilibrium configurations. However, the left equilibrium configuration is unstable so that the electrostatic force can not be balanced by the spring beyond this point \cite{Pamidighantam2002-1, Nemirovsky2005-1}. This is widely known as a pull-in instability. Electrostatic actuators can be easily manufactured at small scales. Furthermore, their force increases nonlinearly with decreasing gaps so that they are particularly suited for the use in micro-electromechanical systems (MEMS). In contrast, the use of magnetostatic actuators is currently limited by incompatibilities \cite{Niarchos2003-1} with existing manufacturing techniques. Furthermore, the ratio between electrostatic and magnetostatic forces increases for decreasing actuator sizes. As a consequence, the subsequent focus is on electrostatic actuators.\\

                \marginnote{
                \begin{center}
                    \includegraphics[width=\marginparwidth]{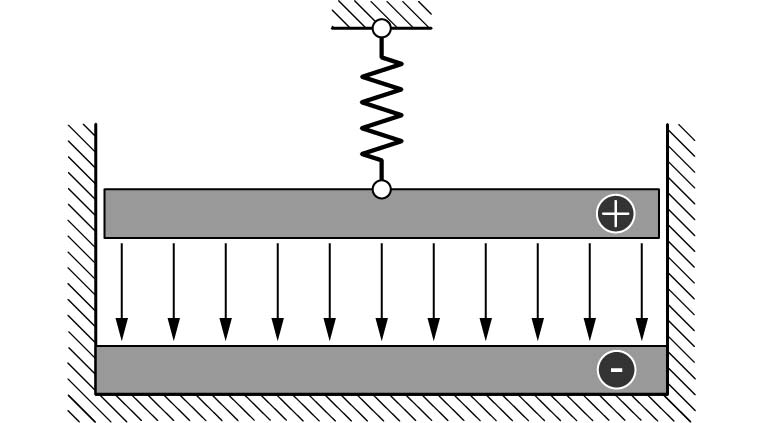}
                    \small(a)\vspace{8mm}
                    \includegraphics[width=\marginparwidth]{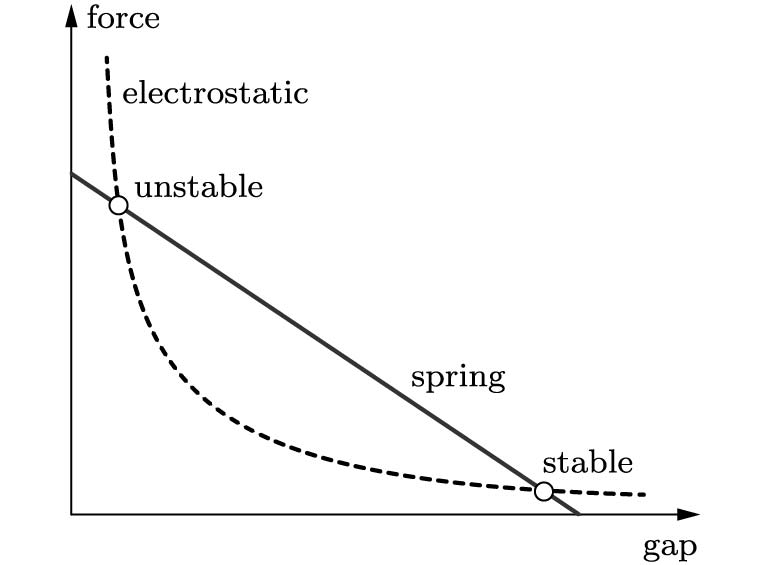}
                    \small(b)
                    \captionof{figure}{(a) Two parallel, separated and electrically charged plates. The lower plate is fixed and the upper plate is connected to a spring. (b) The electrostatic force depends nonlinearly on the gap between both plates so that a stable and an unstable equilibrium configuration exists for any electric field.}
                    \label{pic:Figure_3_4}
                \end{center}}[-99mm]

                MEMS are manufactured by bulk-, surface and high aspect silicon micromachining. Materials such as polymers, metals and ceramics can be deposited in layers on the silicon and patterned, etched to produce the required shapes \cite{Judy2001-1}. Commonly used electrostatic actuators are parallel plates \cite{Yao1992-1}, inchworms \cite{Yeh2002-1} and comp- \cite{Tang1990-1}, scratch drives \cite{Akiyama1993-1}. They are found, for example, in micro mirrors \cite{Hornbeck1989-1} for cinema projectors, television sets and rapid prototyping machines. Electrostatic actuators are mature and well understood. Current developments focus mainly on advanced manufacturing techniques as well as new materials. Furthermore, electrostatically actuated mechanisms for micro-fluidic devices receive a lot of interest due to their potential impact on clinical diagnosis \cite{Sackmann2014-1}.


            \subsection{Ionic}
                The ions in electrostrictive actuators are only slightly displaced whereas their movements in ionic actuators are considerably larger. These kind of movements lead to local volume changes that mostly preserve an actuators total volume. It should be noted that a magnetically actuated version of an ionic actuator is not possible as magnetic charges or monopoles do not exist. Depending on the chosen electrode material, ionic actuators are subsequently divided into two groups as illustrated in Figure~\ref{pic:Figure_3_5}.\\

                An electrode storage actuator that consists of doped polymer electrodes and a separating electrolyte for the ion transport was proposed by Baughman et al \cite{Baughman1996-1} in the early 1990s. This kind of actuator became possible after the discovery of the high conductivity of polyacetylene derivatives at room temperatures by Shirakawa et al \cite{Shirakawa1977-1} in 1977. Current electrodes are often made from polypyrrole \cite{Bolto1963-1}, polyaniline \cite{Syed1991-1} or polythiophene \cite{Roncali1992-1}.\\

                Polymer metal actuators were first investigated by Oguro et al \cite{Oguro1999-1} in 1992. They are based on flexible metal electrodes that are plated on a polyelectrolyte where one ion type is provided by the polymer chains and the opposite type by solvated ions. Gold and platinum are often used as electrode materials whereas styrene/divinylbenzene polymers and perfluorinated alkenes \cite{Shahinpoor2001-1} are commonly used as electrolytes. Current research aims at creating advanced electrode and electrolyte materials where the main focus is on carbon nanotubes \cite{Baughman1999-1,Kosidlo2013-1} and polymers with zwitter ions \cite{Kim2016-1}.\\

                \marginnote{
                \begin{center}
                    \includegraphics[width=\marginparwidth]{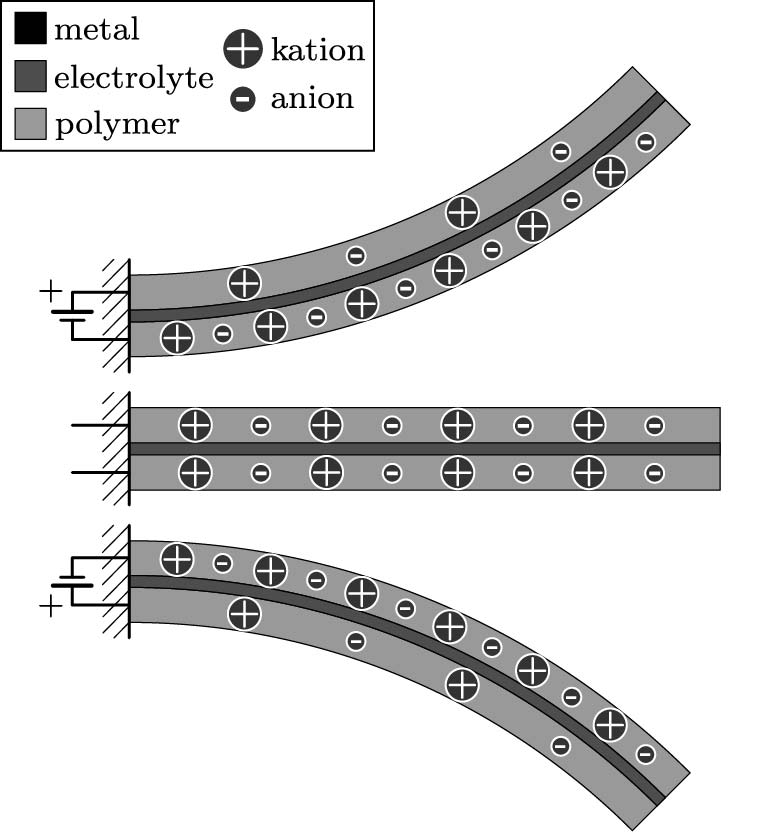}
                    \small(a)\vspace{2mm}
                    \includegraphics[width=\marginparwidth]{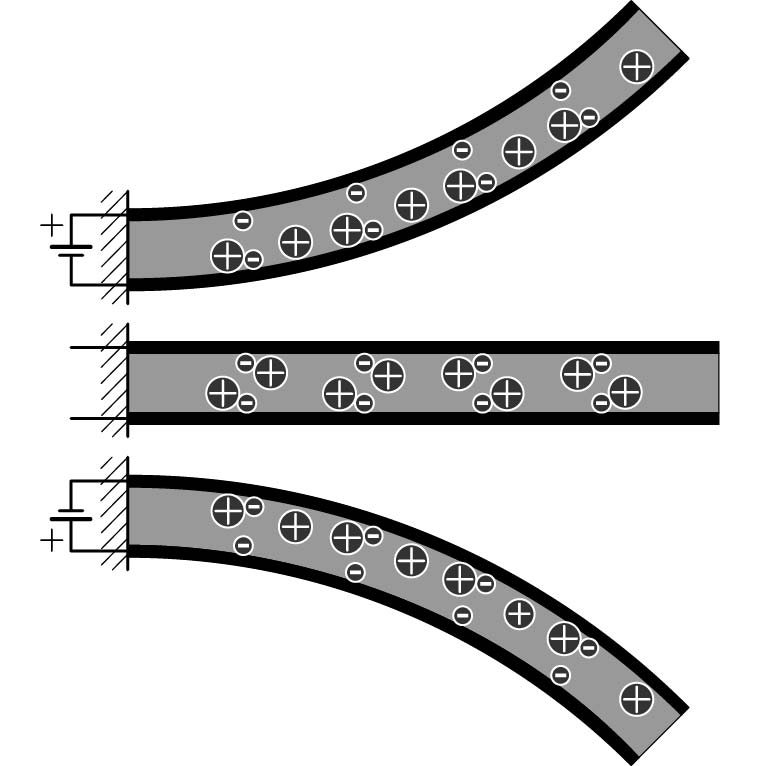}
                    \small(b)
                    \captionof{figure}{(a) Electrode storage actuator with doped polymer electrodes and separating electrolyte. (b) Polymer metal composite actuator with flexible metal electrodes and polyelectrolyte polymer.}
                    \label{pic:Figure_3_5}
                \end{center}}[-73mm]

                Ionic actuators work with low voltages and their deformations depend on the polarity of the electric field. They are flexible, relatively inexpensive and easy to manufacture. However, their industrial use is limited as they can only operate at low frequencies, possess a low endurance and often suffer from irreversible deformations.


            \subsection{Phase Transition}
                A phase is a region of space throughout which all physical properties of a material are uniform. The transition of a material from one phase to another can affect its density, refractive index, magnetization or chemical composition. Phase transitions can be triggered by temperature, pressure and electric or magnetic fields. Materials that reversibly vary their shape or volume during a phase transition can be used as actuators. Commonly utilized are the solid-solid phase transitions of shape memory alloys, the solid-liquid transition of paraffin \cite{Ogden2014-1} and the liquid-gas transition of ethanol \cite{Miriyev2017-1}. Particularly shape memory alloys are widely used since they remain solid. Further advantages are their high strength, biocompatibility and workability.\\

                Shape memory alloys are a group of metals that can repeatedly return from a deformed- to a memorized shape when heated or subjected to a magnetic field \cite{Jani2014-1}. A thermally activated shape memory effect was first observed in 1932 in an Au-Cd alloy by \"Olander \cite{Olander1932-1} and in 1962 in a Ni-Ti alloy by Buehler et al \cite{Buehler1963-1}. The latter stands out of the large number of shape memory alloys that are known today \cite{Wei1998-1} due to its remarkable properties \cite{Huang2001-1}. A subgroup within the thermally actuated shape memory alloys that can also be magnetically actuated was discovered by Ullakko \cite{Ullakko1996-1} in 1996. However, magnetically actuated alloys are rarely used and the only commercially available material is a Ni-Mn-Ga alloy \cite{Faran2016-1}. Shape memory alloys possess three potentially stable crystal structures (twinned martensite, detwinned martensite and austenite). The austenite crystal structure is the preferred material phase at high temperatures. A shape memory alloy returns to its austenite structure upon heating even if it is heavily loaded. Transformations between crystal structures can be categorized into pseudoelastic, one- and two way shape memory effects as illustrated in Figure~\ref{pic:Figure_3_6}.\\

                \marginnote{
                \begin{center}
                    \includegraphics[width=\marginparwidth]{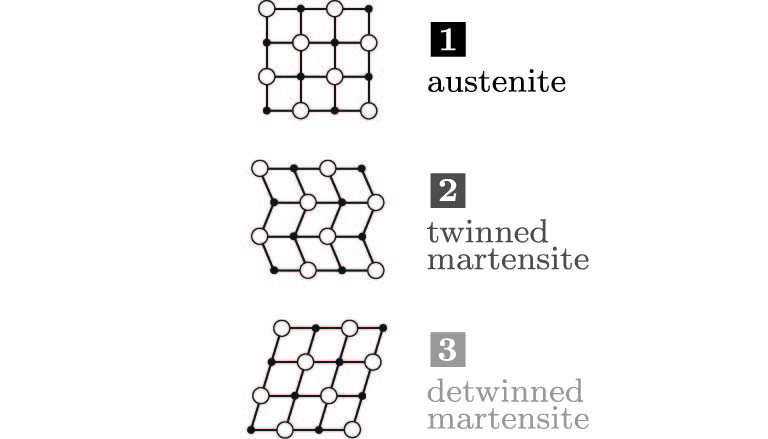}
                    \small(a)\vspace{5mm}
                    \includegraphics[width=\marginparwidth]{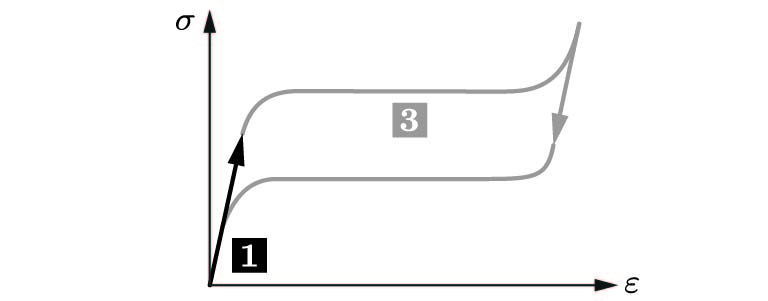}
                    \small(b)\vspace{5mm}
                    \includegraphics[width=\marginparwidth]{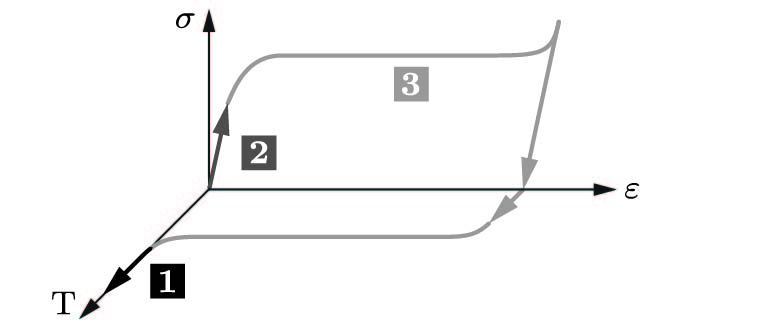}
                    \small(c)\vspace{5mm}
                    \includegraphics[width=\marginparwidth]{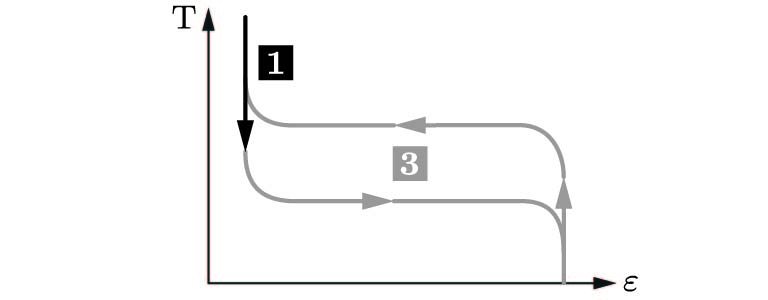}
                    \small(d)
                    \captionof{figure}{(a) Potentially stable crystal structures of shape memory alloys. (b) Pseudoelastic behavior at high temperatures and (c) one way shape memory effect at lower temperatures. (d) Two way shape memory effect of ``trained'' alloys.}
                    \label{pic:Figure_3_6}
                \end{center}}[-40mm]

                Due to instant phase transitions, shape memory alloys can undergo large pseudoelastic strains at high temperatures. External loads lead to a transition from the austenitic to the martensitic phase which is instantaneously reversed upon unloading \cite{Brinson1993-1}. In contrast, a one way shape memory alloy remains in the deformed configuration upon unloading if the environmental temperature is not sufficiently high for a phase transformation. However, it eventually recovers to its initial shape if the temperature is raised.\\

                A one way shape memory alloy can only remember its austenitic shape whereas a two way shape memory alloy can remember its austenitic and martensitic shape. Depending on the temperature, a two way shape memory alloy can repeatedly cycle between both shapes. Hence it does not require external forces that deform the austenite crystal structure into a certain martensite structure upon cooling. However, potential actuation strains of two way alloys are considerably smaller than those of one way alloys. Furthermore, they suffer from strain deteriorations that can be observed within a small number of cycles \cite{Scherngell2002-1}.\\

                A two way shape memory alloy needs to be ``trained'' in order to remember its austenitic and martensitic shapes \cite{Luo2007-1}. This can be done by repeatedly performing the following steps:\vspace{-1mm}
                \begin{enumerate}
                    \item Heating of unloaded alloy to its austenitic phase.\vspace{-2mm}
                    \item Deformation into the desired cold shape.\vspace{-2mm}
                    \item Preservation of cold shape and cooling to its martensitic phase.\vspace{-2mm}
                    \item Unloading of alloy.
                \end{enumerate}\vspace{-1mm}
                The hysteresis of a shape memory alloy is the difference in transition temperatures between heating and cooling. Shape memory alloys with a small hysteresis are used in applications that require fast shape changes \cite{Cui2006-1} whereas alloys with a large hysteresis are used in applications that need to maintain certain shapes within a large temperature range.\\

                The low energy efficiency, fatigue life \cite{Eggeler2004-1} and phase stability \cite{Zarubova2004-1} of shape memory alloys limits their applicability. However, there exists a number of industrially relevant applications that exploit their pseudoelasticity and biocompatibility. For example, Ni-Ti alloys are widely used in eyeglass frames, dental wires, stents, vena cava filters and aortic valves \cite{Petrini2011-1}.


            \subsection{Thermal}
                The density of existing materials depends to a varying degree on the temperature. Thermal expansion coefficients and conductivities as well as Young's moduli of various material classes are illustrated in Figure~\ref{pic:Figure_3_7}. Based on their stress-strain-temperature relationship, thermal actuators can be made from a wide range of materials. Actuators that consist of a single, solid piece of material that is actively heated or cooled can generate large forces but only small strains. Different approaches that can be used to increase the actuation strains at the cost of reduced actuation forces are subsequently discussed.\\

                \marginnote{
                \begin{center}
                    \includegraphics[width=\marginparwidth]{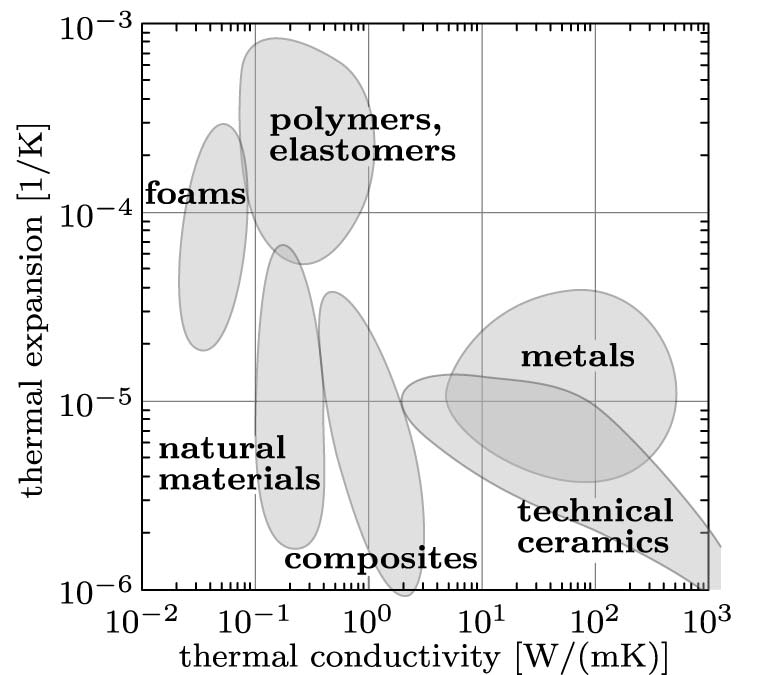}
                    \small(a)\vspace{5mm}
                    \includegraphics[width=\marginparwidth]{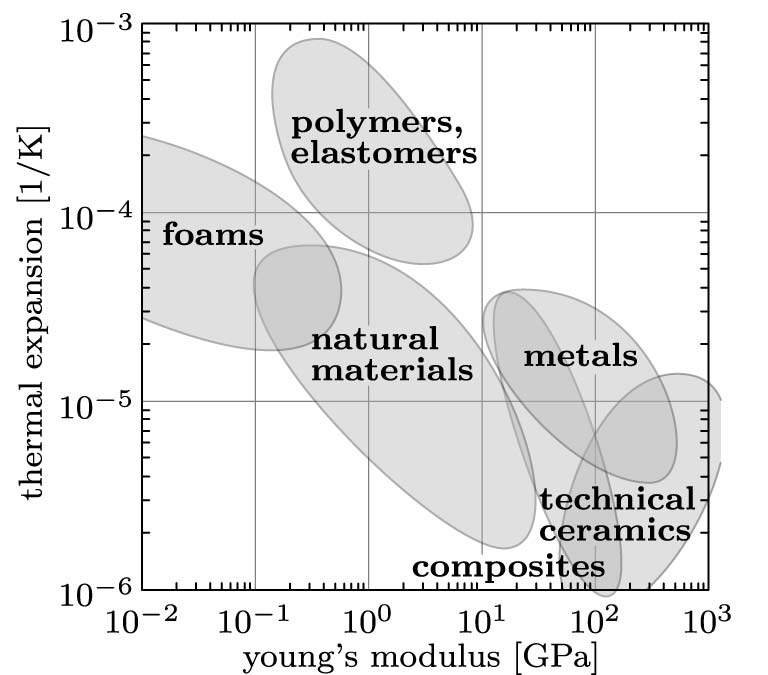}
                    \small(b)
                    \captionof{figure}{Thermal expansion versus (a) thermal conductivity and (b) Young's modulus of various material classes (data from \cite{Ashby2016-1}).}
                    \label{pic:Figure_3_7}
                \end{center}}[-50mm]

                It was shown by Lima et al \cite{Lima2012} in 2012 and subsequently by Haines et al \cite{Haines2014-1} in 2014 that the effective axial strains of anisotropic fibers can be significantly increased by coiling them. The underlying principle is as follows. The thermal expansion coefficient of a straight nylon fiber with axially oriented polymer chains is negative in axial and positive in radial direction. A reorientation of the axially oriented polymer chains into helices thus leads to a torsional deformation of the straight fiber upon heating. An actuator that translates these torsional deformations into a linear motion can be created by fiber coiling. Such a nylon based actuator can generate contractile strains of more than 30\% at the cost of a reduced actuation force \cite{Haines2016-1}.\\

                Microarchitectured materials that consist of two or more material phases can be designed for a wide range of properties. This makes them particularly attractive for the use as thermal actuators since they can occupy regions of the material property space that were previously empty \cite{Fleck2010-1}. Sigmund pioneered the topology optimization of microarchitectured materials with extreme thermal expansion coefficients for two \cite{Sigmund2001-1} and three \cite{Sigmund2001-2} phases. However, large thermal expansion coefficients come at the cost of a low bulk modulus. A simple relationship between the thermal expansion coefficients and bulk modulus of two phase materials \cite{Levin1967-1} is known whereas a corresponding relationship for three phase materials does not exist. However, bounds for the thermal expansion of microarchitectured multiphase materials are known \cite{Gibiansky1997-1}.\\

                The energy efficiency and actuation frequency of thermal actuators is relatively low so that they are particularly suited for applications that need to slowly respond to changes in the environmental temperature. For example, thermal actuators are widely used in greenhouse ventilation systems and thermostats. The steady progress of additive manufacturing techniques for multiple materials \cite{Vaezi2013-1} enables the creation of increasingly sophisticated microarchitectured actuators with desired shape changing properties. This might enable one day the construction of multiphase materials with properties similar to those of shape memory alloys.


            \subsection{Swelling and Shrinking}
                Gels consist of polymer chains that form three-dimensional networks. They differ from soluble materials insofar that their chains are connected by cross links so that they can enclose large amounts of liquids (hydrogels) or gasses (aerogels) as shown in Figure~\ref{pic:Figure_3_8}. For example, aerogels are excellent insulators as more than 99\% of their volume is made up of immobile air\footnote{Graphene aerogels with a density of about 160~g/m$^3$ are the least dense solids known \cite{Sun2013-1}. They are nearly eight times lighter than air.}. In contrast, the water content of hydrogels \cite{Wichterle1960-1} can be varied within a wide range or redistributed with the help of different mechanisms. This leads to large potential volume changes that can be used to create various kind of actuators.\\

                \marginnote{
                \begin{center}
                    \includegraphics[width=\marginparwidth]{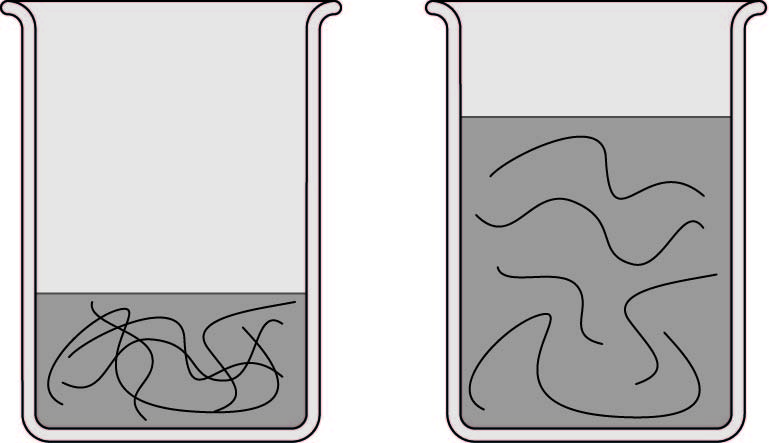}
                    \small(a)\vspace{5mm}
                    \includegraphics[width=\marginparwidth]{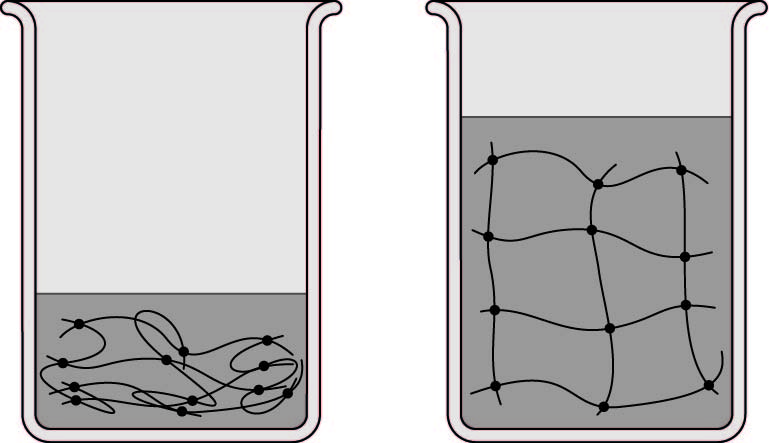}
                    \small(b)
                    \captionof{figure}{Dilution of polymer chains (a) without and (b) with cross links.}
                    \label{pic:Figure_3_8}
                \end{center}}[-49mm]

                Volume changes can be triggered by a wide range of external stimuli. For example, osmotic gradients can be generated by ionizing ultraviolet light \cite{Mamada1990-1} or in ionic hydrogels by electric fields \cite{Tanaka1982-1}. Other mechanisms include a breakdown \cite{Tanaka1978-1} of hydrogel networks or a temperature dependent variation of their hydrophilicity \cite{Matsumoto2018-1}. The speed of volume changes and the mechanical properties of hydrogels are heavily influenced by the network structure. Response times can be significantly reduced by large, interconnected pores \cite{Chen1999-1} whereas the mechanical properties are mostly determined by the nature and density of the cross links. Haraguchi and Takehisa \cite{Haraguchi2002-1} showed in 2002 that cross links can be modified by the inclusion of nanoparticles. Another approach that leads to stiff and ductile hydrogels was proposed by Gong et al \cite{Gong2013-1} in 2003. It is based on two interlaced networks where one is stiff and brittle and the other is soft and ductile. Further improvements seem to be possible by tailoring the network structure with the help of carefully designed, self-assembling building blocks \cite{Petka1998-1}.\\

                The size dependent and relatively slow response times of hydrogel actuators vary between shrinking and swelling. Hence they can be rarely found in commercial applications. However, the working principle and potential of hydrogels is intriguing as it has much in common with the nastic movement of plants.


    \newpage

    \sectionmark{Summary}
    \begin{framed}
        \noindent \textbf{Summary}\\

        \noindent Actuators translate various kind of energy forms into motions so that they are an indispensable part of any machinery. They can be solid state materials or mechanisms themselves. Solid state actuators are usually driven by variations in their crystal or molecular structure whereas mechanical actuators are driven by the translation or rotation of rigid bodies. Compared to their solid counterparts, mechanical actuators can usually generate larger strains or, in terms of motors, unbounded rotations. This makes them particularly suited for the use in heavy machinery such as airplanes and excavators.\\

        Electricity and differential pressures are the two most commonly used energy forms for the operation of mechanical actuators. Electric actuators are often motors whose unbounded rotations can be translated, if required, into linear motions via simple mechanisms. In contrast, pressure based actuators are usually designed for their intended motions where the most commonly used linear actuators are hydraulic and pneumatic cylinders.\\

        The ongoing trend towards the electrification of airplanes led to the development of electro hydraulic actuators. These actuators are often based on an electric motor that locally generates the pressurized fluid for the operation of a hydraulic cylinder. This reduces the required plumbing within an airplane and thus leads to significant weight savings while it avoids some of the safety critical drawbacks of electromechanical actuators.\\

        \noindent \textbf{Conclusion}\\

        \noindent It was argued in the previous chapter that some of the most efficient and lightweight structures avoid compressive forces with the help of pressure gradients by introducing a sufficiently large prestress into the material. The same can be said for hydraulic and pneumatic cylinders that are, due to their large actuation strains and stresses the de facto standard for the construction of heavy machinery.
    \end{framed}

        \cleardoublepage
    \chapter{High Lift Devices}
        The roll movements of early airplanes were controlled with the help of flexible wings that could be twisted in opposite directions. This approach was a viable option as long as flight speeds were relatively low. The rapid development of engines with an increasingly large power-to-weight ratio enabled flight speeds that could not be sustained by flexible wings. This led to the development of rigid wings with hinged ailerons whose airfoils were optimized for cruise flight. The reduced lift at takeoff and landing caused a need for longer runways which was, in return, countered by the development of various high lift devices that were widely used since the 1930s. For example, the deployed high lift devices of a Boeing 747-400 increase the wing area by 21\% and the lift by 90\%.


        \section{Conventional Designs}
            Wings that can seamlessly change their shape were already investigated in 1915 by the Varioplane Company \cite{Varioplane1915-1}. However, their realization within an acceptable weight limit is far from trivial and most likely not possible with existing technologies. Wings are therefore often split into wing boxes and leading, trailing edges. Wing boxes are rigid structures that carry most of the loads whereas the design of leading and trailing edges is dominated by aerodynamic considerations. This section provides a brief overview of the historic development \cite{Smith1975-1} of high lift devices for leading and trailing edges that consist of aerodynamically shaped rigid bodies and various kind of mechanisms.


            \subsection{Leading Edge}
                Fixed leading edge slats or slots were invented by Lachmann in 1917 after he stalled his airplane during training. Slots consist of a small, separate airfoil shaped device that is rigidly placed in front of the wing, Figure~\ref{pic:Figure_4_1}. The corresponding patent application was initially rejected by the German Patent Office in 1918 and retroactively granted \cite{Lachmann1922-1} in 1922 after experiments confirmed an increased lift of about 63\%. Slots can considerably reduce an airplanes stall speed and improve its handling properties at low speeds. However, they are rarely used in modern airplanes since the additional drag becomes prohibitive at higher speeds. As a consequence, slots are mostly found in slow flying, low maintenance bush planes that need to takeoff and land on short runways.\\

                Slots were independently developed in the United Kingdom by Handley Page who applied for a patent in 1919. In order to avoid a patent challenge he joined forces with Lachmann who moved to the United Kingdom in 1929. They continued their work on high lift devices by investigating moveable slats that increase an airplanes lift at low speeds without increasing its drag at high speeds. The first slats were automatically deployed by springs at low speeds and retracted by aerodynamic forces at high speeds. Automatic slats were, due to their simplicity and reliability, widely used during the Second World War. However, they were soon replaced by actively controlled slats that are commonly used today.\\

                The first actively controlled slats were either fully deployed or retracted. Two position slats were widely used in passenger airplanes such as the Boeing B707 that made its maiden flight in 1957. However, it was soon realized that it is advantageous to use different slat positions during takeoff and landing. Three position slats are fully deployed, slotted during landings to maximize lift whereas a partially deployed, sealed configuration is used during takeoffs. Examples of modern airplanes with three position slats include the Airbus A380 that made its maiden flight in 2005 and the Boeing B787 that made its maiden flight in 2009.\\

                Kr\"uger flaps were invented in 1943 at the Aerodynamische Versuchsanstalt in G\"ottingen. Three kind of Kr\"uger flaps with increasingly refined aerodynamic surfaces are commonly used today. Simple Kr\"uger flaps deploy a single, aerodynamically shaped rigid body from the lower wing surface whereas bull nose flaps deploy two rigid bodies that are connected to each other. The latter can be found, for example, in the Boeing B747 that made its maiden flight in 1969. Variable camber flaps are a further development of the bull nose design. They improve the aerodynamic properties with the help of a flexible skin that is deformed in the deployed configuration. Despite their superior aerodynamic properties, variable camber flaps are rarely used since they require relatively complex and heavy mechanisms. Kr\"uger flaps are generally lighter, simpler and less expensive than moveable slats. However, they can only be deployed into one position so that their aerodynamic performance is, particularly during takeoff, inferior to three position slats. Nonetheless, they have recently attracted a lot of attention since they do not require any gaps on the suction side of a wing. This is particularly important for the design of laminar flow airfoils that are considered for the next generation of passenger airplanes.\\

                Rigid droop noses were already investigated by Harris and Bradfield \cite{Harris1920-1} in 1920. They differ from slats insofar that the entire leading edge is rotated downwards without moving away from the wing. A rigid droop nose can not achieve the lift of a slotted slat due to its stall angle that is limited by the large curvatures on the suction side of the wing. Nonetheless, they are easier to manufacture and require a smaller installation space than slats. Hence they are particularly suited for the use in supersonic airplanes such as the F-5 that made its maiden flight in 1959. Furthermore, their use in passenger airplanes seems to be advantageous under certain conditions. For example, rigid droop noses are installed between the fuselage and inboard engines of the Airbus A350 and A380.

                \afterpage{
                    \newgeometry{}
                    \begin{figure}[htbp]
                        \begin{center}
                            \subfloat[]{
                                \includegraphics[width=0.95\textwidth]{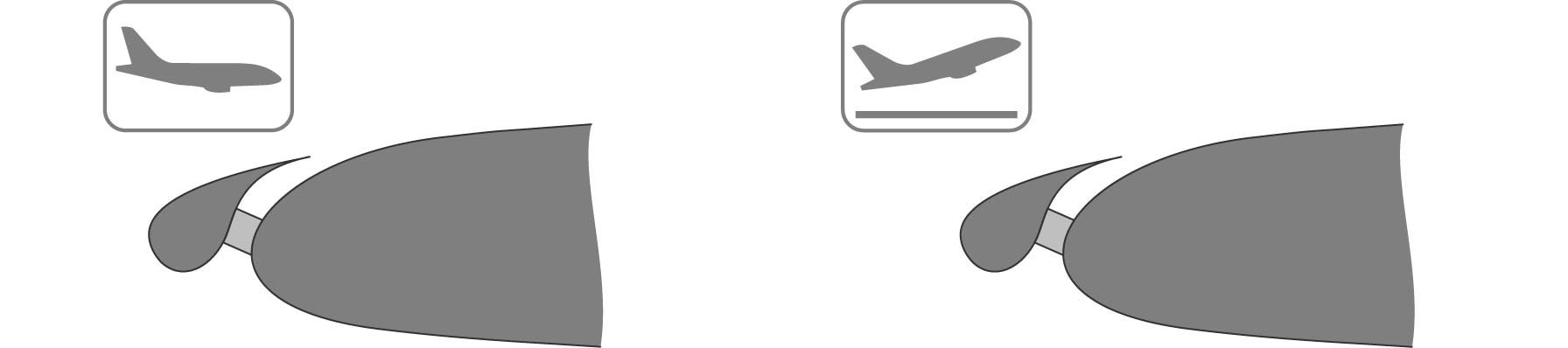}}

                            \subfloat[]{
                                \includegraphics[width=0.95\textwidth]{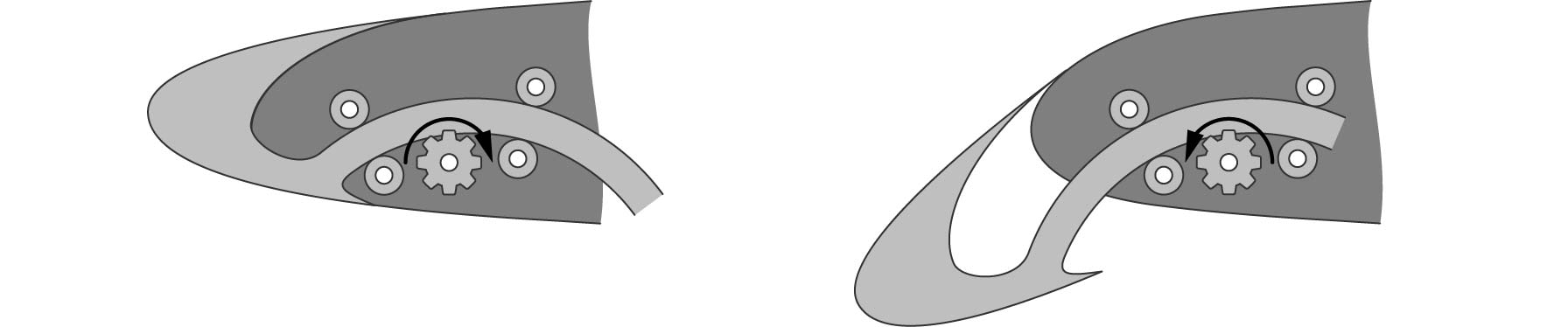}}

                            \subfloat[]{
                                \includegraphics[width=0.95\textwidth]{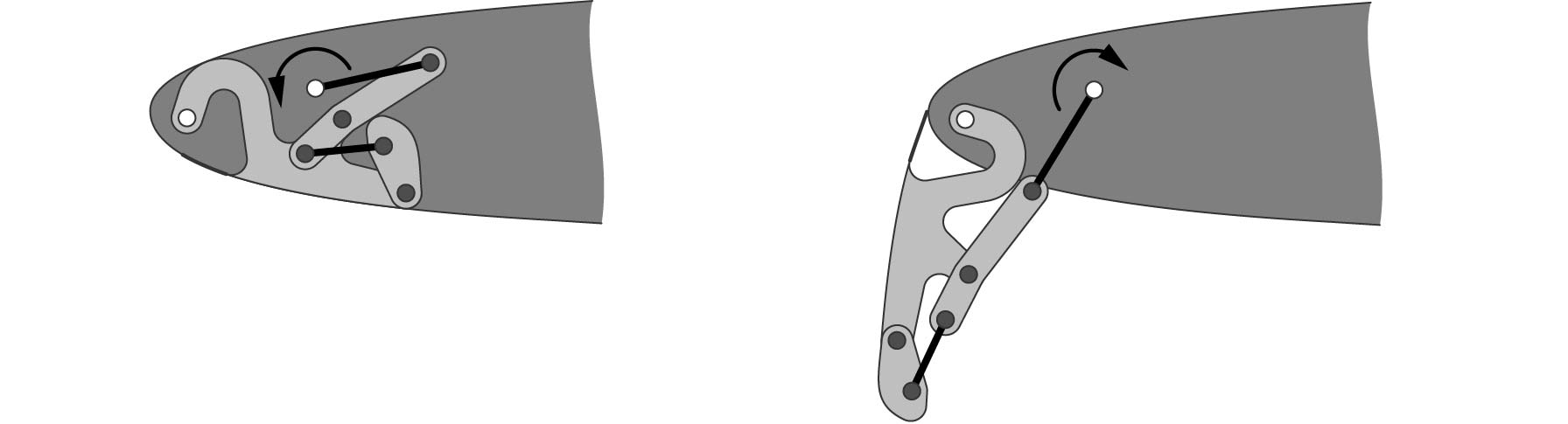}}

                            \subfloat[]{
                                \includegraphics[width=0.95\textwidth]{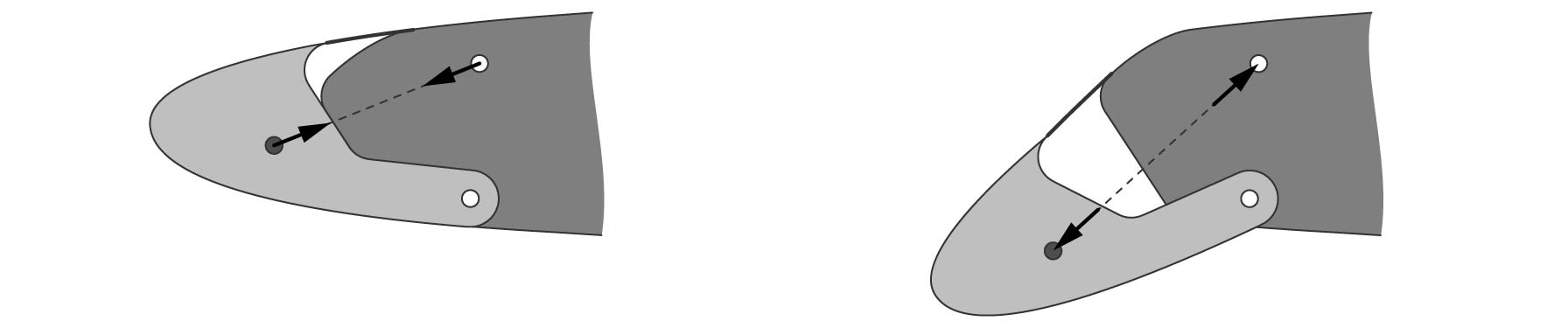}}
                            \caption{Various high lift devices for leading edges. (a) Fixed slat or slot. (b) Retracted and sealed configuration of a three position slat. The sealed configuration is used during takeoffs whereas the fully extended, slotted configuration is used during landings. The underlying mechanism is based on constant radius tracks that possesses a single, rotary degree of freedom. (c) Inboard bull nose Kr\"uger flap with a rotary degree of freedom that consists of two connected 4-bar linkages. (d) Rigid droop nose with a linear degree of freedom that is based on a 2-bar linkage.}
                            \label{pic:Figure_4_1}
                        \end{center}
                    \end{figure}

                    \begin{figure}[htbp]
                        \begin{center}
                            \subfloat[]{
                                \includegraphics[width=0.95\textwidth]{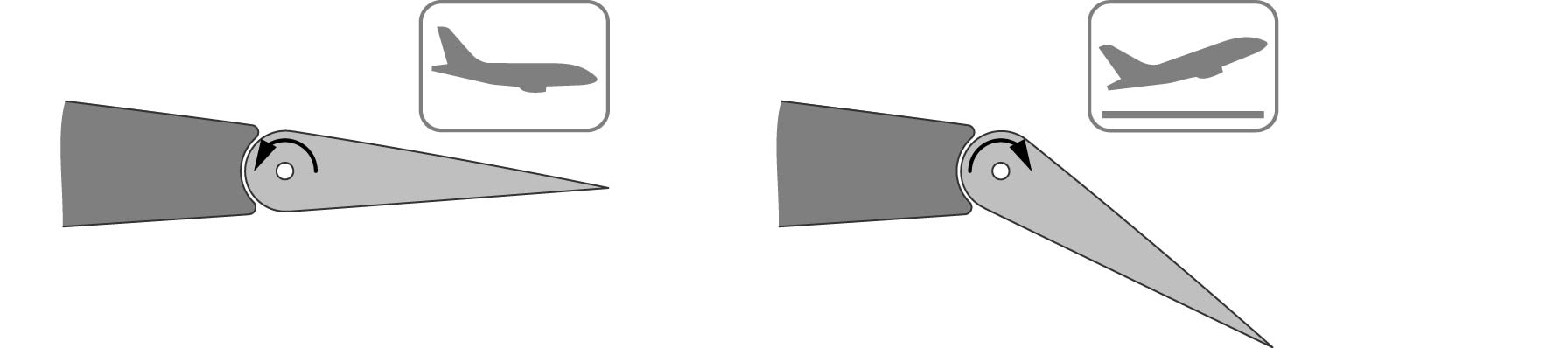}}

                            \subfloat[]{
                                \includegraphics[width=0.95\textwidth]{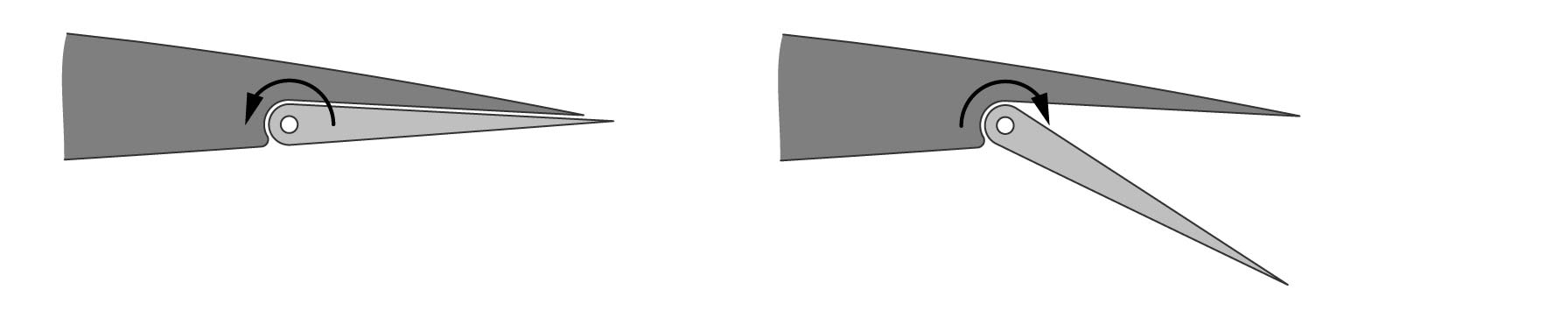}}

                            \subfloat[]{
                                \includegraphics[width=0.95\textwidth]{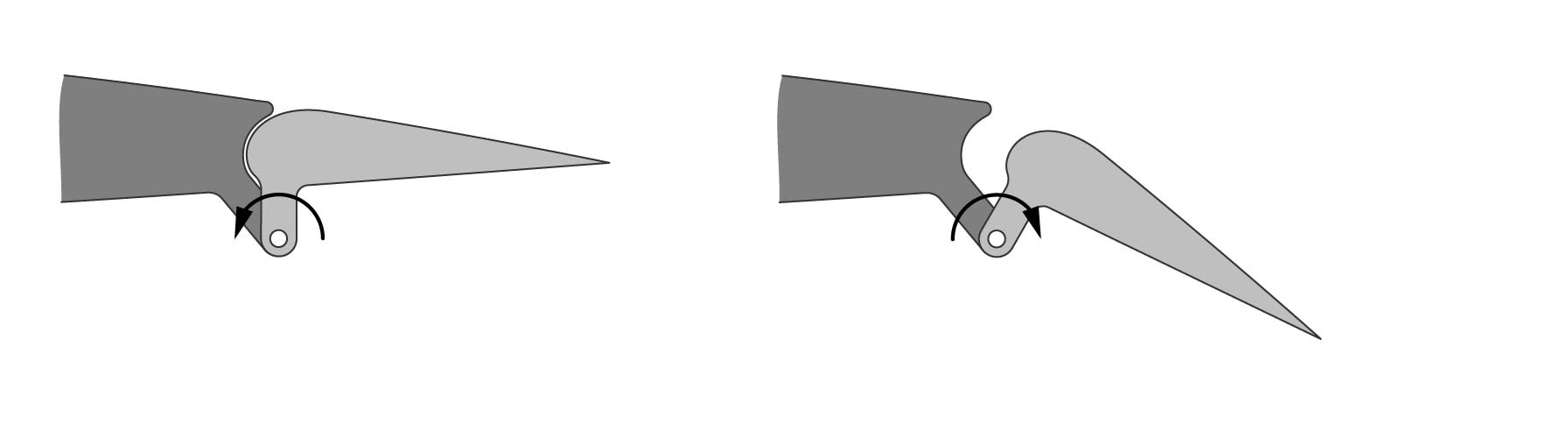}}

                            \subfloat[]{
                                \includegraphics[width=0.95\textwidth]{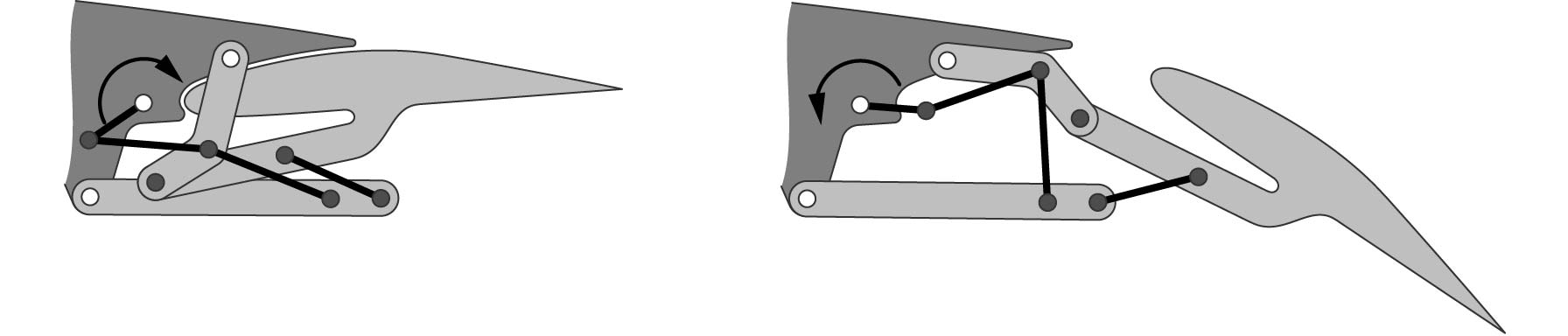}}

                            \caption{High lift devices for trailing edges that possess a single, rotary degree of freedom. Mechanisms consist of an aerodynamically shaped rigid body, struts and beams. (a) Plain flaps are used in small airplanes and fighter jets with relatively thin wings. (b) Split flaps were used in fighter planes during the Second World War. They turned out to be inferior to single or multi slotted flaps. (c) Single slotted and (d) outboard Fowler flap. The latter increases the wing area and thus the lift in the deployed configuration.}
                            \label{pic:Figure_4_2}
                        \end{center}
                    \end{figure}
                    \restoregeometry}


            \subsection{Trailing Edge}
                Ailerons were already used by Esnault-Pelterie in 1904 to control the roll movements of a glider. The major difference between plain flaps, as illustrated in Figure~\ref{pic:Figure_4_2}, and ailerons is that the former can only rotate downwards to increase lift. Plain flaps have been tested in the United Kingdom by The Royal Aircraft Factory and the National Physical Laboratory between 1913 and 1914. They were first used in a floatplane for the Royal Naval Air Service that was build in 1916 by the Fairey Aviation Company. Plain flaps are inexpensive, compact and easy to construct. However, their performance is usually inferior to other kind of flaps so that they are mostly used in small airplanes and fighter jets with relatively thin wings. Examples include the MiG-29 and the F-15E that made their maiden flights in 1977 and 1986, respectively.\\

                Split flaps were invented by Orville Wright and James Jacobs \cite{Wright1921-1} in 1921. They resemble plain flaps that are mounted in a cutout at the bottom of the leading edge. Split flaps were widely used during the 1930s in passenger airplanes such as the DC-3 that made its maiden flight in 1935 and in military airplanes such as the Focke-Wulf Fw 190 that made its maiden flight in 1939. The lift that is generated by split flaps is comparable to plain flaps whereas their drag is considerably larger. This enables steeper approaches during landings but leads to longer takeoff rolls. Modern airplanes avoid this problem by using flaps with a larger lift to drag ratio and additional air breaks that increase the drag during landings.\\

                The lift to drag ratio of plain flaps can be increased with the help of an eccentric rotation center. This leads to a gap between wing and flap in the deployed configuration that stabilizes the boundary layer on the suction side. The working principle of single slotted flaps is similar to those of leading edge slats so that it is not surprising that their development coincided at Handley Page Ltd \cite{Miller1968-1} in 1920. The lift to drag ratio of single slotted flaps was further improved by Fowler in 1924. He proposed the use of larger flaps that are partially enclosed by the wing in the retracted configuration. This increases the wing area in the deployed configuration and thus has a positive effect on the lift. Fowler flaps were first used in the Martin 146 that made its maiden flight in 1935. They are, despite their relatively complex mechanisms, still used in modern airplanes.\\

                The performance of Fowler flaps was further improved by splitting them into two or more segments. These segments separate during deployment so that they increase the number of slots and thus the lift. Double slotted flaps were first used in the Piaggio P.32 and triple slotted flaps in the Boeing B727 that made their maiden flights in 1936 and 1963, respectively. However, the performance gain of double and triple slotted flaps comes at the cost of relatively complex mechanisms and increased maintenance requirements. The availability of powerful computers and simulation tools enabled the design of increasingly refined flaps. This led to a reversal of this trend \cite{Reckzeh2003-1} so that single slotted flaps can be found in many modern airplanes. Examples include the Boeing B787 that made its maiden flight in 2009 and the Airbus A350 that made its maiden flight in 2013.


        \marginnote{
        \begin{center}
            \includegraphics[width=\marginparwidth]{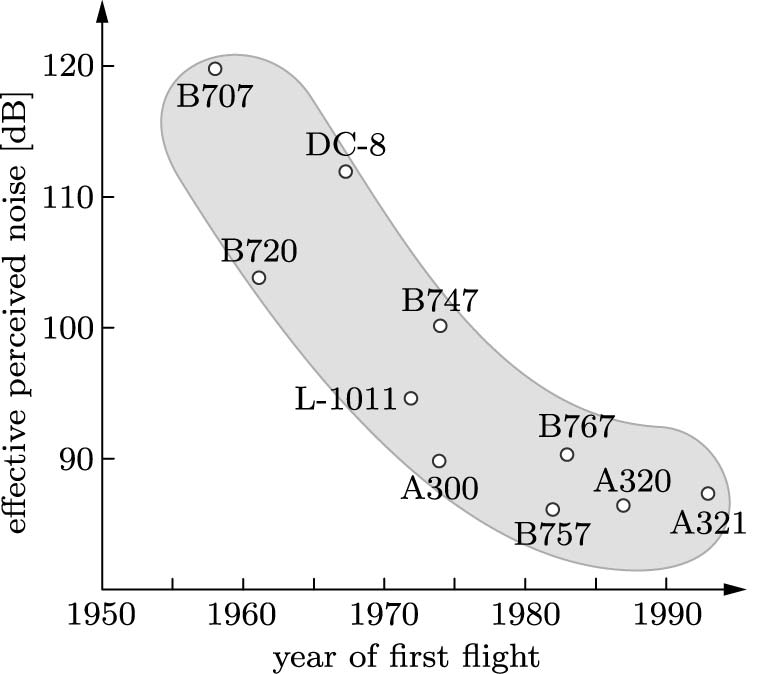}
            \captionof{figure}{Effective perceived takeoff noise and year of maiden flight of various passenger airplanes (data from \cite{Filippone2000-1}).}
            \label{pic:Figure_4_3}
        \end{center}}[13mm]

        \section{Gapless High Lift Devices}
            The effective perceived takeoff noise of various passenger airplanes is illustrated in Figure~\ref{pic:Figure_4_3}. It can be seen that noise emissions\footnote{Noise emissions of passenger airplanes reduced by about 30~dB since the 1950s. This is comparable to the reductions that can be achieved by a silencer on a rifle.} decreased significantly since the 1950s. Furthermore it can be seen that this development started to slow down in the 1990s. Other design aspects that reached a high level of maturity include the energy efficiency and stealth properties\footnote{Reliable data on the radar cross section of military airplanes is usually classified. A detailed study is further complicated by the dependency on the radar frequency and the relative position between the source and the airplane.} of airplanes. As a consequence, significant improvements in these fields can only be achieved by the development of new technologies. A promising approach in this regard is the use of gapless high lift devices. They have the potential to not only reduce an airplanes noise emissions, fuel consumption and radar reflections but also to provide a basis for new technologies such as laminar flow control \cite{Stanewsky2001-1}.\\

            The cruise and low speed configuration of a gapless leading and trailing edge is shown in Figure~\ref{pic:Figure_4_4}. The continuity of the leading edge skin enables the design of mechanisms that preserve its arc-length during shape changes. In contrast, the arc-length of the trailing edge skin can not be preserved so that it needs to undergo stretching and bending deformations. Various mechanisms for gapless leading and trailing edges are subsequently reviewed. It should be emphasized at this point that none of them is currently used in passenger or military airplanes.

            \marginnote{
            \begin{center}
                \includegraphics[width=\marginparwidth]{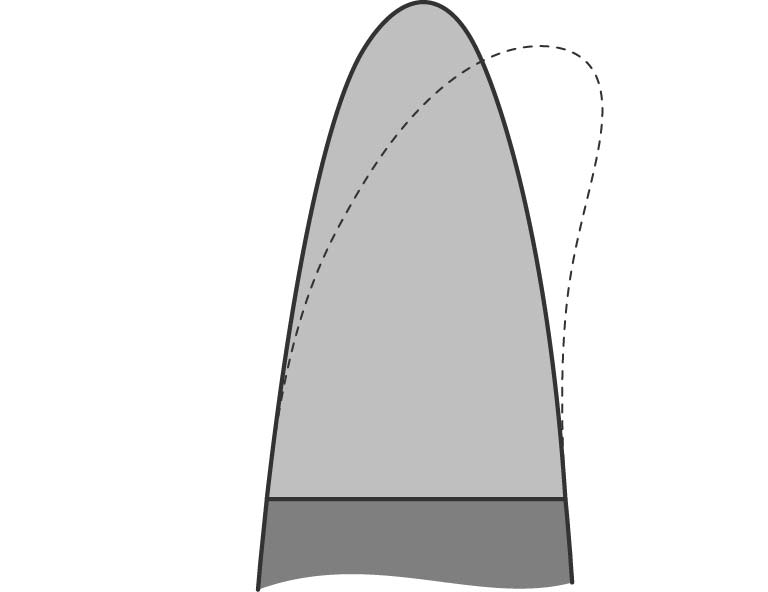}
                \small(a)\vspace{3mm}
                \includegraphics[width=\marginparwidth]{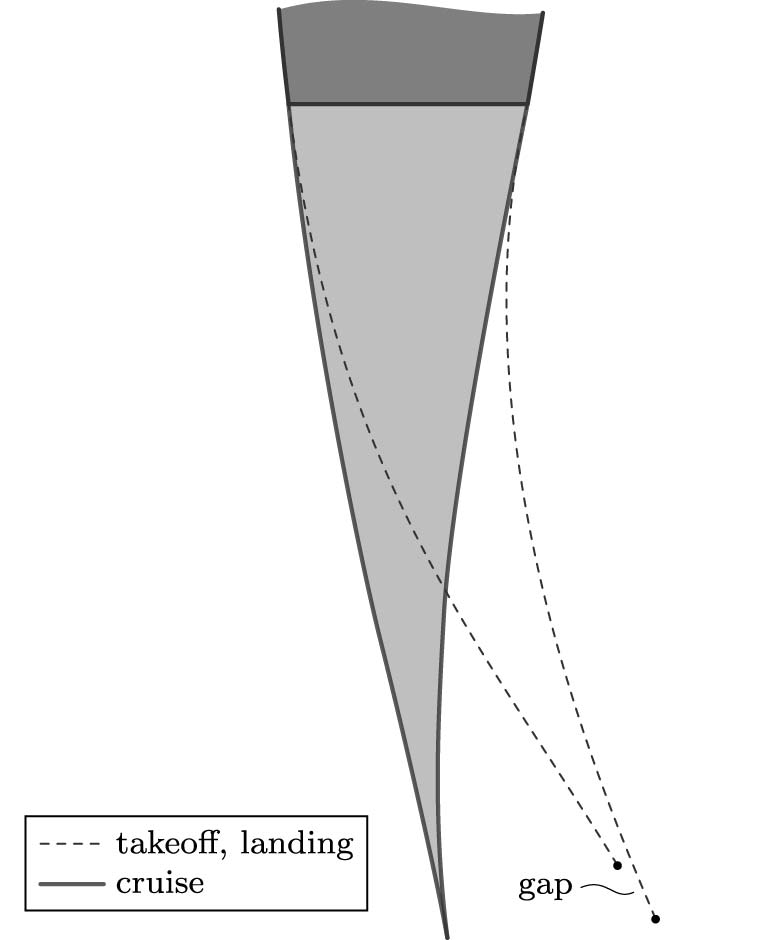}
                \small(b)
                \captionof{figure}{Cruise and low speed configuration of a gapless (a) leading and (b) trailing edge. Leading edge skins need to undergo only bending deformations whereas trailing edge skins need to additionally undergo stretching deformations.}
                \label{pic:Figure_4_4}
            \end{center}}[-40mm]


            \subsection{Leading Edge}
\begin{figure}[htbp]
                    \begin{center}
                        \subfloat[]{
                            \includegraphics[width=\textwidth]{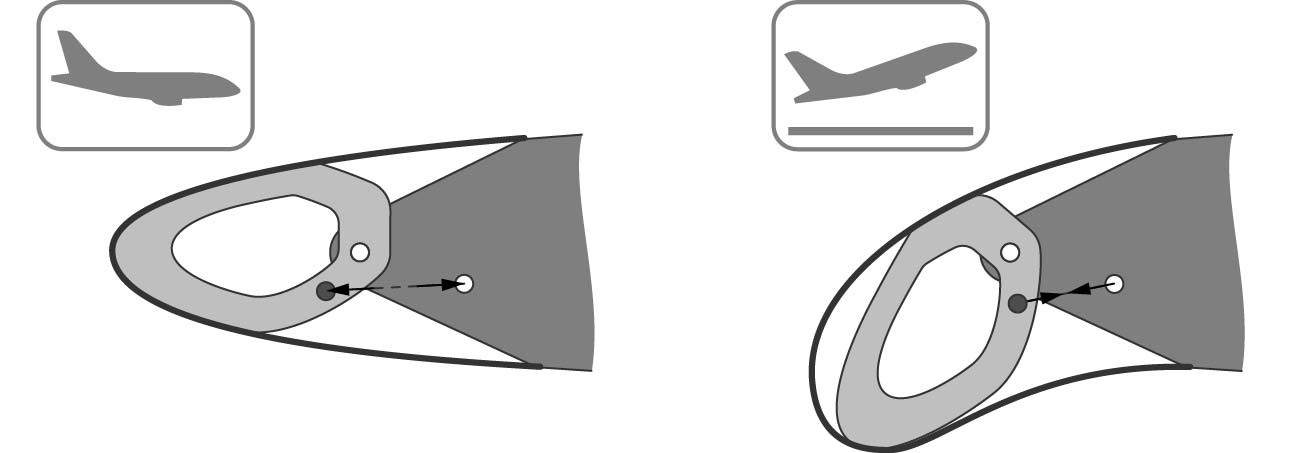}}

                        \subfloat[]{
                            \includegraphics[width=\textwidth]{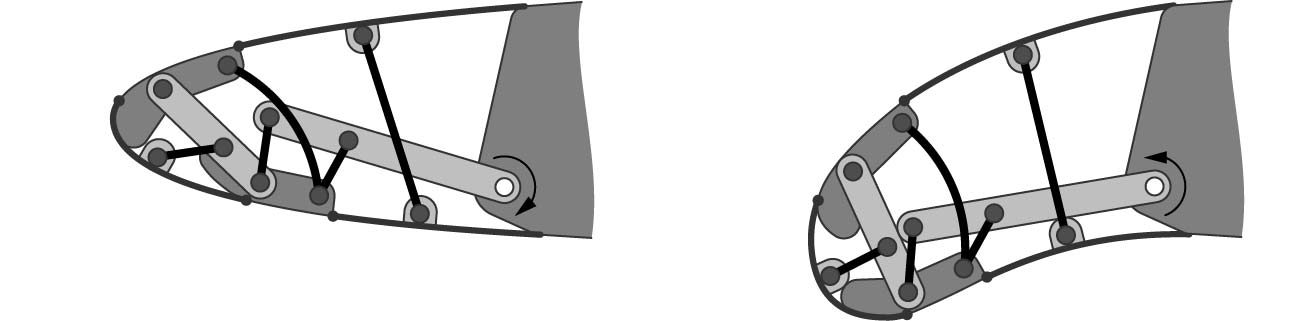}}

                        \subfloat[]{
                            \includegraphics[width=\textwidth]{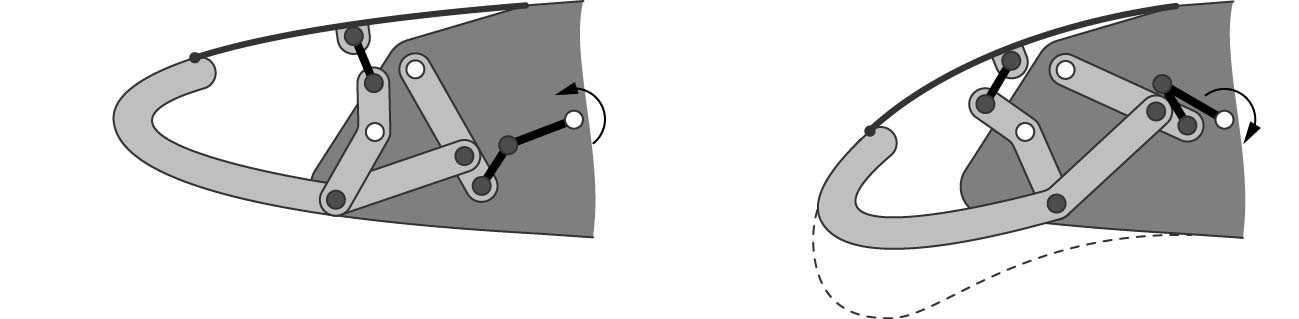}}

                        \subfloat[]{
                            \includegraphics[width=\textwidth]{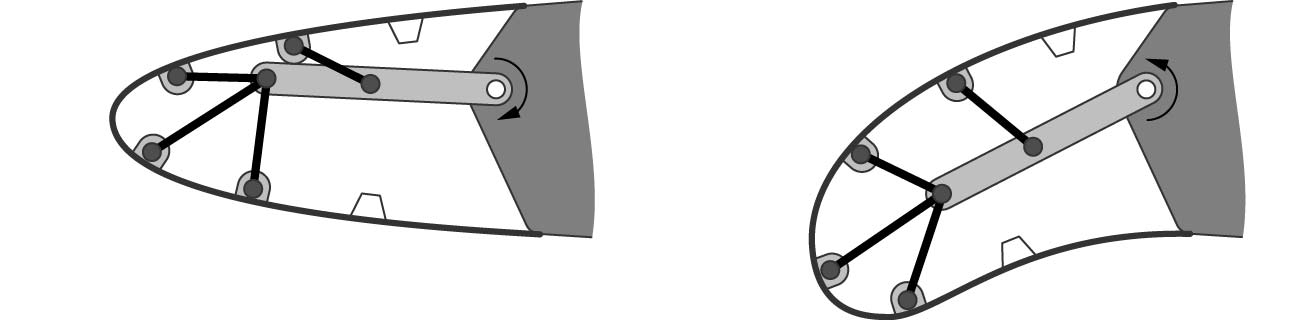}}

                        \subfloat[]{
                            \includegraphics[width=\textwidth]{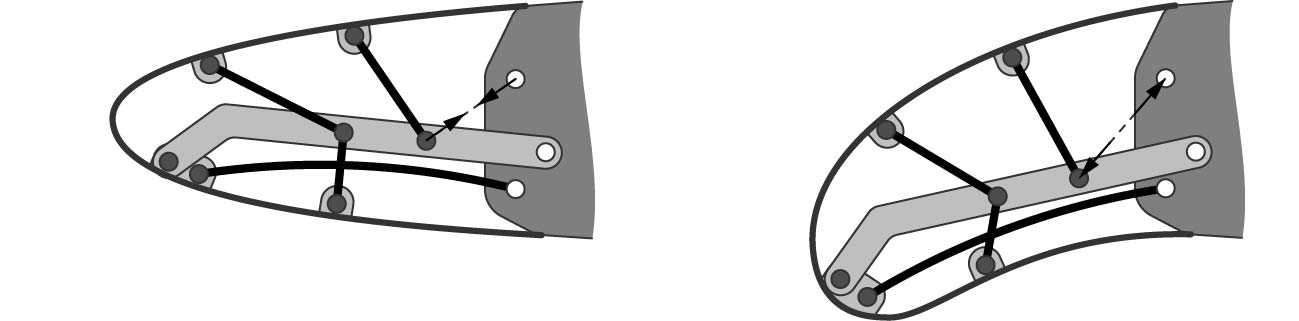}}
                    \end{center}

                    \marginnote{
                        \begin{center}
                            \captionof{figure}{Various mechanisms that preserve the arc-length of the leading edge skin during shape changes. (a) Inextensional, flexible skin with a cam mechanism that possesses a linear degree of freedom. (b) Segmented skin with three flexible and two rigid segments. The underlying mechanism that is based on beams and rigid struts possesses a rotational degree of freedom. (c) A similar approach with a skin that is based on a flexible and rigid segment. The skin is continuous during cruise and discontinuous on the pressure side during takeoff and landing. (d) An unsegmented, flexible skin and a relatively simple mechanism with a rotary degree of freedom that consists of a beam and rigid struts. (e) A variation of the previous mechanism that locally enforces the gradient of the skin.}
                            \label{pic:Figure_4_5}
                        \end{center}}[-200mm]
                \end{figure}

                One of the first gapless leading edges was proposed by Pierce \cite{Pierce1970-1} in 1970. It possesses a linear degree of freedom and consists of a flexible skin and one or more cams that determine its shape, Figure~\ref{pic:Figure_4_5}. An advantage of this approach is that the total arc-length of the skin is preserved as long as the friction forces between  skin and cams are relatively small. However, the sliding motions between them can lead to increased maintenance requirements. Furthermore, large skin deformations on the suction side can not be prevented without additional supports.\\

                A gapless leading edge that avoids any sliding motions was developed by Zimmer \cite{Zimmer1979-1} in 1979. It is based on a segmented skin that consists of three flexible and two rigid, cam based segments. The upper and lower segments are deformed by the underlying mechanism whereas the third segment is deformed by the contact forces that are exerted by both cams. Although rolling contacts are an advancement compared to sliding contacts, their use can still increase the maintenance requirements. Furthermore, the continuity of the skin in the deployed configuration is limited by the rigid segments. A similar, less complex leading edge was subsequently proposed by Cole \cite{Cole1982-1} in 1982. It is based on a skin that consists of a flexible and rigid segment. Contact forces are avoided and the complexity of the mechanism is considerably reduced. However, this comes at the cost of a discontinuity at the bottom of the leading edge in the deployed configuration.\\

                A gapless leading edge with an unsegmented skin that is continuous in the retracted and deployed configuration was proposed by Kintscher et al \cite{Kintscher2016-1} in 2009. It avoids any contact forces and possesses a relatively simple mechanism with a rotary degree of freedom that consists of a beam and rigid struts. The stringers that reinforce the skin in spanwise direction serve as attachment points for the underlying mechanism. A variation of this mechanism that additionally enforces the gradient of the skin at the front of the leading edge was proposed by Grip et al \cite{Grip2014-1} in 2014.\\

                Despite a massive research effort, gapless leading edges are currently not used in commercial or military airplanes. This is mostly due to the relatively heavy mechanisms and complex skins. The construction of skins for gapless high lift devices is particulary challenging as they need to be soft enough to undergo large deformations while being stiff enough to carry the aerodynamic loads. Furthermore, they have to protect the wing against erosion, icing, lightning- and bird strikes. Nonetheless, these problems might eventually be overcome by the availability of increasingly advanced materials and manufacturing techniques.


            \subsection{Trailing Edge}
                There exists an astonishingly large number of different mechanisms for gapless trailing edges. This variety is potentially due to the false assumption that their design is not constrained by the induced skin deformations. A few interesting mechanisms that vary particularly with respect to the stretching of their skins are subsequently reviewed.\\

                \begin{figure}[htbp]
                    \begin{center}
                        \subfloat[]{
                            \includegraphics[width=\textwidth]{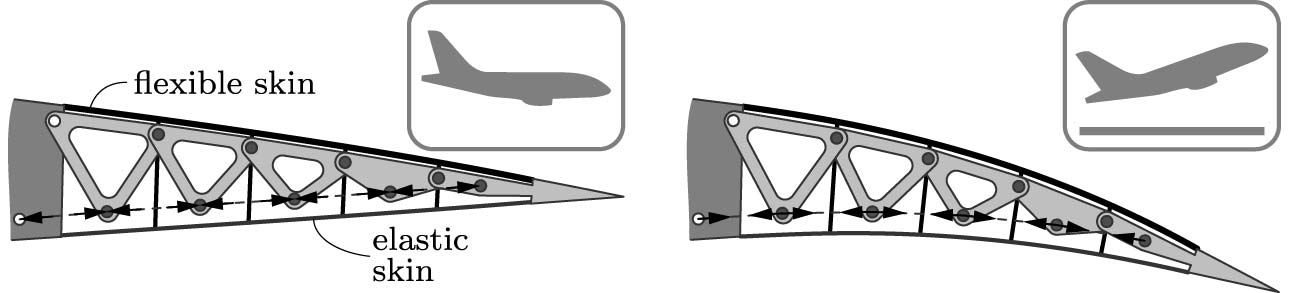}}

                        \subfloat[]{
                            \includegraphics[width=\textwidth]{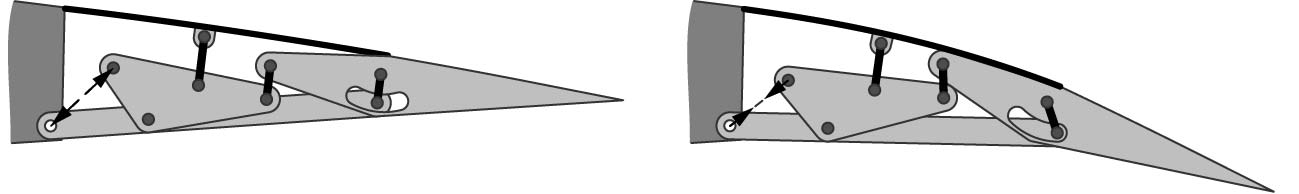}}

                        \subfloat[]{
                            \includegraphics[width=\textwidth]{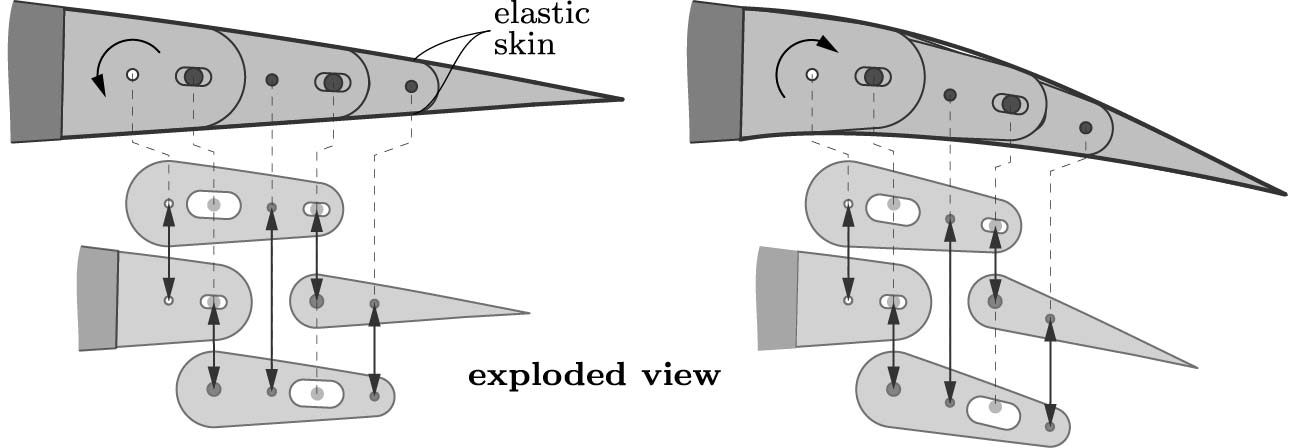}}

                        \subfloat[]{
                            \includegraphics[width=\textwidth]{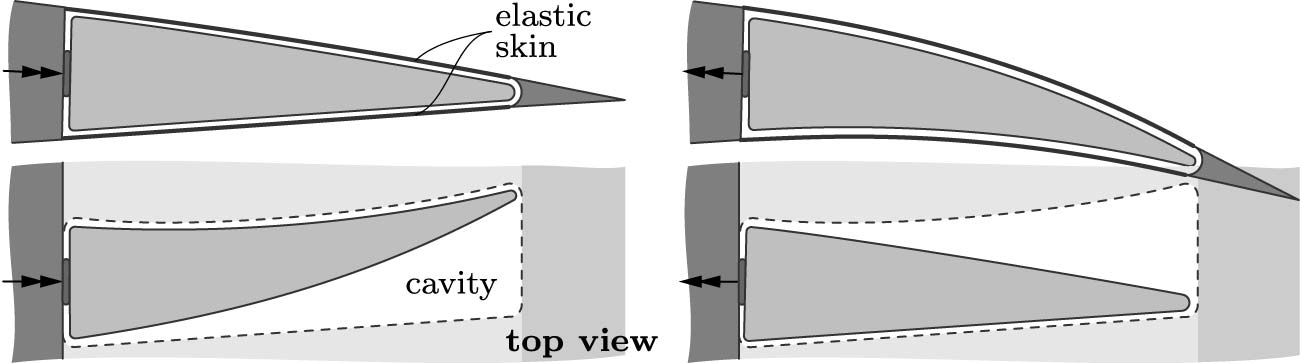}}

                        \subfloat[]{
                            \includegraphics[width=\textwidth]{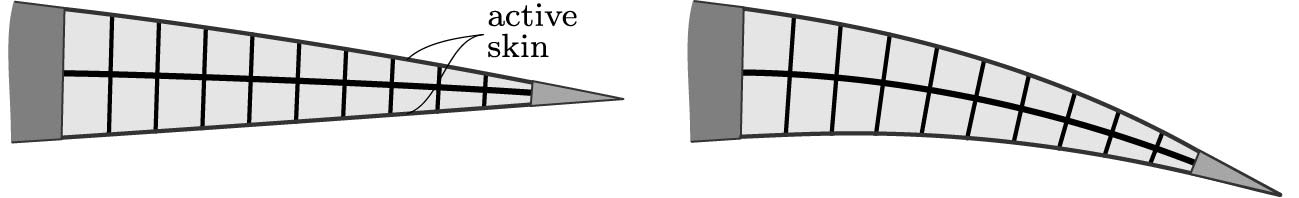}}

                        \subfloat[]{
                            \includegraphics[width=\textwidth]{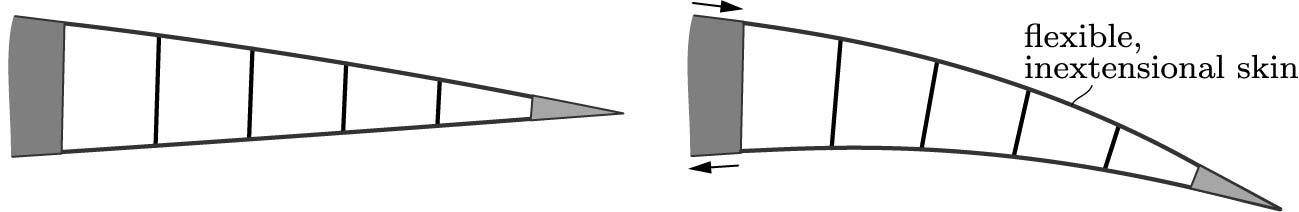}}
                    \end{center}

                    \marginnote{
                        \begin{center}
                            \captionof{figure}{Various mechanisms for gapless trailing edges. (a) Truss based mechanism with multiple degrees of freedom that can be merged with the help of a drive train. The arc-length of the upper skin is preserved whereas the bottom skin needs to undergo large stretching deformations. (b) Mechanism with a single degree of freedom that consists of three rigid bodies and three struts. It preserves the arc-length and continuity of the upper, flexible skin whereas the lower skin becomes discontinuous during takeoff and landing. (c) Mechanism with a single degree of freedom that is based on a chain of rigid bodies that are connected to each other via hinges and sliders. Both, the upper and lower skins are stretched during shape changes. (d) Single degree of freedom mechanism that is based on a three-dimensional cam. It stretches both, the upper and lower skins. (e) Compliant mechanism that consists of a central, flexible plate and two honeycomb cores that act as spacers between plate and skins. The mechanism is actuated by skins that can actively change their lengths. (f) Compliant mechanism that completely integrates a pair of flexible, inextensional skins. A relative movement between both trailing edge ends leads to shape changes.}
                            \label{pic:Figure_4_6}
                        \end{center}}[-213mm]
                \end{figure}

                Truss structures for fuselages and wings were widely used since the early days of aviation. For example, the wings of the Junkers Ju~52 that made its maiden flight in 1932 are based on a relatively complex, three-dimensional truss. It is likely that the idea to change the shape of a truss by varying the length of its members goes back to these early days. Shape changing truss structures are periodically revisited since then although they were never used in commercial or military airplanes. For example, Lyon \cite{Lyon1960-1} investigated a wing with a flexible skin and an internal, shape changing truss structure in 1960. Forty years later, Perez \cite{Perez2000-1} proposed a gapless trailing edge that is based on a similar mechanism as shown in Figure~\ref{pic:Figure_4_6}. It possesses a large number of linear degrees of freedom that can be either separately actuated or merged into a single, rotary degree of freedom with the help of a drive train. A major drawback of this approach is that the lower skin needs to be highly stretched whereas the arc-length of the upper skin is preserved. Furthermore, it requires a large number of actuators or a relatively heavy drive train.\\

                A mechanism with fewer parts and a single, linear degree of freedom was proposed by Cole \cite{Cole1975-1} in 1975. It consists of struts and rigid bodies that form, in conjunction with a flexible segment, the upper and lower skin. The arc-length of the flexible segment is preserved during shape changes whereas the use of rigid bodies causes discontinuities on the pressure side. A similar approach \cite{Smith1992-1} was successfully used by the National Aeronautics and Space Administration during the 1980s to evaluate the performance of an upgraded F-111A fighter jet.\\

                Gapless trailing edges with continuous surfaces require mechanisms that equally distribute the deformations between the upper and lower skins. Such a mechanism with a single rotary degree of freedom was developed by Piening and Monner \cite{Piening1997-1} in 1997. It consists of a chain of overlapping rigid bodies that are connected to each other via hinges and sliders. The mechanism was initially proposed in conjunction with a segmented, discontinuous skin. Nonetheless, it can be used together with elastic, flexible skins that avoid surface gaps. However, potential shape changes are rather small and the overall weight is relatively large due to the overlapping chain segments.\\

                A completely different mechanism was proposed by M\"uller \cite{Mueller1996-1} in 1996. It is based on a three-dimensional, rigid cam with a horn like shape that is rotated within the specifically designed cavity of an elastic trailing edge. The geometries of the line contacts between the cam and the trailing edge vary considerably during shape changes. This makes the use of low friction surfaces that minimize actuation forces and wear indispensable. Another problem is the load transfer between the skins and the horn particularly on the suction side.\\

                A compliant mechanism that avoids any moving parts, friction forces and discontinuities was developed by Maclean et al \cite{Maclean1993-1} in 1993. It consists of a central, flexible plate that is sandwiched between two honeycomb cores that act as spacers between plate and skins. The mechanism is either actuated by internally placed tendons that induce a bending moment into the plate \cite{Beauchamp1993-1} or by skins that can actively change their length \cite{Maclean1993-1}.\\

                Arguably one of the most promising mechanism for a compliant, shape changing wing was developed by Campanile and Hanselka \cite{Campanile1997-1} in 1997. It is based on a closed, flexible skin whose upper and lower surfaces are connected via spokes. The latter divide the wing into prismatic cells with a quadrilateral cross section that possess a single degree of freedom. The stiffness, position and angle of each spoke is chosen such that the desired wing shape is obtained from the accumulated shear deformations of each cell due to one or more internally placed actuators. This concept was taken up by Kota and Hetrick \cite{Kota2003-1} in 2003 for the development of a gapless trailing edge. Their approach differs from previous work only with regards to the actuation principle. Relative movements between the upper and lower trailing edge skins are generated with the help of one or two linear actuators that are placed within the wing box. Such a trailing edge was tested in a modified Gulfstream III that underwent a series of successful flight tests in 2014. These tests indicated potential fuel savings of up to 10\% and noise reductions during landings of up to 40\%. Although not directly related to high lift devices, a similar mechanism for grippers was patented by Bannasch and Kniese \cite{Bannasch2007-1} in 2007.


    \newpage

    \sectionmark{Summary}
    \begin{framed}
        \noindent \textbf{Summary}\\

        \noindent The airfoil shaped cross sections of rigid wings can be either optimized for cruise or takeoff and landing. The former leads to fast airplanes that can cover long distances whereas the latter leads to relatively slow airplanes that can takeoff and land on short runways. Wings that combine the advantages of both designs can be realized with the help of high lift devices. These devices consist of one or more aerodynamically shaped rigid bodies that are tightly integrated into the wing during cruise and deployed via various kind of mechanisms during takeoff and landing. For example, the deployed high lift devices of a Boeing 747-400 increase the wing area by 21\% and the lift by 90\%.\\

        Rigid body based high lift devices cause gaps in a wings surface that increase the fuel consumption, noise emissions and radar reflections. This led to the development of gapless high lift devices that consist of a flexible, continuous skin that is, depending on the current flight regime, deformed via various kind of mechanisms. However, despite a massive research effort, gapless high lift devices are relatively heavy so that they are currently not used in passenger or military airplanes.\\

        \noindent \textbf{Conclusion}\\

        \noindent The most promising approach towards gapless high lift devices so far is based on a closed, flexible skin whose upper and lower surfaces are connected via spokes. The latter divide the wing into prismatic cells with a quadrilateral cross section that possess a single degree of freedom. The stiffness, position and angle of each spoke is chosen such that the desired wing shape is obtained from the accumulated shear deformations of each cell due to one or more internally placed actuators.\\

        It was argued in previous chapters that some of the lightest and most efficient structures and actuators are based on differential pressures. The use of pressurized cells with individually tailored geometries that possess more than one degree of freedom thus seems to be an interesting approach.
    \end{framed} 
        \cleardoublepage
    \chapter{Flexible Airplane Skins}
        Current high lift devices of passenger and military airplanes consist of aerodynamically shaped rigid bodies that are deployed and retracted via various kind of mechanisms. Despite being inexpensive and reliable, their inherent surface gaps are a bottleneck that limits potential noise reductions, fuel savings and stealth properties of future airplanes. This limitation led to the development of various approaches towards gapless high lift devices that can be incorporated into existing wings. The elimination of surface gaps requires skins with soft deformation modes whose equilibrium shapes are fully defined by the displacements of a small number of interface points that are connected to the underlying mechanisms. Unfortunately, the available literature in this field \cite{Thill2008-1} focuses mostly on rigid skins with generically designed deformation modes. Some general approaches for the design of anisotropic skins with a constant and variable in-plane and bending stiffness are subsequently reviewed.


        \section{Tailored Poisson's Ratio}
            The complexity of an airplane skin depends to a large degree on its Gaussian curvature \cite{Gauss1902-1}. Wings with a constant chord length are widely used in small airplanes. This is a reasonable tradeoff between their manufacturing costs and aerodynamic efficiency. In contrast, wings with a double tapered chord are often used in large passenger airplanes \cite{Grant2007-1}. However, the variation of their outer taper is usually small so that the skins of most high lift devices can be considered to be developable.\\

            Developable surfaces with an infinitesimally small thickness have the benefit that their in-plane and bending strains are decoupled. However, skins of gapless high lift devices possess a finite thickness. This leads to a coupling between in-plane and bending strains if they are made from a material with a non-zero Poisson's ratio. This coupling either causes anticlastic curvatures or, if out-of-plane deformations are constrained, large in-plane stresses and support forces. Hence it is advantageous to use skin materials with a Poisson's ratio that is very close to zero. Various design approaches for materials with a zero or negative Poisson's ratio\footnote{Auxetic is another commonly used term to designate materials with a negative Poisson's ratio. It was coined by Ken Evans and stems from the Greek word \textit{auxesis} which means ``increase".} are subsequently reviewed. It is shown that a zero or negative Poisson's ratio can be introduced on a material and geometric level.


            \subsection{Material Level}
                Negative Poisson's ratios in single crystals were already observed by Voigt \cite{Voigt1966-1} in 1920. Many natural and biological materials that exhibit this property were discovered since then \cite{Lim2015-1}. It is well known from the linear theory of elasticity that the Poisson's ratio of isotropic materials is bounded \cite{Greaves2011-1} between -1 and 0.5. In contrast, the theoretically unbounded Poisson's ratio of anisotropic materials \cite{Ting2005-1} enables the design of fiber reinforced composites with extreme properties.\\

                Fiber reinforced composites \cite{Evans2004-1} with a negative Poisson's ratio in the in-plane direction were studied by Tsai and Hahn \cite{Tsai1980-1} in 1980 and in the out-of-plane direction by Herakovich \cite{Herakovich1984-1} in 1984. The achievable auxeticity of conventional laminates depends on the stacking sequence, fiber volume fraction and fiber stiffness. However, the negative Poisson's ratios that were realized thus far are moderate. For example, Miki and Morotsu \cite{Miki1989-1} manufactured an unbalanced, bi-directional laminate with a negative in-plane Poisson's ratio of -0.4 in 1989 and Hine et al \cite{Hine1997-1} manufactured an angle-ply laminate with a negative out-of-plane ratio of -0.5 in 1997.\\

                \marginnote{
                \begin{center}
                    \includegraphics[width=\marginparwidth]{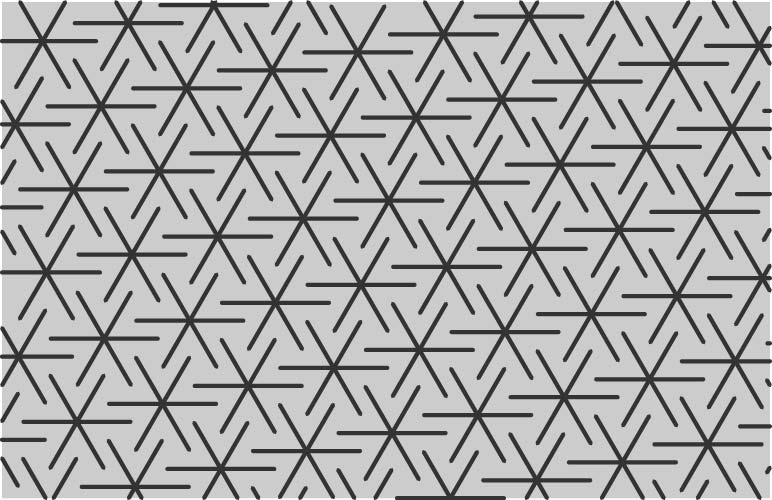}
                    \small(a)\vspace{7mm}
                    \includegraphics[width=\marginparwidth]{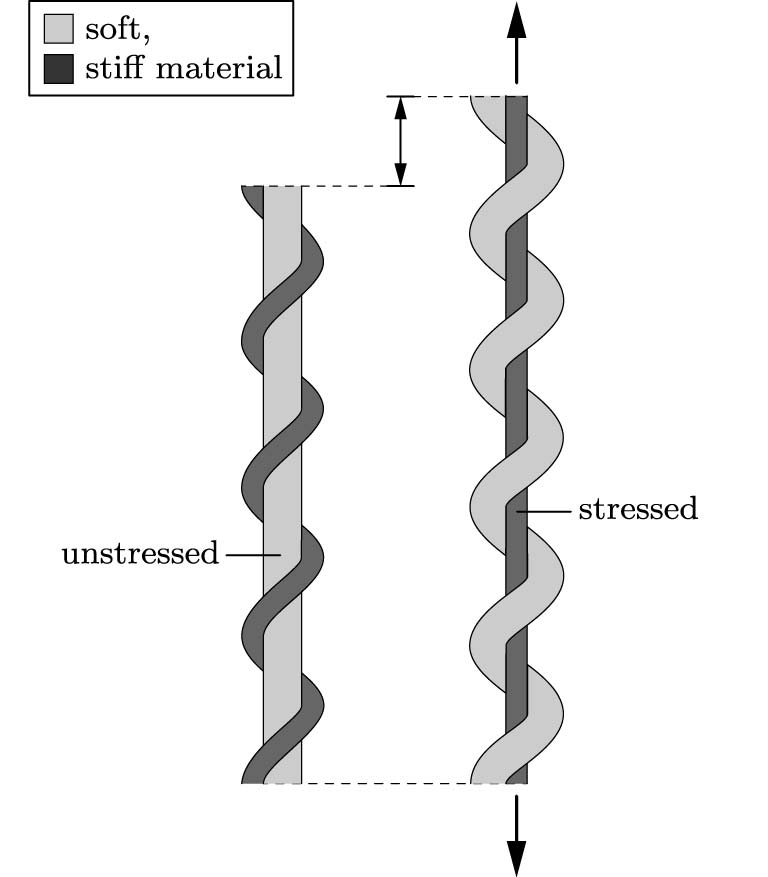}
                    \small(b)
                    \captionof{figure}{(a) The Poisson's ratio of a composite matrix can be reduced by rigid, star shaped inclusions. (b) An auxetic fiber that consists of a helical and a straight thread. The helical thread is relatively stiff and thin whereas the straight thread is soft and thick in the unstressed configuration. The fiber thickness increases in the stressed configuration as the helical thread becomes straight and the straight thread becomes helical.}
                    \label{pic:Figure_5_1}
                \end{center}}[-120mm]

                Composites with a considerably smaller Poisson's ratio can be created with the help of auxetic fibers and/or an auxetic matrix \cite{Nkansah1993-1}. Milton \cite{Milton1992-1} showed in 1992 that rigid inclusions can be used to lower the Poisson's ratio of a composite matrix. However, these inclusions need to be numerous and carefully placed as shown in Figure~\ref{pic:Figure_5_1}. Furthermore, they might have a detrimental effect on other composite properties so that they are currently not used.\\

                The fiber volume fraction of round fibers is usually between 50-65\% whereas the maximum achievable fraction is about 90\%. As a consequence, due to the rule of mixtures, the gain from using auxetic fibers clearly exceeds that of an auxetic matrix. It was found by Caddock and Evans \cite{Caddock1989-1} in 1989 that the Poisson's ratio of polymers can be considerably lowered by a three stage process that consists of compaction, sintering and extrusion. Polypropylene fibers with a Poisson's ratio of -0.6 and 0.3 that are otherwise identical were used in various composite structures by Alderson et al \cite{Alderson2002-1} in 2002. They found that the pull-out force of auxetic fibers is twice as large as those of conventional fibers \cite{Simkins2005-1}.\\

                A fiber with a remarkably small Poisson's ratio was studied by Miller et al \cite{Miller2009-1} in 2006. The unstressed fiber consists of a straight thread that is soft and thick and a helical thread around it that is relatively stiff and thin. The fiber thickness increases upon loading as the helical thread becomes straight and the straight thread becomes helical. This geometric effect led to a Poisson's ratio of -2.1 during first tests. An auxetic knit structure with a similar mechanism was subsequently developed by Ugbolue et al \cite{Ugbolue2007-1} in 2007.


        \subsection{Geometric Level}
            Evans et al \cite{Evans2000-1} pointed out that structures with a single, soft deformation mode and a Poisson's ratio of -1 were already developed during the 1950s for the graphite cores of nuclear reactors \cite{Poulter1963-1}. Each core consisted of about 44,000 rigid bricks that were connected at their corners and sides via sliding joints so that they could sustain large earthquakes and non-uniform temperature distributions. However, it took another three decades until structures with a geometrically induced negative Poisson's ratio were systematically investigated. Closed form expressions for the material properties of regular, compliant honeycombs were developed by Gibson and Ashby \cite{Gibson1982-1} in 1982. They showed that the material properties of a honeycomb cell can be tailored within a wide range by varying its geometry. A honeycomb with reentrant cell corners that possesses a small Poisson's ratio is shown in Figure~\ref{pic:Figure_5_2}. Various cellular structures with a negative or zero Poisson's ratio that were mostly found by trial and error were published since then. Two approaches that can be used to systematically design these structures are subsequently discussed.\\

            \marginnote{
                \begin{center}
                    \includegraphics[width=\marginparwidth]{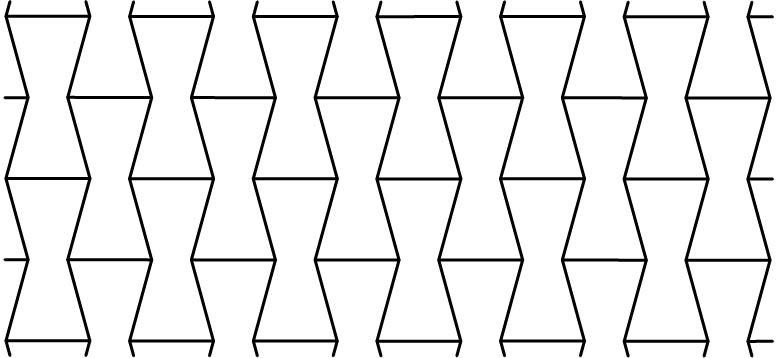}
                    \small(a)\vspace{2mm}
                    \includegraphics[width=\marginparwidth]{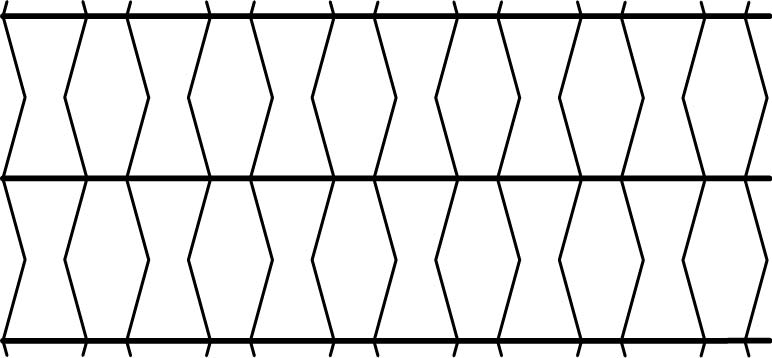}
                    \small(b)\vspace{0mm}
                    \includegraphics[width=\marginparwidth]{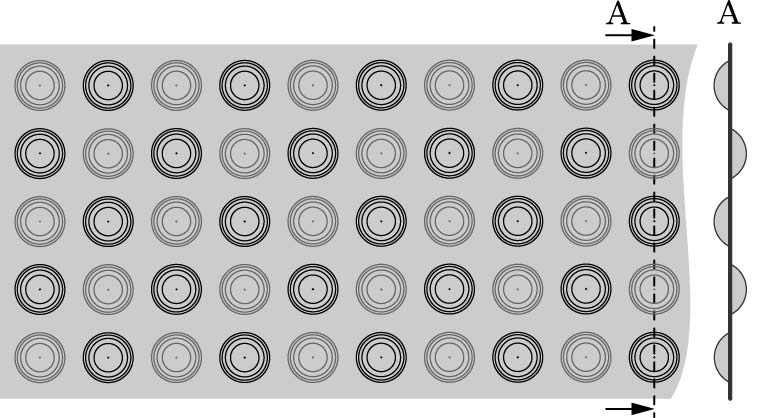}
                    \small(c)\vspace{2mm}
                    \includegraphics[width=\marginparwidth]{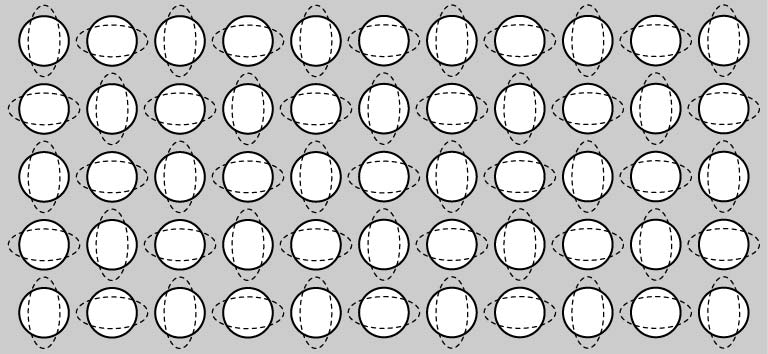}
                    \small(d)
                    \captionof{figure}{Different structures with a negative or zero Poisson's ratio. (a) Reentrant honeycomb and (b) a variation with a large, uniaxial in-plane and bending stiffness. (c) Sheet with spherical dimples that are arranged on both sides in a checkerboard pattern. (d) Sheet with a lattice of circular holes and superimposed deformation modes.}
                    \label{pic:Figure_5_2}
                \end{center}}[-141mm]

                Linear topology optimization was used by Sigmund in 1994 for the design of cellular structures with desired material properties in two- \cite{Sigmund1994-1} and three-dimensions \cite{Sigmund1995-1}. A corresponding nonlinear approach for finite deformations was published by Clausen et al \cite{Clausen2015-1} in 2015. Various structures with Poisson's ratios between $\pm 0.8$ were designed and manufactured. Experiments confirmed a nearly constant Poisson's ratio for strains of up to 20\%.\\

                Cellular structures with a regular, two-dimensional geometry are ultimately based on the triangular, rectangular or hexagonal tiling of a plane. K\"orner and Liebold-Ribeiro \cite{Koerner2015-1} showed in 2015 that the simple eigenvector analysis of a unit cell with straight sides is sufficient to find a large number of auxetic structures. Furthermore, they showed that this approach can be generalized to three-dimensions and semiregular structures. This enables a systematic classification of previously published results. For example, the eighth eigenvector of a triangular unit cell has close similarities with the chiral honeycomb that was published by Prall and Lakes \cite{Prall1997-1} in 1997.\\

                A completely different approach towards auxetic structures was developed by Golabchi and Guest \cite{Golabchi2009-1} in 2009. They observed that the positive Poisson's ratio of a homogeneous and isotropic sheet can be lowered by regularly denting its surface on both sides in a checkerboard pattern. This effect, that stems from the out-of-plane deformations of its dimples, was rediscovered by Javid et al \cite{Javid2015-1} in 2015. A similar effect that is driven by elastic instabilities was discovered by Bertoldi et al \cite{Bertoldi2010-1} in 2010. They found that the otherwise positive Poisson's ratio of an elastic sheet with regularly placed holes can become negative under compression loads. This work was extended to three-dimensions by Babaee et al \cite{Babaee2013-1} in 2013. However, the nonlinear relationship between the applied loads and the effective material properties limits the usefulness of these approaches.


        \section{Leading Edge Skins}
            An airplanes leading edge skin needs to protect the wing against erosion, icing, lightning- and bird strikes. For example, commercial passenger jets are hit about once a year by a lightning strike \cite{Gagne2014-1}. In addition, more than 65,000 bird strikes that led to nine human fatalities were reported \cite{Cleary2006-1} in the United States during a fifteen year period. The design of lightweight skins that can safely operate under these conditions is difficult. Leading edge skins of modern passenger airplanes such as the Boeing 787 are made from carbon fiber metal laminates. Their deicing system is based on an integrated metal layer that is sandwiched between two insulating glass fiber fabrics as illustrated in Figure \ref{pic:Figure_5_3}.\\

            \marginnote{
            \begin{center}
                \includegraphics[width=\marginparwidth]{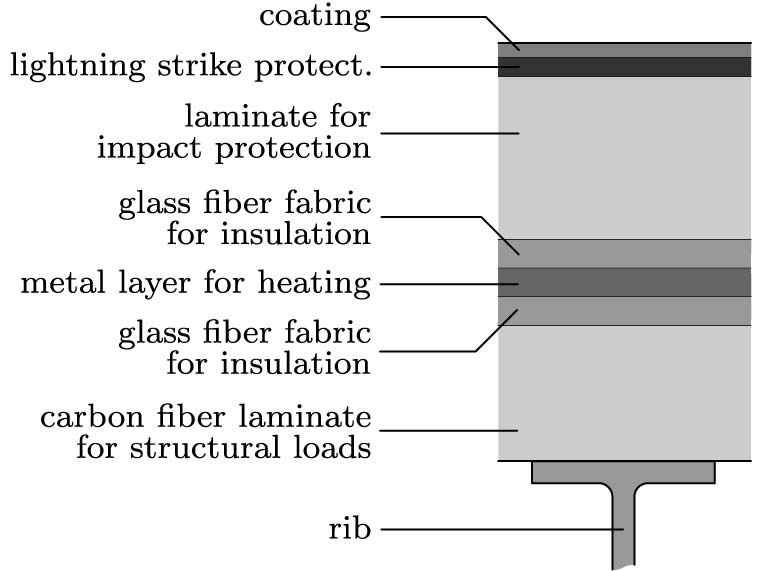}
                \small(a)\vspace{5mm}
                \includegraphics[width=\marginparwidth]{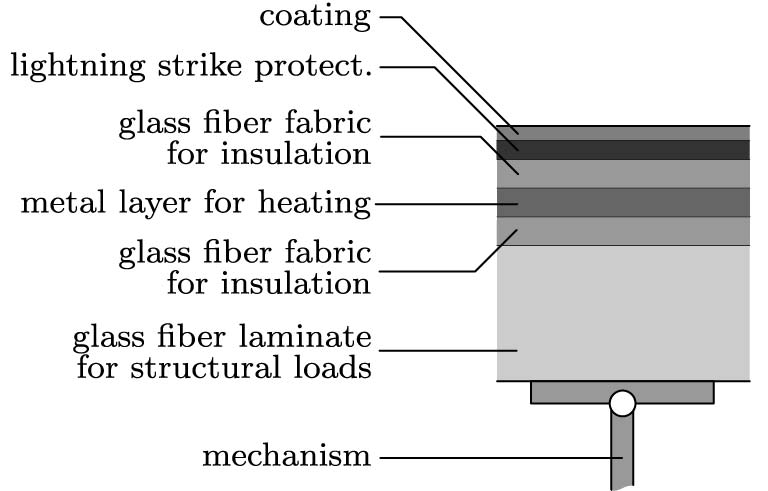}
                \small(b)\vspace{5mm}
                \includegraphics[width=\marginparwidth]{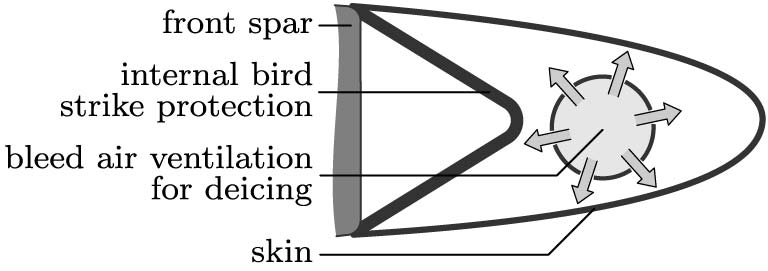}
                \small(c)
                \captionof{figure}{(a) Impact resistive leading edge skin of a Boeing 787. It consists of a carbon fiber metal laminate and a deicing system that is sandwiched between two insulating glass fiber fabrics. (b) A similar approach for a shape changing, non-impact resistive leading edge skin that consists of a glass fiber metal laminate and a deicing system. (c) Internally placed impact protection and deicing system of a gapless leading edge.}
                \label{pic:Figure_5_3}
            \end{center}}[-145mm]

            Gapless leading edge skins need to be stiff enough to carry external loads and soft enough to undergo large shape changes. These conflicting demands can be relaxed by using mechanisms that preserve their arc-length in chordwise direction. An optimal skin thus possesses a large in-plane stiffness and a small, large bending stiffness in chordwise, spanwise direction, respectively. In order to achieve such an anisotropy it is advantageous to remove the impact protection and deicing system from the skin. The front spar, fuel reservoir of a wing can be protected by a secondary structure against bird strikes \cite{Chary2016-1} and the deicing can be realized by a ventilation system that circulates hot bleed air \cite{Goraj2004-1} inside the leading edge.\\

            A relatively thin skin with a tailored, varying stiffness in chordwise direction that breaches upon impact was investigated by Kintscher et al \cite{Kintscher2016-1} in 2009. It consists of an anisotropic glass fiber laminate that is reinforced by stringers in spanwise direction. The latter increase the anisotropy of the skins bending stiffness and serve as attachment points for the underlying mechanism. An integrated deicing system for such a skin was investigated by Geier et al \cite{Geier2016-1}. They found that the outer bending strains can be reduced if a thin rubber layer with a low shear stiffness is placed between the glass fiber laminate and the deicing system. The use of a rubber layer has the additional benefit that it acts as an insulator that decreases the energy consumption during deicing and increases, although only slightly, the impact resistance.


        \section{Trailing Edge Skins}
            Skins for shape changing leading- and trailing edges differ insofar that the latter need to additionally undergo large in-plane strains in chordwise direction. These strains, that are caused by the discontinuity of the trailing edge, have a severe impact on the overall design complexity. However, trailing edge skins do not require a deicing- or impact protection system. Furthermore, they are less susceptible to erosion.\\

            Leading edge skins usually consist of thin laminates that are based on low modulus fibers. However, these kind of laminates can not undergo the large in-plane strains in chordwise direction that are required by trailing edge skins. As a consequence, the anisotropy of trailing edge skins needs to be even more extreme so that their design has many similarities with the development of rigid body mechanisms.


            \subsection{Sandwich Structure}

                \marginnote{
                \begin{center}
                    \includegraphics[width=\marginparwidth]{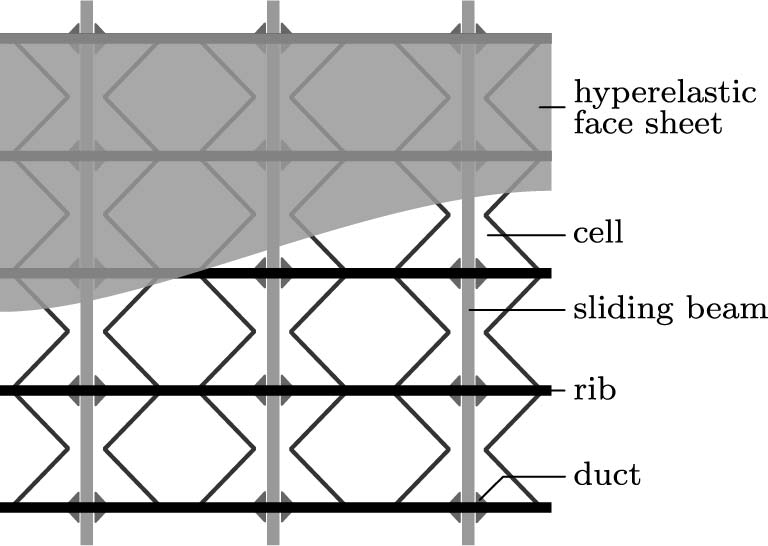}
                    \captionof{figure}{Top view of a trailing edge skin that consists of a nonregular cell core and hyperelastic face sheets with a zero Poisson's ratio. Additional sliding beams can be incorporated into the core to vary the bending stiffness in chordwise direction.}
                    \label{pic:Figure_5_4}
                \end{center}}[-18mm]

                Some promising approaches towards trailing edge skins are based on sandwich structures. For example, Olympio and Gandhi \cite{Olympio2007-1} proposed the use of a nonregular cell core with a zero Poisson's ratio that is covered by hyperelastic face sheets as shown in Figure~\ref{pic:Figure_5_4}. It can be seen that the face sheets are only connected to the straight cell sides or ribs in spanwise direction since the accordion shaped sides in chordwise direction need to undergo incompatible motions during shape changes.\\

                This approach has the drawback that additional supports in chordwise direction are only provided to the face sheet on the pressure side of the trailing edge via contact forces. Another problem is the cores relatively small bending stiffness. Bubert et al \cite{Bubert2010-1} demonstrated that this problem can be overcome with the help of guided beams. However, the required sliding mechanisms have a detrimental impact on the skins overall weight, complexity and maintenance requirements.


            \subsection{Corrugated Shell}
                Other design approaches for trailing edge skins are based on corrugated shells that possess, due to their wavy geometry, a large anisotropy with regards to their in-plane and bending stiffness. Trailing edge skins that are solely based on corrugated shells were first investigated by Yokozeki et al \cite{Yokozeki2006-1} in 2006. A large number of different geometries and various combinations of shells as shown in Figure~\ref{pic:Figure_5_5} were investigated since then. For example, the use of two symmetrically stacked shells was studied by Thill et al \cite{Thill2010-1} in 2007. This arrangement decouples the in-plane and bending deformations and leads to closed cells that possess a large torsional stiffness. A similar approach that is based on a nested combination of two corrugated shells was proposed by Previtali et al \cite{Previtali2014-1} in 2014. This arrangement increases the outer thickness and thus the bending stiffness in spanwise direction while it preserves the large torsional stiffness and the decoupling between in-plane an bending deformations.\\

                \marginnote{
                \begin{center}
                    \includegraphics[width=\marginparwidth]{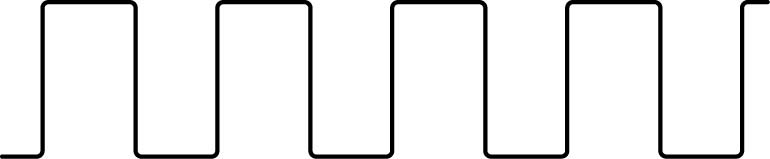}
                    \small(a)\vspace{3mm}
                    \includegraphics[width=\marginparwidth]{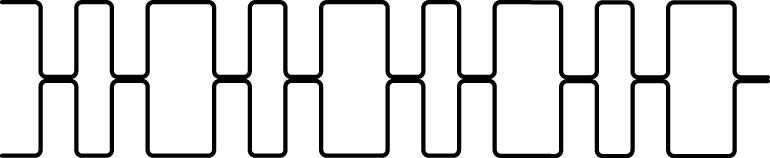}
                    \small(b)\vspace{3mm}
                    \includegraphics[width=\marginparwidth]{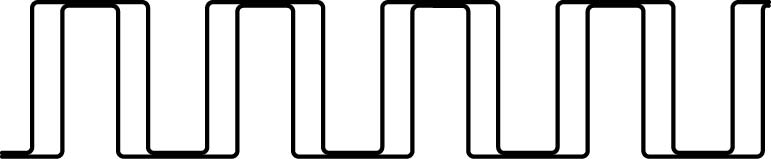}
                    \small(c)
                    \captionof{figure}{Side views of corrugated trailing edge skins. (a) A single, (b) symmetrically stacked and (c) nested shell.}
                    \label{pic:Figure_5_5}
                \end{center}}[-70mm]

                \marginnote{
                \begin{center}
                    \includegraphics[width=\marginparwidth]{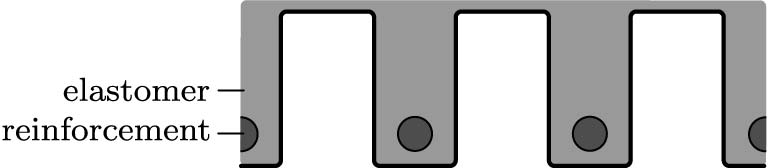}
                    \small(a)\vspace{3mm}
                    \includegraphics[width=\marginparwidth]{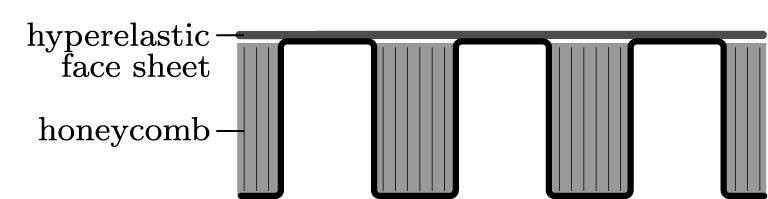}
                    \small(b)\vspace{3mm}
                    \includegraphics[width=\marginparwidth]{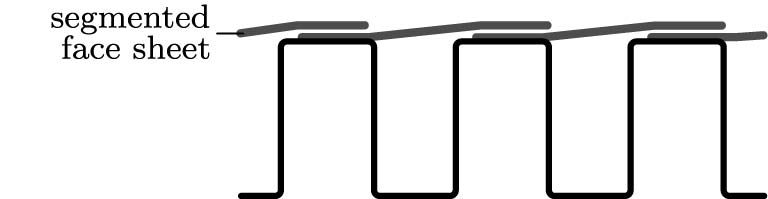}
                    \small(c)
                    \captionof{figure}{Smooth shell surfaces. (a) Elastomer filled corrugations with additional reinforcements. (b) Elastic face sheet that is supported by honeycomb cores against out-of-plane deformations. (c) Segmented face sheet.}
                    \label{pic:Figure_5_6}
                \end{center}}[21mm]

                The use of corrugated skins is, at least from an aerodynamic point of view, disadvantageous. Their wavy surfaces can be smoothed by filling one side of the shells with an elastomer that might or might not be reinforced in spanwise direction as illustrated in Figure~\ref{pic:Figure_5_6}. Unfortunately, the weight penalty of such an approach is usually prohibitive. Airoldi et al \cite{Airoldi2017-1} proposed the incorporation of lightweight honeycomb cores with a zero Poisson's ratio that act as fillers between the corrugations and thin, elastic face sheets. However, as for the previously introduced sandwich structures, it is not possible to directly connect the face sheet to the honeycombs due to different relative motions. Thill et al \cite{Thill2010-1} avoided these problems by using segmented face sheets instead. Two neighboring segments can be connected via simple sliding mechanisms to reduce out-of-plane deformations. This approach is widely used in nature and can be found, for example, in shark skins and bird wings \cite{Bechert2000-1}.


        \section{Variable Stiffness Skins}
            The carefully tailored deformation modes of otherwise stiff skins enable large shape changes. However, their deformations can only be controlled by the underlying mechanisms at a few locations so that large, undesired displacements can generally not be avoided. As a consequence, it is either necessary to tightly support the skins with complex mechanisms or to actively control their stiffness. The weight penalty of closely spaced supports is usually prohibitive so that different approaches for variable stiffness structures were developed. Some of the most promising approaches that are based on a variation of the material or geometric properties of substructures or the friction forces between them are subsequently reviewed.


            \subsection{Material Properties}
                Feedback controlled actuators that utilize force and position sensors as well as microprocessors \cite{Salisbury1980-1} can be used to create programmable structural elements that can switch between different nonlinear stiffness functions, Figure~\ref{pic:Figure_5_7}. Programmable elements are widely used in robotics as they enable a fast, software centered development. Furthermore, they are currently investigated \cite{Weidner2018-1} for the compensation of wind induced deformations in extremely lightweight and slender buildings. The complexity of a programmable element can be significantly reduced if it only needs to switch between two constant stiffness values. Such an element that functions without sensors, microprocessors and complex algorithms was developed by Kobori et al \cite{Kobori1993-1} in 1993. It consists of a double acting, hydraulic cylinder whose ports are connected via a valve. A closed valve leads to a large stiffness that is only limited by the compressibility of the fluid and the elasticity of the cylinder. In contrast, an open valve enables a flow between both ports so that the elements stiffness becomes negligible. This relatively simple and robust approach has a fast response time and requires only a small amount of energy. Hence, it is widely used in buildings that need to vary their stiffness during earthquakes. However, the one-dimensional nature of these elements renders them unsuitable for an application in airplane skins.\\

                \marginnote{
                \begin{center}
                    \includegraphics[width=\marginparwidth]{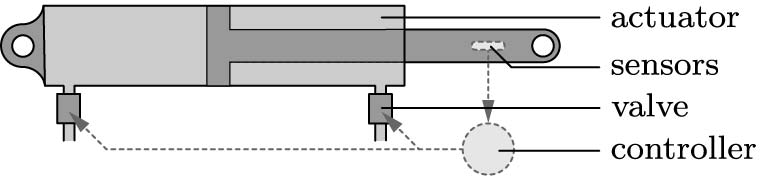}
                    \small(a)\vspace{2mm}
                    \includegraphics[width=\marginparwidth]{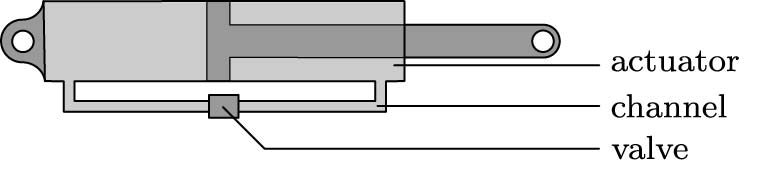}
                    \small(b)\vspace{2mm}
                    \includegraphics[width=\marginparwidth]{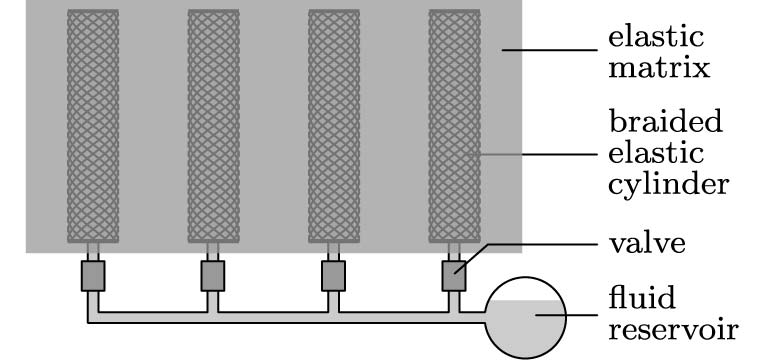}
                    \small(c)\vspace{2mm}
                    \includegraphics[width=\marginparwidth]{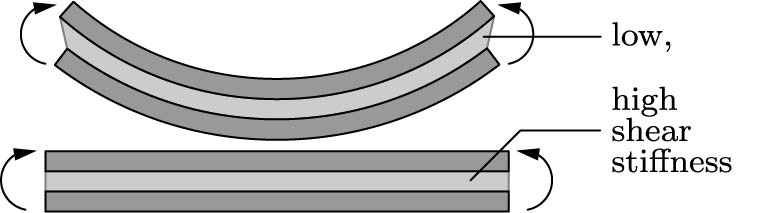}
                    \small(d)\vspace{2mm}
                    \includegraphics[width=\marginparwidth]{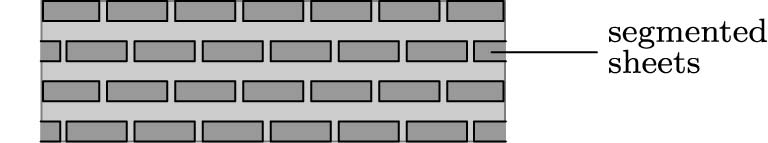}
                    \small(e)
                    \captionof{figure}{Variable stiffness structures. (a) Feedback controlled actuator that can switch between arbitrary force displacement relationships. (b) Simplified actuator with connected ports that can switch between two constant stiffness values. (c) Similar approach that is based on a number of compliant, pneumatic muscles that are embedded in an elastic matrix. (d) Sandwich structure with continuous face sheets and a variable stiffness core. (e) Sandwich structure with a large number of layers that alternately consist of hexagonal platelets and a variable stiffness polymer.}
                    \label{pic:Figure_5_7}
                \end{center}}[-95mm]

                A plate like element with a variable stiffness that is based on a similar approach was proposed by Shan et al \cite{Shan2009-1} in 2009. It consists of a large number of pneumatic muscles that are embedded in an elastic matrix. These actuators can be arranged in several layers and their orientations might vary between layers or even within a single layer. Actuators can be combined into groups of varying sizes such that their in- and outflow is controlled by a single valve. This reduces the control complexity and enables tailored stiffness variations within a plate.\\

                A sandwich structure with a variable bending stiffness was proposed by Martin et al \cite{Martin1998-1} in 1998. It consists of a thermoplastic polymer core\footnote{High melting thermoplastics such as the semi-crystalline polyether ether ketones (PEEK) and the amorphous polyether sulfones (PES) are commonly used. The shear modulus of a thermoplastic polymer can decrease by up to two orders of magnitude if it is heated above its glass transition temperature.} and continuous, flexible face sheets. The core material is chosen such that its glass transition temperature is well above the structures operating temperatures. The nonlinear relationship between the polymers temperature and shear modulus around the glass transition temperature is exploited to vary the bending stiffness of the sandwich. This can be done by using heating and/or cooling elements that are embedded in the sandwich core. The performance of these structures is limited by the continuous face sheets that allow only developable deformations. Inspired by bio-composites \cite{Wegst2015-1,Ji2004-1} such as nacre and dentin, McKnight et al \cite{McKnight2004-1} overcame this problem by using a large number of layers that alternately consist of a thermoplastic polymer and hexagonal platelets.\\

                The response times of these structures is limited by the heat capacity and thermal conductivity of their polymer layers. Furthermore, the use of heating and/or cooling elements is energy intensive. Alternative materials \cite{Kuder2013-1} with a variable stiffness that improve upon some of these aspects include phase transition, magnetorheological and electrorheological materials. For example, shape memory alloys such as Ni-Ti can undergo a solid-solid phase transition. The Young's modulus of their high temperature phase is about 80~GPa whereas their low temperature modulus is about 30~GPa.

                \marginnote{
                \begin{center}
                    \includegraphics[width=\marginparwidth]{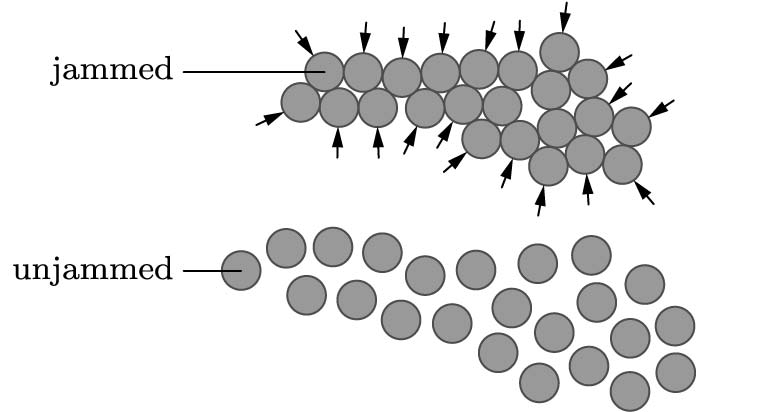}
                    \small(a)\vspace{5mm}
                    \includegraphics[width=\marginparwidth]{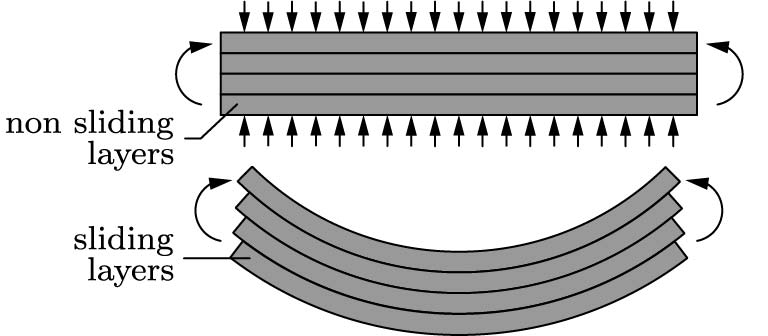}
                    \small(b)\vspace{5mm}
                    \includegraphics[width=\marginparwidth]{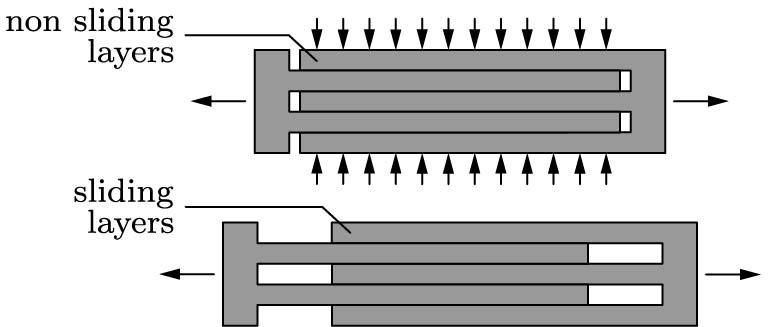}
                    \small(c)
                    \captionof{figure}{(a) Jamming induced stiffness variation of a granular material. Friction induced stiffness variations of a layered (b) beam and (c) bar.}
                    \label{pic:Figure_5_8}
                \end{center}}[1mm]


            \subsection{Friction Forces}
                Granular materials can exhibit properties of fluids and solids \cite{Liu1998-1}. The dominant phase depends mainly on their loading, temperature, density and geometric, material properties. For example, the vacuum packaging of coffee beans that is done since the early 1900s leads to rigid bodies whereas the unpacked beans can flow from a storage tank into the grinder of a coffee machine, Figure~\ref{pic:Figure_5_8}. This effect was exploited by Suits \cite{Suits1952-1} in 1952 for portable shelters and by Campanaro et al \cite{Campanaro1962-1} in 1962 for shapeable structures.\\

                Granular materials possess a large number of degrees of freedom so that they can be formed into a wide range of shapes. However, this is usually neither required nor desirable so that simpler systems with only a few degrees of freedom are used instead. For example, rotary clutches possess only a single degree of freedom. Their torsional stiffness is commonly varied by changing the normal force induced friction forces between two layers. These normal forces can be generated with the help of pressure, mass inertia, electromagnetism or electric charges.\\

                A similar approach can be used for layered bars, beams and plates. The available literature in this field differs mainly with regards to the layer geometry and normal force generation. For example, Tabata et al \cite{Tabata2001-1} proposed the use of one-dimensionally stacked plates with patterned electrodes in 2001. Electrostatic forces between the plates are generated by alternately connecting them to a high voltage source. The use of two-dimensionally stacked beams that enable biaxial stiffness variations and vacuum induced pressure forces was subsequently proposed by Kawamura et al \cite{Kawamura2002-1} in 2002. It was pointed out by Kim et al \cite{Kim2012-1} in 2012 that the previous approach can be easily extended towards beams with a varying in-plane and bending stiffness.


            \subsection{Geometry}
                The second moment of area of a layered structure is indirectly affected by the friction forces between its substructures. Instead of using friction forces, it is possible to directly vary a structures geometric properties by applying translations or rotations to its otherwise rigidly connected substructures. For example, Griffin et al \cite{Griffin1998-1} filed a patent in 1998 where they describe a rotating, elliptic trailing edge spar that varies a wings second moment of area, Figure~\ref{pic:Figure_5_9}.
                \marginnote{
                \begin{center}
                    \includegraphics[width=\marginparwidth]{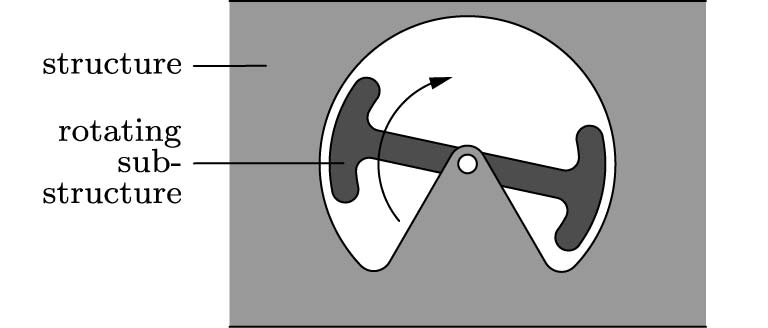}
                    \captionof{figure}{Overall stiffness is varied by the rotation of a substructure.}
                    \label{pic:Figure_5_9}
                \end{center}}[-33mm]


    \newpage

    \sectionmark{Summary}
    \begin{framed}
        \noindent \textbf{Summary}\\

        \noindent Gapless high lift devices need to withstand significant external loads while being capable of large shape changes. These conflicting demands can only be satisfied by stiff skins with specifically tailored, soft deformation modes in chordwise direction. Leading edge skins usually require a relatively soft bending mode and trailing edge skins an additionally soft in-plane mode.\\

        Skin deformations during shape changes are relatively large so that parasitic stresses or anticlastic curvatures due to a non-zero Poisson's ratio are problematic. Leading edge skins with a highly anisotropic bending stiffness and a zero Poisson's ratio are commonly based on fiber reinforced composites with tailored material properties. The design of trailing edge skins is, due to their additionally large in-plane deformations, much harder. Promising approaches in this regard are either based on sandwich structures with hyperelastic face sheets and cell cores or corrugated shells.\\

        The soft deformation modes of gapless high lift devices can only be controlled by closely spaced supports or, alternatively, skins with an adjustable stiffness. The former approach leads to mechanisms that are often prohibitively heavy whereas the latter is not a viable option with existing technologies. Furthermore, gapless leading edges require, due to their relatively thin skins, additional structures that protect their front spar from impacts.\\

        \noindent \textbf{Conclusion}\\

        \noindent Stiff skins with a highly anisotropic bending stiffness and a zero Poisson's ratio can be easily manufactured. This is in great contrast to skins that additionally need to undergo large in-plane deformations. However, their use can be avoided by changing a relatively large portion of the airfoil around the leading edge instead of a small, isolated portion around the discontinuous trailing edge.\\

        Another major problem is the need for closely spaced skin supports as it often leads to heavy mechanisms. A potential remedy is the use of mechanisms that are based on pressurized cellular structures with individually tailored cell geometries. Their relatively large, pressure induced stiffness might be even sufficient to withstand required impact loads. These kind of structures are widely used in the plant kingdom so that a thorough study of their working principles is advantageous.
    \end{framed} 
        \cleardoublepage
    \chapter{Plant Movements}

        \marginnote{
        \begin{center}
            \includegraphics[width=\marginparwidth]{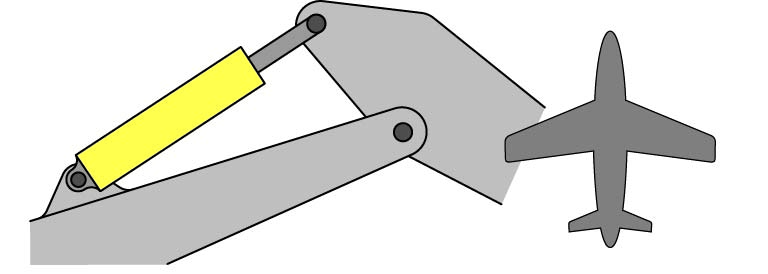}
            \small(a) rigid body actuator and mechanism\vspace{4mm}
            \includegraphics[width=\marginparwidth]{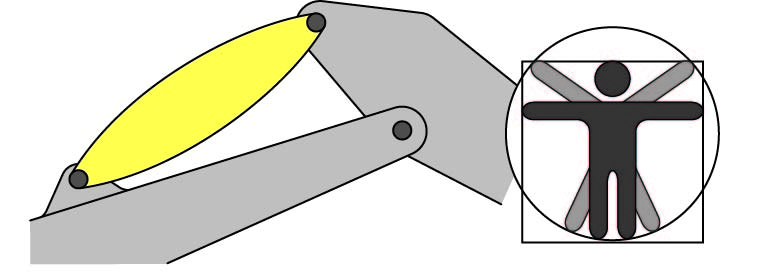}
            \small(b) compliant actuator and rigid body mechanism\vspace{4mm}
            \includegraphics[width=\marginparwidth]{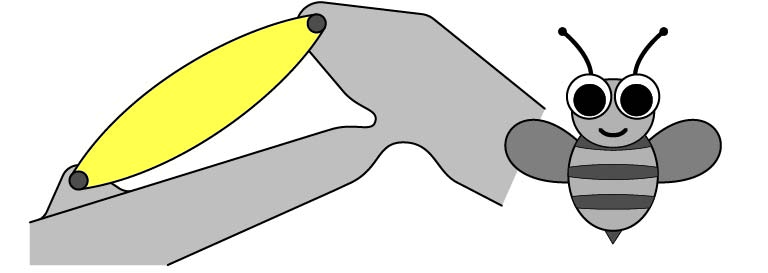}
            \small(c) compliant actuator and mechanism\vspace{0mm}
            \includegraphics[width=\marginparwidth]{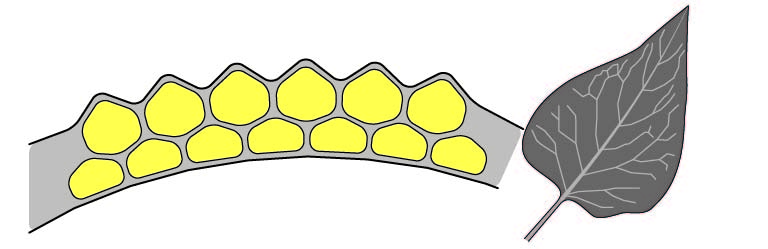}
            \small(d) complete fusion of actuator and mechanism\vspace{0mm}
            \captionof{figure}{Implementation of a single degree of freedom mechanism and actuator in (a) heavy machinery, (b) humans (c) bees and (d) plants.}
            \label{pic:Figure_6_1}
        \end{center}}[-28mm]

        Despite their clear advantages, gapless high lift devices can not be found in the latest generation of commercial or military airplanes. Their lack of success can be explained by the weight penalty that mostly stems from an insufficient integration of actuators, mechanisms and structures. Inspiration for new, potentially superior approaches with different integration levels can be found in nature \cite{Ball2001-1}. For example, humans tend to build heavy machinery from rigid bodies, frictionless hinges and hydraulic cylinders as illustrated in Figure~\ref{pic:Figure_6_1}. A similar approach is used in humans and large animals where the hydraulic cylinders are substituted by compliant muscles. This approach is further refined in smaller animals such as bees where the segmented joints are substituted by compliant hinges. Nonetheless, their integration level is still low in comparison to plants \cite{Guo2015-1}. They utilize a large number of compliant, fluid filled cells so that their overall shape is highly coupled to the individual cell geometries and pressures. It is thus not possible to clearly distinguish between their actuators, mechanisms and structures. Furthermore, unlike humans and animals, plants do not possess a central control system although some species are capable of large, reversible movements. If well understood, their working principle might enable the creation of gapless high lift devices with a reduced part count, weight and complexity.\\

        Depending on the species and intended purpose, the complexity and speed of plant movements can vary significantly. A comprehensive selection of automatic and stimulated movements is subsequently provided. It is pointed out that the interactions between cell geometries, cell materials, cytoskeletons and osmotic hydration motors are far-reaching and well beyond the relatively simple plant movements.


        \section{Automatic Movements}
            Plants are bound to the place where they germinate. Nonetheless, they are far from being immobile as they can adapt to their environment by changing the shape of their roots, stems and leaves. These movements are either automatically initiated or triggered by directional or non-directional external stimuli. Fully automatic movements are commonly used by plants to disperse their pollen and seeds or to find mechanical support \cite{Stolarz2009-1}. Their speed varies between the slow growth of shoots and roots to the explosive rupture of seed pods. Automatic movements are either passive or active. Active movements depend on the plant metabolism whereas passive movements are driven by external energy sources such as temperature and humidity gradients. Furthermore, the energy can be continuously translated into movements or stored inside the plant to enable rapid, discontinuous shape changes.


            \marginnote{
            \begin{center}
                \includegraphics[width=\marginparwidth]{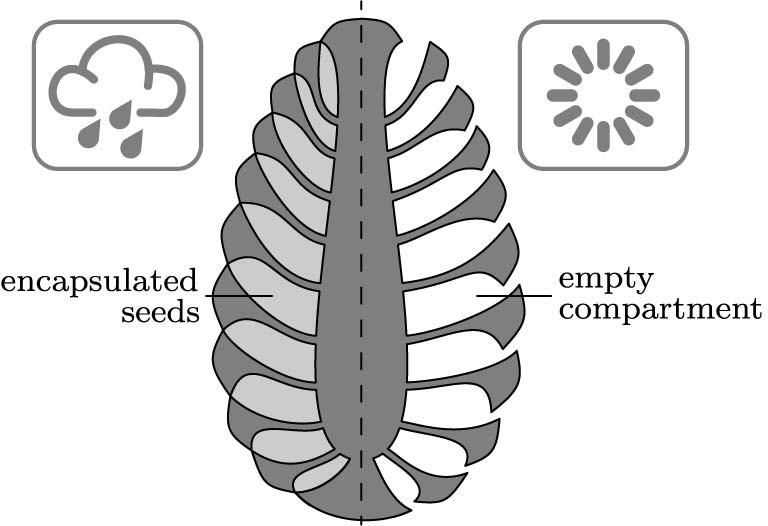}
                \small(a)\vspace{0mm}
                \includegraphics[width=\marginparwidth]{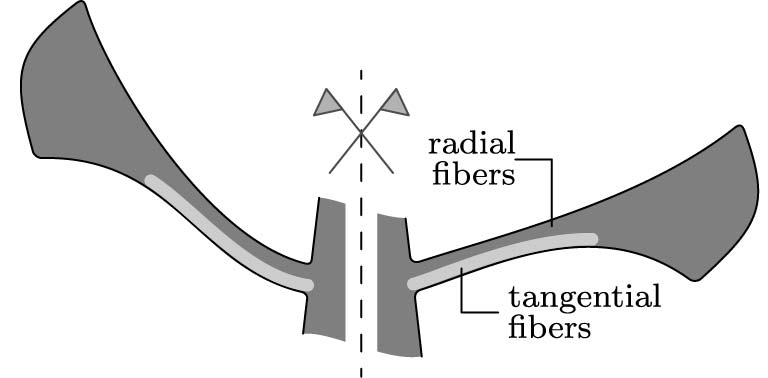}
                \small(b)\vspace{10mm}
                \includegraphics[width=\marginparwidth]{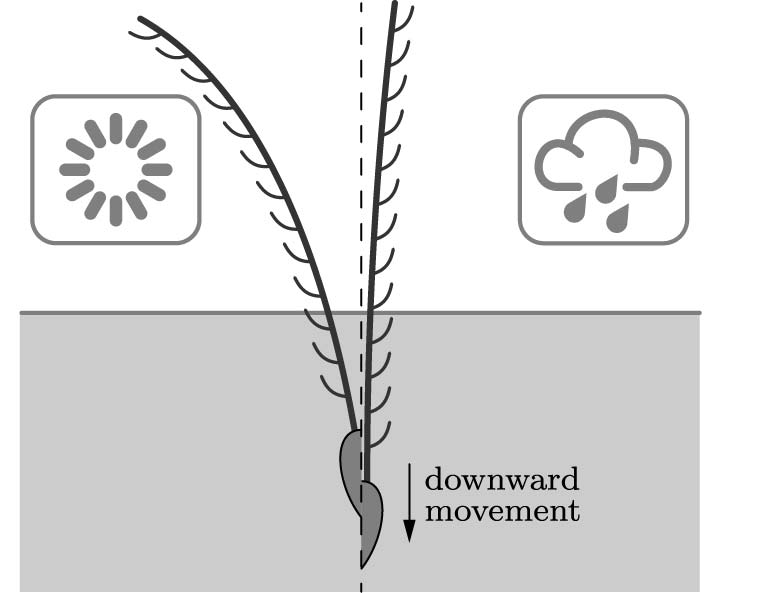}
                \small(c)\vspace{2mm}
                \captionof{figure}{Plant movements that are driven by the continuous swelling and shrinking of cell walls. (a) Pine cone and (b) single scale in a humid and dry environment. Pine cones open in a dry environment to release seeds. (c) Self burial of wheat due to awn movements. Seed depth increases during each dry-humid cycle.}
                \label{pic:Figure_6_2}
            \end{center}}[-65mm]

            \subsection{Passive Movements}
                \subsubsection{Continuous}
                    Some of the simplest plant movements are driven by the swelling and shrinking of cell walls. The matrix of these fiber reinforced composites undergoes large volume changes whereas fiber lengths remain unchanged. A tailored composition of these constituents thus enables plant movements that are solely driven by humidity gradients. This principle is widely used by plants to disperse their seeds and to improve the conditions for germination. For example, pine seeds are protected within a cone during their growth phase that can take up to three years. Upon maturation, they are automatically released if the weather is warm and dry \cite{Dawson1997-1} so that they can be carried away by the wind. The cone opening is accomplished by the bending of a large number of scales as illustrated in Figure~\ref{pic:Figure_6_2}. Scales are bi-layered structures where the fibers of the upper, lower layers are in the radial, tangential directions, respectively. This leads to different axial strains in each layer as the swelling and shrinking deformations are orthogonal to the fiber directions. Poppinga et al \cite{Poppinga2017-1} showed that this simple approach works reliably even in 15 million year old conifer cones.\\

                    Another example for humidity driven passive movements can be found in wheat awns \cite{Elbaum2007-1}. Each seed is attached at its rear to two awns that help to improve its germination conditions. Awns stabilize the seeds free fall after the separation from the ear and propel it along and into the ground. They are curved towards the outside during daytime when the humidity is relatively low and nearly straight during nighttime when the humidity is high. The seeds mobility is further improved by a large number of hairs along the awns that point towards their tip. These hairs act as hooks and support the seeds movement into the ground. A similar approach that is based on a single awn that undergoes helical instead of bending movements is used by Erodium cicutarium \cite{Stamp1984-1}. These awns can bury their seed into a favourable, porous ground over the course of five day-night cycles. A further increase in the complexity of passive plant movements can be found in ice plant seed capsules \cite{Harrington2011-1}. Their valves are shell like structures that translate the swelling and shrinking of cell walls via curved hinge lines into two-dimensional shape changes.


                \subsubsection{Discontinuous}
                    Continuous plant movements that are driven by the swelling and shrinking of cell walls are relatively slow. They are thus often used for the release of winged seeds that can independently travel over long distances. In contrast, the ballistic dispersal of reproduction units requires considerably faster movements. Based on the same actuation principles, a significant speedup can only be achieved by the storage and sudden release of elastic energy.\\

                    \blfootnote{\vspace{-3mm}
                        \captionof{figure}{Discontinuous, rapid plant movements that are driven by water evaporation. (a) Prestressed Cardamine parviflora (Brassicaceae) valves coil after fracture between valves and replum. Seeds are accelerated by adhesion forces that decrease with increasing valve curvatures. (b) Air and spore filled capsule of genus Sphagnum (Sphagnaceae) is pressurized by radial contractions. Spore dispersal is triggered by a fracture between capsule and lid. Turbulent vortex rings near the ejected lid increase the ejection range. (c) Seamless pollen dispersal of Phlebodium aureum (Polypodiaceae). Water evaporation from cell row leads to cavitation and rapid movements. Spore dispersal around midway is triggered by a sudden retardation due to an internal damping system.}
                        \label{pic:Figure_6_3}}
                    \marginnote{
                        \begin{center}
                            \includegraphics[width=\marginparwidth]{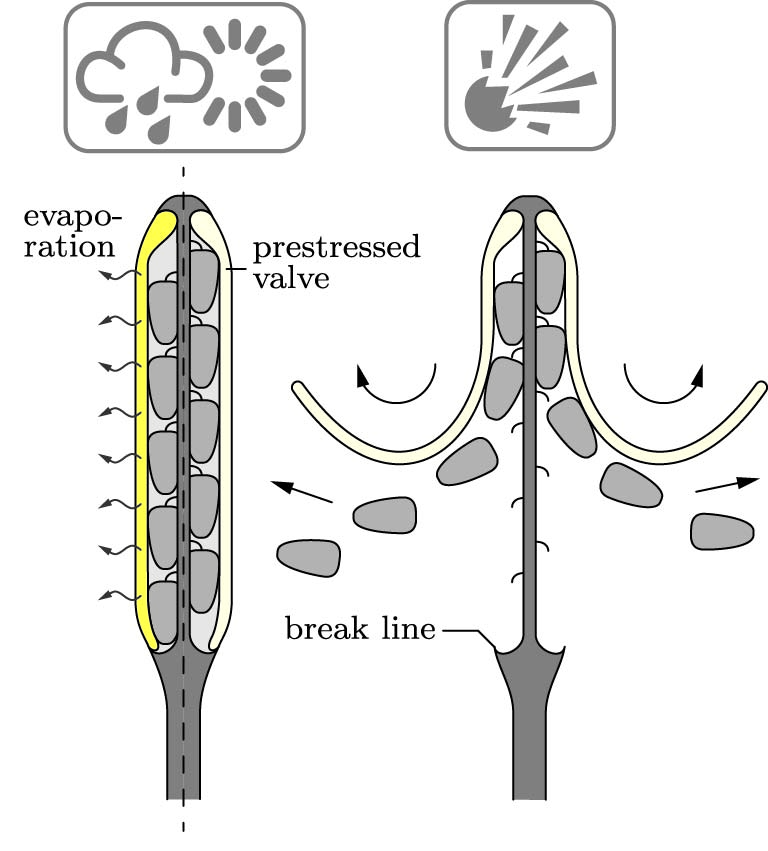}
                            \small(a)\vspace{2mm}
                            \includegraphics[width=\marginparwidth]{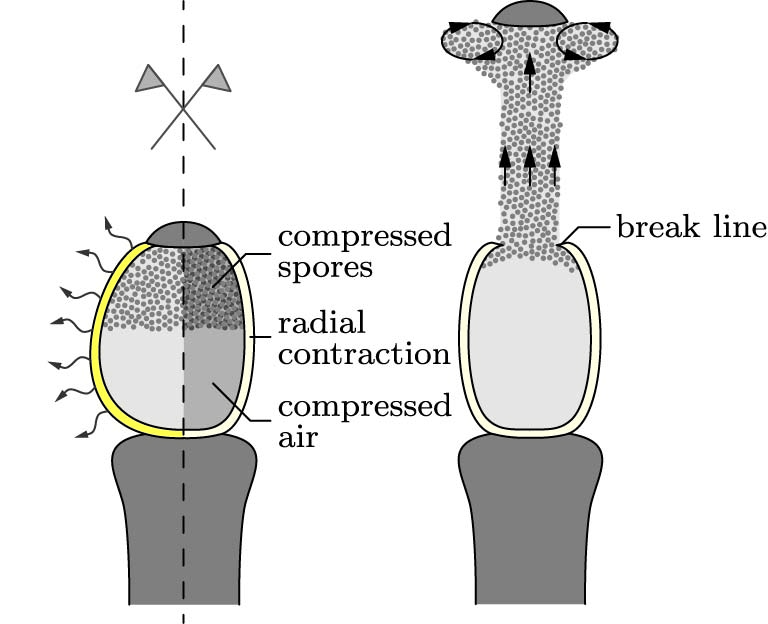}
                            \small(b)\vspace{2mm}                        \includegraphics[width=\marginparwidth]{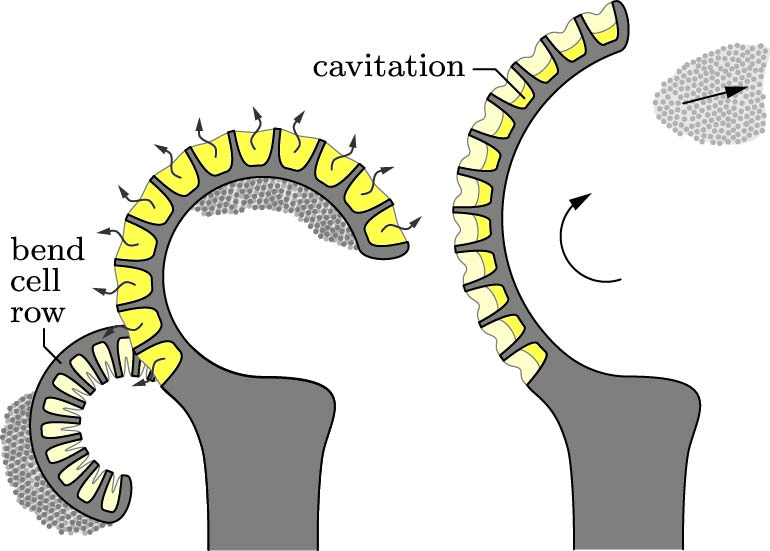}
                            \small(c)
                        \end{center}}[-85mm]
                    \vspace{-3mm}

                    The fruit of Cardamine parviflora from the mustard family Brassicaceae contains about 300 seeds that are attached to a replum and protected by two valves as illustrated in Figure~\ref{pic:Figure_6_3}. The elastic bending energy that is stored in the bilayered valves through water evaporation is suddenly released upon reaching a critical stress between valves and replum \cite{Vaughn2011-1}. Subsequent coiling motions accelerate the seeds, each weighs about 150~$\mu$g, to velocities of around 12~m/s which is sufficient for distances of up to 2~m. An interesting property of this ejection mechanism is that the seeds are accelerated by adhesion forces that decrease with increasing valve curvatures. However, their dependency on the valve moisture leads to a high variability of the launch performance.\\

                    A similar approach that is based on contact instead of adhesion forces can be found in Tetraberlinia moreliana from the family Leguminosae that is native in Gabon and the south-west of Cameroon \cite{Burgt1997-1}. Its fruit consists of two connected valves that cover up to four seeds with an average weight of about 2.5~g. Drying of the valves leads to the buildup of elastic energy that is suddenly released upon reaching a critical stress between both valves. The subsequent helical valve motions catapult the seeds to distances of up to 60~m by accelerating them to velocities of around 37~m/s. Even greater speeds are achieved by the evergreen tree Hura crepitans from the family Euphorbiaceae that is domiciled in the tropical regions of North and South America and the Amazon Rainforest. Its pumpkin shaped fruit contains 7-16 seeds that each weighs about 1~g. Water evaporation from the fruit tensions the carpels and compresses the seeds. The carpels split at a critical stress into two halves \cite{Swaine1977-1} and accelerate the seeds via squeeze forces to velocities of around 70~m/s. Despite the relatively small seed mass, ejection distances of up to 45~m were observed. Hura crepitans is sometimes known, due to the emitted noise at dispersal, as the monkey's dinner bell. However, the raw seeds are poisonous for humans and most animals.\\

                    Unlike previous approaches, a compressible fluid is used for the energy storage by plants such as Sphagnum fimbriatum of the family Sphagnaceae. Their spherical capsules are divided into two compartments where the lower is filled with air and the upper with up to 240,000 spores. The four to five layer thick capsule walls are airtight and relatively dark on the outside. Water evaporation from the walls at sunny days leads to cylindrically shaped capsules with a smaller volume and a nearly identical height \cite{Leeuwen2010-1}. The spores with a size of 22 to 45~$\mu$m are ejected after a breaking line failure between the capsule and the lid at a differential pressure of about 400~kPa. They are accelerated by the compressed air to velocities of up to 30~m/s. This initial speed is sufficient to reach altitudes of nearly 20~cm where they can be carried away by the wind within the turbulent boundary layer. These impressive altitudes are enabled by the relatively low drag coefficients of turbulent vortex rings that form in the vicinity of the ejected lid \cite{Whitaker2010-1}.\\

                    The water potential of humid air is a measure for the work that needs to be done to add a unit volume of water to a plant. For example, air with a humidity of 50\% has a potential of about \mbox{-94~MPa}. As a consequence, water evaporation from cells with rigid walls can cause negative pressures that break up the cohesion of the remaining cell water. This leads to a sudden vaporization and thus to large volume changes that release most of the cell walls elastic energy. For example, cavitation is used for the spore dispersal of ferns such as Polypodium aureum from the family Polypodiaceae \cite{Llorens2016-1}. They consist of a row of 12 to 25 cells that enclose the spores in their hydrated configuration. Loss of water through evaporation leads to a slow bending of the row and a collapse of lateral cell walls. The cavitation that occurs at a pressures of about -10~MPa initiates rapid movements towards the fully hydrated shape during which the spores are ejected at velocities of up to 10~m/s. The ejection angle is predetermined by an internal damping system that causes a sudden retardation around the midpoint. This catapults the spores, that possess a size of about 45~$\mu$m, over distances of 20~mm. Such a cavitation based approach is outstanding as it does not require the failure of any component.


            \subsection{Active Movements}
                \subsubsection{Continuous}
                    Unlike passive movements, active movements are independent of the environmental conditions. This is achieved by using energy from the plants metabolism via pressure variations or the directed growth of cells. As for passive movements, the energy can be stored and rapidly released or continuously translated into shape changes \cite{Sakes2016-1}.\\

                    \marginnote{
                    \begin{center}
                        \includegraphics[width=\marginparwidth]{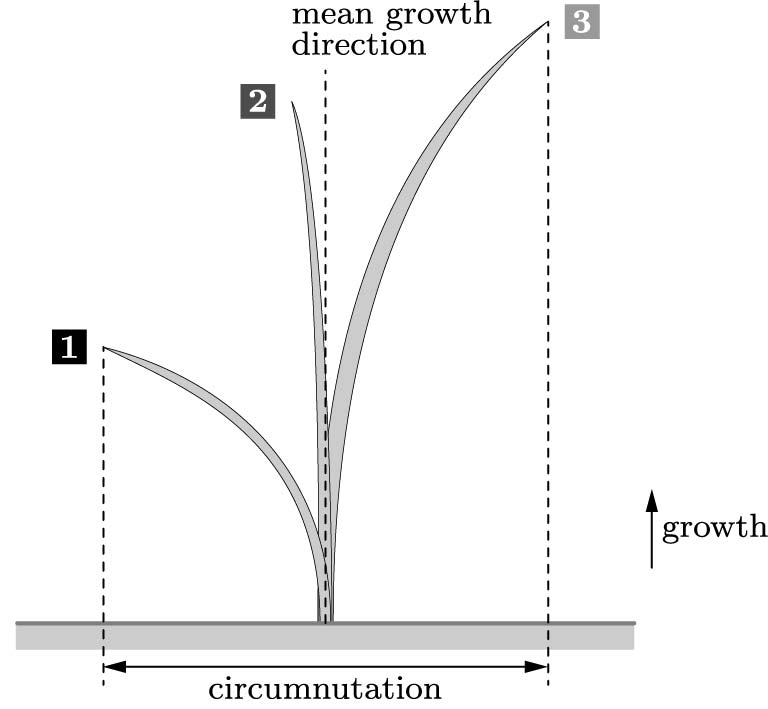}
                        \small(a)\vspace{7mm}
                        \includegraphics[width=\marginparwidth]{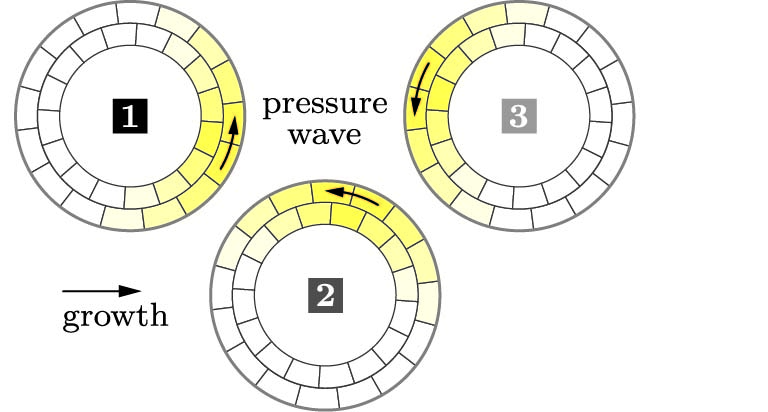}
                        \small(b)\vspace{7mm}                        \includegraphics[width=\marginparwidth]{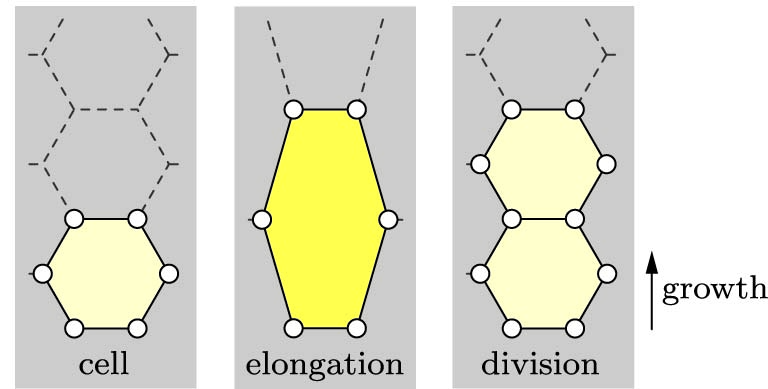}
                        \small(c)\vspace{1mm}
                        \captionof{figure}{Circumnutation of plants due to local cell elongations and cell divisions. (a) Oscillation around the mean growth direction. (b) Cross section of stem and pressure wave that travels in counterclockwise direction. (c) Longitudinal growth through pressure driven cell elongation and cell division.}
                        \label{pic:Figure_6_4}
                    \end{center}}[-80mm]

                    Plants rarely grow in one direction. Instead they oscillate around a mean growth direction \cite{Brown1993-1} as shown in Figure~\ref{pic:Figure_6_4}. These movements that are known as circumnutation follow an elliptical orbit if viewed from the growth direction and thus resemble a helix as the plant grows. The amplitude, period and shape of circumnutations can vary greatly and depends on the species, the involved plant organs and the developmental stage. For example, the shoots and tendrils of many climbing plants move regularly on circular orbits whereas other, non-climbing plants may exhibit highly irregular movements that hardly resemble an elliptical orbit. Furthermore, the purpose of circumnutations can differ. Climbing plants use regular movements to find mechanical supports whereas the benefit of circumnutations in many non-climbing plants remains unclear.\\

                    Plant growth is, to a varying degree, driven by hormones that are known as auxins \cite{Friml2003-1}. They influence the cell elongations and divisions that lead to plant movements. Elongations are directed by local cell wall reinforcements and initiated by increased cell pressures. A simplified mechanical model for cell divisions was proposed by Frei Otto \cite{Thywissen1979-1} in 1979. He observed that, for a critical deformation, the strain energy of a constricted soap bubble is minimized by forming two separate spheres. Although not fully understood, it is likely that circumnutations are caused by rotating turgor waves during growth \cite{Hejnowicz1995-1}. The period of these waves differs and ranges between twenty and three-hundred minutes in most plants. For example, cell walls in Phycomyces can grow in longitudinal direction with strain rates of more than fifty per day \cite{Jordan2010-1}. Nonetheless, circumnutations are considerably slower than the speeds that can be achieved by the sudden release of energy that is initiated by either fracture or cavitation. Furthermore, they are far more complex and less understood. Particularly the interaction of auxins with other plant processes and their detailed involvement in the cell elongation and division is still an open question.\\


                \blfootnote{\vspace{-3mm}
                    \captionof{figure}{Pressure driven, discontinuous plant movements. (a) Valves of Impatiens glandulifera (Balsaminaceae) separate after reaching critical prestress. Their rapid coiling motions accelerate the seeds via momentum exchange. (b) Seeds of the genus Arceuthobium (Santalaceae) are embedded in a sticky layer of pressurized viscin cells that are surrounded by an elastic, energy storing exocarp. They are accelerated by squeeze forces after a separation between pedicel and fruit. (c) Pollen filled anther of Morus alba (Moracea) is attached to a curved, energy storing filament. The anther is accelerated on a circular trajectory after slippage past a mechanical restraint.}
                    \label{pic:Figure_6_5}}
                \marginnote{
                    \begin{center}
                        \includegraphics[width=\marginparwidth]{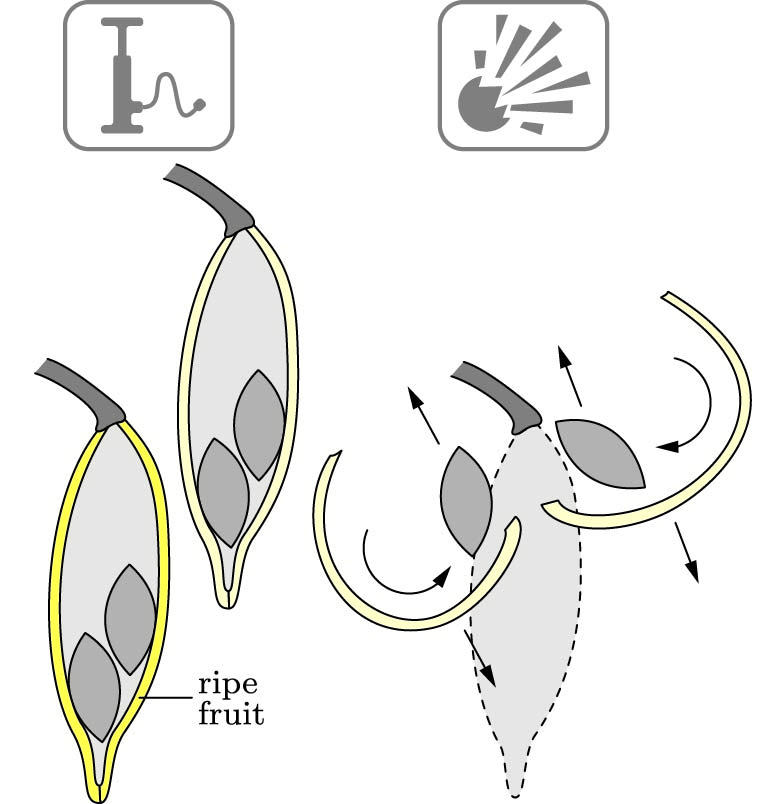}
                        \small(a)\vspace{7mm}
                        \includegraphics[width=\marginparwidth]{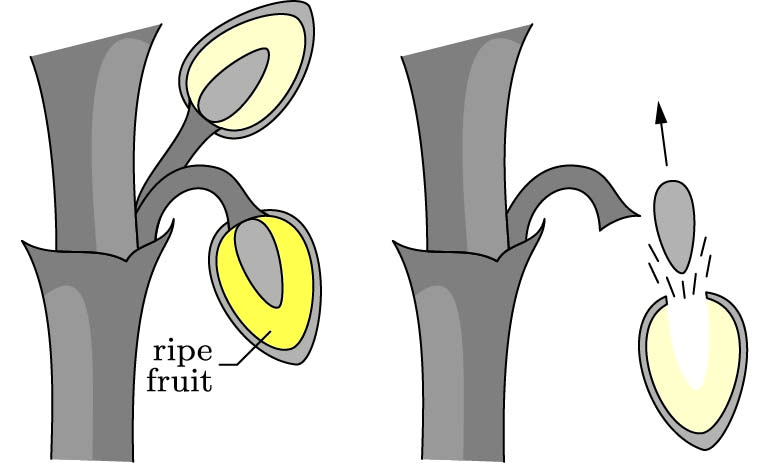}
                        \small(b)\vspace{7mm}
                        \includegraphics[width=\marginparwidth]{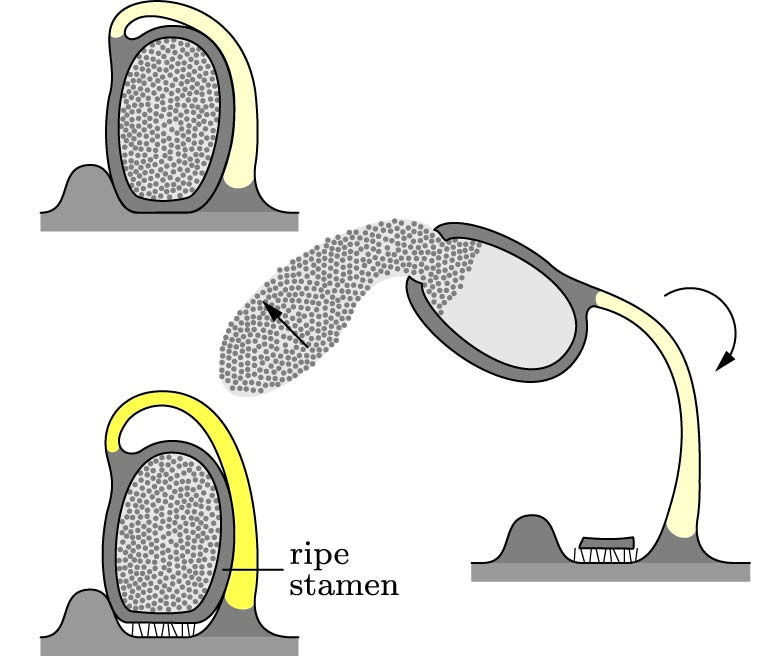}
                        \small(c)
                    \end{center}}[-19mm]
                \vspace{-3mm}

                \subsubsection{Discontinuous}
                    Impatiens glandulifera from the family Balsaminaceae is a highly invasive plant that conquered much of the Northern Hemisphere. This success can be partially attributed to its efficient seed dispersal mechanism. Each of I. glandulifera's seed pods consists of five, bilayered valves that enclose 5 to 10 seeds. The valves, that are connected to each other, are prestressed by the swelling of the outer layers. It is speculated that the stored energy is at least partially generated by the plants metabolism via cell pressure variations \cite{Hayashi2009-1}. Hence it seems plausible that the fracture between the valves and thus the seed dispersal can be actively triggered. About 45\% of the elastic energy that is released during valve coiling is transferred into the kinetic energy of the seeds \cite{Deegan2012-1}. Initial velocities of 12~m/s were observed that catapult the seeds, that weigh about 7~mg each, to distances of up to 10~m. This is remarkable as the less invasive Impatiens capensis from the same genus utilizes only about 0.5~\% of the elastic valve energy.\\

                    Plants of the genus Arceuthobium from the family Santalaceae are commonly known as dwarf mistletoes. They cause considerable economic damages by parasitizing species of Pinaceae and Cupressaceae in North and Central America, Asia, Europe and Africa. The ellipsoidal fruits of dwarf mistletoes contain a single seed with a weight of about 2 to 3~mg that is embedded in a sticky layer of viscin cells and surrounded by an exocarp as illustrated in Figure~\ref{pic:Figure_6_5}. Ripe fruits develop an abscission zone around their pedicel and alter their position by pointing downwards. The energy for the fruit dispersal is stored in the elastic exocarp and generated by the increasing cell pressures of the viscin layer. A remarkable property of these plants is that they can actively trigger their seed dispersal by initiating a failure in the abscission zone via thermogenesis \cite{Bruyn2015-1}. The rapid contraction of the exocarp after its separation from the pedicel causes squeeze forces that accelerate the aerodynamically shaped seeds to velocities of up to 25~m/s \cite{Hawksworth1959-1}. The vertically ejected seeds can reach altitudes of 20~m. Some of them stick to upright needles during their descent at which they slide down during rainfall to reach the bark.\\

                    \marginnote{
                    \begin{center}
                        \includegraphics[width=\marginparwidth]{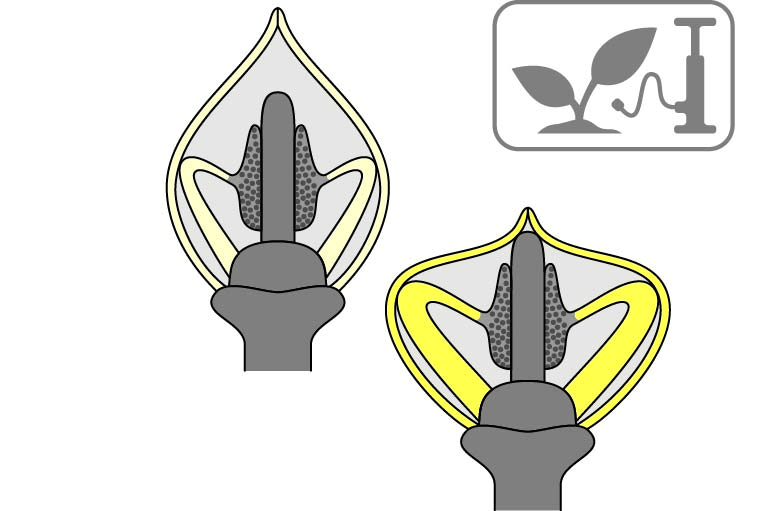}
                        \small(a)\vspace{0mm}
                        \includegraphics[width=\marginparwidth]{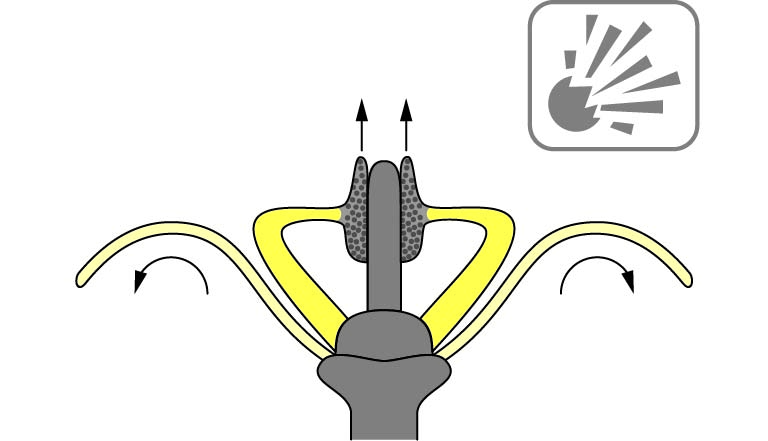}
                        \small(b)\vspace{3mm}
                        \includegraphics[width=\marginparwidth]{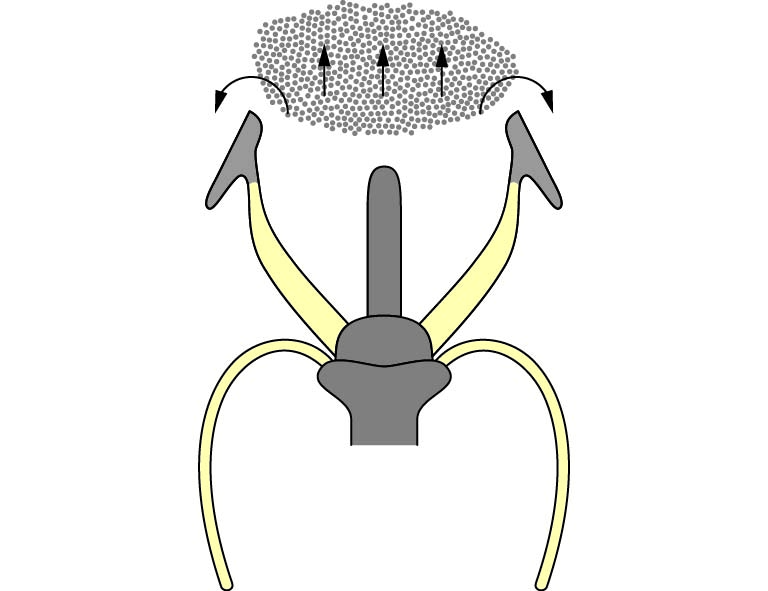}
                        \small(c)\vspace{0mm}
                        \captionof{figure}{Pollen dispersal of Cornus canadensis (Cornaceae). (a) Elastic energy is stored in the petals and filaments through differential growth and cell pressures. (b) Petal opening is triggered by cell pressures or large insects. Anthers start to accelerate vertically. (c) Separation from the inset leads to anther rotations that increase tip velocities.}
                        \label{pic:Figure_6_6}
                    \end{center}}[-36mm]

                    Morus alba from the family Moracea is a small to medium-sized tree that is native to North China. Its male inflorescence possesses about 60 individual flowers where each bud consists of four stamens that resemble manmade catapults \cite{Taylor2006-1}. Their pollen filled anther is connected to a curved, elastic filament that can be prestressed via cell pressure variations. Anther movements are prevented by fine threads and a mechanical restraint. Slippage past the restraint, that is triggered by increasing cell pressures, opens up the anther and accelerates the pollen via centrifugal forces to velocities of up to 240~m/s. These speeds, that approach the physical limit for plant movements, are among the fastest biological movements that were recorded so far. The pollens with a size of about 25~$\mu$m are dispersed by this mechanism over distances of up to 65~mm.\\

                    A slightly different approach can be found in Cornus canadensis from the family Cornaceae that is native to Eastern Asia. Its dispersal mechanism differs from M. alba insofar that it resembles more a trebuchet than a classical catapult \cite{Edwards2005-1}. Each of its flowers consists of an inset and four anthers that are flexibly connected to filament tips as shown in Figure~\ref{pic:Figure_6_6}. A characteristic of C. canadensis is that the elastic energy for the pollen dispersal is partially generated through differential growth. This is achieved by the relatively fast growing filaments that are restrained by the petals during flower development. The rapid opening of the prestressed petals is triggered by cell pressure variations or the touch of large insects such as bumblebees. This leads to an initially vertical acceleration of the anthers that start to rotate after the separation from the inset. The pollens that weigh about 24~$\mu$g are accelerated by this mechanism to velocities of up to 7.5~m/s \cite{Whitaker2007-1}. Their relatively moderate speed is sufficient to reach altitudes of 25~mm where they can be carried away by the wind or a pollinating insect.


        \section{Stimulated Movements}
            Plant movements that occur in response to an external stimulus can be divided into predetermined, nastic movements that do not depend on the direction of the stimulus and the more sophisticated tropisms that depend on the magnitude and direction of the stimulus. The required movements of gapless high lift devices are known in advance so that tropisms are of little interest in this context. Nonetheless, their major working principles are subsequently discussed.

            \marginnote{
            \begin{center}
                \includegraphics[width=\marginparwidth]{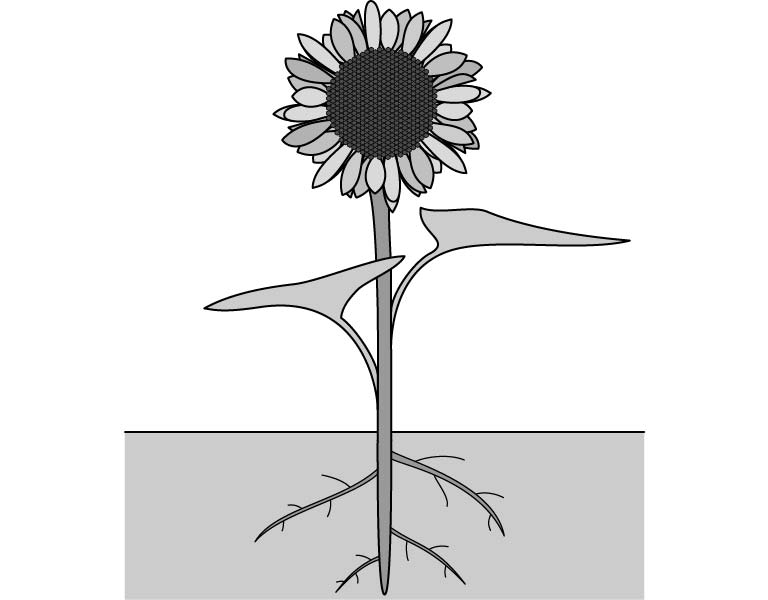}
                \small(a)\vspace{1mm}
                \includegraphics[width=\marginparwidth]{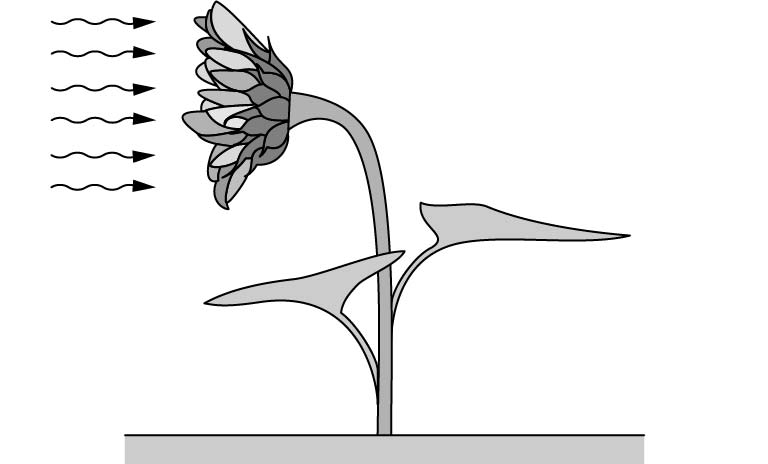}
                \small(b)\vspace{3mm}
                \includegraphics[width=\marginparwidth]{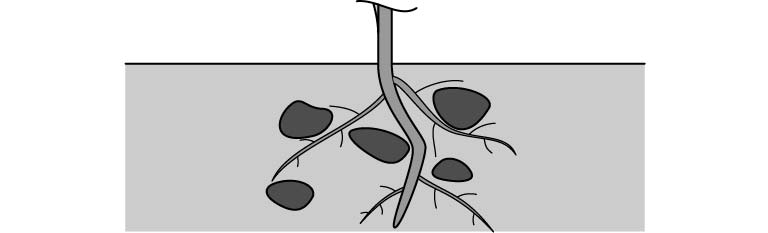}
                \small(c)\vspace{3mm}
                \includegraphics[width=\marginparwidth]{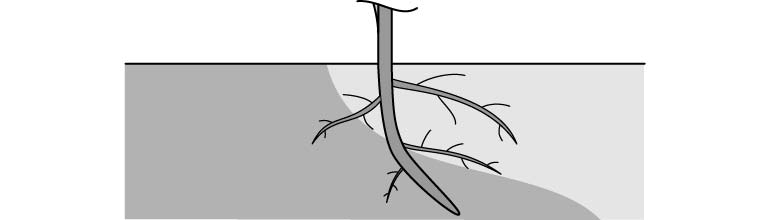}
                \small(d)\vspace{2mm}
                \captionof{figure}{Various forms of tropism. (a) \textit{Geotropism} drives the vertical growth of stems and roots. (b) \textit{Phototropism} optimizes the irradiation of plant organs. (c) \textit{Thigmotropism} avoids obstacles during root growth. (d) \textit{Hydrotropism} directs root growth towards nutrient rich soil.}
                \label{pic:Figure_6_7}
            \end{center}}[-1mm]


            \subsection{Tropisms}
                Circumnutations and tropisms are growth driven movements. The former happen automatically whereas the latter are driven by various kind of external stimuli. Common stimuli include irradiation \textit{(phototropism, heliotropism)}, gravity \textit{(geotropism)}, nutrients \textit{(hydrotropism)} and touch \textit{(thigmotropism)} as illustrated in Figure~\ref{pic:Figure_6_7}. The close link between circumnutations and tropisms is highlighted by many climbing plants. They minimize their supportive tissue during vertical growth by utilizing external supports. Circumnutations are used to randomly search for new supports whereas touch induced growth movements are initiated upon contact.\\

                A more sophisticated approach is used by climbing plants such as Epipremnum giganteum from the family Araceae that is native to the tropical forests of Indochina. These plants avoid random searches by using negative phototropism \textit{(skototropism)} that directs them towards dark, solid objects \cite{Strong1975-1}. This is remarkable as they initially impair their light exposure before they can start to climb towards the canopy. Phototropism that directly improves a plants exposure to light can be observed at different timescales. Most plants optimize their orientation with respect to average light conditions that occur over a longer period of time. Other plants such as juvenile sunflowers directly track the motion of the sun during a diurnal cycle \textit{(heliotropism)} and reorient them after sunset to face the next sunrise \cite{Atamian2016-1}. The latter movement, that is not driven by phototropism, seems to be triggered by the circadian clock \cite{McClung2006-1}.\\

                The study of tropisms is often complicated by their mutual interactions. For example, the root growth direction depends mostly on three different tropisms. Geotropism aligns the mean growth direction to the gravitational field, thigmotropism uses touch responses to avoid obstacles and hydrotropism directs the growth towards nutrient and water rich soil. It is thus difficult to quantify the influence of each tropism on the growth direction \cite{Eapen2005-1}. Some of these interactions can be switched off with the help of mutants or microgravity environments that can be found on space stations. However, despite a considerable research effort, a detailed understanding of the mechanisms behind tropisms remains elusive.


            \subsection{Nastic Movements}
                Nastic movements occur, unlike tropisms, in response to non-directional stimuli. Common stimuli include irradiation \textit{(photonastic)}, temperature \textit{(thermonastic)}, touch \textit{(thigmonastic)}, vibration \textit{(seismonastic)} and the circadian clock \textit{(nyctinastic)}. Nastic movements are usually hardwired into plants so that they are identical within a certain species. This reduces their overall complexity and response times. Hence they are of particular interest for the design of gapless high lift devices.\\

                \blfootnote{\vspace{-3mm}
                    \captionof{figure}{Nastic movements of carnivorous plants (side and top view). (a) Drosera capensis uses nastic growth movements to clutch its tentacles around a prey and tropic movements to bend its leaf. (b) Utricularia vulgaris prestresses its bladders by pumping out water via osmosis. Prey is sucked in by rapid water inflow after trigger hairs are touched. (c) Dionea muscipula prestresses its leaves by varying their cell pressures. Prey is catched between leaves after trigger hairs are touched.}
                    \label{pic:Figure_6_8}}
                \marginnote{
                    \begin{center}
                        \includegraphics[width=\marginparwidth]{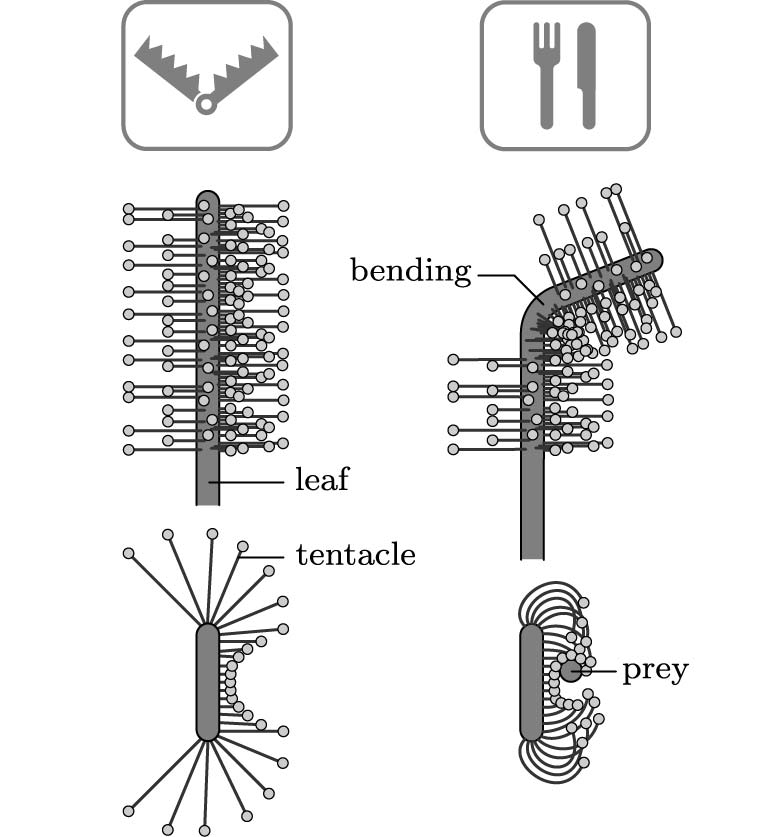}
                        \small(a)\vspace{7mm}
                        \includegraphics[width=\marginparwidth]{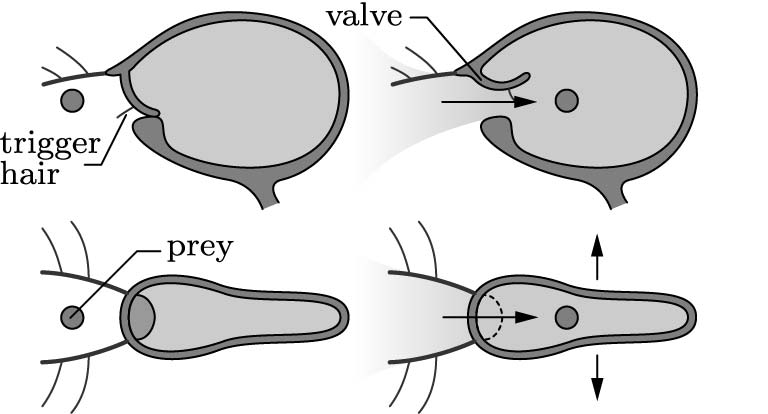}
                        \small(b)\vspace{7mm}
                        \includegraphics[width=\marginparwidth]{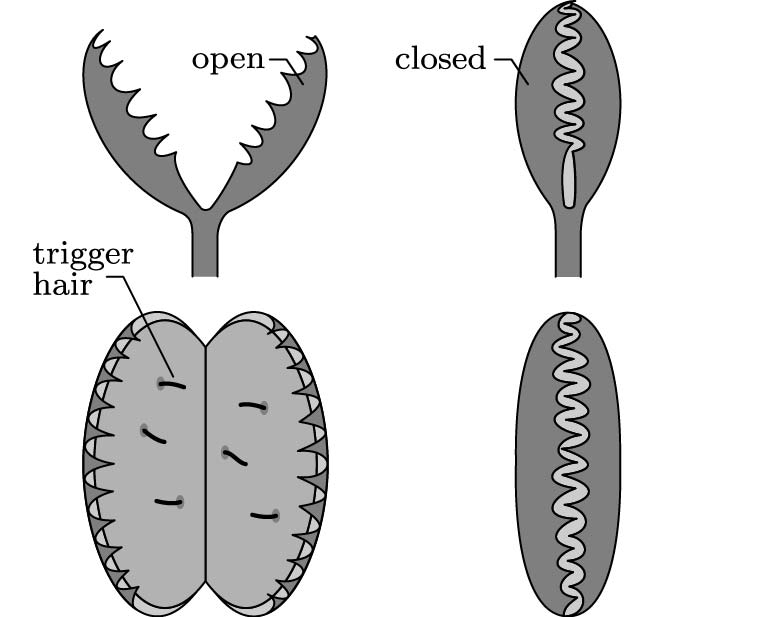}
                        \small(c)
                    \end{center}}[-32mm]
                \vspace{-3mm}

                Photonastic and thermonastic movements play a major role in the opening and closing of flowers. Some species open their flowers only once whereas others open and close them regularly \cite{Doorn2003-1}. The latter approach has the advantage that it protects the flowers against environmental conditions during times when a pollination is unlikely. Diurnal flowers such as Iridacea are open during the day and closed during the night whereas nocturnal flowers such as Cactaceae or Bromeliaceae are open during the night and closed during the day. Diurnal flowers are often pollinated by butterflies or bees whereas nocturnal flowers are pollinated by moths and bats. The opening and closing of most flowers is based on growth movements. Opening times can vary significantly. Flowers that open only once can store elastic energy in bracts or sepals to accelerate their opening after abscission. For example, Hedera helix from the family Araliaceae requires about five minutes whereas Oenothera biennis from the family Onagraceae that opens its flowers every evening requires about twenty minutes.\\

                Plants that are capable of fast, reversible movements have several advantages over their conspecific, static rivals. For example, carnivorous plants survive in places where the soil is extremely poor in nutrients by catching insects and arachnids \cite{Ellison2006-1}. Drosera capensis from the family Droseraceae that is native to South Africa \cite{LLoyd1942-1} possesses strap like leaves that are covered by long tentacles on the sides and by short tentacles on the central part of the leaf. The glands at the top of the tentacles secret a sticky, sugar containing mucilage that attracts prey. A struggling animal that sticks to a tentacle is transported towards the center of the leaf and covered by the remaining lateral tentacles. These growth driven, thigmonastic movements are followed by the thigmotropic bending of the leaf at the location of the prey as shown in Figure~\ref{pic:Figure_6_8}. A remarkable property of many carnivorous plants is that they can avoid unnecessary movements by recognizing stimuli that are unrelated to food.\\

                \marginnote{
                \begin{center}
                    \includegraphics[width=\marginparwidth]{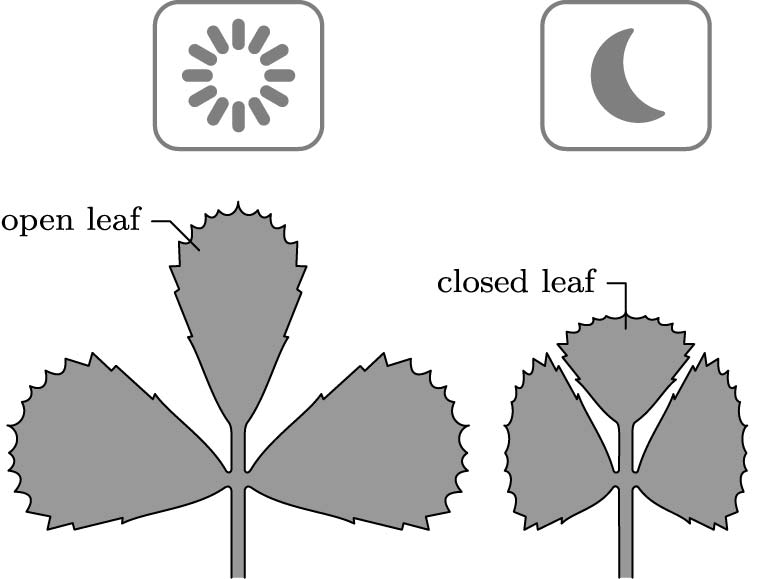}
                    \small(a)\vspace{7mm}
                    \includegraphics[width=\marginparwidth]{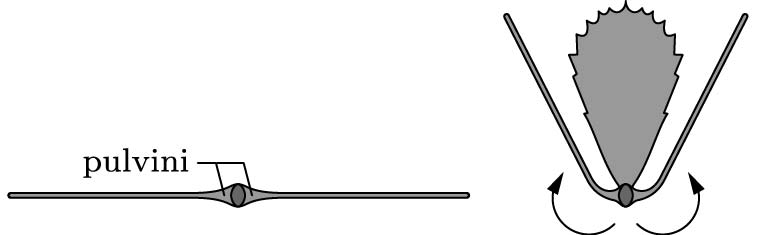}
                    \small(b)\vspace{7mm}
                    \includegraphics[width=\marginparwidth]{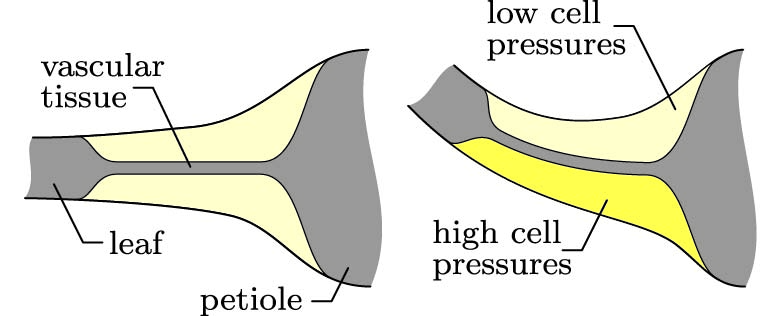}
                    \small(c)
                    \captionof{figure}{Seismonastic, thigmonastic or nyctinastic closure, opening of Medicago truncatula (Fabaceae) leaves. (a) Top and (b) side view of a single leaf. (c) Simplified cross section of pulvinus that translates cell pressure variations into movements.}
                    \label{pic:Figure_6_9}
                \end{center}}[-6mm]

                A trapping mechanism that functions without complex growth movements is used by Utricularia vulgaris from the family Lentibulariaceae. The aquatic plant that can be found in Asia and Europe captures small animals with bladder like traps. These traps are armed by an underpressure of up to 15~kPa that is generated within the bladder via osmosis. Their in- and outflow is controlled by a compliant, inward opening valve that is sealed by an adhesive. Animals are attracted towards the bladder by a sugar containing mucilage that is excreted around the valve. Trigger hairs that are connected to the valve serve as levers that break the seal upon touch. Driven by the elastic energy of the bladder, the rapid inflow of water sucks in sufficiently small prey that is dissolved by digestive secretions within a few hours. The trap continuously pumps out water during this process to increase the nutrient concentration so that it can be reused within half an hour after digestion.\\

                Dionaea muscipula from the family Droseraceae is native to the subtropical wetlands of North and South Carolina. Its leaves are divided into a heart shaped petiole for the photosynthesis and a pair of terminal lobes that form a trap. The lobes secret, like many other carnivorous plants, mucilage to attract prey. The upper, red dyed surface of each lobe contains three hair like trichomes that trigger rapid thigmonastic movements. D. muscipula exploits a geometric instability to close its leaves within a fraction of a second \cite{Forterre2005-1}. This discontinuous, repeatable movement is particularly interesting as it does not require complex growth processes or subcomponents such as valves. Instead it is based on cells with tailored geometries whose total elastic energy is generated and partially released by cell pressure variations \cite{Pagitz2013-1}. The edges of the trap are seamed by stiff hairs that interlock as the leaves close. Gaps between the hairs allow small animals to escape whereas larger prey is digested if the trigger hairs are further stimulated. The digestion takes about ten days and the reopening up to twelve hours.\\

                The discontinuous, rapid movements of Dionaea muscipula are fascinating. However, they can rarely serve as an inspiration for gapless high lift devices as they require fully controllable, continuous movements. Plants such as Mimosa pudica or Medicago truncatula from the family Fabaceae can continuously close their leafs within seconds, Figure~\ref{pic:Figure_6_9}. These movements are driven by a pulvinus at the base of each petiole that consist of a stiff core of vascular tissue that is surrounded by thin walled cells. Leaf closure minimizes a plants radiative heat losses during nighttime \textit{(nyctinastic)} and creates the illusion of being a less voluminous meal during daytime if touched by a predator \textit{(thigmonastic)}. Furthermore, the defensive capabilities of M. pudica improve during closure as the relative angles between protective thorns and predators vary \cite{Eisner1981-1}. Leaf closure during daytime is energetically costly as the plants photosynthesis reduces drastically. Hence there is a tradeoff between the protection against potential threads and the harnessing of vital energy. An impressive property of M. pudica is its ability to habitually learn which kind of seismonastic and thigmonastic stimuli require leaf closure \cite{Gagliano2014-1}. This property is more pronounced and persistent for plants that grow in challenging environments as they have a smaller margin for superfluous movements.


        \marginnote{
        \begin{center}
            \includegraphics[width=\marginparwidth]{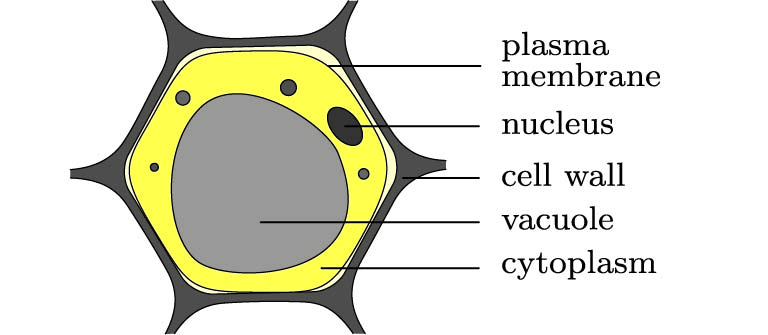}
            \small(a)\vspace{4mm}
            \includegraphics[width=\marginparwidth]{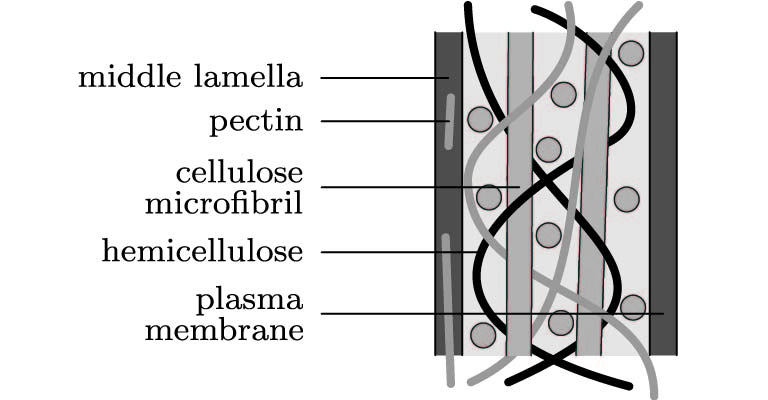}
            \small(b)\vspace{5mm}
            \includegraphics[width=\marginparwidth]{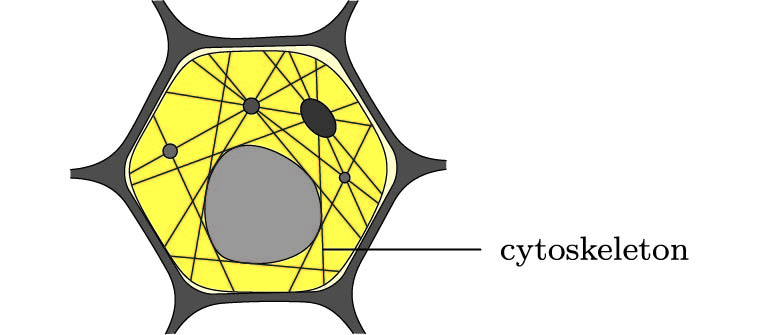}
            \small(c)\vspace{5mm}
            \includegraphics[width=\marginparwidth]{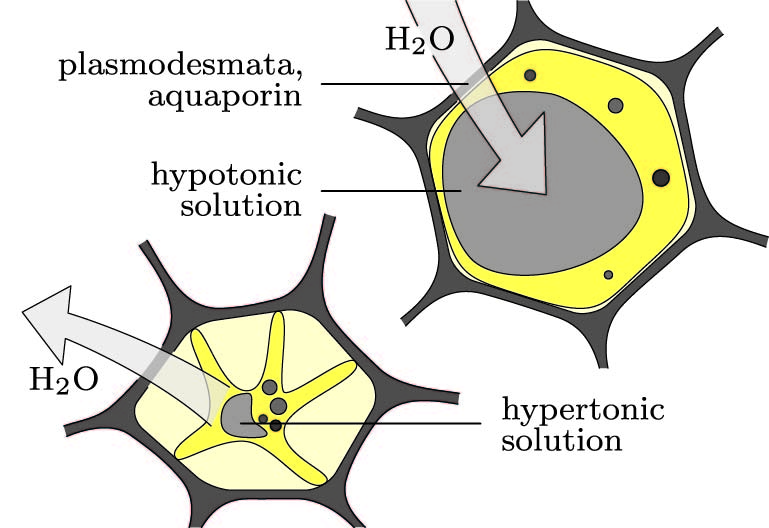}
            \small(d)\vspace{2mm}
            \captionof{figure}{Cell components that are involved in nastic movements. (a) Cell organization. (b) Middle lamella, primary cell wall and plasma membrane. (c) Cytoskeletons and (d) osmotic hydration motors.}
            \label{pic:Figure_6_10}
        \end{center}}[-36mm]

        \section{Cell Components}
            Nastic plants such as Mimosa pudica are highly integrated, lightweight and energy efficient. Their continuous, repeatable movements that are hardwired into the cell geometries are driven by cell pressure variations. The main components and processes that are involved in these movements include cell walls, cytoskeletons and subcellular osmotic hydration motors as shown in Figure~\ref{pic:Figure_6_10}.


            \subsection{Geometry}
                The Weaire-Phelan structure possesses the smallest total surface area observed thus far \cite{Weaire1994-1} for a packing of equal volume cells. It utilizes two kind of cell geometries where one is a dodecahedron with pyritohedral symmetry and the other is a truncated hexagonal trapezohedron. This structure is used, for example, in the Beijing Aquatic Center. Plant cells rarely form this ideal structure as their geometries are driven by more than one objective. However, like the Weaire-Phelan structure, they always pack their cells such that only three edges meet at a single point. The diversity of cell geometries can be highlighted with the help of epidermal plant cells whose basic geometry depends on the plants habitat \cite{Koch2009-1}. Plants that are exposed to intense irradiation and heat are often protected by fine hairs against water evaporation whereas wetland plants are usually equipped with self cleaning surfaces that reduce contaminations and pathogen attacks.\\

                The periclinal cell walls that form a plants outer surface can be seen with a light microscope. Their tetragonal or polygonal outlines can be isodiametric (constant cell side lengths) or elongated as illustrated in Figure~\ref{pic:Figure_6_11}. The basic geometry of these cell walls can vary even within a single leaf. For example, cells of the lower leaf side of Poa annua (Poaceae) are tetragonal elongated whereas cells of the upper side are polygonal. The boundaries between periclinal cell walls can be straight or undulated with U, V, $\Omega$ and S shapes. These undulations can be regular or non-regular and their amplitudes may vary. It is widely believed that the mechanical properties of the epidermis are improved by the interlocking between undulated cell boundaries.\\

                \blfootnote{\vspace{-3mm}
                    \captionof{figure}{Basic epidermal plant cell geometries. (a) Plant surfaces are formed by periclinal cell walls. Their tetragonal or polygonal outlines possess sides with identical (isodiametric) or varying lengths (elongated). (b) Boundaries between periclinal cell walls can be straight or undulated with U, V, $\Omega$ and S shapes. (c) Periclinal cell walls can be flat, concave or convex. (d) Anticline cell walls can be flush, sunken or exposed.}
                    \label{pic:Figure_6_11}}
                \marginnote{
                    \begin{center}
                        \includegraphics[width=\marginparwidth]{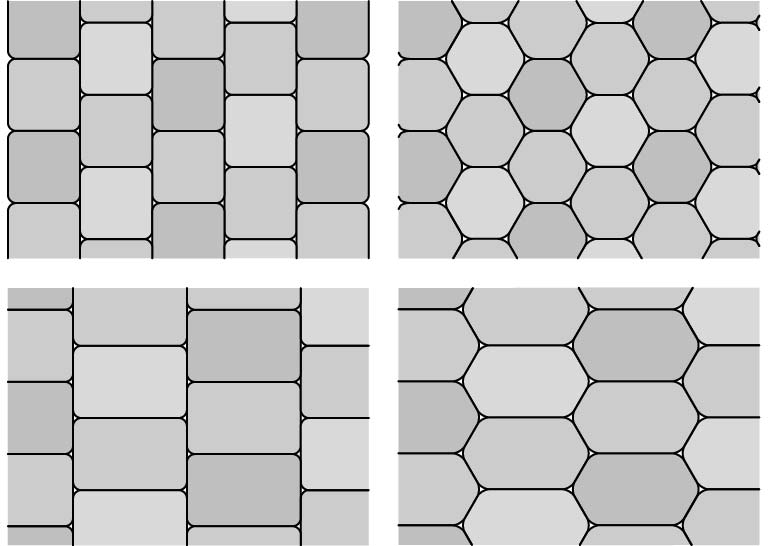}
                        \small(a)\vspace{7mm}
                        \includegraphics[width=\marginparwidth]{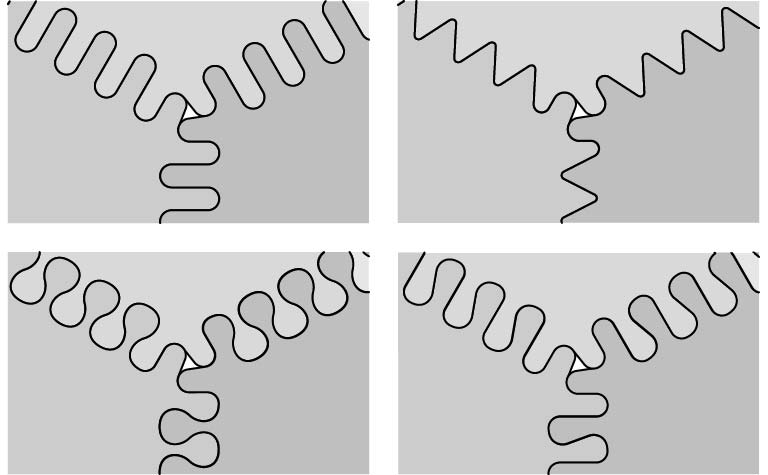}
                        \small(b)\vspace{7mm}
                        \includegraphics[width=\marginparwidth]{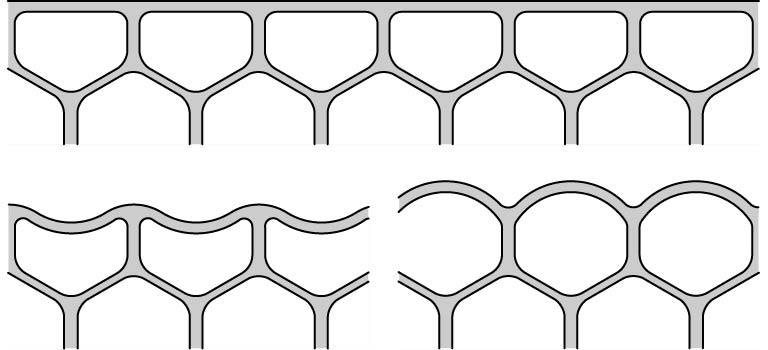}
                        \small(c)\vspace{7mm}
                        \includegraphics[width=\marginparwidth]{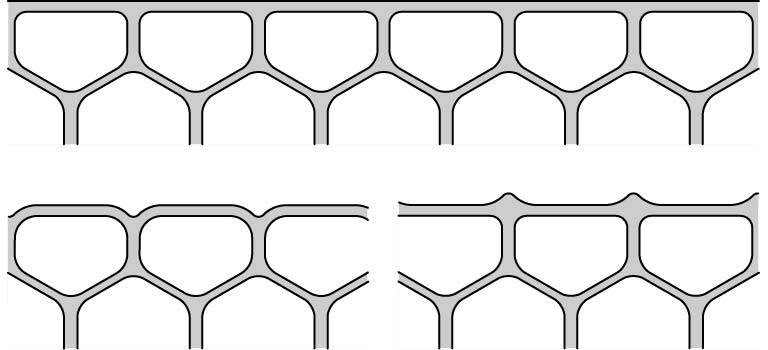}
                        \small(d)
                    \end{center}}[-15mm]
                \vspace{-3mm}

                Anticline cell walls are orthogonal to the periclinal cell walls. With regards to the plant surface they can be flush, sunken or exposed. The periclinal cell walls that are bounded by the anticlines can be tabular (flat), concave or convex. Tabular surfaces can be often found in submerged and floating plants such as Anubias barteri from the family Araceae. In contrast, concave periclinal cells are rarely present in living cells as the concavity is usually caused by water evaporation. Hence they are commonly found in dry seeds such as the wind dispersed seeds of Aeginetia indica from the family Orobanchaceae. Convex periclinal cell walls are the most common. Underwater plants are usually hydrophilic whereas emerged (above the water) and wetland plants are often hydrophobic. Plants such as Nelumbo nucifera from the family Nelumbonaceae possess convex walls with optimized shapes that improve their self cleaning capabilities \cite{Neinhuis1997-1}. This reduces N. nucifera's susceptibility to contaminations and pathogen attacks that are promoted by the high, constant moisture of its habitat.\\

                Plant hairs are formed by extremely elongated convex cell walls. They can be dead or living and based on single or multiple periclinal cell walls. Many plants use hairs as a protection against intense irradiation. For example, the straight, air filled hairs of the South African tree Leucadendron argenteum consist of a single cell wall. They form a dense layer that appears to be white as it reflects most of the visible light. Other densely covered plants such as Encelia farinosa reflect mainly the near infrared radiation (700-3000~nm) whereas the photosynthetically useful radiation (400-700~nm) can reach the leaves \cite{Ehleringer1978-1}. Although hairs are absent in underwater plants (tubular surfaces), they can be commonly found in floating plants. An example are the rough, hairy leaves of the water ferns of genus Salvinia. Their water repelling, multicellular hairs increase the buoyancy by retaining an air layer for several days even if fixed underwater. Other multicellular hairs with glands that secrete adhesives and enzymes for the capture and digestion of prey can be found in carnivorous plants such as Drosera capensis.\\

                The study of internal cell geometries is far more difficult as they can not be directly observed. Hence it is not surprising that the available literature in this field is relatively sparse. Irrespective of the limited knowledge, a direct translation of cell geometries into technical applications is unrealistic as their shapes are driven by various objectives whose interactions are little understood. In contrast, cell geometries for engineering structures are potentially simpler as they need to be optimized solely for the shape changing requirements.


            \subsection{Material}
                Unlike animal cells, plant cells are surrounded by stiff cell walls with a thickness that varies between a tenth to several $\mu m$. This is necessary as plants do not possess a centralized skeleton that carries most of the loads. Another consequence of their design is that they can not utilize muscles in the same way as animals do. Instead they can generate large cell pressures that are contained by the surrounding cell walls. Commonly used turgor pressures are between 0.4-0.8~MPa but can reach 5~MPa in guard cells that control a plants gas exchange. Cell walls are a living material that can grow and reorganize its constituents. This enables them to undergo large plastic strains that are driven by pressure induced tension forces during cell division. It is thus difficult to quantify the mechanical properties of their composite material on a microscopic level \cite{Ambrosi2011-1}.\\

                \marginnote{
                \begin{center}
                    \includegraphics[width=\marginparwidth]{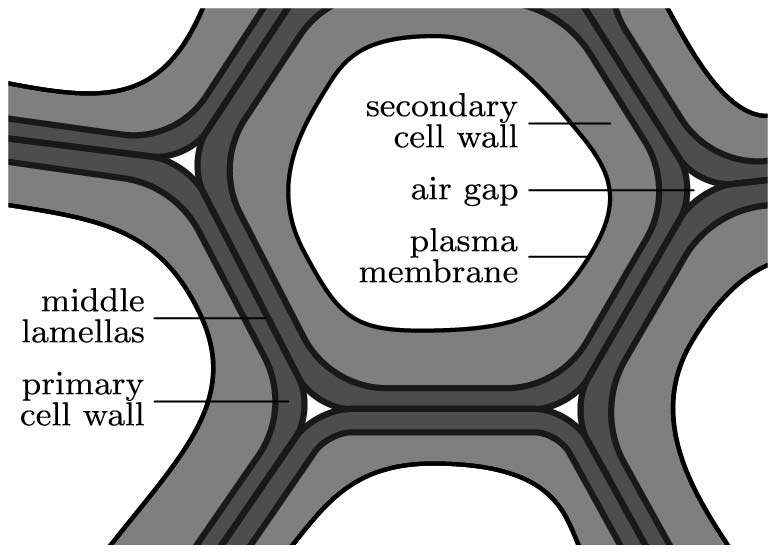}
                    \captionof{figure}{Neighbouring cells are glued together by middle lamellas where the air filled gaps at cell corners are known as intercellular spaces. Fully grown cells of woody plants develop a secondary cell wall that increases their load carrying capacity and resistance against pathogens at the cost of a reduced permeability.}
                    \label{pic:Figure_6_12}
                \end{center}}[-6mm]

                A cell wall consist of a primary cell wall that is sandwiched between the plasma membrane and the middle lamella. Some fully grown cells additionally possess a secondary cell wall that is located between the plasma membrane and the primary cell wall as illustrated in Figure~\ref{pic:Figure_6_12}. The primary cell wall is a thin, flexible and extensible layer that is formed during growth. It consists of a network of cellulose microfibrils that are connected by hemicellulose tethers. This network is embedded in a pectin matrix, a hydrated gel that pushes the microfibrils apart, thus easing their sideways motion \cite{Cosgrove2005-1}. For example, primary cell walls in dicotyledonous plants consist of approximately 35\% pectin, 30\% cellulose, 30\% hemicellulose and 5\% protein on a dry weight basis \cite{Fry1988-1}. The cellulose fibers possess a Young's modulus of about 130~GPa and a tensile strength of about 1~GPa \cite{Gibson2012-1}. In contrast, the Young's modulus of hemicellulose is between 5-8~GPa and thus considerably smaller. As a consequence, the mechanical properties of a primary cell wall depends mainly on the density and orientation of its microfibrils. Cellulose fibers are widely used in natural fiber composites as their mechanical properties and biodegradability make them a serious competitor to glass fiber reinforced plastics \cite{Pickering2016-1}.\\

                The plasma membrane is a semipermeable lipid bilayer that separates the cytoplasm from the cell wall. It has a thickness between 7-10~nm and forms through the process of self assembly. The lipid bilayer possesses a hydrophobic internal phase and a hydrophilic external phase on both sides. The hydrophobic phase creates a barrier between the interior and exterior of the cell that can not be easily passed by water whereas the hydrophilic surfaces on both sides are in contact with aqueous fluids. A plasma membrane possesses passive and active mechanisms that transport molecules in both directions so that it can influence the osmotic gradient. Furthermore, it contains sensors that receive signals from the environment and hormones that control the plants growth and cell division.\\

                Pectin based middle lamellas are the outmost cell wall layers that glue neighbouring cells together. The finite curvature of cell walls usually leads to air filled gaps at cell corners that are known as intercellular spaces. It is often difficult to distinguish between the middle lamellas and the primary cell walls so that they are often referred to as a compound middle lamella. The splitting of cells is accomplished by enzymes that dissolve the middle lamellas. This happens for example during the separation of petals and leaves.\\

                The secondary cell walls of woody plants are relatively thick layers between the primary cell walls and plasma membranes. They are formed in fully grown cells that need to carry large external loads so that they usually account for most of the biomass. Similar to primary cell walls, secondary cell walls contain cellulose, hemicellulose and pectin. However, their composition differs as their cellulose fraction and thus their strength, stiffness and durability is considerably higher. Furthermore, their permeability is lower so that they can efficiently protect a cell against pathogen attacks. Attributed to their large stiffness, secondary cell walls can carry tension as well as compression stresses. In contrast, primary cell walls carry mostly tension forces due to cell pressures. However, the large stiffness of secondary cell walls and their low permeability hinders nastic movements so that they are of little interest for the work that is presented in this thesis.


            \marginnote{
            \begin{center}
                \includegraphics[width=\marginparwidth]{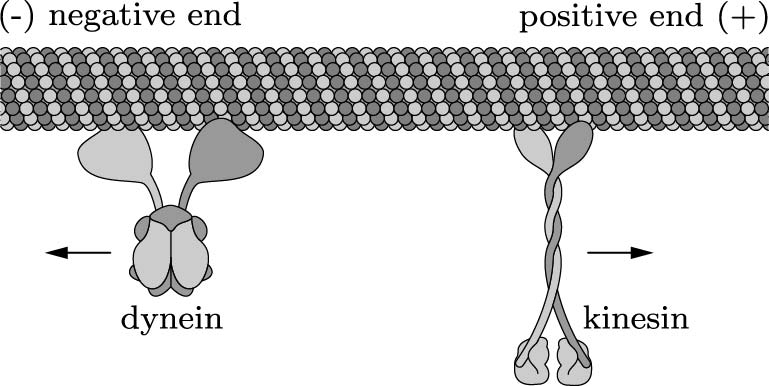}
                \small(a)\vspace{7mm}
                \includegraphics[width=\marginparwidth]{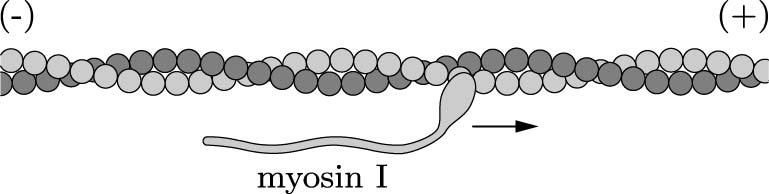}
                \small(b)\vspace{7mm}
                \includegraphics[width=\marginparwidth]{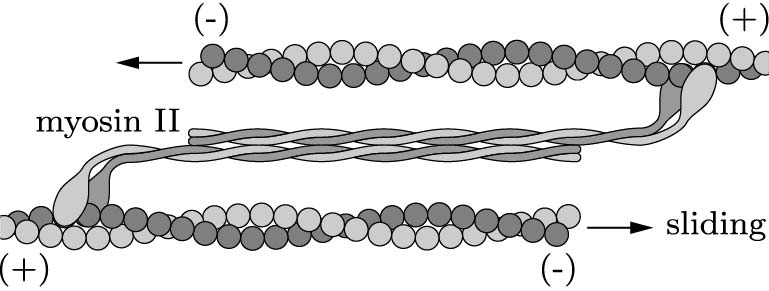}
                \small(c)\vspace{1mm}
                \captionof{figure}{(a) Microtubules and motor proteins for intracellular transports. Dynein transports towards the negative and kinesin towards the positive end. (b) Microfilaments can serve as tracks for myosin I motors. (c) Muscle like fibers that consist of myosin II motors and microfilaments are used to tension the cytoskeleton and to vary the cell shape.}
                \label{pic:Figure_6_13}
            \end{center}}[-92mm]

            \subsection{Cytoskeletons}
                Cytoskeletons are complex networks that can rapidly adapt their structure to varying cell requirements through rapid growth and disassemblies. They extend throughout the cytoplasm and support the cell shape, anchor subcellular components and participate in intracellular transports and communications \cite{Fletcher2010-1}.\\

                Cytoskeletons consist of interconnected microfilaments, intermediate filaments and microtubules that are anchored in the plasma membrane. Microtubules are highly dynamic, polar structures with an outer diameter of about 24~nm and an inner diameter of about 12~nm. They create a scaffold for the cell and serve as tracks for intracellular transports. These transports are accomplished by motor proteins that are powered by adenosine triphosphate. Kinesin transports towards the positive and dynein towards the negative pole of a microtubule as illustrated in Figure~\ref{pic:Figure_6_13}. Most microtubules possess a half life of five to ten minutes whereas others remain stable for hours. Microtubules grow faster at the positive and slower at the negative end. It is believed that their dynamic growth and disassembly is used to probe a cells three-dimensional space \cite{Kirschner1986-1}.\\

                Microfilaments are less dynamic than microtubules as they can grow, shrink or remain stable. They consist of two helical, interlaced strands that form a polarized structure with a diameter of about 7~nm. Much like microtubules, microfilaments grow faster at the positive and slower at the negative end. Different myosin motors that are powered by adenosine triphosphate can bind to microfilaments. The myosin I protein that moves towards the positive end of a microfilament is particularly used for transports in regions that are poor in microtubules. Muscle like fibers are created with the help of myosin II proteins that connect two or more microfilaments. Fiber contractions are accomplished by the relative sliding motions between them. The tension forces that result from these contractions are used to rigidify the cytoskeleton or to vary the cell shape. For example, a contracting ring of microfilaments is used to divide the cytoplasm into two parts during cell division.\\

                Intermediate filaments with a diameter of about 10-13~nm are thicker than microfilaments and thinner than microtubules. They are less dynamic and more elastic than microfilaments and microtubules. It is estimated that they can be stretched between three to five times their initial length. Elastic strains of these magnitude are enabled by a hierarchical structure where deformation mechanisms vary with strain levels. A common occurrence of intermediate filaments in plant cells has not been unequivocally demonstrated \cite{Kost2002-1} so that little is known about their function. However, it seems likely that they support the cytoskeletal structure.\\

                The involvement of cytoskeletons in the nastic movement of plants is mostly unknown \cite{Morillon2001-1}. It is known that local tensile stresses of up to 0.3~MPa can be exerted by muscle like fibers \cite{Bray2000-1} whereas turgor pressures that act on the whole cell wall can reach up to 5~MPa. This discrepancy makes it unlikely that microfilaments are actively involved in plant movements. However, it is shown in a subsequent chapter that subcellular structures with inherent mechanisms are capable of carrying a significant portion of cell pressures. This reduces the required cell wall thicknesses and thus the rigidity of cells. Furthermore, cytoskeletons might be used to constrain cell geometries prior to shape changes. This hypothesis is supported by the observations of Kanzawa et al \cite{Kanzawa2006-1}. They found that the highly organized cytoskeletons in motor cells of Mimosa pudica become fragmented after shape changes. Hence it seems that they support plant movements by a rapid disintegration.


            \marginnote{
            \begin{center}
                \includegraphics[width=\marginparwidth]{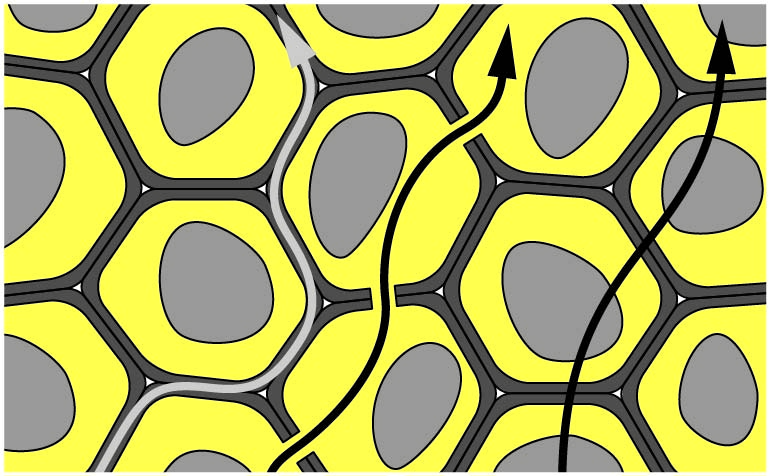}
                \small(a)\vspace{1mm}
                \includegraphics[width=\marginparwidth]{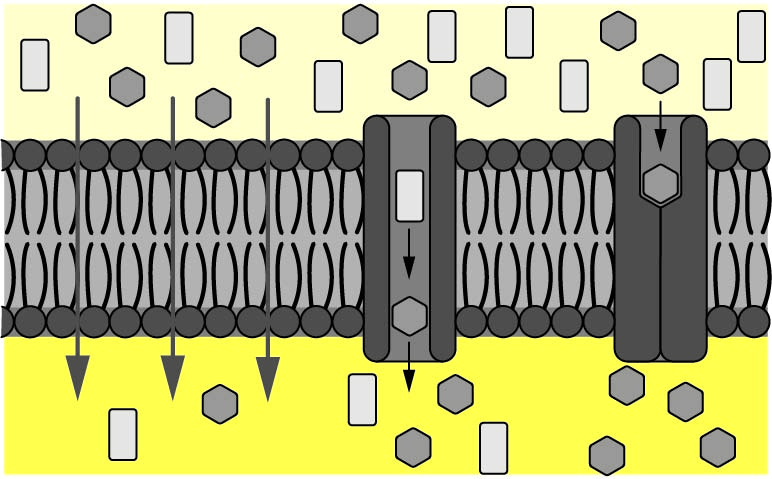}
                \small(b)\vspace{1mm}
                \includegraphics[width=\marginparwidth]{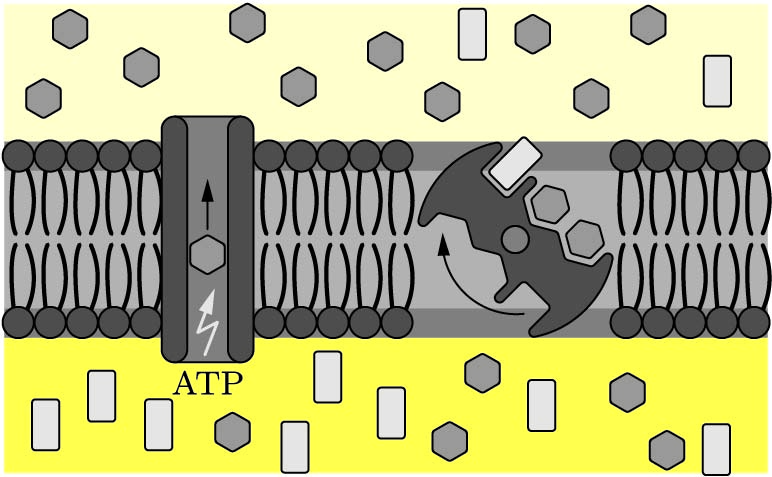}
                \small(c)\vspace{0mm}
                \captionof{figure}{(a) Pathways for water transport in plants. Apoplastic pathway through porous cell walls and intercellular spaces. Symplastic pathway through cytoplasm continuum via plasmodesmata. Transmembrane pathway through cell walls, cytoplasm and vacuoles. (b) Passive transport across plasma membrane through diffusion and non-specific, specific transporters. (c) Active transport against concentration gradients through primary and secondary transporters.}
                \label{pic:Figure_6_14}
            \end{center}}[-23mm]

            \subsection{Subcellular Motor}
                Plant movements are driven by cell pressures that can be as low as $-10$~MPa in Phlebodium aureum and as high as 5~MPa in guard cells \cite{Franks2001-1}. This is impressive as the hydraulic pressures of airplanes that were constructed in the 1940s\footnote{The need for weight savings led to hydraulic pressures that reached 21~MPa (3,000~psi) in the 1950s, 28~MPa (4,000~psi) in the 1970s and 34~MPa (5,000~psi) in the early 1980s. Pressures will most likely increase to 56~MPa (8,000~psi) in the near future.} are between 3.5-7~MPa. The osmotic hydration motors that generate these large pressures within each cell are of particular interest for engineering applications. Although their working principles are well known, a detailed understanding of their functionality remains elusive.\\

                As illustrated in Figure~\ref{pic:Figure_6_14}, the water transport in plants occurs through three different pathways \cite{Dumais2012-1}. The apoplastic pathway is a flow through cell walls and intercellular spaces. It depends on the cell wall porosity that is driven by the pectin deposition in the primary, secondary cell wall and the middle lamella. The symplastic pathway is a direct flow through the cytoplasm continuum via plasmodesmata that cross the cell walls of neighbouring cells. In contrast, the transcellular pathway is a flow through cell walls and plasma, vacuolar membranes. It needs to pass two plasma membranes at each crossing so that its solute conductivity is relatively low.\\

                Osmotic hydration motors are based on a semipermeable plasma membrane whose expansion is restrained by adjacent cell walls. A flow through the cell walls (apoplastic pathway) is mostly driven by pressure gradients as the cell wall continuum is permeable to most solutes. In contrast, a flow through the plasma membrane requires an osmotic gradient that is the difference in solute concentrations on both sides of the membrane. Water flows towards the solution with the higher concentration until an equilibrium with the pressure gradient is reached. The osmotic gradient is controlled by passive and active transports through the plasma membrane. Passive transports are directly driven by gradients and thus do not require any additional energy. They can be based on diffusion or non-specific and specific transporters such as aquaporins. Active transports against a gradient require additional energy. Primary active transporters use adenosine triphosphate as an energy source whereas secondary active transporters gain energy from the transport of two substances where one is moved against and the other in the direction of its osmotic gradient.


    \newpage

    \sectionmark{Summary}
    \begin{framed}
        \noindent \textbf{Summary}\\

            \noindent Despite their clear advantages, gapless high lift devices can not be found in the latest generation of commercial or military airplanes. Their lack of success can be explained by the weight penalty that mostly stems from an insufficient integration of actuators, mechanisms and structures. Inspiration for new, potentially superior approaches with different integration levels can be found in nature. Large animals and humans possess a nervous system, a skeleton and muscles whereas plants consist only of a large number of compliant, fluid filled cells that function without a central control system. Irrespective of their relative simplicity, plants are capable of a wide range of movements that vary with respect to their energy source, speed and repeatability.\\

            Plants that are capable of continuous, nastic movements are of particular interest for the design of gapless high lift devices. Their shape changes are hardwired into the cell geometries and driven by cell pressure variations. It is thus virtually impossible to distinguish between their actuators, mechanisms and structures. However, a detailed understanding of the interactions between cell geometries, materials and hydration motors remains elusive.\\

        \noindent \textbf{Conclusion}\\

            \noindent The general working principles of nastic plant movements are well known albeit a detailed understanding remains out of reach. Based on sufficiently large abstractions, a translation of these principles into engineering applications seems to be feasible within current technological boundaries. These abstractions can be based on the following observations.\\

            Cell wall materials with dynamically varying properties are not fundamental to nastic movements so that they can be substituted by common engineering materials such as fiber reinforced composites. Shape changing structures with developable surfaces can be realized with prismatic cells and thus do not require a large number of complexly shaped, three-dimensional cells. Individual cell pressure variations are not required so that groups of cells with identical pressures can be connected via pressure lines. This enables an external, outsourced pressure generation via conventional compressors. Based on these observations it is possible to create a surprisingly large number of relatively abstract engineering models that need to be evaluated for their potential.
    \end{framed}

        \cleardoublepage
    \chapter{Previous Work}
        Previous attempts towards gapless high lift devices were, despite a massive research and development effort, of little success. As it appears today, it is unlikely that this will change without a novel approach that significantly reduces their weight by a tighter integration of actuators, mechanisms and structures. A role model for such an integration can be found in plants that are capable of nastic movements. Their overall shape depends solely on the geometries and pressures of a large number of coupled cells so that it is virtually impossible to distinguish between their components.\\

        A translation of the working principles of plants into engineering applications is difficult and requires considerable abstractions. For example, cell wall materials with dynamically varying properties need to be substituted by engineering materials such as carbon fiber reinforced composites. The complexly shaped plant cells are commonly represented by prismatic cells with polygonal cross sections and the number of distinct cell pressures is reduced to only a few. The latter enables a connection between equally pressurized cells and the use of an external compressor instead of subcellular osmotic hydration motors.\\

        Various approaches that mimic at least some of the working principles of  plants were developed in fields that range from architecture\footnote{Particularly noteworthy in this regard is the work \cite{Bach1977-1,Bubner1976-1,Thywissen1979-1,Otto1995-1} by Frei Otto and coworkers at the University of Stuttgart in the 1970s. They investigated a wide range of pressurized cellular structures and their potential applications to architecture. Frei Otto is well known for his tensile and membrane structures that include the stadium roof for the 1972 Summer Olympics in Munich. He was awarded the Pritzker Architecture Prize in 2015 which is one of the highest honors in this field.} and civil engineering to aerospace and robotics. For example, many weight critical structures such as balloons, rockets and large stadium roofs can be considered to be pressure stiffened cells with extremely thin cell walls. However, this chapter reviews only structures that can significantly alter their shape via cell pressure variations. These structures are subsequently grouped, according to their shape changing capabilities, into beam, plate and shell like structures.


        \section{Beams}

            \blfootnote{\vspace{-5mm}
                \captionof{figure}{Various pressure actuated structures with uniform, prismatic cells that are capable of bending deformations. (a) Two separately pressurized cell rows are connected to a stiff membrane. The shape of this agonist-antagonist pair depends on the pressure ratio and its stiffness on the pressure magnitude. (b-d) Different assemblies of equally pressurized cells that are connected to an inextensional skin. Their shape depends on the skin's bending stiffness and pressure magnitude. (b) Structure that consists of a curved skin and a large number of layered cells (third layer is only partially shown). (c) Structure with a straight skin and a single layer of elastic cells with a rectangular cross section that contribute to the skins bending stiffness. (d) Similar approach with gaps that reduce the cell wall strains.}
                \label{pic:Figure_7_1}}
            \marginnote{
                \begin{center}
                    \includegraphics[width=\marginparwidth]{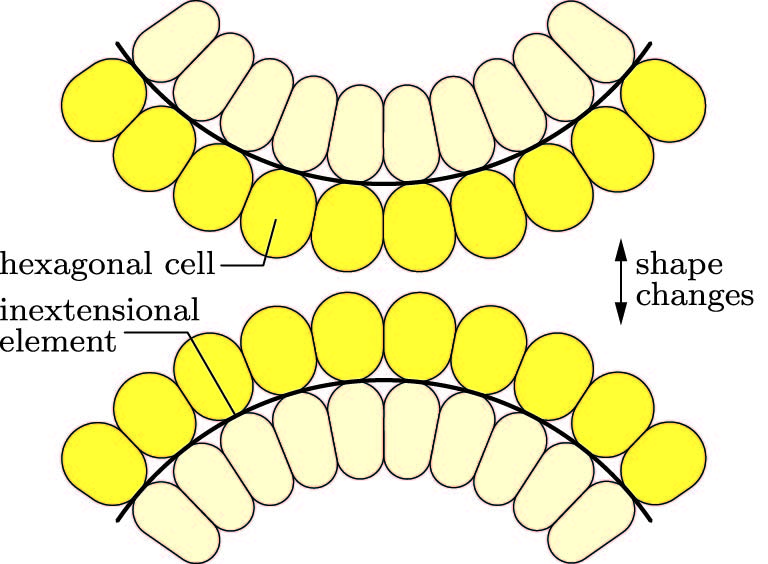}
                    \small(a)\vspace{7mm}
                    \includegraphics[width=\marginparwidth]{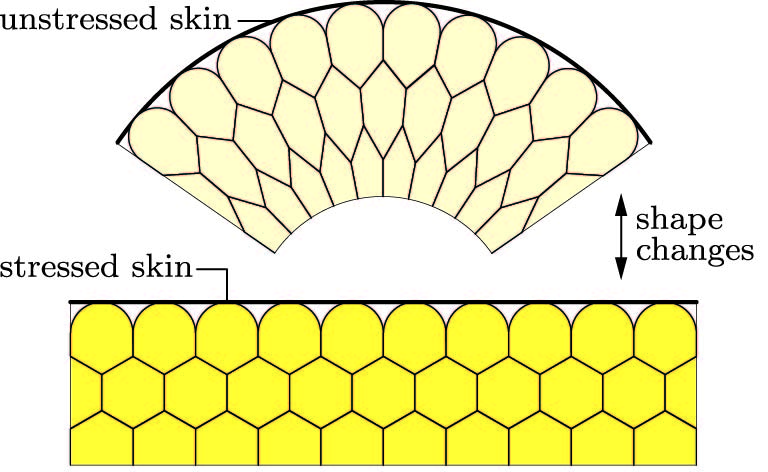}
                    \small(b)\vspace{7mm}
                    \includegraphics[width=\marginparwidth]{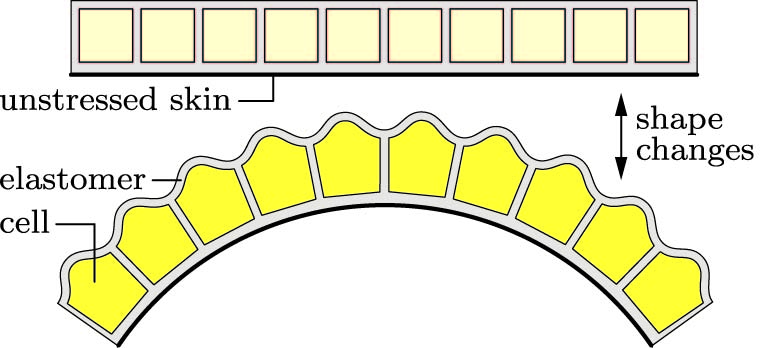}
                    \small(c)\vspace{7mm}
                    \includegraphics[width=\marginparwidth]{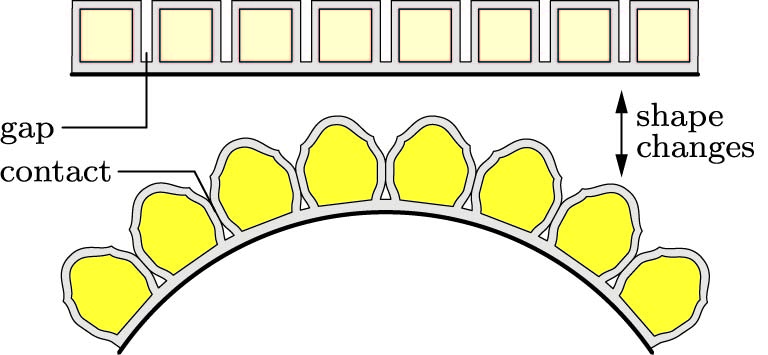}
                    \small(d)
                \end{center}}[-38mm]

            Cellular structures that can undergo large, uniaxial bending deformations can be split into two groups. The first group comprises structures with two or more cell rows that, although connected to each other, are independently pressurized. This enables a separate variation of their shape and stiffness. The deformability of these structures might be additionally constrained by an inextensional element as shown in Figure~\ref{pic:Figure_7_1}(a). The second group consists of structures that are based on equally pressurized cells and an energy storing element such as a flexible skin or compliant cells, Figure\ref{pic:Figure_7_1}(b-d). The shape of these structures depends on the stiffness of the energy storing element and the magnitude of the cell pressures. Hence, there exists a unique, invariant relationship between their shape and stiffness.\\

            A shape changing structure that consists of two independently pressurized rows of flexible tubes that are welded to both sides of an inextensional membrane \cite{Bubner1976-1} was proposed by Frei Otto in 1971\footnote{At least it was build that year by Wellesley-Miller at MIT.}. The tubes, as illustrated in Figure~\ref{pic:Figure_7_1}(a), are spaced such that the structure deforms into the desired shapes for given cell row pressures. The separate pressurization of cell rows has the advantage that its shape and stiffness can be independently varied. A nearly identical approach was patented about three decades later by Vos and Barrett \cite{Vos2010-1}. The main difference is the use of a flexible skin instead of a second, independently pressurized cell row as shown in Figure~\ref{pic:Figure_7_1}(b). As a consequence, these kind of structures are not capable of separately varying their shape and stiffness. Furthermore, Vos and Barrett proposed the use of a large number of equally pressurized cell layers. This increases the manufacturing complexity and leads to large deformations in the outer layers that prohibit the use of compliant cells.\\

            A similar concept that is based on a straight skin and a single cell row with rectangular cross sections \cite{Correll2010-1} was investigated by Correll et al in 2010. The use of rectangular instead of hexagonal cells is rather unusual and was motivated by the casting based manufacturing process. Rectangular cells that are made from rigid sides and frictionless hinges possess a single degree of freedom whereas pentagonal and hexagonal cells possess two and three degrees of freedom, respectively. Hence, rectangular cells that share a common side can only be used in shape changing structures if they are made from an elastic material that indirectly increases their degrees of freedom. However, this leads to relatively large strains as the cell geometry needs to be transformed during pressurization. An improved version with small gaps between neighbouring cells was subsequently developed by Polygerinos et al \cite{Polygerinos2013-1} in 2013. Nonetheless, it would be best to use cells with a tailored pentagonal or hexagonal geometry instead that can be constructed from rigid sides and frictionless hinges.\\

            Particularly interesting is the work by Conolly, Fisher and Neumark \cite{Bubner1976-1} that was undertaken in the 1970s. They proposed an airfoil that consists of a large number of independently pressurized cells where the laminarity of the airflow on the suction side is improved by blowouts. However, structures with a large number of independently pressurized cells require extensive piping which increases the overall weight and control complexity so that this approach is not of practical use.


        \section{Plates}
            Plate like structures differ from beams insofar that their in-plane dimensions are usually of the same order of magnitude. It is subsequently assumed that the plane strain compatibility condition \cite{Calladine1983-1} is satisfied so that they undergo only in-plane deformations. Plate like structures are thus of potential interest for applications that require continuous shape changes along a closed boundary. However, the same capability can be achieved by beam like structures where both ends are connected. This approach has the benefit that it frees up internal space that can be used to accommodate other components or active, passive mechanisms that support the boundary.\\

            The general design of plate like structures is, due to the increased coupling between cells, relatively difficult. This can be highlighted by the following thought experiment. As previously discussed, a beam can be used to create a closed boundary with desired shape changing capabilities. A transition from such a structure to a plate can be realized by an onion like arrangement of additional beams that are connected to each other and that do not affect the boundaries shape changing capabilities. The alternative to this approach is even harder and requires the search for feasible topologies of cell groups with identical pressures. Hence, relatively little work was published so far about shape changing plates as they do not provide a clear benefit with regards to beam like structures. Nonetheless, three relatively simple approaches are subsequently reviewed.\\

            \marginnote{
            \begin{center}
                \includegraphics[width=\marginparwidth]{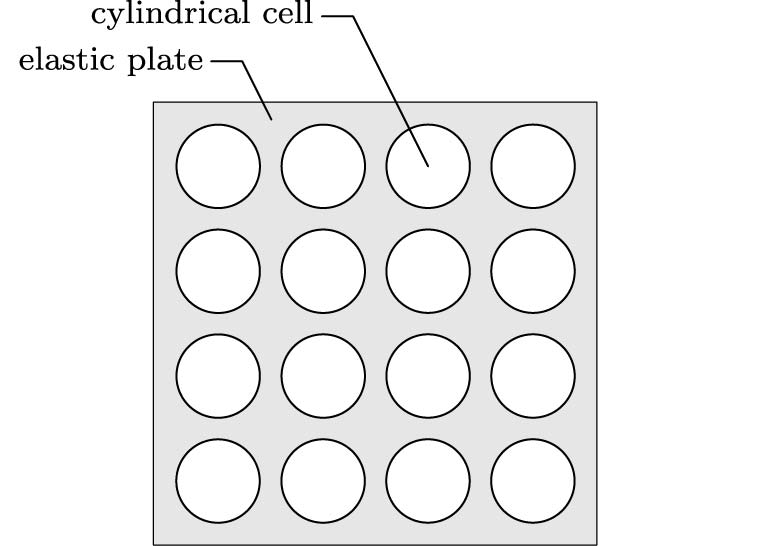}
                \small(a)\vspace{7mm}
                \includegraphics[width=\marginparwidth]{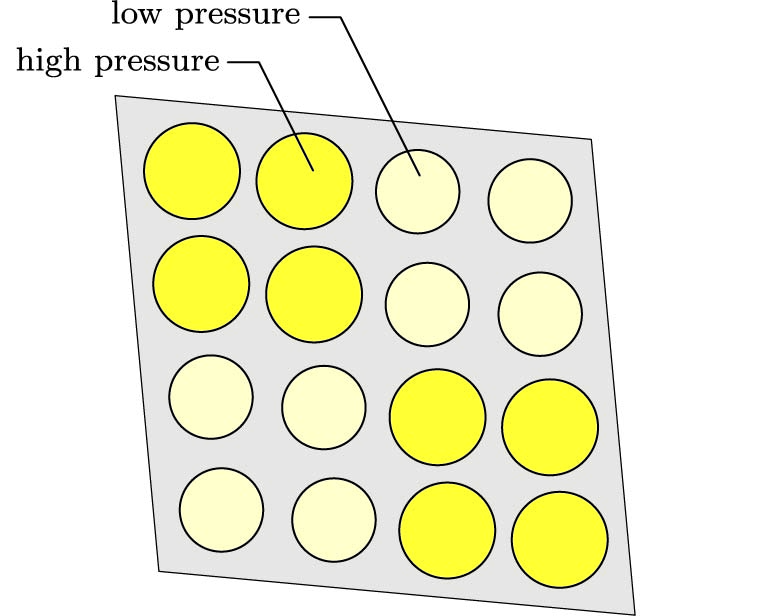}
                \small(b)\vspace{1mm}
                \captionof{figure}{(a) Pressure actuated structure that consists of a monolithic, elastic plate with a lattice of cylindrical cutouts. (b) Pressurization of one or more sets of cells leads to local in-plane deformations that accumulate throughout the plate.}
                \label{pic:Figure_7_2}
            \end{center}}[-57mm]

            A significant amount of conceptual and experimental work on pressure actuated plates was undertaken in the 1970s. For example, a single layer of connected and independently pressurized spherical cells was investigated by Prada and Aroca \cite{Bubner1976-1}. However, as for the related beam like structures, this approach requires extensive piping and complex control strategies so that it is not of practical use. Nonetheless, a similar approach was published by Matthews et al \cite{Matthews2006-1} in 2006. It is based on a monolithic, elastic plate with a lattice of cylindrical cutouts. Pressurization of one or more cells leads to local deformations that accumulate into global shape changes. The desired equilibrium shapes are not preprogrammed into the individual cell geometries so that it can be difficult, if not impossible to group the single cutouts into sets of equally pressurized cells. Furthermore, large material strains are another problem that limit potential shape changes so that this approach is merely an interesting thought experiment.\\

            \begin{figure}[htbp]
                \begin{center}
                    \subfloat[]{
                        \includegraphics[width=\textwidth]{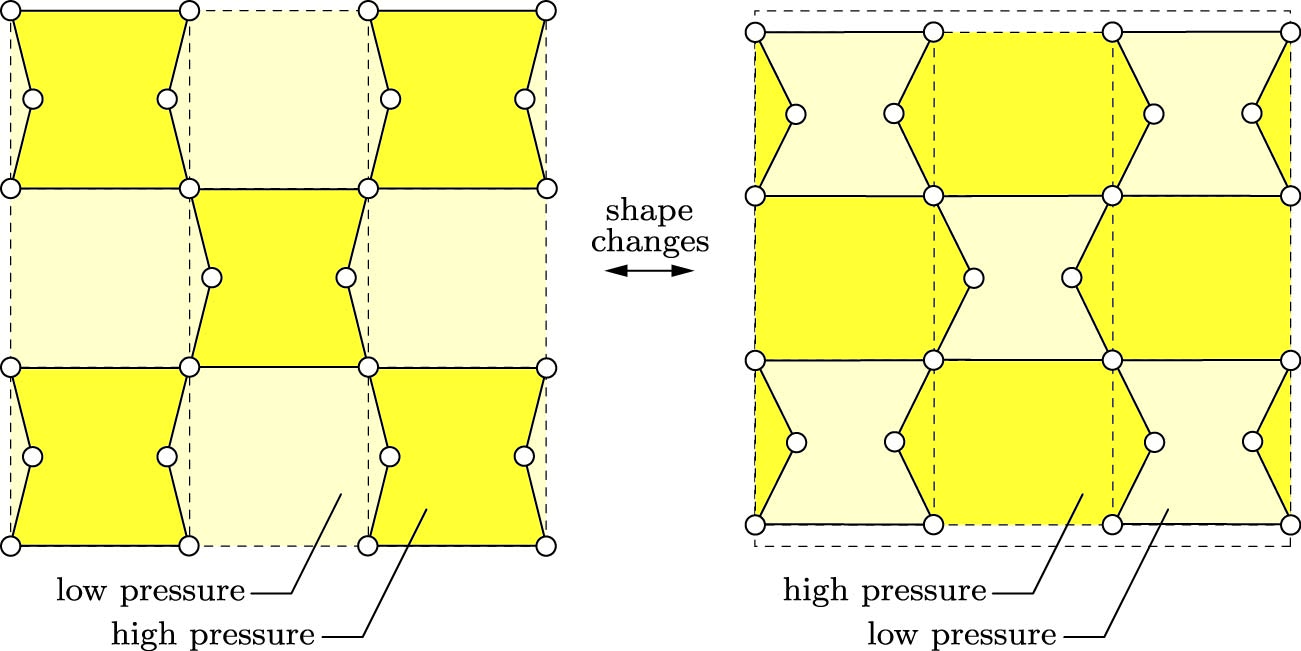}}
                        \vspace{5mm}

                    \subfloat[]{
                        \includegraphics[width=\textwidth]{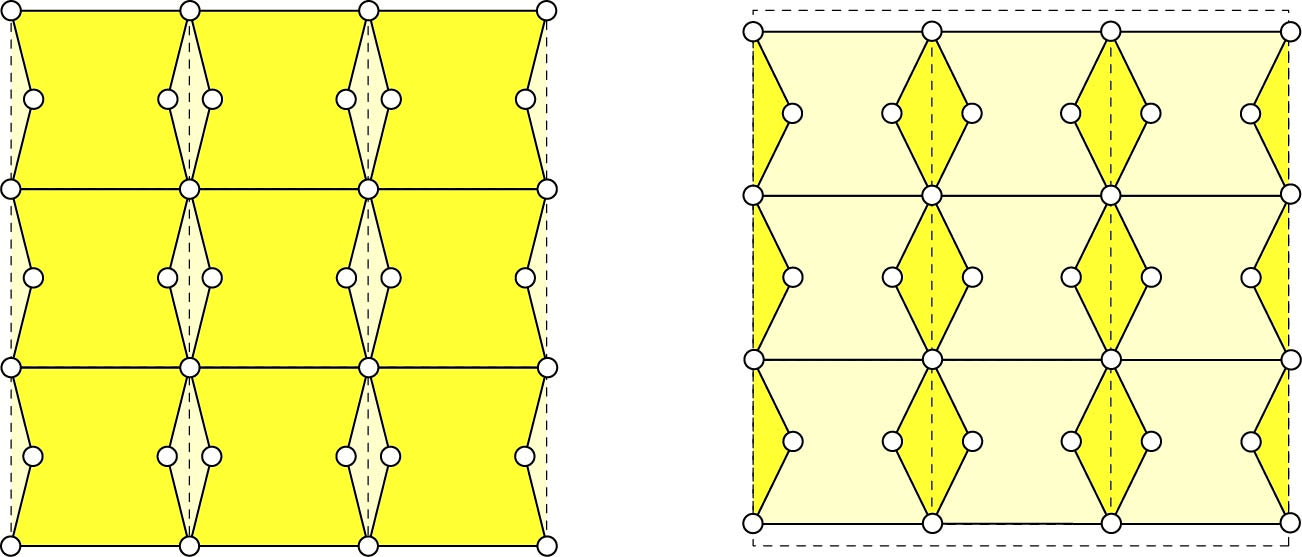}}
                        \vspace{5mm}

                    \subfloat[]{
                        \includegraphics[width=\textwidth]{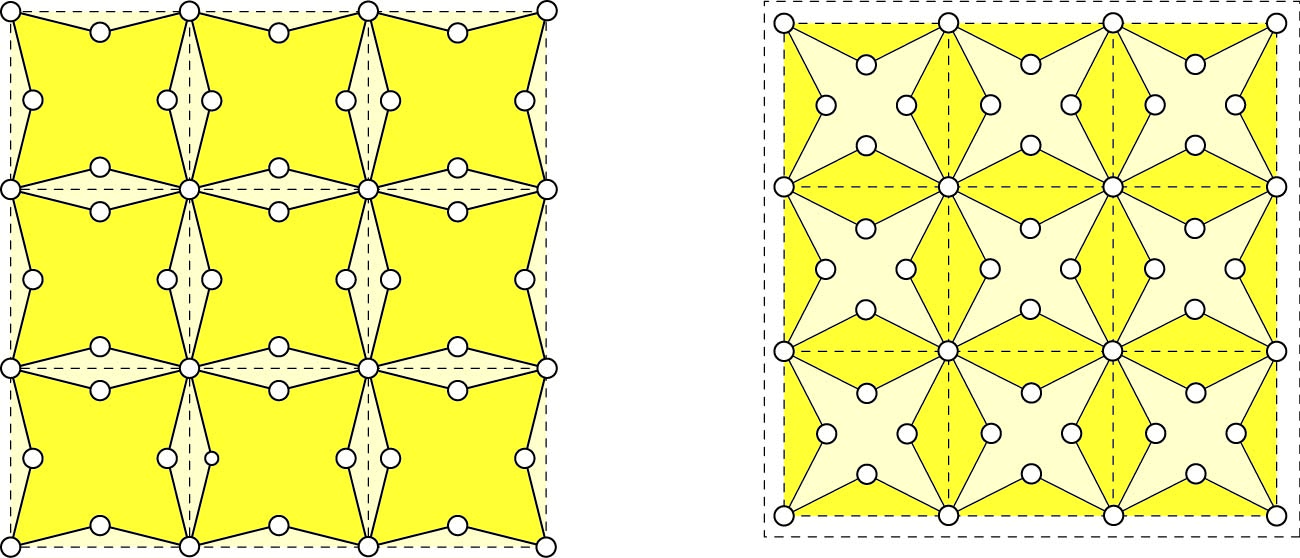}}
                \end{center}

                \marginnote{
                    \begin{center}
                        \captionof{figure}{Shape changing plate like structure that is generated from a quadrilateral tessellation of a plane. (a) Vertical cell sides are cut at their center and connected by a hinge. (b) Vertical cell sides are split along their axis, cut at their center and connected by two hinges. (c) All cell sides are split along their axis, cut at their center and connected by two hinges.}
                        \label{pic:Figure_7_3}
                    \end{center}}[-195mm]
            \end{figure}

            A concept for plate like structures that comprises different topologies and two sets of equally pressurized cells was patented by Dittrich \cite{Dittrich2003-1} in 2003. It is based on the regular tessellation of a plane where rigid cell sides are connected at cell corners via frictionless hinges. Some or all of the cell sides are then modified to increase the structures total number of degrees of freedom. This is done by cutting the sides at their center and splitting them along their axial direction. Three common combinations are
            \begin{enumerate}
                \item Vertical or horizontal sides are cut at their center...\vspace{-2mm}
                \item ...and additionally split in their axial direction.\vspace{-2mm}
                \item All sides are cut at their center and split in axial direction.
            \end{enumerate}
            as illustrated in Figure~\ref{pic:Figure_7_2} for a quadrilateral tessellation. These kind of structures can vary their stiffness and are capable of large, uniform shape changes so that they can be used instead of double acting hydraulic or pneumatic cylinders. However, this approach suffers from the occurrence of cell corners where only two cell sides meet. Their axial forces can vary greatly so that they are prone to snap through buckling. Hence, this concept is of little interest for practical applications.


        \section{Shells}
            The shape changing capabilities of plates can be fully described on the basis of their one-dimensional, closed boundary so that they can be substituted by beams. In contrast, shell like structures can additionally undergo out-of-plane deformations that violate the plane strain compatibility condition so that their shape changing capabilities need to be described with respect to their surface. This considerably complicates their design and explains the limited number of publications in this field. An early contribution to shape changing shells by Wellesley-Miller \cite{Bubner1976-1} dates back to the 1970s. He investigated structures that consist of three layers of connected, spherical cells. The central layer is constantly pressurized to provide a base stiffness whereas the remaining cells can be individually pressurized to vary the shells curvature. As for the equivalent beam and plate like structures, the required piping and control complexity is usually prohibitive.\\

            Pagitz and Bold \cite{Pagitz2012-2,Pagitz2013-1} developed a novel approach for shape changing shell like structures in 2012. It is based on two separately pressurized layers of prismatic cells with individually tailored hexagonal cross sections as illustrated in Figure~\ref{pic:Figure_7_5}~(a). Cell sides consist of bars that are connected at cell corners via frictionless hinges. Bars on the top and bottom surface are elastic whereas all other bars are rigid so that the undeformed structure possesses a mirror symmetry with regards to its geometric and mechanical properties. Cell layers are sealed at their boundaries by plates that possess a negligible in-plane and a large out-of-plane stiffness. As a consequence, the hexagonal plates between both layers can only change the angles between their sides whereas all other plates can additionally change the length of at least one of their sides.\\

            \afterpage{
                \begin{center}
                    \includegraphics[width=\textwidth]{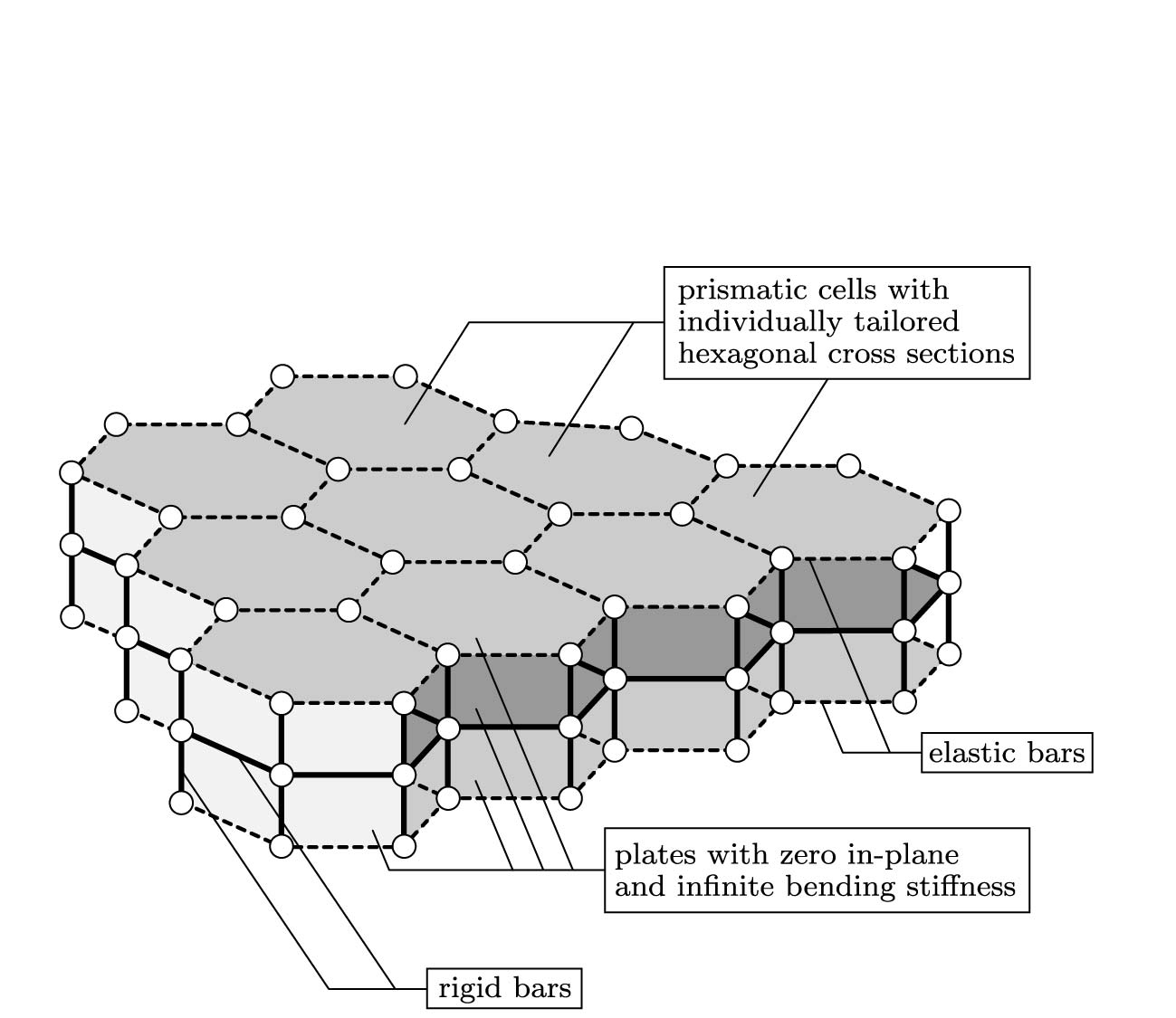}
                    \small(a)
                \end{center}
                \vspace{10mm}

                \blfootnote{\vspace{-3mm}
                    \captionof{figure}{Shape changing shell like structure consists of two separately pressurized layers of prismatic cells with individually tailored hexagonal cross sections. (a) Geometric and mechanical properties of the undeformed structure are mirror symmetric. The otherwise rigid cell sides are elastic on the top and bottom surface. Cell layers are sealed at their boundaries by plates with a zero in-plane and an infinite bending stiffness. The central hexagonal plates can thus vary the angles between their sides whereas the top and bottom plates can additionally vary their side lengths. (b) Hemispherical equilibrium shape of an annulus shaped shell after pressurization of the upper compartment. The hemisphere is hardwired into the structure as individual cell side lengths are tailored to maximize its potential energy. (c) Shell develops regions of negative curvatures if the pressure in the upper, lower layer is decreased, increased respectively. (d) Rapid shape changes into a hemisphere with opposite curvature are triggered upon reaching a critical pressure ratio between both layers.}
                    \label{pic:Figure_7_5}}
                \marginnote{
                    \begin{center}
                        \includegraphics[width=\marginparwidth]{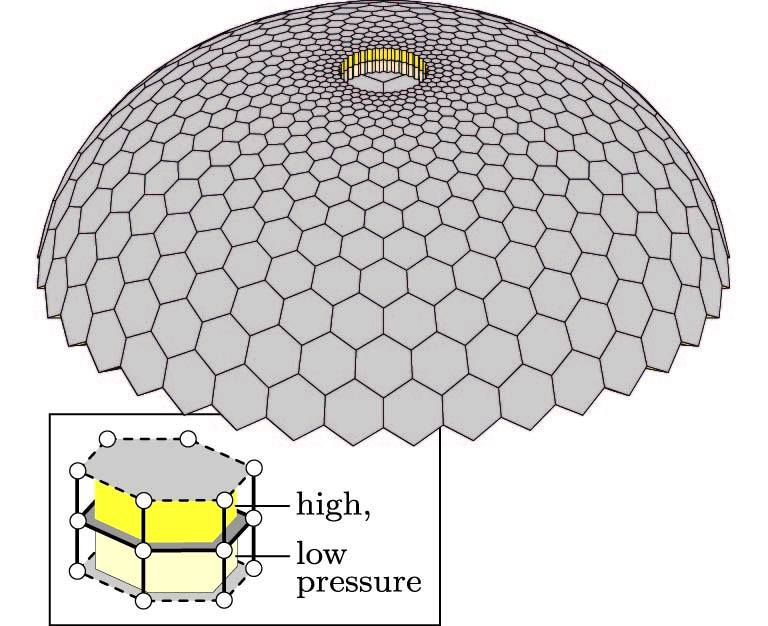}
                        \small(b)\vspace{11mm}
                        \includegraphics[width=\marginparwidth]{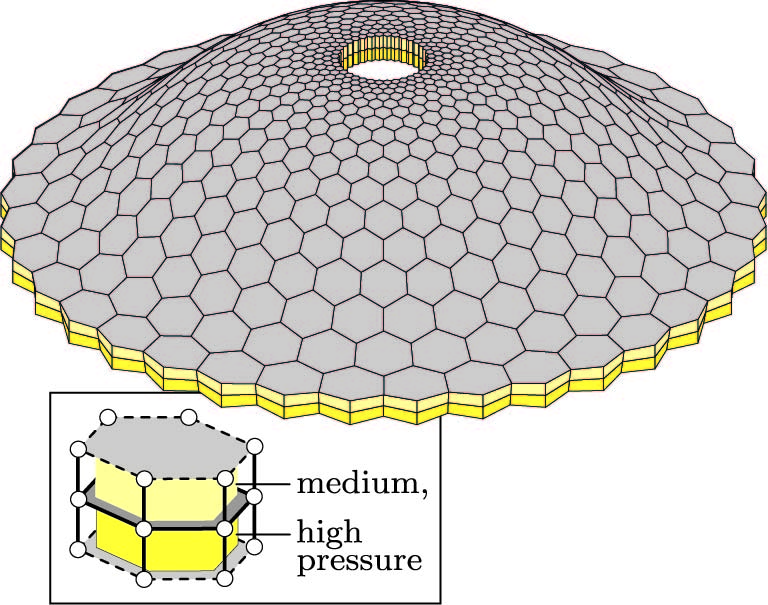}
                        \small(c)\vspace{11mm}
                        \includegraphics[width=\marginparwidth]{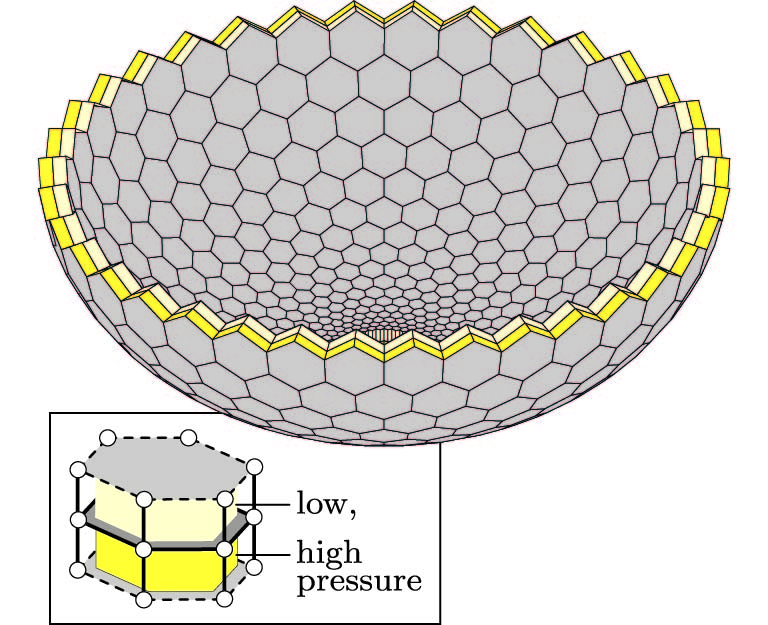}
                        \small(d)
                    \end{center}}[-121mm]
            }

            Two symmetric equilibrium configurations can be hardwired into the structure by tailoring the side lengths that define the cells cross sectional geometry. This can be done in a first approximation by maximizing the area of the central hexagonal plates upon deformation into the desired target shapes. A circular shell\footnote{The circular opening at its center is required as the assembly of hexagonal cells needs to satisfy the Euler characteristic \cite{Richeson2012-1}.} that is designed to change its shape between two hemispheres with an opposite curvature is shown in Figure~\ref{pic:Figure_7_5}~(b-d). It can be seen that the structure deforms into the desired target shapes upon pressurization of the upper or lower cell layer. Regions with opposite curvatures develop if the other layer is subsequently pressurized. Rapid shape changes that are driven by the release of strain energy occur if a critical pressure ratio between both layers is reached. This is remarkable as it shows that the fast nastic movements as known from the trapping leaves of Dionea muscipula can be triggered by a pressure release in a single layer of cells. Hence, these kind of plants do not necessarily have to overcome an energy barrier that would have to be sufficiently large to prevent unintended closures.\\

            This design approach can be extended towards structures with two or more layers of cells that are not necessarily prismatic. Another potential improvement is the fusion of shell and beam like structures into gridshells. However, the optimization and manufacturing of shape changing shell like structures is quite complex so that it is uncertain if sufficiently powerful tools and manufacturing techniques can be developed in the foreseeable future. Nonetheless, the corresponding work does not have to start from scratch as many of the underlying geometric problems are similar to those of freeform architectural designs that are based on hexagonal grids \cite{Pottmann2015-1}.


    \newpage

    \sectionmark{Summary}
    \begin{framed}
        \noindent \textbf{Summary}\\

        \noindent Plants that are capable of predetermined, repetitive shape changes exhibit an unparalleled integration of actuators, mechanisms and structures. Despite their attractive properties, relatively little work can be found in the literature that mimics these movements. Corresponding publications can be grouped according to the structural dimensions and shape changing capabilities into beam, plate and shell like structures.\\

        Previous work on beams focuses on structures that consist of one or more layers of prismatic cells with an identical cross sectional geometry. Their shape and stiffness are either controlled by an energy storing element or the interactions between separately pressurized cell groups. However, none of the existing approaches enables the design of beams with a smooth surface that can vary their shape between a given set of continuous target shapes by separately adjusting the pressures of a minimal number of cell groups.\\

        The design and manufacturing complexity of shape changing plate and shell like structures exceeds those of beams. This is mostly due to their increased number of relatively compact cells and the tight coupling between them. It is thus not surprising that the literature in this field is relatively sparse and more of a conceptual nature.\\

        \noindent \textbf{Conclusion}\\

        \noindent The state of the art of plant inspired shape changing structures is, despite a considerable research and development effort, surprisingly lacking. This is astonishing since many engineering applications would benefit from relatively simple, beam like structures. The remainder of this thesis focuses on the development of a novel concept for plant inspired, shape changing beams that fill this gap.
    \end{framed}

    \newpage
    \thispagestyle{empty} 
        \cleardoublepage
    \chapter{Novel Approach}
        Existing approaches that mimic the working principles of nastic plants are poorly conceived and usually not applicable to the design of engineering structures. This becomes apparent if one considers the corresponding plate and shell like structures whose design is undoubtedly more complex than that of beams. The realization of gapless high lift devices requires beam like structures with a smooth surface and cell geometries that are tailored for an arbitrary set of continuous target shapes. In order to reduce the overall complexity, their shape and stiffness needs to be fully controllable by varying the pressures of a small number of cell groups. A novel approach towards pressure actuated cellular structures \cite{Pagitz2012-1,Pagitz2012-3} that satisfies these requirements was first proposed by Pagitz et al in 2011. The main ideas behind this approach are subsequently presented where a particular focus is put on the following considerations:\\

        $\bullet$~The cell topology of pressure actuated cellular structures needs to be chosen such that their total perimeter is small and their number of inherent degrees of freedom is large. $\bullet$~Groups of equally pressurized cells must be defined that enable large shape changes and reduce the required plumping. $\bullet$~The usefulness of these structures depends to a large degree on their shape changing capabilities, load carrying capacity and weight. A basic overview of these properties is thus indispensable. $\bullet$~Pressure actuated cellular structures can be manufactured from a single piece of material with the help of compliant hinges that are located around cell corners. This introduces a dependency between cell sizes, pressures and material properties that needs to be taken into account. $\bullet$~Prismatic cells can be reinforced by subcellular structures and sealed by compliant end caps without significantly affecting their shape changing capabilities.


        \section{Working Principle}
            \subsection{Topologies}
                Pressure actuated cellular structures need to be lightweight and capable of a wide range of shape changes. This can only be achieved if their total perimeter is small and their inherent number of degrees of freedom large. It is well known that a monohedral tiling of a plane with convex polygons is possible for $n_v \in \left[3,4,5,6\right]$ where $n_v$ is the number of sides. Arbitrary polygonal geometries can be used for $n_v \in \left[3,4\right]$ whereas only fifteen \cite{Rao2017-1} polygon types can be used for $n_v=5$ and three for $n_v=6$. Furthermore, only eight \cite{Bagina2011-1} out of the fifteen pentagonal types are edge-to-edge and thus of potential interest for the use in pressure actuated cellular structures.\\

                \marginnote{
                \begin{center}
                    \includegraphics[width=\marginparwidth]{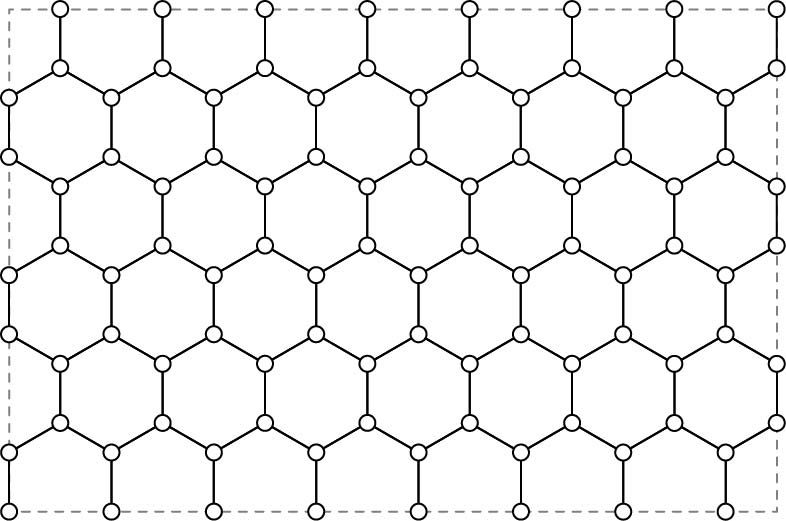}
                    \small(a)\vspace{9mm}
                    \includegraphics[width=\marginparwidth]{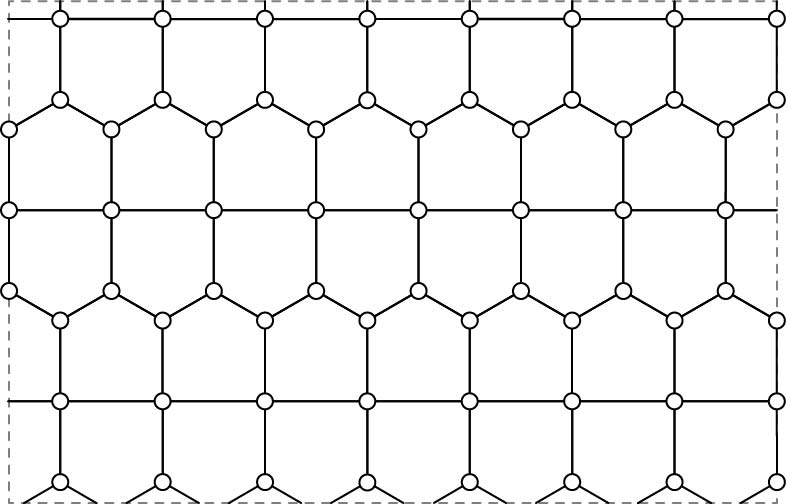}
                    \small(b)\vspace{9mm}
                    \includegraphics[width=\marginparwidth]{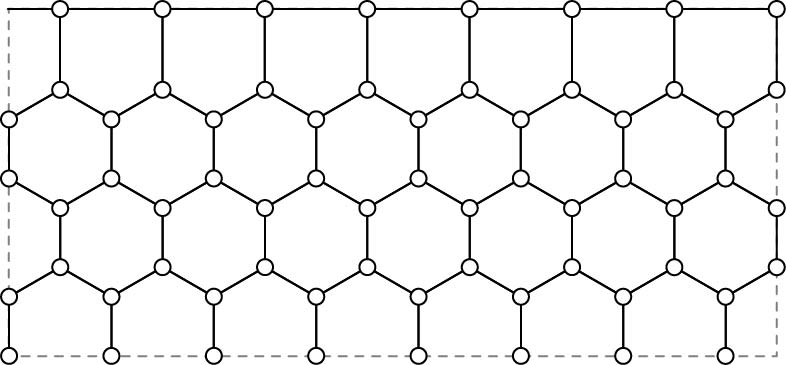}
                    \small(c)
                    \captionof{figure}{Division of a plane into regions of equal area with the least total perimeter. Monohedral division with (a) hexagonal and (b) pentagonal tiles. (c) Dihedral division of a half plane with pentagonal and hexagonal tiles.}
                    \label{pic:Figure_8_1}
                \end{center}}[-53mm]

                It was speculated for a very long time that a plane is divided by regular hexagons into regions of equal area with the least total perimeter, Figure~\ref{pic:Figure_8_1}~(a). However, it took more than two millennia until this conjecture was proven \cite{Hales2001-1} by Thomas Hales in 1999. Chung et al \cite{Chung2012-1} showed in 2012 that the prismatic pentagonal tiling, as shown in Figure~\ref{pic:Figure_8_1}~(b), possesses the least total perimeter of all pentagonal tilings. Although not unique\footnote{The Cairo pentagonal tile possesses an identical perimeter for an equal area.}, it is of particular interest as it possesses straight, intercellular lines. This is important as applications such as gapless high lift devices require smooth boundaries. A solution for the division of a half plane into regions of equal area with the least total perimeter is thus most likely based on regular hexagonal tiles that are terminated at the boundary by a row of pentagonal tiles as shown in Figure~\ref{pic:Figure_8_1}~(c).\\

                Another important aspect is the total number of degrees of freedom. A polygon that consists of $n_v$ rigid sides that are connected at vertices via frictionless hinges possesses $n_v-3$ intrinsic degrees of freedom. Hence it would be best to use only cells with a hexagonal cross section. If this is not possible it is advantageous to minimize the number of additional pentagonal cells. This point of view leads to the same topology for a half plane as previous considerations. Hence, the topology of a half plane that is based on hexagonal cells that are terminated by pentagonal cells at the boundary is subsequently used as a basis for the development of pressure actuated cellular structures.


            \subsection{Cell Groups}
                The simplest way to minimize the piping of a pressure actuated cellular structure is to separately pressurize its cell rows. The required number of rows is thus identical to the number of target shapes that need to be considered. This is highlighted in the following by a structure that consists of two rows of pentagonal and hexagonal cells with identical cross sectional geometries. Each row can be separately assembled from rigid sides and frictionless hinges as illustrated in Figure~\ref{pic:Figure_8_2}. Their pressurized shape is a circular arc with a radius that solely depends on the cell side lengths. Both cell rows can be connected if the opposite cell sides are of equal length. This leads to a structure that can independently vary its stiffness and shape within the bounds that are defined by the individual cell geometries and admissible cell row pressures.\\

                \blfootnote{\vspace{-5mm}
                    \captionof{figure}{Working principle of pressure actuated cellular structures. (a) Two separate rows of identical pentagonal and hexagonal cells are assembled from rigid sides and frictionless hinges. (b) Cell rows deform into circular arcs upon pressurization. Their radii depends solely on the cell geometries. (c) Cell rows can be connected if opposite sides are of equal length. (d) The shape of a structure that consists of two cell rows depends on the pressure ratio between both rows and its stiffness on the pressure magnitudes.}
                    \label{pic:Figure_8_2}}
                \marginnote{
                    \begin{center}
                        \includegraphics[width=\marginparwidth]{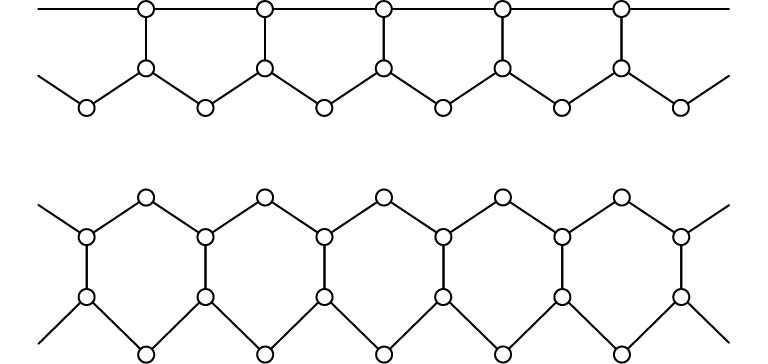}
                        \small(a)\vspace{10mm}
                        \includegraphics[width=\marginparwidth]{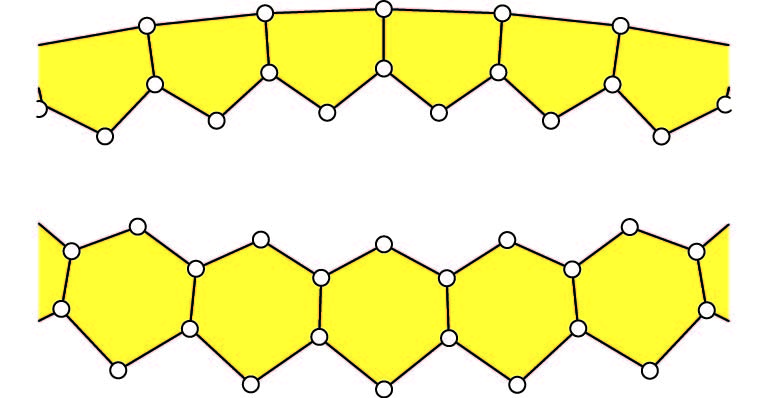}
                        \small(b)\vspace{10mm}
                        \includegraphics[width=\marginparwidth]{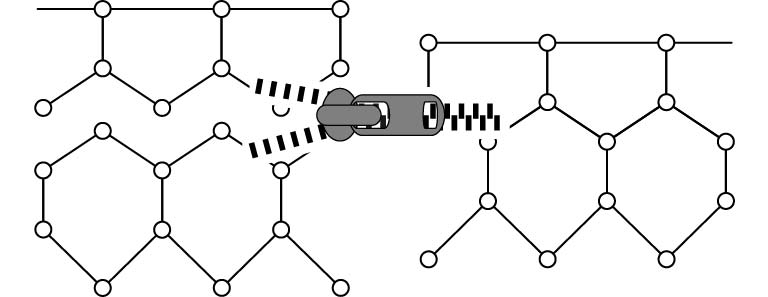}
                        \small(c)\vspace{10mm}
                        \includegraphics[width=\marginparwidth]{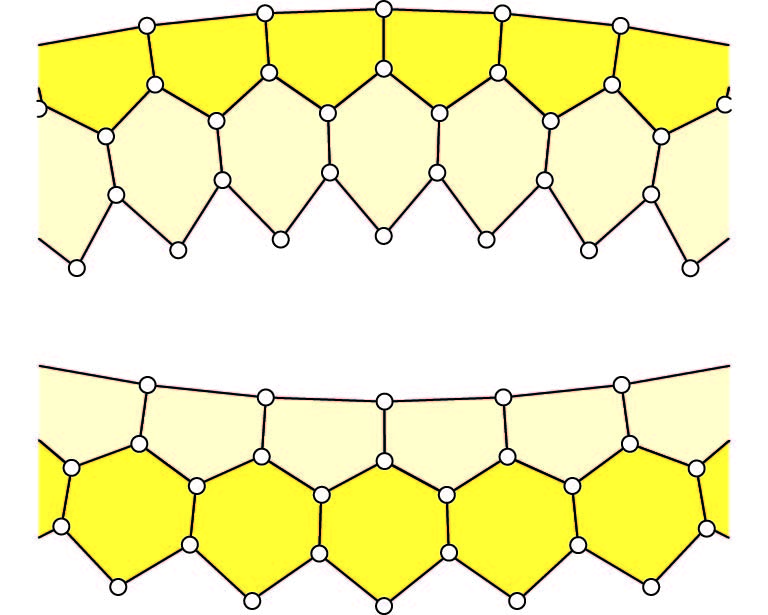}
                        \small(d)
                    \end{center}}[-45mm]
                    \vspace{-3mm}

                Equilibrium shapes for given cell row pressures can be hardwired into the cell geometries of pressure actuated cellular structures. An efficient algorithm for this is presented in the following three chapters. The remainder of this chapter focuses on their basic properties that can be demonstrated with rows of identical cells.


        \section{Potential}
            \subsection{Load Carrying Capacity}
                The potential of pressure actuated cellular structures depends on their shape changing capabilities, stiffness and weight. A one meter long cantilever as shown in Figure~\ref{pic:Figure_8_3}~(a) is subsequently used to provide a first insight. It consists of two cell rows that are assembled from pentagonal and hexagonal cells with identical geometries. The inherent cell stiffness is negligible as they consist of rigid sides and frictionless hinges. The cell corner movements at both ends are constrained by supports or a rigid endplate such that edge effects are eliminated. The structure thus deforms into a circular arc upon pressurization. Cell row pressures of either 0.4 or 2.0~MPa lead to subtended angles of about $60^\circ$ and $110^\circ$ which corresponds to a shape changing capability of $50^\circ/$m. Angle changes between adjacent pentagonal cells are thus about $5^\circ$.\\

                The equilibrium shapes of the cantilever for tip loads of 5 and 10~kN/m are shown in Figure~\ref{pic:Figure_8_3}~(b-c) where the unit ``kN/m" refers to a depth of one meter. It can be seen that the cantilever is capable of moving a tip load of 10~kN/m over a vertical distance that nearly equals its own length. A remarkable property of pressure actuated cellular structures is that their cell side forces are mostly driven by cell pressures. This makes them highly resilient against unexpectedly large loads and deformations. For example, the maximum axial cell side force for all load cases is less than 250~kN/m. The required cell side thickness, due to the maximum normal force, is therefore less than half a millimeter if modern materials such as carbon fiber reinforced polymers are used. The nodal coordinates and cell side forces of the equilibrium shapes for a tip load of 5~kN/m can be found in the appendix. This enables a verification of the presented results via simple hand calculations.\\

                \blfootnote{\vspace{-3mm}
                    \captionof{figure}{Shape changing capability and load carrying capacity of a pressure actuated cellular structure. The one meter long cantilever consists of two cell rows that are assembled from identical pentagonal and hexagonal cells. Cell sides are rigid and connected at cell corners via frictionless hinges. Edge effects are eliminated by constraining the displacements of end nodes to a common plane so that the unloaded structure deforms into a circular arc upon pressurization. Cell row pressures are either $0.4$ or $2.0$~MPa. Equilibrium shapes of (a) an unloaded and (b-c) loaded cantilever with a tip load of (b) 5 and (c) 10~kN/m.}
                    \label{pic:Figure_8_3}}
                \marginnote{
                    \begin{center}
                        \includegraphics[width=\marginparwidth]{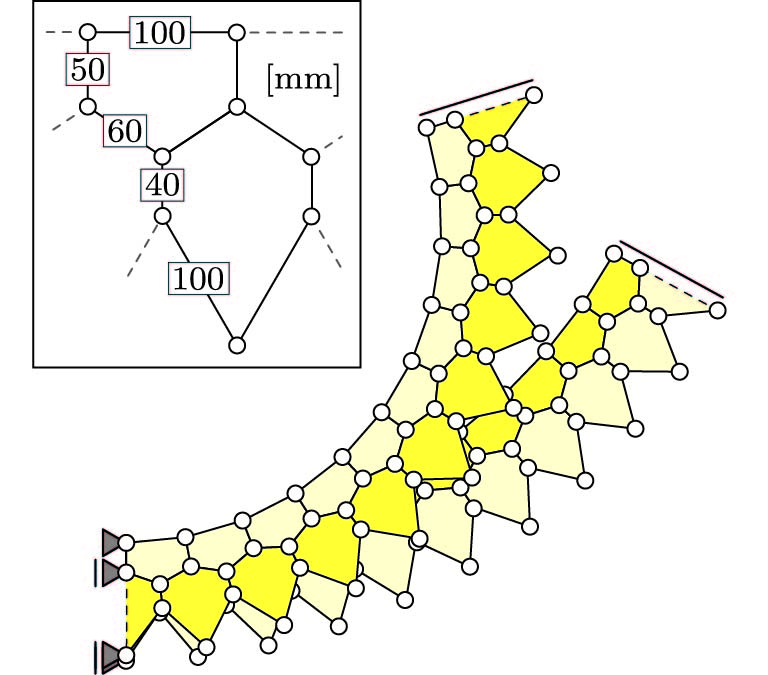}
                        \small(a)\vspace{2mm}
                        \includegraphics[width=\marginparwidth]{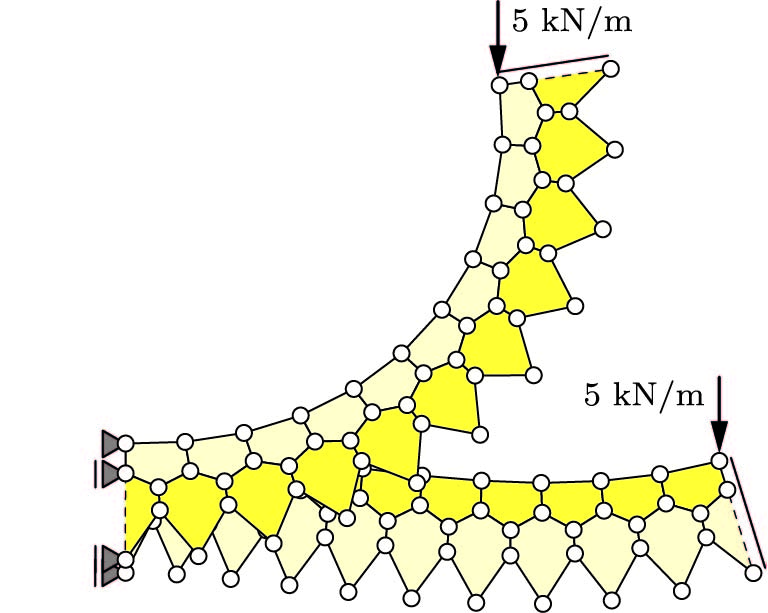}
                        \small(b)\vspace{2mm}
                        \includegraphics[width=\marginparwidth]{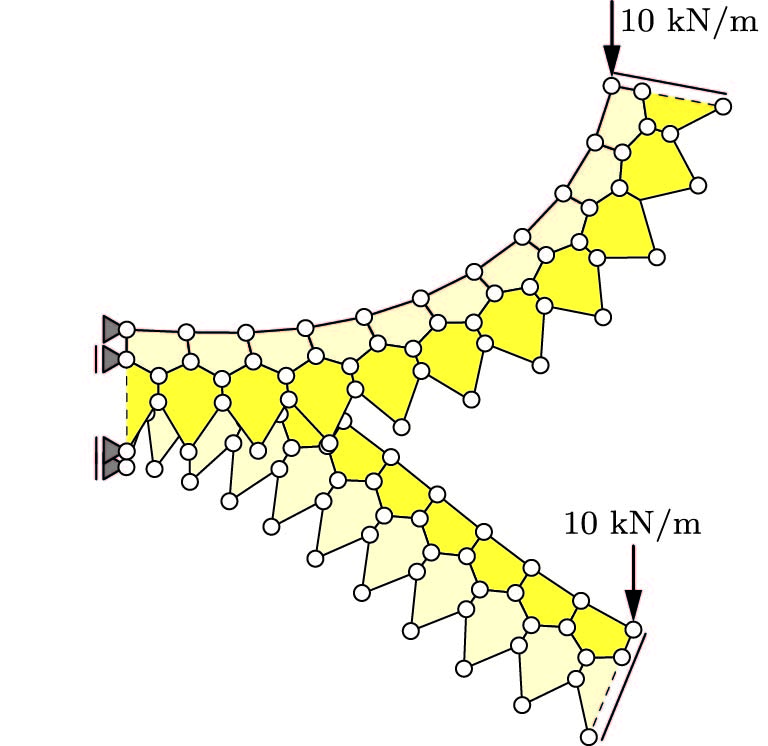}
                        \small(c)
                    \end{center}}[-52mm]
                \vspace{-3mm}

                Cell sides that are located at the boundary or between cell rows are subjected to differential pressures. These additional loads lead to bending moments that considerably increase the required cell side thicknesses and thus the overall weight. Bending moments can be minimized with the help of subcellular structures or efficiently carried by cell sides with a variable thickness. Furthermore, the stiffness and load carrying capacity of the cantilever increases linearly and its weight sublinearly with increasing cell row pressures. This is due to the superlinear relationship between the bending stiffness and cell side thicknesses. Commonly used pressures of pneumatic systems reach values of up to 5~MPa so that the chosen cell row pressures for this example are relatively moderate. A preliminary sizing of the cantilever indicates that a selfweight of as low as 10~kg/m$^2$ seems to be possible.


            \subsection{Shape Changing Capabilities}
                \subsubsection{Considerations}
                    Trusses and pressure actuated cellular structures have much in common as they both consist of cells that can be pressurized to increase their stiffness. Hence, they differ only with respect to their inherent degrees of freedom that depend on their topology. It is generally pointless to use pentagonal and hexagonal cells in regions of pressure actuated cellular structures that need to undergo only small shape changes. A tight integration of triangular and pentagonal, hexagonal cells can thus be advantageous. A reversible transition between different cell topologies can be achieved with the help of cytoskeletons. These subcellular structures are relatively thin so that they can carry large tension but only minor compression forces. They can thus be used to fully or partially subdivide a cell into triangles upon reaching limit deformations that put them into tension \cite{Pagitz2014-1}.\\

                    \marginnote{
                    \begin{center}
                        \includegraphics[width=\marginparwidth]{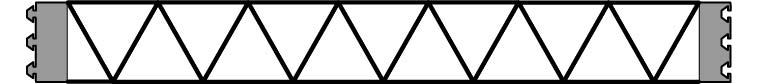}
                        \small(a)\vspace{10mm}
                        \includegraphics[width=\marginparwidth]{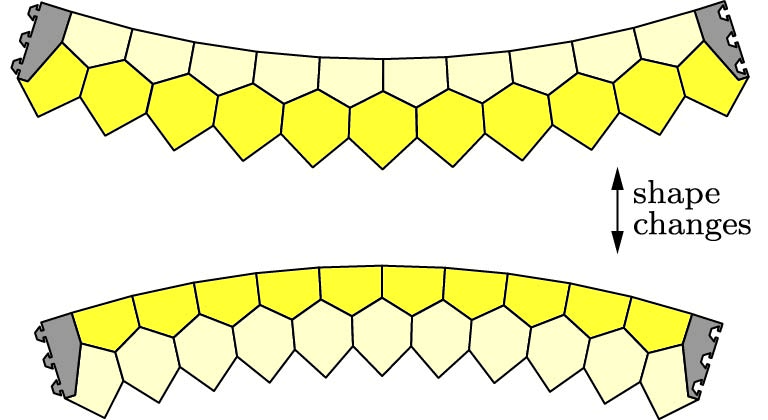}
                        \small(b)\vspace{10mm}
                        \includegraphics[width=\marginparwidth]{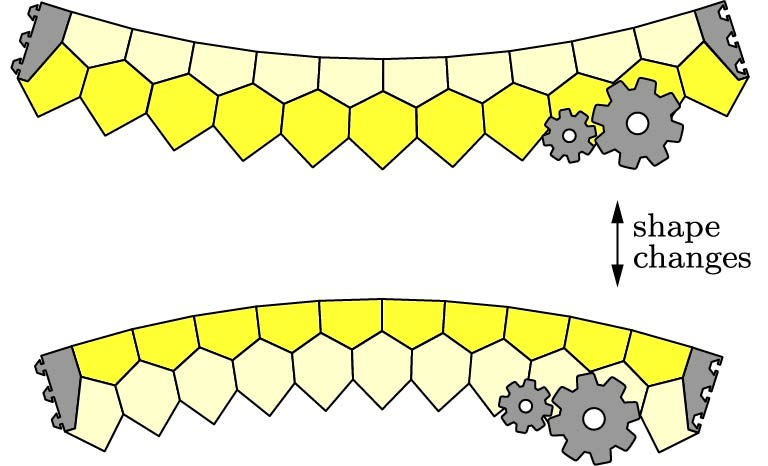}
                        \small(c)
                        \captionof{figure}{Three kind of modules with an increasing shape changing capability and weight. Modules possess connectors at both ends so that they can be arbitrarily combined. (a) Rigid module that is based on a truss structure. (b) Pressure actuated cellular module whose equilibrium shapes resemble circular arcs. Its potential shape changing capabilities are limited by stiffness requirements. (c) A mechanical module that combines a pressure actuated cellular structure and a passive or active mechanisms to increase its shape changing capabilities and stiffness.}
                        \label{pic:Figure_8_4}
                    \end{center}}[-28mm]

                    Potential shape changing capabilities of pressure actuated cellular structures increase with the number of pentagonal cells whereas their stiffness decreases. The achievable change of angle between adjacent pentagonal base sides is limited so that structures with demanding stiffness and shape changing requirements can not be realized with pressure actuated cellular structures. This problem can be overcome with the help of additional supports that are provided by passive or active mechanisms. As a consequence, depending on the required shape changing capabilities and stiffness, there exists a smooth transition between trusses, pressure actuated cellular structures and flexible skins that are supported by active mechanisms. It is remarkable that classical high lift devices are based on rigid structures whereas gapless high lift devices are commonly based on flexible skins and active mechanisms. However, both approaches are rather extreme and most likely not optimal. Pressure actuated cellular structures can be used instead to create a wide range of intermediate solutions that might even function without additional mechanisms.\\

                    A detailed investigation of pressure actuated cellular structures with different cell topologies, cytoskeletons and rigid body mechanisms is beyond the scope of this thesis. However, three kind of modules with an increasing shape changing capability as illustrated in Figure~\ref{pic:Figure_8_4} are subsequently used to outline the potential of an integral approach. These modules possess connectors at both ends so that they can be arbitrarily combined. Furthermore, their outer surfaces resemble circular arcs with a constant arc-length and a constant or variable radius. Rigid modules are based on truss structures so that their surface shapes are invariant. In contrast, the shape changing capabilities of modules that are based on pressure actuated cellular structures are only limited by stiffness constraints. These constraint can be overcome with the help of passive or active mechanisms that provide additional supports at the cost of an increased weight, complexity and construction space. Similar to Lego bricks \cite{Bartneck2018-1}, modules can be manufactured with different, standardized lengths, surface radii and shape changing capabilities. Their equilibrium shapes can then be adjusted within these bounds by cell pressure variations.\\

                    Different one-dimensional geometries that optimize certain properties for a given set of environmental conditions can be found in the literature. For example, optimal airfoil geometries for high lift and flight configurations are well known. Their geometries can be broken down into $C^1$ continuous sequences of circular arcs with different radii and arc-lengths. The properties of these arcs can be optimized such that they approximate the airfoil geometries in a least square sense. This requires a discontinuous optimization approach as the arc-lengths need to be chosen from a set of admissible values. The optimization results can then be used to assemble the corresponding structure with the help of shape changing modules. This is subsequently demonstrated by means of two examples.


                \subsubsection{Car Seat}
                    The first example deals with a passenger seat that can adapt to persons with a body height between 1.7-1.9~m. Boundary geometries of humans with different sizes and postures are derived from a model where the head, torso and extremities are assumed to be rigid bodies that are connected via hinges. Details about the dummy can be found in the appendix. Nine different boundary geometries are taken into account that correspond to 1.7, 1.8 and 1.9~m tall persons with a torso inclination of 14, 24 and $34^\circ$. The resulting curves are interpolated by four modules with arc-lengths $\mathbf{L}$ that are chosen from the set $L\in\left[5,10,\ldots,80\right]$~cm. The optimized arc-lengths of the modules from top to bottom are
                    \begin{align*}
                        \mathbf{L} =
                        \left[
                        \begin{array}{cccc}
                            40 & 20 & 25 & 10
                        \end{array}
                        \right]~\textrm{cm}
                    \end{align*}
                    and the corresponding minimum and maximum subtended angles of all considered positions are
                    \begin{align*}
                        \left[
                        \begin{array}{c}
                            \boldsymbol{\alpha}_{\textrm{min}}\\
                            \boldsymbol{\alpha}_{\textrm{max}}
                        \end{array}
                        \right]=
                        \left[
                        \begin{array}{rrrr}
                            -19 & -42 & 42 & -80\\
                              6 & -33 & 45 & -48
                        \end{array}
                        \right]^\circ.
                    \end{align*}
                    Therefore, the required shape changing capabilities per unit length $\hat{\alpha} = \left(\alpha_\textrm{max}-\alpha_\textrm{min}\right)/L$ result in
                    \begin{align*}
                        \hat{\boldsymbol{\alpha}} =
                        \left[
                        \begin{array}{cccc}
                            63 & 45 & 12 & 320
                        \end{array}\right]~\frac{\circ}{\textrm{m}}.
                    \end{align*}
                    The shape changes of the third module are relatively small whereas those of the fourth module are large. Hence it is best to use pressure actuated modules for the first two arcs, a rigid module for the third arc and a mechanical module for the fourth arc as illustrated in Figure~\ref{pic:Figure_8_5}. This result is rather unsurprising as most passenger seats are based on a mechanism that connects a rigid backrest to its base. The only difference is thus the use of shape changing modules in the upper half of the backrest. It can be seen that the approximation errors are, despite the use of only four modules and the consideration of nine boundary geometries, remarkably small.

                    \afterpage{
                        \newgeometry{}
                        \begin{figure}[htbp]
                            \begin{center}
                                \subfloat[]{
                                    \includegraphics[width=\textwidth]{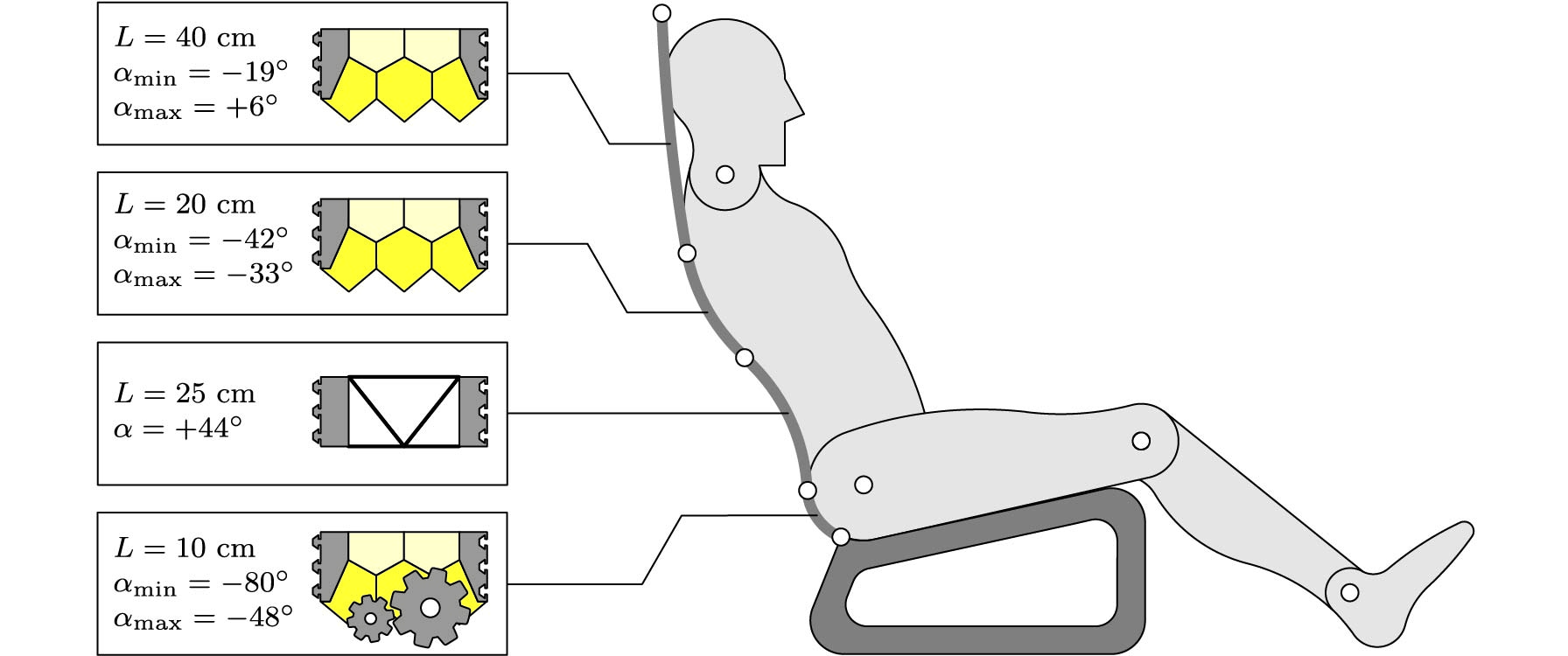}}
                                \vspace{10mm}

                                \subfloat[]{
                                    \includegraphics[width=\textwidth]{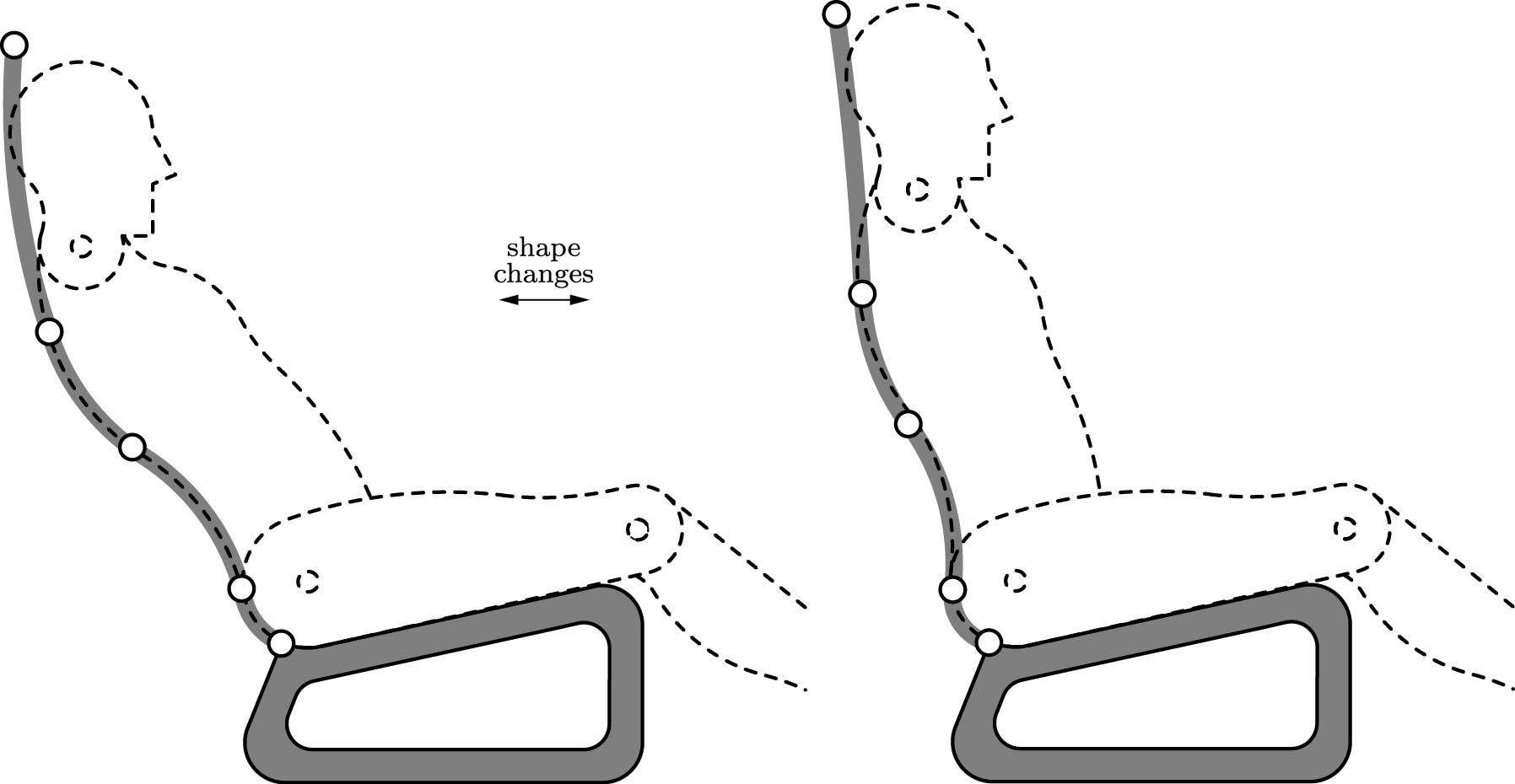}}
                                \caption{Passenger seat that can adapt to 1.7-1.9~m tall persons. The superimposed dummy is 1.8~m tall. (a) Seat consists of one rigid, one mechanical and two pressure actuated modules. (b) Backward and forward position that correspond to a 14 and $34^\circ$ inclination of the torso.\vspace{3.5mm}}
                                \label{pic:Figure_8_5}
                            \end{center}
                        \end{figure}
                        \restoregeometry}


                \subsubsection{Gapless Leading and Trailing Edge}
                    The second example deals with gapless high lift devices that can continuously vary their shapes between a high lift and flight configuration. Details about the geometries of the corresponding airfoils can be found in the appendix. Starting from the base in counterclockwise direction, the optimized arc-lengths $\mathbf{L}$ of the leading edge modules are
                    \begin{align*}
                        \mathbf{L} =
                        \left[
                        \begin{array}{cccc}
                            50 & 10 & 20 & 60
                        \end{array}
                        \right]~\textrm{cm}
                    \end{align*}
                    and the corresponding minimum and maximum subtended angles of both positions are
                    \begin{align*}
                        \left[
                        \begin{array}{c}
                            \boldsymbol{\alpha}_{\textrm{min}}\\
                            \boldsymbol{\alpha}_{\textrm{max}}
                        \end{array}
                        \right]=
                        \left[
                        \begin{array}{rrrr}
                            -18 &  5 &  80 & 15\\
                             13 & 75 & 140 & 35
                        \end{array}
                        \right]^\circ.
                    \end{align*}
                    Therefore, the required shape changing capabilities per unit length $\hat{\alpha} = \left(\alpha_\textrm{max}-\alpha_\textrm{min}\right)/L$ result in
                    \begin{align*}
                        \hat{\boldsymbol{\alpha}}=
                        \left[
                        \begin{array}{cccc}
                            62 & 700 & 300 & 33
                        \end{array}
                        \right]~\frac{\circ}{\textrm{m}}.
                    \end{align*}
                    The shape changes of the second and third module are relatively large whereas those of the first and fourth module are moderate. The leading edge can thus be assembled from two pressure actuated and two mechanical modules as shown in Figure~\ref{pic:Figure_8_6}.\\

                    Geometric data for the high lift configuration of the trailing edge is not available. Starting from the base in counterclockwise direction, it is subsequently assumed that their arc-lengths $\mathbf{L}$ are
                    \begin{align*}
                        \mathbf{L} =
                        \left[
                        \begin{array}{cccc}
                            80 & 30 & 60 & 50
                        \end{array}
                        \right]~\textrm{cm}
                    \end{align*}
                    and the minimum and maximum subtended angles for both positions are
                    \begin{align*}
                        \left[
                        \begin{array}{c}
                            \boldsymbol{\alpha}_{\textrm{min}}\\
                            \boldsymbol{\alpha}_{\textrm{max}}
                        \end{array}
                        \right]=
                        \left[
                        \begin{array}{rrrr}
                             7 & -2 & -15 & -21\\
                            27 & -2 & -15 &  -3
                        \end{array}
                        \right]^\circ.
                    \end{align*}
                    The required shape changes per unit length thus result in
                    \begin{align*}
                        \hat{\boldsymbol{\alpha}}=
                        \left[
                        \begin{array}{cccc}
                            25 & 0 & 0 & 36
                        \end{array}
                        \right]~\frac{\circ}{m}.
                    \end{align*}
                    The first and fourth module need to undergo only small shape changes so that the suction and compression side of the trailing edge can be assembled from a rigid and pressure actuated module as shown in Figure~\ref{pic:Figure_8_6}. It should be noted that the gap between the rigid modules in the high lift configuration is inevitable as their arc-lengths are invariant. Hence, a closed boundary would either require modules that can vary their arc-length or a different approach where the complete airfoil is changed instead.\\

                    \afterpage{
                        \newgeometry{}
                        \begin{figure}[htbp]
                            \begin{center}
                                \subfloat[]{
                                    \includegraphics[width=\textwidth]{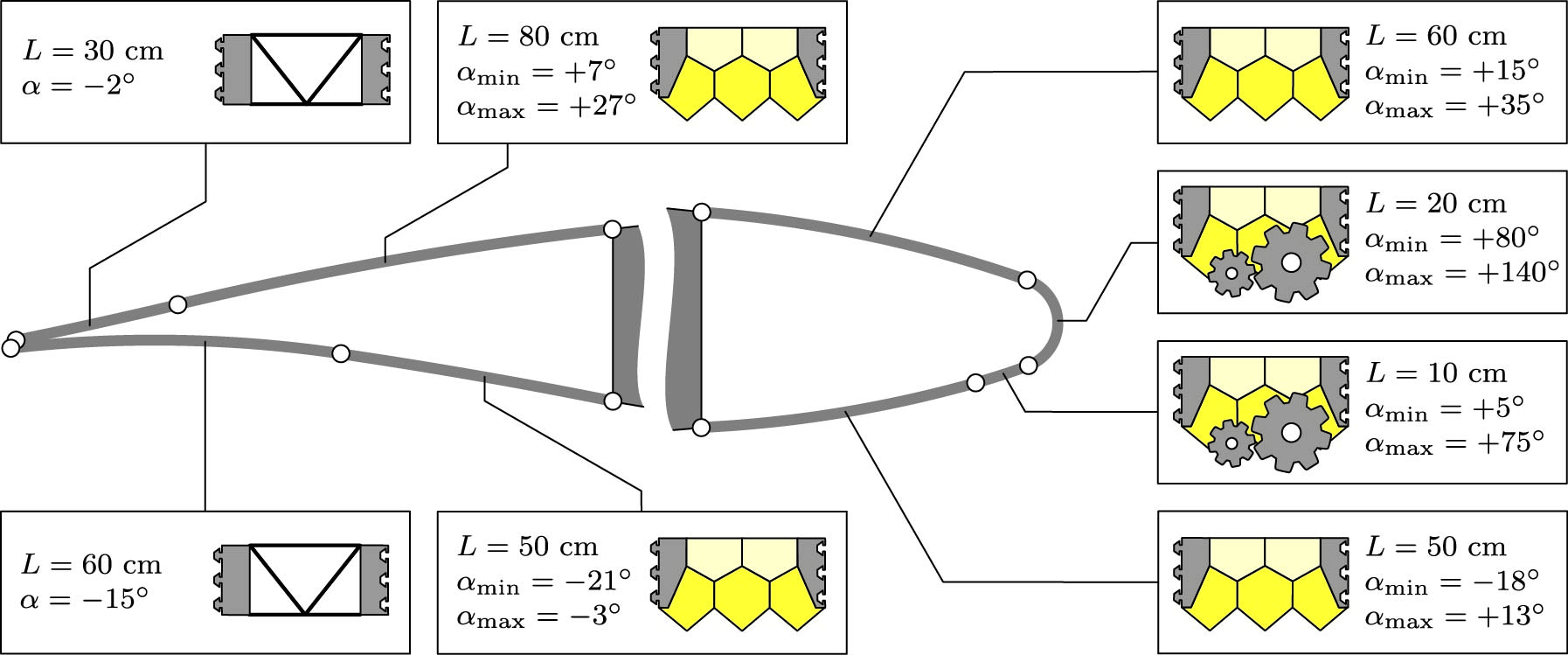}}
                                \vspace{10mm}

                                \subfloat[]{
                                    \includegraphics[width=\textwidth]{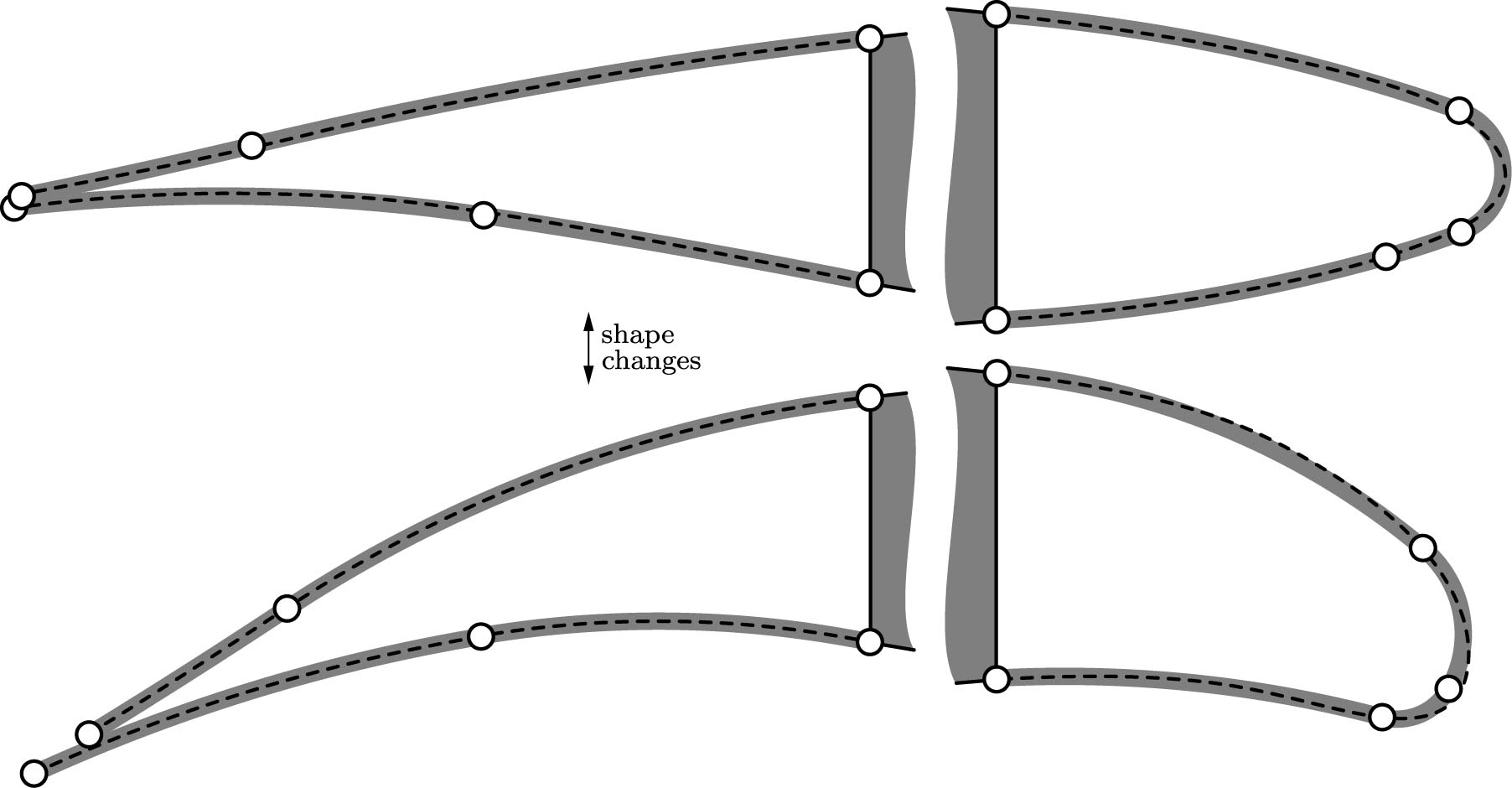}}
                                \caption{Gapless leading and trailing edge. (a) Leading edge consists of two mechanical and two pressure actuated modules whereas trailing edge consists of two rigid and two pressure actuated modules. (b) Flight and high lift configurations with superimposed target shapes. The gap between the rigid modules in the high lift configuration is due to their invariant arc-length.}
                                \label{pic:Figure_8_6}
                            \end{center}
                        \end{figure}
                        \restoregeometry}

                    It is rather surprising that a wide range of shape changing structures can be realized with the help of a few standardized modules that resemble circular arcs. However, such an approach leads to an increased weight, plumping complexity and a reduced stiffness. Furthermore, compromises are made with respect to approximation errors so that it is often best to use structures with tailored cell geometries instead. It is shown in the following three chapters how the cell corner geometries, cell side lengths and cell side thicknesses of pressure actuated cellular structures can be efficiently optimized for arbitrary sets of target shapes.


        \section{Compliant Structures}
            \label{sec:CompliantStructures}
            \subsection{Hinges}
                Pressure actuated cellular structures that are assembled from a large number of rigid cell sides and frictionless hinges are relatively heavy and maintenance intensive. The use of compliant structures that localize bending deformations in regions around cell corners is therefore often advantageous. Their susceptibility to fatigue \cite{Lobontiu2003-1} depends heavily on the hinge geometry. A hinge with a lumped compliance \cite{Hopkins2010-1} possesses a unique load-displacement relationship at the cost of a reduced fatigue life. In contrast, hinges with a distributed compliance \cite{Ananthasuresh1995-1} possess a long fatigue life at the cost of a non-unique load-displacement relationship. For the sake of simplicity, only hinges with a lumped compliance are subsequently considered.\\

                \marginnote{
                \begin{center}
                    \includegraphics[width=\marginparwidth]{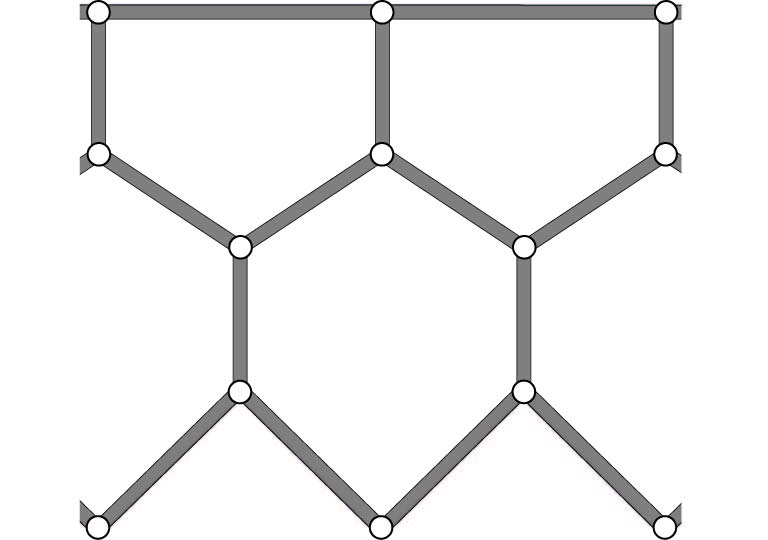}
                    \small(a)\vspace{10mm}
                    \includegraphics[width=\marginparwidth]{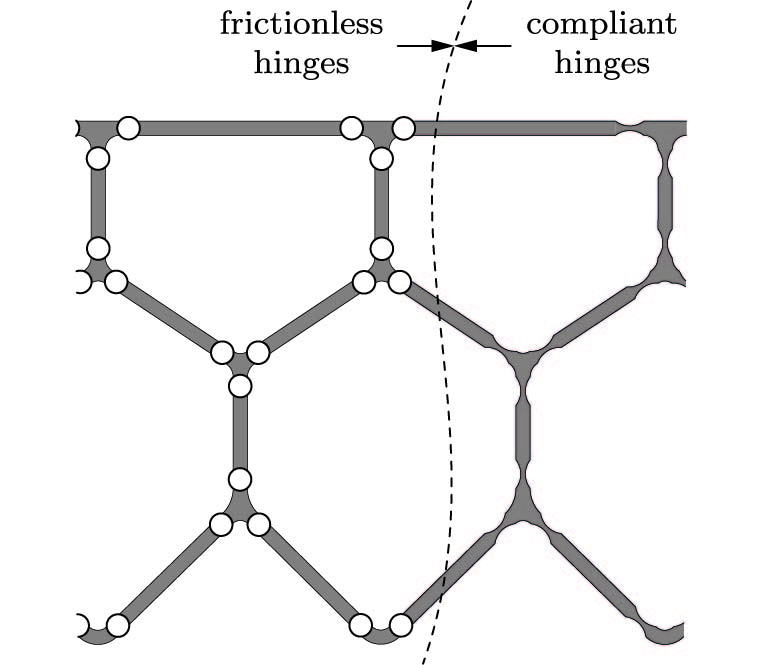}
                    \small(b)\vspace{10mm}
                    \includegraphics[width=\marginparwidth]{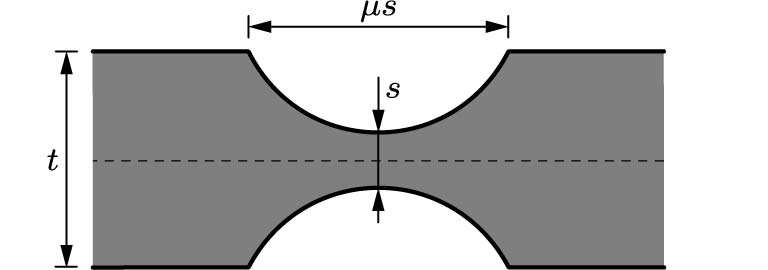}
                    \small(c)
                    \captionof{figure}{Pressure actuated cellular structure with frictionless, (a) centric and (b) eccentric hinges. (c) The latter can be replaced by compliant hinges that are created with circular cutouts.}
                    \label{pic:Figure_8_7}
                \end{center}}[-125mm]

                A hinge with a lumped compliance can be created within a rectangular cell side by two circular cutouts as shown in Figure~\ref{pic:Figure_8_7}. The resulting hinge possesses two reflection symmetry planes so that its geometry can be described by the cell side thickness $t$, central hinge thickness $s$ and an aspect ratio $\mu$. The equivalent rotational stiffness $n$ can be analytically derived with the help of the Euler-Bernoulli beam theory. However, the corresponding solution is rather bulky so that it is advantageous to express it on the basis of a rectangular beam with a unit length and a thickness $s$ so that
                \begin{align}
                    n = \frac{E^\textrm{eff} s^2}{12} \chi\left(t,s,\mu\right)
                \end{align}
                where $\chi$ is a numerically determined correction factor for the hinge geometry. The effective Young's modulus $E^\textrm{eff}$ for the plane strain condition of a material with a Young's modulus $E$ and Poisson's ratio $\nu$ is
                \begin{align}
                    E^\textrm{eff} = \frac{E}{1-\nu^2}.
                \end{align}


            \subsection{Material Selection}

                \marginnote{
                \begin{center}
                    \includegraphics[width=\marginparwidth]{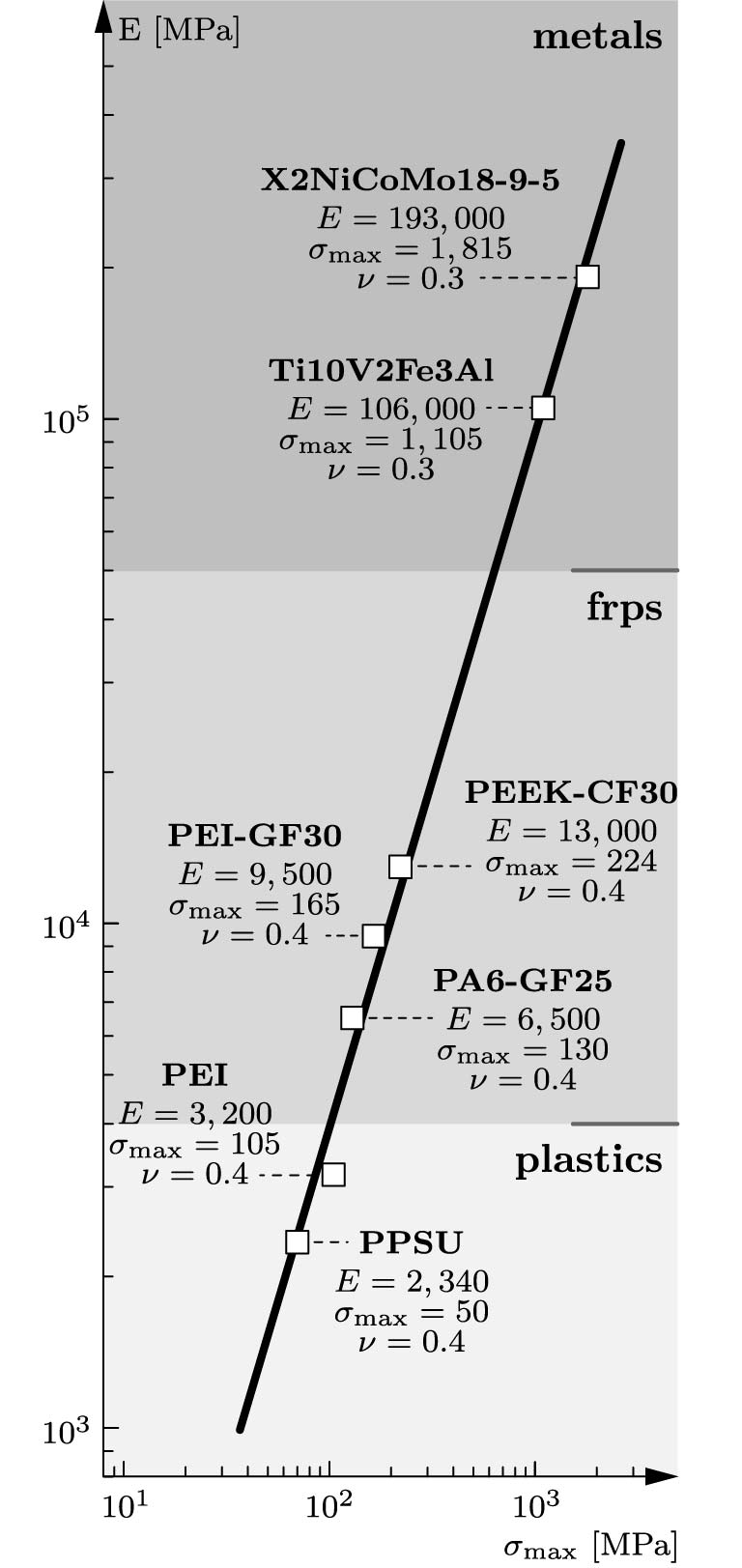}
                    \captionof{figure}{Yield strength $\sigma_\textrm{max}$ versus Young's modulus $E$ of materials that range from plastics over fiber reinforced plastics (frps) to metals.}
                    \label{pic:Figure_8_8}
                \end{center}}[-6mm]

                \noindent The many advantages of compliant pressure actuated cellular structures come at the cost of an increased coupling between material properties, cell sizes and pressures. It is thus important to select their material such that the overall bending energy of the compliant hinges is minimized. Compliant structures can be made from materials that range from plastics over fiber reinforced plastics to metals. A selection of materials with a particularly large ratio between yield strength $\sigma_\textrm{max}$ and Young's modulus $E$ is shown in Figure~\ref{pic:Figure_8_8}. It can be seen that their yield strengths obey a power law
                \begin{align}
                    \sigma_\textrm{max} = a E^b
                \end{align}
                where the parameters $a$ and $b$ are computed from a least square interpolation so that
                \begin{align}
                    a = 0.243
                    \hspace{5mm}\textrm{and}\hspace{5mm}
                    b = 0.727.
                \end{align}
                It should be noted that the creep properties of the considered materials differ significantly. Their time dependent impact on the admissible stresses are not taken into account in the following. Bending angles of compliant hinges are driven by shape changing requirements whereas their axial forces are driven by cell pressures. Based on a constant correction factor\footnote{$\chi\left(t,s,\mu\right)$ is constant for a fixed aspect ratio $\mu$ and a fixed thickness ratio $t/s$.} $\chi$, the required central hinge thickness $s$ for a given axial cell side force $F$ and bending angle $\varphi$ is
                \begin{align}
                    s = \frac{2 \rho \left|F\right|}{2 \sigma_\textrm{max} - \rho \chi E^\textrm{eff} \left|\varphi\right|}
                \end{align}
                where $\rho$ is the stress reduction factor for the von Mises yield criterion and $\sigma_\textrm{max}$ the admissible stress. Unlike the yield strength, the Poisson's ratio can not be expressed with the help of a power law. For example, a wide range of plastics and metals possess a similar Poisson's ratio \cite{Greaves2011-1}. Hence, its variation is not considered. The bending energy
                \begin{align}
                    \Pi_n = \frac{\chi E^\textrm{eff}}{24} s^2  \varphi^2
                \end{align}
                of a compliant hinge is minimal for
                \begin{align}
                    \tilde{E} = \left( \frac{\rho \chi |\varphi|}{2 a \left(2b-1\right)\left(1-\nu^2\right)} \right)^{\frac{1}{b-1}} \propto |\varphi|^{-3.7}
                \end{align}
                so that the optimal Young's modulus $\tilde{E}$ decreases exponentially for increasing bending angles. Therefore, the minimum bending energy $\tilde{\Pi}_n$ of a compliant hinge is proportional to
                \begin{align}
                    \tilde{\Pi}_n \propto |F|^2|\varphi|^{\frac{1}{1-b}} \approx |F|^2|\varphi|^{3.7}.
                \end{align}
                Hinges throughout a compliant pressure actuated cellular structure are usually subjected to different bending angles and axial cell side forces. As a consequence, it would be optimal if they are made from different materials. If only one material is used throughout a structure it is best to select it such that it minimizes the overall bending energy.


            \subsection{Cell Sizes}

                \marginnote{
                \begin{center}
                    \includegraphics[width=\marginparwidth]{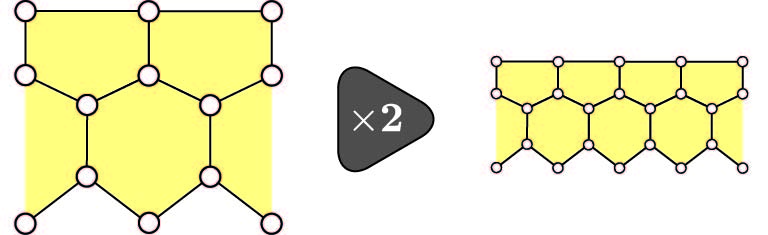}
                    \small(a)\vspace{7mm}
                    \includegraphics[width=\marginparwidth]{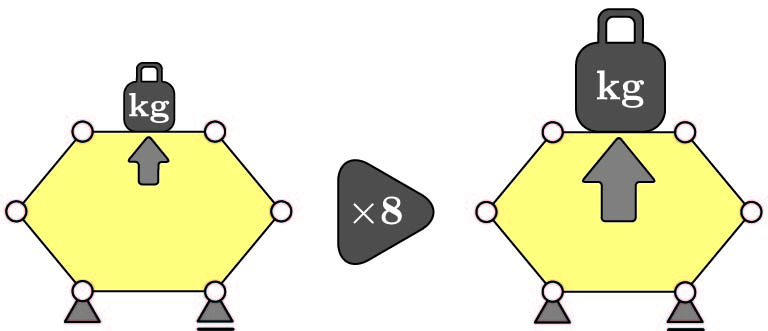}
                    \small(b)\vspace{15mm}
                    \includegraphics[width=\marginparwidth]{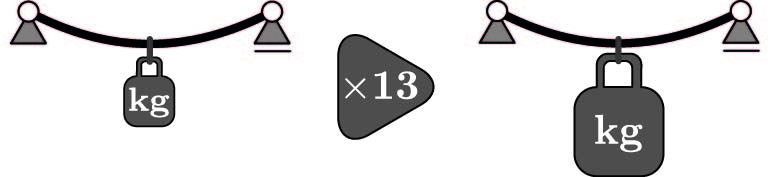}
                    \small(c)
                    \captionof{figure}{Modifications that preserve the shape changing capabilities and stiffness of a compliant pressure actuated cellular structure. (a) Doubling the number of cells reduces the stiffness. (b) This can be compensated by an eightfold increase of cell pressures. (c) The reduced hinge rotations and increased axial cell side forces affect the material selection. Structures with twice as many cells require a material with a thirteen times larger Young's modulus.}
                    \label{pic:Figure_8_9}
                \end{center}}[-4mm]

                \noindent The stiffness and shape changing requirements of a compliant pressure actuated cellular structure are known prior to its design whereas the optimal material and cell sizes are unknown. Their dependencies are subsequently outlined with the help of a structure that consists of a varying number of cells as illustrated in Figure~\ref{pic:Figure_8_9}. The scaled cell side lengths $\mathbf{v}^\textrm{s}$ and bending angles $\boldsymbol{\varphi}^\textrm{s}$ are
                \begin{align}
                    \mathbf{v}^\textrm{s} = \eta^{-1} \mathbf{v}
                    \hspace{5mm}\textrm{and}\hspace{5mm}
                    \boldsymbol{\varphi}^\textrm{s} = \eta^{-1} \boldsymbol{\varphi}
                \end{align}
                where the scaling factor $\eta$ is proportional to the number of cells. The expression for the optimal, scaled Young's modulus $\tilde{E}^\textrm{s}$ of a compliant hinge can be written as
                \begin{align}
                    \tilde{E}^\textrm{s} = \eta^\frac{1}{1-b} \tilde{E} \propto \eta^{3.7}.
                \end{align}
                As a consequence, a material with a thirteen times larger Young's modulus is required if the number of cells is doubled ($\eta=2$). The corresponding expression for the bending energy $\tilde{\Pi}_n^\textrm{s}$ is harder to find as it additionally depends on the axial cell side force $F^\textrm{s}$. The stiffness of pressure actuated cellular structures with twice as many cells and identical cell pressures reduces by a factor of eight so that the scaled pressure $p^\textrm{s}$ that needs to act on a cell side is
                \begin{align}
                    p^\textrm{s} = \eta^3 p.
                \end{align}
                The scaled cell side force $F^\textrm{s}$ becomes
                \begin{align}
                    F^\textrm{s} = \eta^2 F
                \end{align}
                and the bending energy of a hinge results in
                \begin{align}
                    \tilde{\Pi}_n^\textrm{s} = \eta^\frac{3-4b}{1-b} \tilde{\Pi}_n \propto \eta^{0.3}
                \end{align}
                since $F \propto p v$. The proportionality between the number of cells and the ratio between the total pressure $\Pi_p$ and bending $n_v \tilde{\Pi}_n$ energy of a compliant pressure actuated cellular structure follows from the previous equation \footnote{The number of cell sides $n_v$ is proportional to the scaling factor $\eta$ so that the total bending energy is proportional to $n_v^\textrm{s} \tilde{\Pi}_n^\textrm{s} \propto \eta^{1.3}$. Furthermore, the scaled pressure energy is proportional to $\Pi_p^\textrm{s} \propto \eta^2$.}
                \begin{align}
                    \frac{\Pi_p^\textrm{s}}{n_v^\textrm{s} \tilde{\Pi}_n^\textrm{s}} = \eta^\frac{3b-2}{1-b} \frac{\Pi_p}{n_v \tilde{\Pi}_n} \propto \eta^{0.7}
                \end{align}
                where $n_v$ is the number of cell sides. Hence, the influence of compliant hinges on the equilibrium shapes decreases for an increasing number of cells. This might, at least partially, explain the small cell sizes and relatively large cell pressures that can be observed in nastic plants. However, structures with a very large number of cells are difficult to manufacture and therefore not of practical interest. Furthermore, safety and energy efficiency considerations limit the maximum cell pressures for most applications to about 5~MPa if gaseous pressure mediums are used. Cell pressures beyond that limit need to be generated with the help of comparatively heavy hydraulic fluids.


        \section{Cytoskeletons}

            \blfootnote{\vspace{-5mm}
                \captionof{figure}{Cytoskeletons with different member lengths can be used to increase the stiffness of a shape changing module. (a) Cell geometries for equilibrium shapes with a subtended angle of $\alpha_\textrm{min}=60^\circ$ and $\alpha_\textrm{max}=100^\circ$ for cell row pressures of 0.4 and 2.0~MPa. (b) The vertical and horizontal cytoskeleton members are taut at $\alpha_\textrm{min}$ and $\alpha_\textrm{max}$, respectively. (c) Cytoskeleton with vertical and horizontal members that are prestressed at $\alpha_\textrm{min}+10^\circ$ and $\alpha_\textrm{max}-10^\circ$, respectively. (d) Subtended angle of equilibrium shapes as a function of the bending moment $M$. Depending on the cytoskeleton, deformations beyond $\left|\alpha_\textrm{min}\right|$ and $\left|\alpha_\textrm{max}\right|$ or $\left|\alpha_\textrm{min}+10^\circ\right|$ and $\left|\alpha_\textrm{max}-10^\circ\right|$ are restrained. The prestress of the shorter cytoskeleton increases the stiffness at both equilibrium shapes for bending moments with an arbitrary sign.}
                \label{pic:Figure_8_10}}
            \marginnote{
                \begin{center}
                    \includegraphics[width=\marginparwidth]{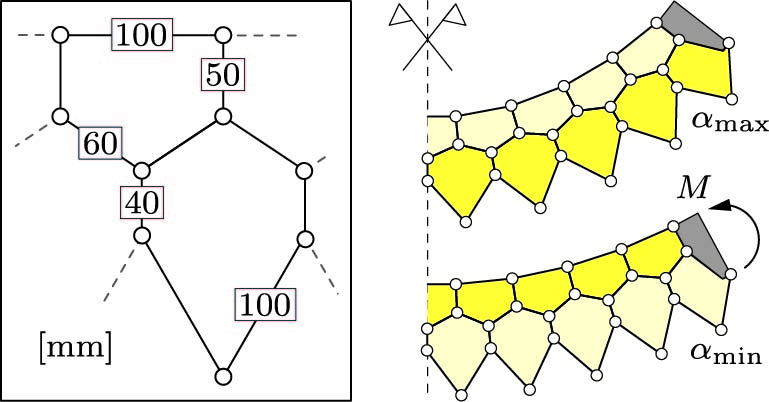}
                    \small(a)\vspace{8mm}
                    \includegraphics[width=\marginparwidth]{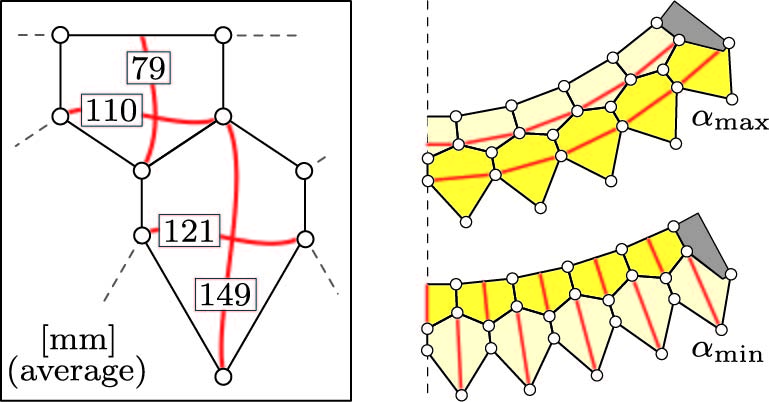}
                    \small(b)\vspace{8mm}
                    \includegraphics[width=\marginparwidth]{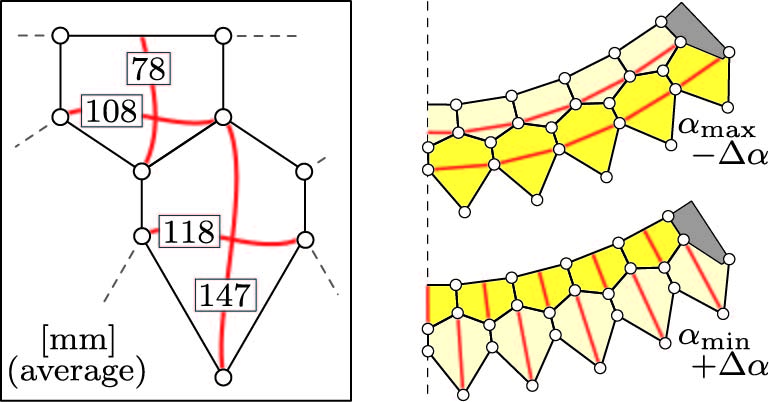}
                    \small(c)\vspace{8mm}
                    \includegraphics[width=\marginparwidth]{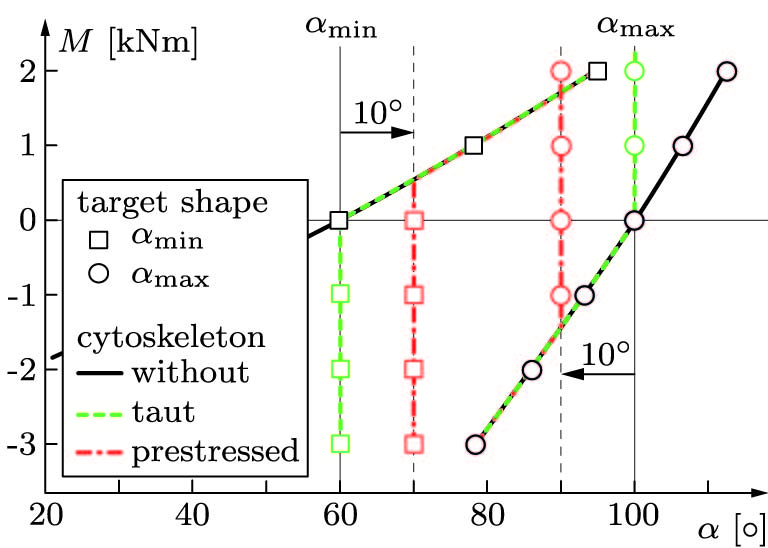}
                    \small(d)
                \end{center}}[-72mm]

            \noindent Pressure actuated cellular structures suffer from two drawbacks. First, their stiffness depends mostly on the cell geometries and pressures. Larger stiffness requirements thus lead to increased cell pressures or structures with a smaller number of larger cells. Second, cell sides along the boundary or between cell layers are subjected to differential pressures. This causes bending moments that can only be carried by relatively thick and heavy cell sides. It is subsequently shown how the weight and stiffness of pressure actuated cellular structures can be improved with the help of cytoskeletons. These subcellular structures are relatively thin so that they can carry large tension but only minor compression forces. This is done on the basis of a shape changing module that is assembled from rigid sides and frictionless hinges as illustrated in Figure~\ref{pic:Figure_8_10}~(a). The side lengths are chosen such that the subtended angles of the equilibrium shapes are $\alpha_\textrm{min}=60^\circ$ and $\alpha_\textrm{max}=100^\circ$ for cell row pressures of 0.4 and 2.0~MPa.


            \subsection{Stiffness}
                The module stiffness can be enhanced by cytoskeletons with horizontal and vertical members. Two cytoskeletons with different member lengths are subsequently considered. The first cytoskeleton, Figure~\ref{pic:Figure_8_10}~(b), is designed such that its vertical and horizontal members are taut at subtended angles of $\alpha_\textrm{min}=60^\circ$ and $\alpha_\textrm{max}=100^\circ$, respectively. The second cytoskeleton, Figure~\ref{pic:Figure_8_10}~(c), possesses shorter members so that it restrains the shape changing capabilities of the module. Its vertical and horizontal members are prestressed at subtended angles of $\alpha_\textrm{min}+10^\circ$ and $\alpha_\textrm{max}-10^\circ$, respectively. It should be noted that member lengths are slightly affected by edge effects so that only average values are given in Figure~\ref{pic:Figure_8_10}(b-c).\\

                Module deformations due to an external bending moment $M$ are summarized in Figure~\ref{pic:Figure_8_10}(d) for both pressure sets. It can be seen that the module stiffness is, depending on the cell pressures, between $7-19^\circ$/kNm. This corresponds to a steel plate with a thickness of 5-8~mm. The use of a taut, unstressed cytoskeleton greatly increases the stiffness against deformations beyond $\left|\alpha_\textrm{min}\right|$ and $\left|\alpha_\textrm{max}\right|$. However, it does not affect the stiffness against deformations inbetween as its members become slack upon compression. In contrast, the prestressed cytoskeleton increases the module stiffness against arbitrary deformations at the cost of reduced shape changing capabilities. Bending moments of up to 0.5-2.0~kNm can be carried at subtended angles of $\alpha_\textrm{min}+10^\circ$ and $\alpha_\textrm{max}-10^\circ$ without significant deformations. This enables the use of smaller cell pressures and thinner cell sides if the module stiffness is not a critical design parameter during shape changes.


            \marginnote{
            \begin{center}
                \includegraphics[width=\marginparwidth]{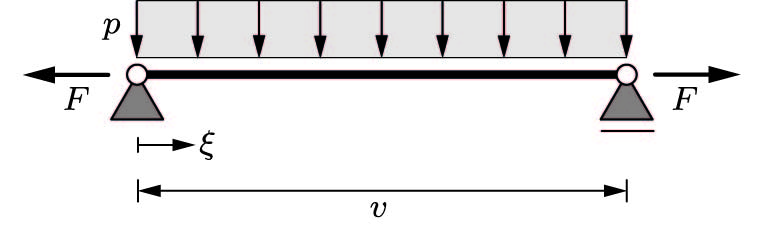}
                \small(a)\vspace{10mm}
                \includegraphics[width=\marginparwidth]{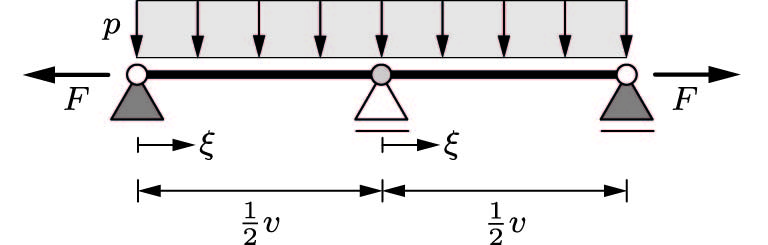}
                \small(b)\vspace{10mm}
                \includegraphics[width=\marginparwidth]{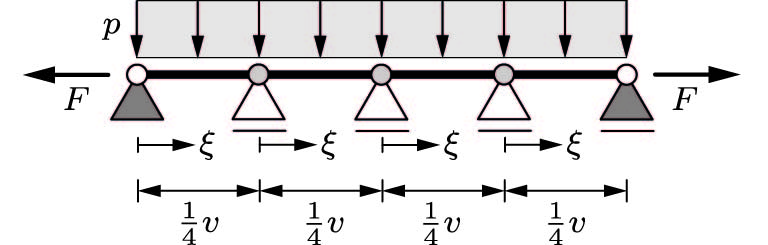}
                \small(c)
                \captionof{figure}{(a) Mechanical model of a cell side that is subjected to an axial force $F$ and pressure $p$. (b-c) Equally spaced supports are provided by a cytoskeleton with a refinement level (b) $q=1$ and (c) $q=2$.}
                \label{pic:Figure_8_11}
            \end{center}}[-49mm]

            \subsection{Weight}
                Unlike axial tension forces, differential pressures can not be efficiently carried by straight cell sides so that it is often advantageous to use curved sides instead. However, their applicability is limited to a few boundary locations. The reason for this is twofold. First, the outer pentagonal sides need to approximate smooth target shapes. Second, sides between cell layers are exposed to differential pressures with varying signs. The snap through buckling of curved sides would increase their hinge deformations and thus reduce the fatigue life of compliant pressure actuated cellular structures.\\

                An alternative to curved cell sides is the use of subcellular structures that provide additional supports as shown in Figure~\ref{pic:Figure_8_11}. For the sake of simplicity, it is subsequently assumed that the cell side segments are connected at intermediate supports via frictionless hinges. This is a conservative assumption that leads to larger bending moments in comparison to a continuous beam. The required cell side thickness $t$ for an axial force $F$ and bending moment $M$ is
                \begin{align}
                    t\left(F,p,\xi,q\right) = \frac{\left|F\right| + \sqrt{F^2 + 24 \left|M\left(F,p,\xi,q\right)\right| \sigma_\textrm{max}}}{2\sigma_\textrm{max}}
                \end{align}
                where $\sigma_\textrm{max}$ is the yield strength of the cell side material. The bending moment
                \begin{align}
                    M\left(F,p,\xi,q\right) = \frac{p v^2}{2 \cdot 4^q} \left(1-\xi\right) \xi
                \end{align}
                depends on the position $\xi\in\left[0,1\right]$ along the cell side and the number of supports
                \begin{align}
                    n_s = 1 + 2^q
                \end{align}
                that follows from the refinement level $q$ of the cytoskeleton. The cell side weight $w$ is proportional to the thickness $t$ so that the weight ratio for different refinement levels $\left(q_2 > q_1\right)$
                \begin{align}
                    2^{q_1-q_2} \leq \frac{w_{q2}}{w_{q1}} \leq 1
                \end{align}
                is bounded by $F=0$ and $p=0$.\\

                A cytoskeleton that supports the cell sides of a shape changing module, as introduced in Figure~\ref{pic:Figure_8_10}~(a), is shown in Figure~\ref{pic:Figure_8_12} for different refinement levels. It is based on four-bar linkages that provide additional supports without hindering shape changes. It can be seen that the cell side thicknesses are greatly reduced by a cytoskeleton with a refinement level $q=1$. The uniformly stressed cytoskeleton members are relatively thin so that, depending on the cell material, a total weight reduction of up to 35~\% can be achieved \cite{Pagitz2014-1}. However, pressure actuated cellular structures are difficult to manufacture. The additional integration of a cytoskeleton with a high refinement level is therefore, at least from that perspective, not helpful.

                \marginnote{
                \begin{center}
                    \includegraphics[width=\marginparwidth]{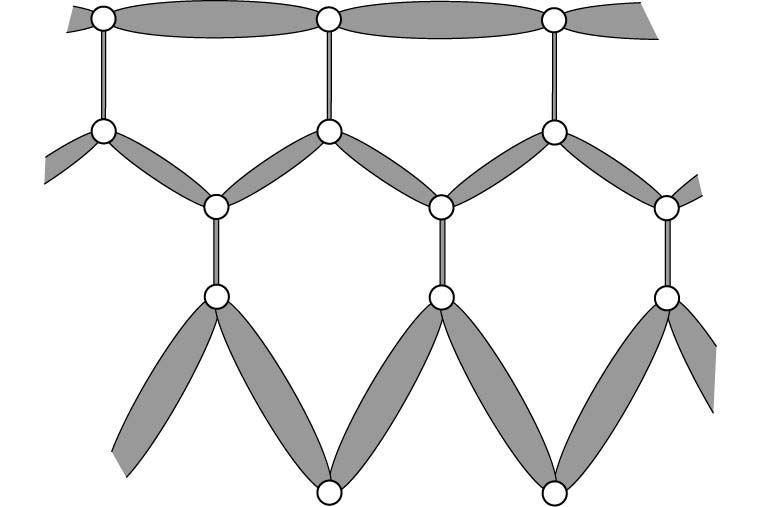}
                    \small(a)\vspace{12mm}
                    \includegraphics[width=\marginparwidth]{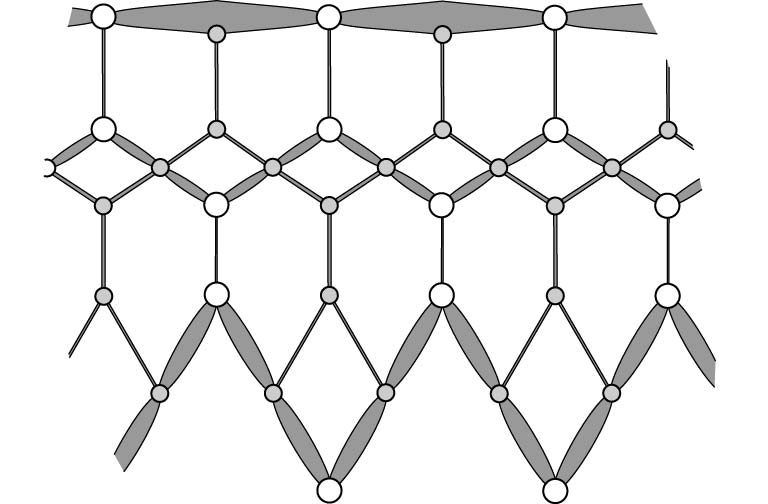}
                    \small(b)\vspace{12mm}
                    \includegraphics[width=\marginparwidth]{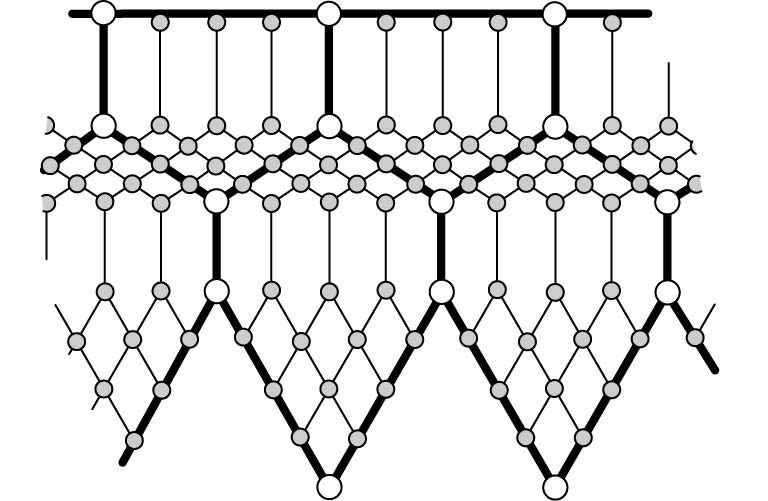}
                    \small(c)
                    \captionof{figure}{(a-b) Cell side thicknesses of a shape changing module that is supported by a cytoskeleton with a refinement level of (a) $q=0$ and (b) $q=1$. (c)~Cytoskeleton can be further refined $\left(q=2\right)$ to increase the number of supports.}
                    \label{pic:Figure_8_12}
                \end{center}}[-150mm]


        \section{Manufacturing}
            Pressure actuated cellular structures are currently not used in engineering applications. One reason for this is the immaturity of previously developed approaches and another their inherent manufacturing complexity. A particular problem in this regard is the pressure tight sealing of prismatic cells. It is subsequently shown how they can be sealed without significantly affecting their shape changing capabilities. Furthermore, various processes are reviewed that can be used for the manufacturing of compliant pressure actuated cellular structures.


            \subsection{End Caps}

                \marginnote{
                \begin{center}
                    \includegraphics[width=\marginparwidth]{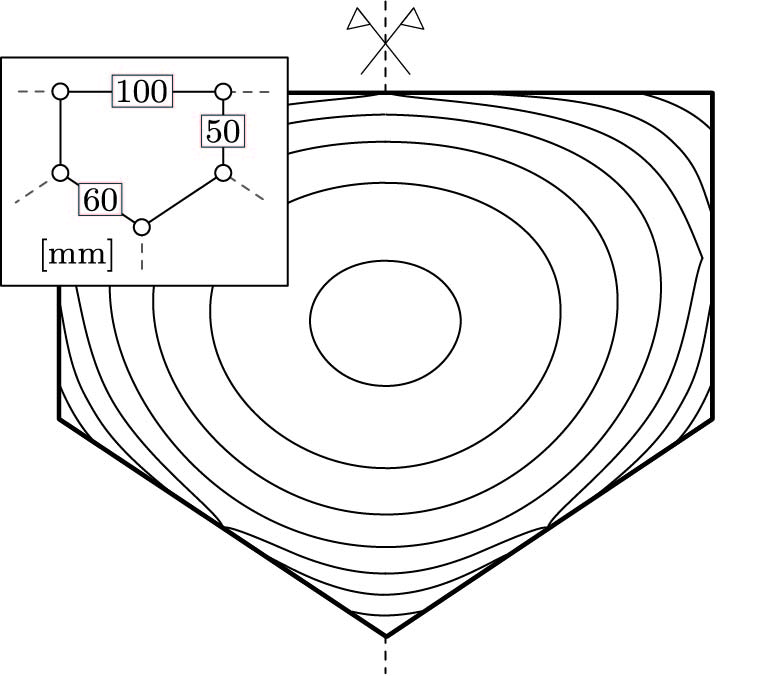}
                    \small(a)\vspace{7mm}
                    \includegraphics[width=\marginparwidth]{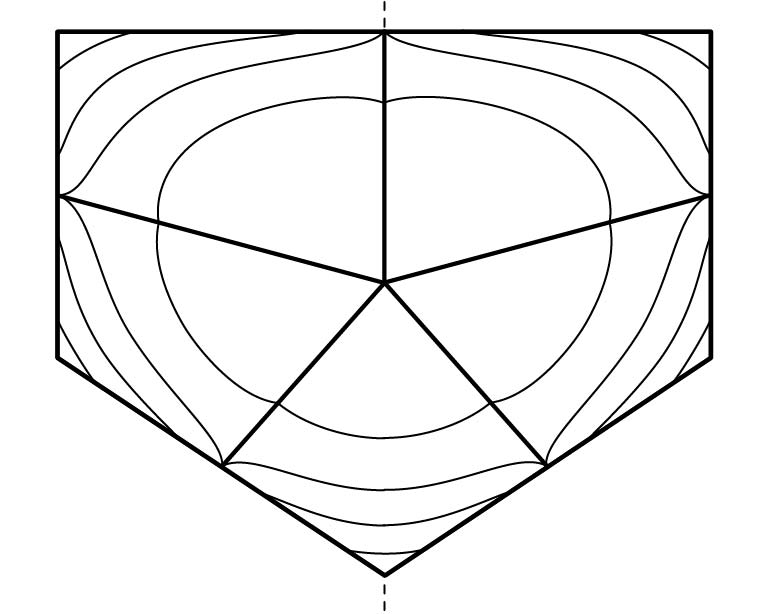}
                    \small(b)
                    \captionof{figure}{Contour lines of two pentagonal end caps that possess a minimal surface. The out of plane spacing of the contour lines is 5~mm. (a) End cap without and (b) with tendons that emerge from the central cell sides.}
                    \label{pic:Figure_8_13}
                \end{center}}[-10mm]

                \noindent The prismatic cells of pressure actuated cellular structures need to be sealed at both ends such that they can sustain the differential pressures while being flexible enough to enable large shape changes. Similar to many bicycle tires, this can be done with the help of flexible inner tubes. Their integration into structures that lack a cytoskeleton is straightforward. However, elastic tubes increase the overall weight and add potential failure modes so that they are of particular interest for prototype structures. Unlike elastic tubes, compliant end caps can be tightly integrated into cells so that they can be used in conjunction with a cytoskeleton. Their reliability and weight advantages come at the cost of an increased design and manufacturing complexity. Optimal end caps possess, similar to soap films, an isotropic stress state for all possible cell deformations. Unfortunately, it is not possible to construct such an end cap from a solid material. For the sake of simplicity, it is subsequently assumed that stress variations are small so that end caps can be designed with the help of minimal surfaces \cite{Bletzinger1999-1} for the reference, undeformed configuration.\\

                Two end cap designs are subsequently considered. The first consists of a thin membrane whereas the second is additionally reinforced by elastic tendons that emerge from the central cell sides and meet at the cell center. The geometry of the pentagonal cell\footnote{It coincides with the geometry of the previously used shape changing module.} and the contour lines of the end caps are shown in Figure~\ref{pic:Figure_8_13} for a membrane thickness $t=1$~mm and a tendon stiffness of $EA=50$~kN. Both, the membrane and tendons are assumed to be made from polyphenylensulfon so that the considered material properties and pressure load are

                \vspace{1mm}
                \begin{center}
                    \begin{tabular}{lll}
                        \textbf{loading}\\
                        \hspace{4mm} $p = 2.0$~MPa                & \hspace{3mm} & pressure\vspace{5mm}\\
                        \textbf{material}\\
                        \hspace{4mm} $E=2,340$~MPa                && Young's modulus\\
                        \hspace{4mm} $\nu=0.4$                    && Poisson's ratio\\
                        \hspace{4mm} $\sigma_\textrm{max}=50$~MPa && yield strength.
                    \end{tabular}
                \end{center}
                \vspace{1mm}

                \noindent It can be seen from the contour lines that uneven cell sides are used to improve the end cap geometries. The altitude $z$ along each cell side is described by a cubic polynomial with a point reflection symmetry
                \begin{align}
                    z =
                    \begin{cases}
                        2 z_\textrm{max} \left(\psi + 2\left(\psi-1\right)\left(4\xi-3\right)\xi\right) \xi & 0 \leq \xi \leq \frac{1}{2}\\
                        2 z_\textrm{max} \left(\psi - 2 \left(\psi-1\right)\left(4\xi-1\right) \left(1-\xi\right)\right) \left(1-\xi\right) & \frac{1}{2} < \xi \leq 1.
                    \end{cases}
                \end{align}
                Herein, $\xi\in\left[0,1\right]$ is the dimensionless coordinate along the cell side, $z_\textrm{max}$ is the altitude at the cell side center $\left(\xi=1/2\right)$ and $\psi$ is the gradient of the polynomial at $\xi=\left[0,0.5,1\right]$. It is assumed in the following that $z_\textrm{max}=15$~mm and $\psi=2$.\\

                \marginnote{
                \begin{center}
                    \includegraphics[width=\marginparwidth]{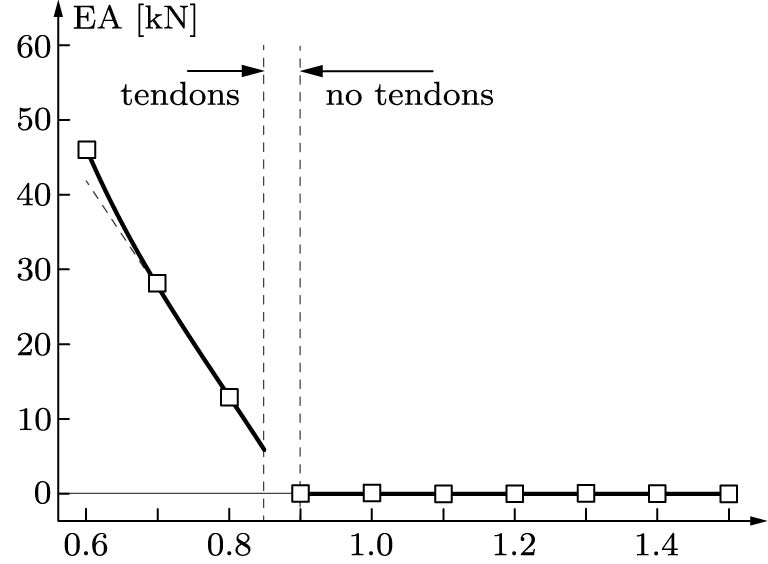}
                    \small(a)\vspace{6mm}
                    \includegraphics[width=\marginparwidth]{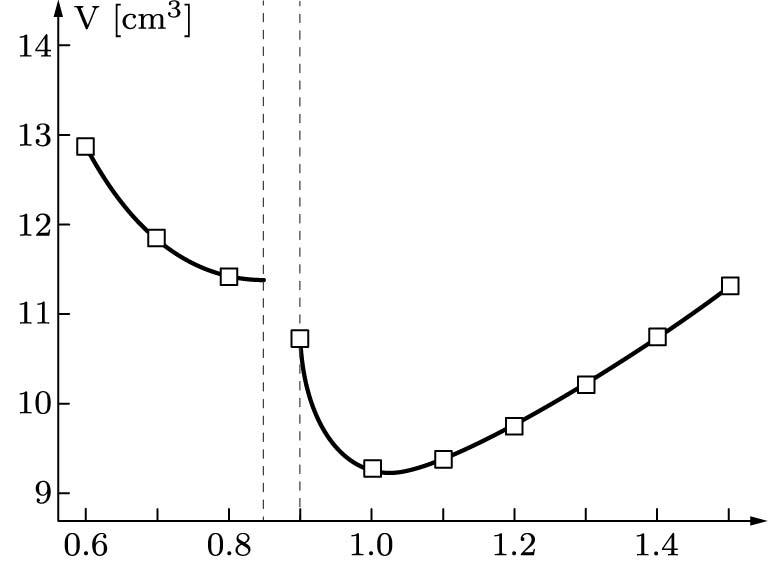}
                    \small(b)\vspace{6mm}
                    \includegraphics[width=\marginparwidth]{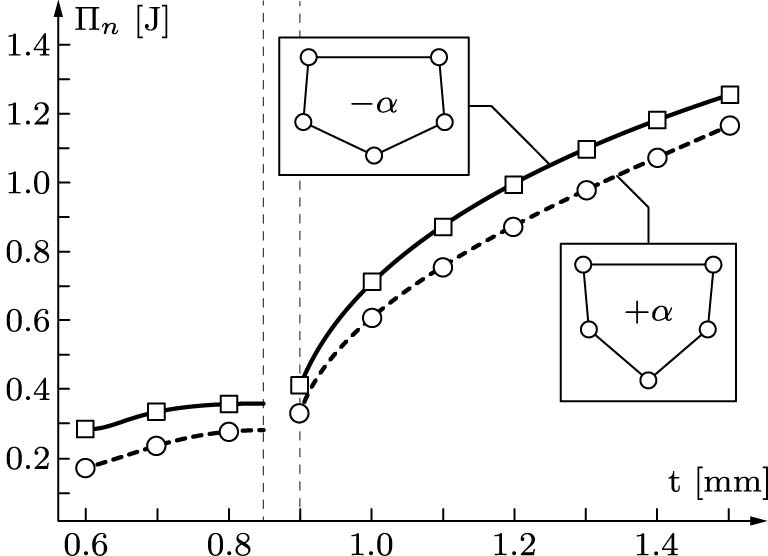}
                    \small(c)
                    \captionof{figure}{End cap properties as a function of the membrane thickness $t$. (a)~Smallest possible tendon stiffness $EA$, (b)~material volume $V$ and (c)~strain energy $\Pi_n$ for a deformation angle $\alpha=0.75^\circ$.}
                    \label{pic:Figure_8_14}
                \end{center}}[-19mm]

                The material volume $V$ and strain energy $\Pi_n$ of the end cap are illustrated in Figure~\ref{pic:Figure_8_14} for the smallest possible tendon stiffness $EA$ as a function of the membrane thickness $t$. The deformations, that preserve the symmetry and side lengths of the end cap, are due to a rotation of the vertical sides by an angle of $\alpha=\pm 3/4^\circ$. It can be seen that the smallest possible membrane thickness that can be achieved without tendons is about $t=0.9$~mm. A further thickness reduction is only possible if local reinforcements are used. However, tendons can carry only uniaxial loads so that the lightest possible end cap possesses a membrane thickness of about $t=1$~mm. Although tendons increase the overall weight of an end cap, they can reduce the required deformation energy as they enable the use of membranes with a higher curvature and a smaller thickness. Furthermore, their weight penalty can be avoided if local reinforcements are provided by a cytoskeleton that additionally supports the cell sides.\\

                The influence of end caps on the shape changing capabilities of pressure actuated cellular structures depends on their strain energy and the variation of the pressure potential for a given deformation. The latter results for a one meter long cell in
                \begin{align}
                    \Pi_p =
                    \begin{cases}
                        -14.4~\textrm{J/m} & \textrm{for}\ \ \alpha=+3/4^\circ\\
                        \hspace{4.5mm} 5.8~\textrm{J/m} & \textrm{for}\ \ \alpha=-3/4^\circ
                    \end{cases}
                \end{align}
                which is considerably larger than the energy needed to deform a well designed end cap. This is remarkable as the undeformed cell geometry is already close to its maximum cross sectional area. As a consequence, the influence of well designed end caps on the simulation and optimization of pressure actuated cellular structures can be neglected if cells are sufficiently long. In general it can be said that it is advantageous to use end cap materials with a small Young's modulus to yield strength ratio\footnote{Thermoplastic polyurethane possesses a smaller $E/\sigma_\textrm{max}$ ratio than titan.}. The argument for this is as follows. The required membrane thickness of an end cap is proportional to
                \begin{align}
                t \propto \frac{1}{\sigma_\textrm{max}} = \frac{1}{\chi a E^b}.
                \end{align}
                The energy density of a membrane with a unit area due to an unidirectional axial strain $\varepsilon$ is
                \begin{align}
                    \Pi_n = \frac{1}{2} E t \varepsilon^2\ \ \ \textrm{so that}\ \ \
                    \Pi_n \propto E^{1-b} = E^{0.27}.
                \end{align}
                Therefore, although only slightly, the energy required to deform an end cap increases with an increasing Young's modulus.


            \subsection{Processes}
                Compliant pressure actuated cellular structures can be manufactured from a wide range of materials with the help of different processes. The most promising approaches for prototyping and production are subsequently discussed.\\

                \textit{\textbf{Rapid Prototyping}} is highly attractive as it minimizes the time between design and manufacturing \cite{Chua2008-1}. This enables short turnaround times during an early development stage. Structures with good mechanical properties can be created by the selective laser sintering of plastics or the laser melting of metals. However, the smallest possible cell side thicknesses that can be realized\footnote{Current constraint of digital manufacturing services such as Shapeways that deliver within a few days.} is 0.7~mm for plastics and 1.0~mm for metals. This leads to relatively large cell sizes so that only structures with a small number of cells can be printed within the $650\times350\times550$~mm construction space for plastics and the $250\times250\times200$~mm construction space for metals.\\

                A \textit{\textbf{Waterjet}} or \textit{\textbf{Laser}} can be used to cutout the cross section of a compliant pressure actuated cellular structure from a plate. A large number of cutouts can be stacked and connected via tension rods through holes in their cell sides. This relatively inexpensive approach, that can be used for all kind of materials, enables the manufacturing of nearly arbitrarily sized prototypes within a short time frame. Its only drawback is the large number of parts and the weight penalty that stems from the increased cell side thicknesses. The use of stacked segments was proposed by the author at the German Aerospace Agency in 2012 and implemented, published by Gram\"uller et al \cite{Gramueller2014-1} in 2014. This article makes a number of wrong claims that were partially corrected in an addendum \cite{Gramueller2015-1} that was published in 2015.\\

                \textit{\textbf{Wire Electrical Discharge Machining}} is probably the best process for cutting out prismatic structures with a complex geometry from a wide range of solid metals \cite{Jameson2001-1}. The contact free machining leads to a high accuracy and enables the creation of compliant structures with small cell sizes. The process is relatively slow and expensive so that it is not a viable option for an early development stage. Furthermore, the production time and cost are mostly independent of the lot size so that it is reserved for relatively small parts that make a large performance difference.\\

                The \textit{\textbf{Extrusion}} of plastics \cite{Rauwendaal2014-1} and certain metals \cite{Saha2000-1} is an interesting option for the inexpensive, large scale production of compliant pressure actuated cellular structures. A heated material is pushed with great force through a specifically shaped die so that it flows like a fluid. This enables prismatic structures with nearly arbitrary lengths. However, large thickness variations within a cell side and thicknesses smaller than 0.8~mm can be problematic. Furthermore, the required die is relatively complex and therefore expensive. Particularly the use of polymers can cause shape deviations that are difficult to predict. An iterative die design can thus be necessary that is hard to justify for a small scale production.\\

                \textit{\textbf{Fiber Reinforced Plastics}} possess many attractive properties that destine them for the use in compliant pressure actuated cellular structures. The simplest approach is to manufacture individual cells\footnote{This can be done by roll wrapping, filament winding, pultrusion and pullbraiding. The latter is of particular interest for the large scale production of prismatic cells with tailored mechanical properties.} that are subsequently glued together. However, this leads to delaminations at cell corners that are driven by large intercellular tension forces. It is thus necessary to use three-dimensionally woven fabrics \cite{Chen2015-1} instead that can efficiently carry these forces. The construction of a weaving machine that is specifically designed for compliant pressure actuated cellular structures can only be justified if their large scale use in passenger or military airplanes is likely.


    \newpage

    \sectionmark{Summary}
    \begin{framed}
        \noindent \textbf{Summary}\\

        \noindent Pressure actuated cellular structures consist of separately pressurized rows of prismatic cells with pentagonal or hexagonal cross sections. They can be assembled from rigid cell sides and frictionless hinges or manufactured from a single piece of material with the help of various processes. The waterjet cutting of cross sectional segments enables fast turnaround times during an early development stage whereas the three-dimensional weaving of composites or the wire electrical discharge machining of metals enable high quality production parts.\\

        Compliant pressure actuated cellular structures can be manufactured from materials that range from plastics over reinforced plastics to metals. The optimal material choice depends on the shape changing requirements, cell sizes and pressures. The weight and stiffness of these structures can be further improved with the help of cytoskeletons that provide additional cell side supports or alter the cell topology at target shapes. Individual cells can be sealed at both ends by caps that do not significantly affect their shape changing capabilities.\\

        \noindent \textbf{Conclusion}\\

        \noindent The load carrying capacity and shape changing capabilities of compliant pressure actuated cellular structures are, in conjunction with their low selfweight and simplicity, remarkable. It is therefore not surprising that they are of interest for the realization of gapless high lift devices. Although not trivial, their prototyping and production is possible with existing manufacturing processes. The major problem that remains is the computation of individual cell geometries for given target shapes and cell row pressures. A novel algorithm for the efficient simulation and optimization of compliant pressure actuated cellular structures with an arbitrary number of cell rows is presented in the following three chapters.
    \end{framed}


    \newgeometry{}
    \thispagestyle{empty}

        \noindent \Large \textbf{Nomenclature} \normalsize
        \vspace{10mm}

            \noindent
            \begin{tabular}{llll}
                \textbf{Superscripts}\hspace{5mm}
                & $\tilde{\square}$       & \hspace{12mm} & optimal\\
                & $\hat{\square}$         &               & normalized\\
                & eff                     &               & effective\\
                & $s$                     &               & scaled
            \end{tabular}

            \vspace{5mm}
            \noindent
            \begin{tabular}{llll}
                \textbf{Subscripts}\hspace{8mm}
                & min, max                & \hspace{1mm}  & minimum, maximum\\
                & $n$, $p$                &               & reference to strain, pressure energy
            \end{tabular}

            \vspace{5mm}
            \noindent
            \begin{tabular}{llll}
                \textbf{Numbers}\hspace{10mm}
                & $n_s$                   & \hspace{12mm} & supports\\
                & $n_v$                   &               & cell sides
            \end{tabular}

            \vspace{5mm}
            \noindent
            \begin{tabular}{llll}
                \textbf{Greek Letters}\hspace{2mm}
                & \multirow{2}{*}{$\alpha$} &               & subtended angle of adaptive module\\
                &                           &               & \textbf{AND} deformation angle of pentagonal cell\\
                &$\varepsilon$              &               & strain\\
                &$\psi$                     &               & gradient of cell side altitude\\
                &$\xi$                      & \hspace{13mm} & local coordinate\\
                &$\eta$                     &               & scaling factor\\
                &$\mu$                      &               & thickness ratio of compliant hinge\\
                &$\nu$                      &               & Poisson's ratio\\
                &$\rho$                     &               & stress reduction factor (von Mises)\\
                &$\sigma$                   &               & stress\\
                &$\varphi$                  &               & bending angle of hinge\\
                &$\chi$                     &               & correction factor for hinge geometry\\ \vspace{-2.5mm}\\
                &$\Pi$                      &               & potential energy
            \end{tabular}

            \vspace{5mm}
            \noindent
            \begin{tabular}{llll}
                \textbf{Roman Letters}
                & $a$, $b$ & \hspace{10mm} & interpolation parameters for material properties\\
                & $n$      &               & rotational spring stiffness\\
                & $p$      &               & pressure\\
                & $q$      &               & refinement level of cytoskeleton\\
                & $s$      &               & central hinge thickness\\
                & $t$      &               & cell side \textbf{AND} membrane thickness\\
                & $v$      &               & cell side length\\
                & $w$      &               & weight\\
                & $z$      &               & cell side altitude\\\vspace{-2.5mm}\\
                & $A$      &               & cross sectional tendon area\\
                & $C$      &               & continuity\\
                & $E$      &               & Young's modulus\\
                & $F$      &               & axial cell side force\\
                & $L$      &               & arc length of module\\
                & $M$      &               & bending moment\\
                & $V$      &               & volume
            \end{tabular}
            \restoregeometry

            \thispagestyle{empty} 
        \cleardoublepage
    \chapter{Geometric Model}

        \blfootnote{\vspace{-5mm}
            \captionof{figure}{Generation of cross sectional geometry. (a) Compliant pressure actuated cellular structures are assembled from rectangular cell sides with individual lengths and thicknesses for given cell corner angles. (b) Pointed cell corners are rounded out with circular arcs and (c) cutouts are made in cell sides to create compliant hinges that localize bending deformations.}
            \label{pic:Figure_9_1}}
        \marginnote{
            \begin{center}
                \includegraphics[width=\marginparwidth]{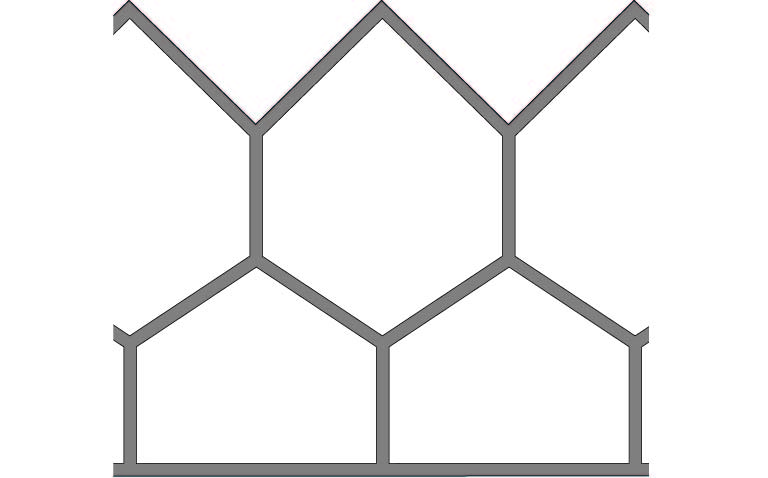}
                \small(a)\vspace{9mm}
                \includegraphics[width=\marginparwidth]{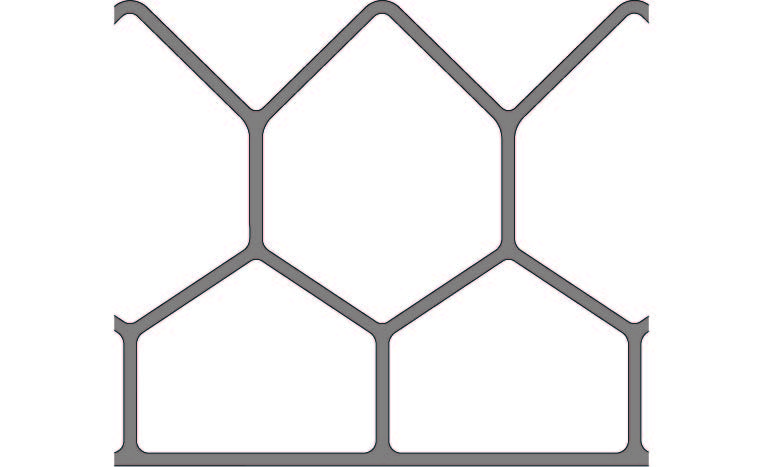}
                \small(b)\vspace{9mm}
                \includegraphics[width=\marginparwidth]{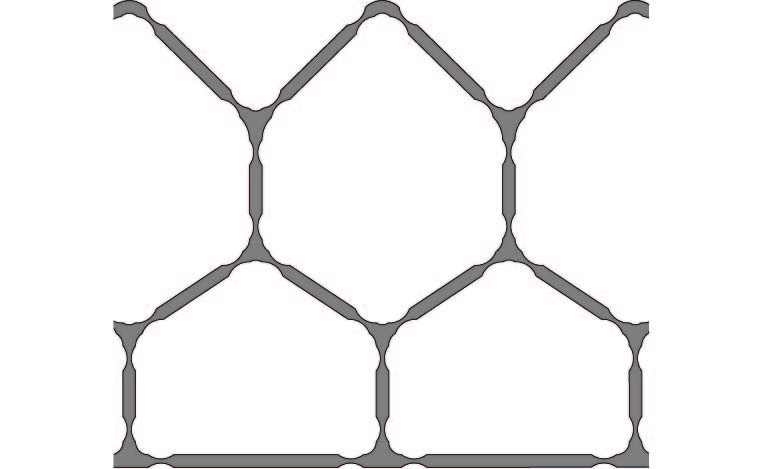}
                \small(c)
            \end{center}}[-31mm]

        \noindent It was argued in the previous chapter that the total perimeter of a half plane with regions of equal area is likely minimal for hexagonal cells that are terminated by a single row of pentagonal cells along the boundary. From that perspective it is rather surprising that some authors approach the design of compliant pressure actuated cellular structures with the help of topology optimization \cite{Maute2005-1,Vasista2012-1}. The structures that are considered in this thesis consist of at least two rows of prismatic cells with pentagonal or hexagonal cross sections so that their geometry can be fully described in two-dimensions. Despite a dimensional reduction and a predetermined topology, an efficient description of their geometry is of utmost importance for an optimization.\\

        It is assumed in the following that the cross sectional geometry of compliant pressure actuated cellular structures is generated in three steps as illustrated in Figure~\ref{pic:Figure_9_1}. Rectangular cell sides with individual lengths and thicknesses are joined at cell corners with individual angles. The pointed cell corners are rounded out and circular cutouts are made in cell sides to create compliant hinges that localize bending deformations. The sole purpose of this chapter is to introduce a minimal set of global state variables that fully describe the cross sectional geometry of compliant pressure actuated cellular structures. This is done such that the coupling between different cross sectional regions is minimized.


        \section{Corner Angles}
            \subsection{Global State Variables}

                \blfootnote{\vspace{-5mm}
                    \captionof{figure}{A minimal set of global state variables that fully describe the cell corner angles of pressure actuated cellular structures for an arbitrary number of cell rows. (a) Cell numbering and effective cell side lengths $\mathbf{v}$ between fictitious, centric cell corners. (b) Cell side inclinations are defined by global state variables $\mathbf{u}$. An exploded view is used to increase the clarity of the presentation. (c) Pressure actuated cellular structures with more than one cell row can be cut into strips of triangular and/or pentagonal cells. Based on an updated cell numbering, global state variables $\mathbf{u}$ can be transformed into local state variables for the pentagonal $\mathbf{u}^\textrm{P}$ and triangular $\mathbf{u}^\textrm{T}$ cells of each strip. Local variables serve as a basis for the computation of cell corner angles.}
                    \label{pic:Figure_9_2}}
                \marginnote{
                    \begin{center}
                        \includegraphics[width=\marginparwidth]{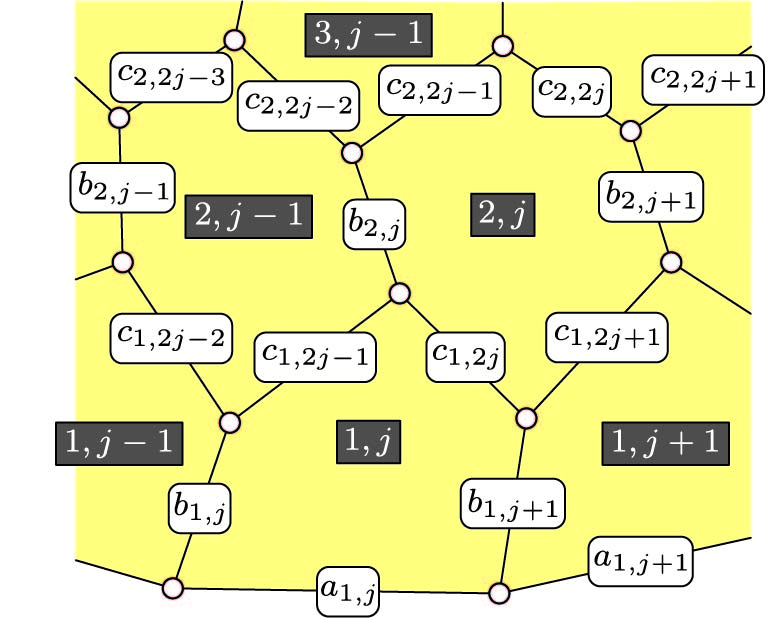}
                        \small(a)\vspace{7mm}
                        \includegraphics[width=\marginparwidth]{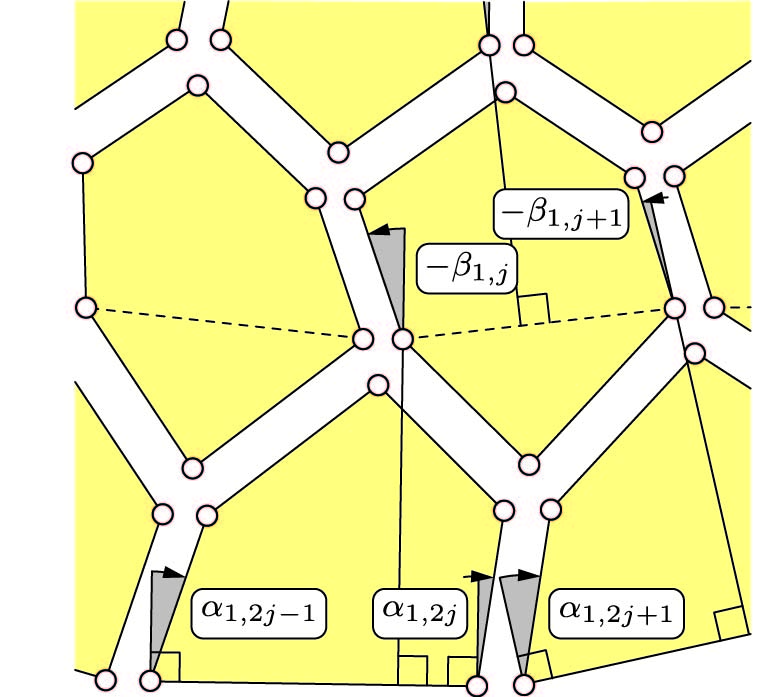}
                        \small(b)\vspace{7mm}
                        \includegraphics[width=\marginparwidth]{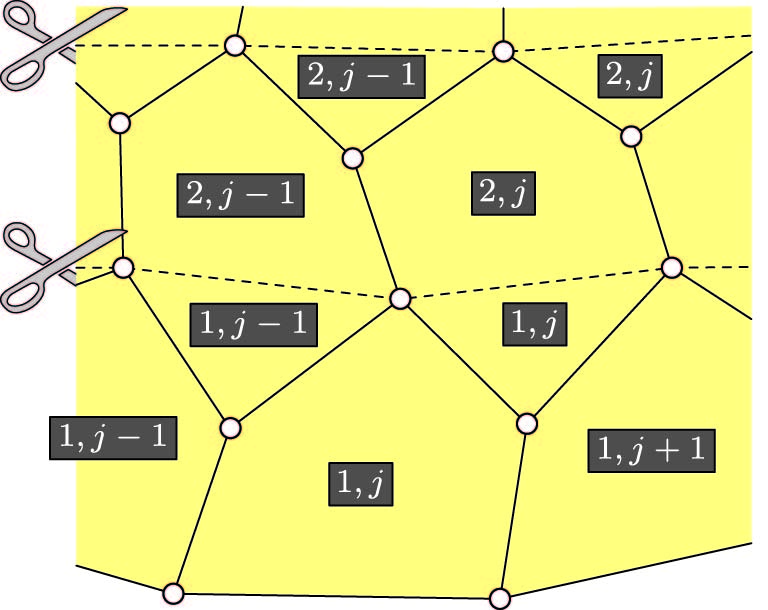}
                        \small(c)
                    \end{center}}[-22mm]

                \noindent The cell corner angles of pressure actuated cellular structures can be expressed in terms of global state variables $\mathbf{u}$ and effective cell side lengths $\mathbf{v}$ as illustrated in Figure~\ref{pic:Figure_9_2}. The cell side lengths are defined with regards to fictitious, centric cell corners where
                \begin{align}
                    \mathbf{v} =
                    \left[
                    \begin{array}{cccccc}
                        {\mathbf{b}_1}^\top &
                        {\mathbf{c}_1}^\top & \ldots &
                        {\mathbf{b}_{nR}}^\top &
                        {\mathbf{c}_{nR}}^\top &
                        {\mathbf{a}}^\top
                    \end{array}
                    \right]^\top\in\mathbb{R}^{nv}
                \end{align}
                and, for example $\mathbf{b}_i = \left[\begin{array}{ccc}b_{i,1} & \ldots & b_{i,nP+2-i}\end{array}\right]^\top$. Herein, $n_P$ denotes the number of base pentagons and $n_R$ the number of cell rows. It can be seen from the vector $\mathbf{b}_i$ that it is assumed that each additional cell row contains one cell less than the previous row. This is not a limitation of the proposed framework since arbitrary topologies at both ends can be modeled with the help of constraints. Therefore, the total number of hexagonal cells $n_T$ and cell sides $n_v$ of a pressure actuated cellular structures are
                \begin{align}
                    n_T = \frac{1}{2}\left(2 n_P-n_R\right)\left(n_R-1\right)
                \end{align}
                and
                \begin{align}
                    n_v = 4 n_P + 3 n_T + n_R.
                \end{align}
                The global state variables $\mathbf{u}$ that define the inclinations of the cell sides $\mathbf{b}$ are
                \begin{align}
                    \mathbf{u} =
                    \left[
                    \begin{array}{cccccc}
                        \alpha_{1,1} & \ldots & \alpha_{1,2 nP} &
                        \beta_{1,1}  & \ldots & \beta_{nR-1,nP-nR+2}
                    \end{array}
                    \right]^\top\in\mathbb{R}^{nu}
                \end{align}
                where the number of state variables $n_u$ is
                \begin{align}
                    n_u = 2 n_P + n_T + n_R - 1.
                \end{align}


            \subsection{Pentagonal Cells}
                \subsubsection{Local State Variables}

                \marginnote{
                    \begin{center}
                        \includegraphics[width=\marginparwidth]{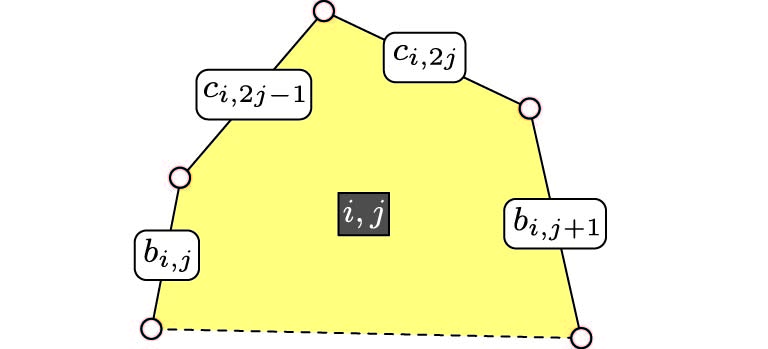}
                        \small(a)\vspace{8mm}
                        \includegraphics[width=\marginparwidth]{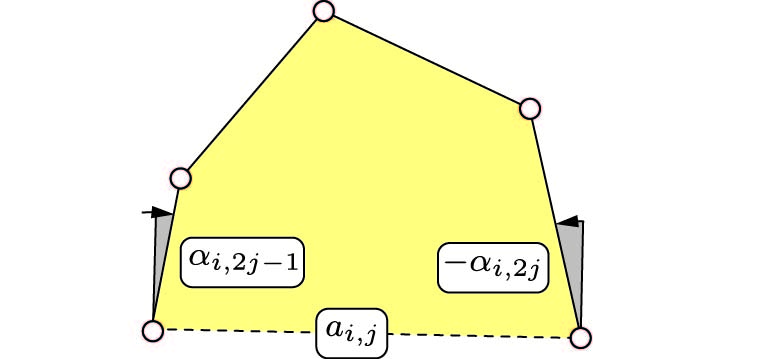}
                        \small(b)\vspace{8mm}
                        \includegraphics[width=\marginparwidth]{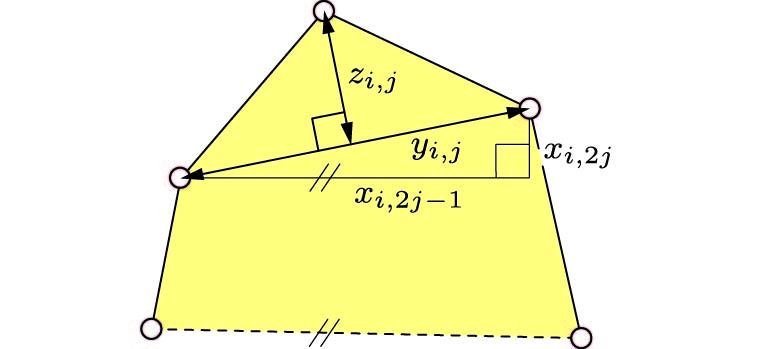}
                        \small(c)\vspace{8mm}
                        \includegraphics[width=\marginparwidth]{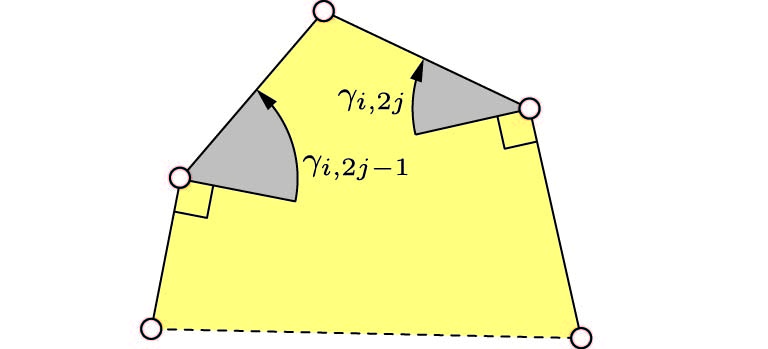}
                        \small(d)
                        \captionof{figure}{Variables of the $j$-th pentagonal cell in the $i$-th cell row. Local state variables (a) $\mathbf{v}^\textrm{P}$ for the cell side lengths and (b) $\mathbf{u}^\textrm{P}$ for the cell side inclinations and base length. (c) Internal lengths $\mathbf{x}$, $y$, $z$ and (d) internal angles $\boldsymbol{\gamma}$.}
                        \label{pic:Figure_9_3}
                    \end{center}}[-20mm]

                    \noindent Pressure actuated cellular structures with more than one cell row can be cut into strips that consist of triangular and/or pentagonal cells. The internal lengths $\mathbf{x}$, $y$, $z$ and cell corner angles $\boldsymbol{\gamma}$ of the $j$-th pentagonal cell in the $i$-th cell row can be expressed in terms of the effective cell side lengths
                    \begin{align}
                        \mathbf{v}^\textrm{P}_{i,j}\left(\mathbf{v}\right) = \left[
                        \begin{array}{cccc}
                            b_{i,j} & b_{i,j+1} & c_{i,2j-1} & c_{i,2j}
                        \end{array}\right]^\top
                    \end{align}
                    and the local state variables
                    \begin{align}
                        \mathbf{u}_{i,j}^\textrm{P}\left(\mathbf{u},\mathbf{v}\right) = \left[
                        \begin{array}{ccc}
                            \alpha_{i,2j-1} & \alpha_{i,2j} & a_{i,j}
                        \end{array}
                        \right]^\top
                    \end{align}
                    as shown in Figure~\ref{pic:Figure_9_3}. It can be seen that a superscript ``P" is used to denote the local state variables of pentagonal cells. The base side $a$ is a part of $\mathbf{u}^\textrm{P}$ since it is an abstract term for hexagonal cells that are cut into a triangular and pentagonal part.


                \subsubsection{Dependent Variables}
                    The local state variables of a pentagonal cell can be used to derive a number of dependent variables. The length $y$ that divides the pentagonal cell into a triangular and quadrilateral part is
                    \begin{align}
                        y_{i,j} &= \sqrt{{x_{i,2j-1}}^2 + {x_{i,2j}}^2}\\\nonumber
                        &= \sqrt{\left(a + \sin\left(\alpha_2\right)b_2 - \sin\left(\alpha_1\right)b_1\right)^2 + \left(\cos\left(\alpha_2\right)b_2 - \cos\left(\alpha_1\right)b_1\right)^2}
                    \end{align}
                    and the altitude $z$ of the triangular part can be expressed as
                    \begin{align}
                        z_{i,j} = \sqrt{{c_1}^2 - \frac{1}{4{y_{i,j}}^2}\left({y_{i,j}}^2 + {c_1}^2 - {c_2}^2\right)^2}.
                    \end{align}
                    Previous expressions are based on the abbreviations
                    \begin{align}
                        \mathbf{v}_{i,j}^\textrm{P} =
                        \left[
                        \begin{array}{cccc}
                            b_1 & b_2 & c_1 & c_2
                        \end{array}\right]^\top
                    \end{align}
                    and
                    \begin{align}
                        \mathbf{u}_{i,j}^\textrm{P} =
                        \left[
                        \begin{array}{ccc}
                            \alpha_1 & \alpha_2 & a
                        \end{array}\right]^\top
                    \end{align}
                    that are used for the sake of clarity. The internal angle $\gamma_{i,2j-1}$ of a pentagonal cell from which the cell corner angles can be derived is completely determined by the local state variables through
                    \begin{align}
                        \gamma_{i,2j-1} &=
                        \begin{cases}
                            \displaystyle\alpha_1 + \arcsin\left(\frac{x_{i,2j}}{y_{i,j}}\right) + \arcsin\left(\frac{z_{i,j}}{c_1}\right)
                            & {c_2}^2 \leq {y_{i,j}}^2 + {c_1}^2\\
                            \displaystyle\alpha_1 + \arcsin\left(\frac{x_{i,2j}}{y_{i,j}}\right) - \arcsin\left(\frac{z_{i,j}}{c_1}\right) + \pi
                            & {c_2}^2 > {y_{i,j}}^2 + {c_1}^2.
                        \end{cases}
                    \end{align}
                    The corresponding expressions for the internal angle $\gamma_{i,2j}$ are derived in a similar manner. The previous equation could be written without a distinction of cases. However, this would result in lengthier expressions. The cross sectional area $A^\textrm{P}$ of a pentagonal cell that is required for the simulation and optimization of pressure actuated cellular structures that is outlined in the following chapters results in
                    \begin{align}
                        A_{i,j}^\textrm{P} =
                        \frac{1}{2} \left(\left(\cos\left(\alpha_1\right) b_1 + \cos\left(\alpha_2\right) b_2\right) a + \sin\left(\alpha_2 - \alpha_1\right) b_1 b_2 + y_{i,j} z_{i,j}\right)
                    \end{align}
                    where the superscript ``P" is used to distinguish it from the cross sectional area of triangular cells. However, despite its notation, it is not a state variable.


            \marginnote{
                \begin{center}
                    \includegraphics[width=\marginparwidth]{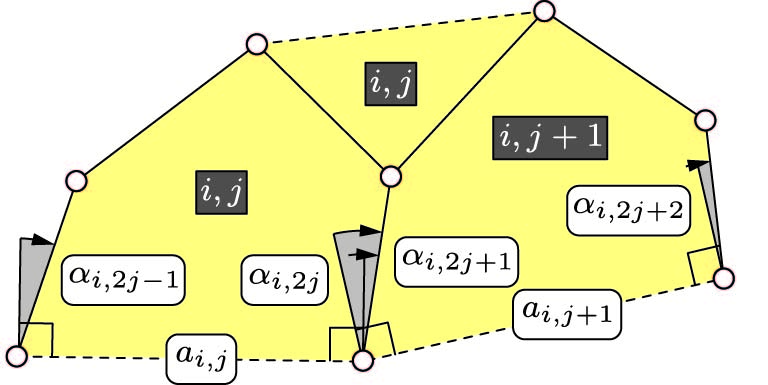}
                    \small(a)\vspace{6mm}
                    \includegraphics[width=\marginparwidth]{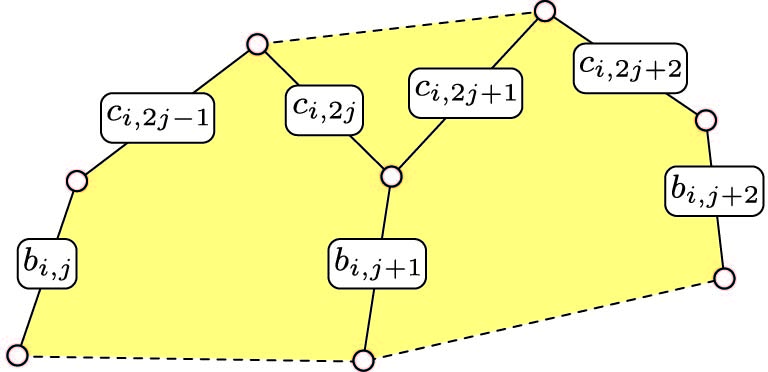}
                    \small(b)\vspace{6mm}
                    \includegraphics[width=\marginparwidth]{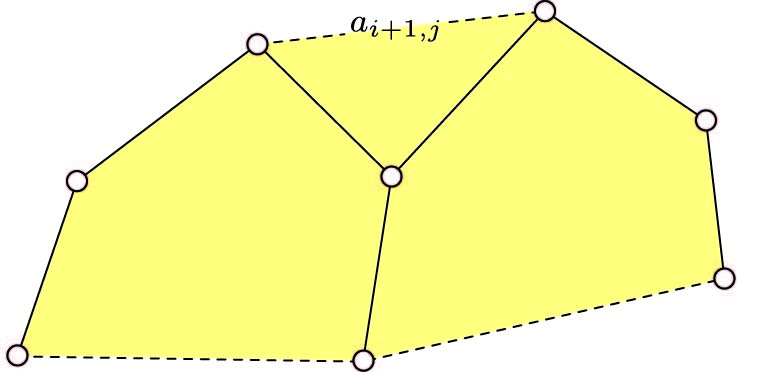}
                    \small(c)\vspace{6mm}
                    \includegraphics[width=\marginparwidth]{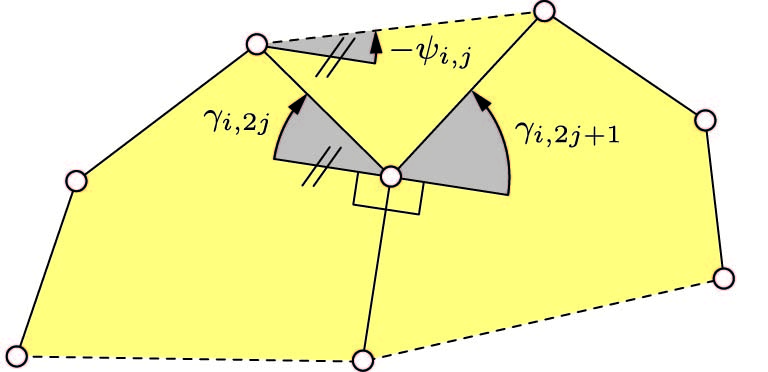}
                    \small(d)
                    \captionof{figure}{Variables of the $j$-th triangular cell in the $i$-th cell row. Local state variables (a) $\mathbf{v}^\textrm{T}$ for the cell side lengths and (b) $\mathbf{u}^\textrm{T}$ for the cell side inclinations and base lengths. (c) Internal length $a$ and (d) internal angles $\boldsymbol{\gamma}$, $\psi$.}
                    \label{pic:Figure_9_4}
                \end{center}}[-63mm]

            \subsection{Triangular Cells}
                \subsubsection{Local State Variables}
                    A triangular cell is obtained by cutting a hexagonal cell into a triangular and pentagonal part. The total number of triangular cells $n_\textrm{T}$ is therefore equal to the total number of hexagonal cells. The cell corner angles of the $j$-th triangular cell in the $i$-th cell row can be expressed in terms of the effective cell side lengths
                    \begin{align}
                        \mathbf{v}^\textrm{T}_{i,j}\left(\mathbf{v}^\textrm{P}\right) =
                        \left[
                        \begin{array}{ccccccc}
                            b_{i,j} & b_{i,j+1} & b_{i,j+2} & c_{i,2j-1} & c_{i,2j} & c_{i,2j+1} & c_{i,2j+2}
                        \end{array}\right]^\top
                    \end{align}
                    and the state variables
                    \begin{align}
                        \mathbf{u}_{i,j}^\textrm{T}\left(\mathbf{u}^\textrm{P}\right) =
                        \left[
                        \begin{array}{cccccc}
                            \alpha_{i,2j-1} & \alpha_{i,2j} & \alpha_{i,2j+1} & \alpha_{i,2j+2} & a_{i,j} & a_{i,j+1}
                        \end{array}\right]^\top.
                    \end{align}
                    The local state variables of a triangular cell that are denoted by a superscript ``T" can thus be fully assembled from the local state variables of its neighbouring pentagonal cells as illustrated in Figure~\ref{pic:Figure_9_4}.


                \subsubsection{Dependent Variables}
                    The length $a$ that divides a hexagonal cell into a triangular and pentagonal part can be derived from the local state variables of a triangular cell so that
                    \begin{align}
                        a_{i+1,j} = \sqrt{{c_{i,2j}}^2 + {c_{i,2j+1}}^2 + 2c_{2j} c_{2j+1} \cos\left(\gamma_{i,2j} + \gamma_{i,2j+1}\right)}.
                    \end{align}
                    Similarly, the internal angle $\psi$ of the triangular cell results in
                    \begin{align}
                        \psi_{i,j} = \gamma_{i,2j} - \arccos\left(\frac{{a_{i+1,j}}^2+{c_{i,2j}}^2-{c_{i,2j+1}}^2}{2 a_{i+1,j} c_{i,2j}}\right).
                    \end{align}
                    Finally, the cross sectional area $A^\textrm{T}$ of a triangular cell is
                    \begin{align}
                        A^\textrm{T}_{i,j} = \frac{1}{2} c_{i,2j} c_{i,2j+1} \sin\left(\gamma_{i,2j}+\gamma_{i,2j+1}\right).
                    \end{align}
                    It can be seen that the length $a$, the angle $\psi$ and the cross sectional area $A^\textrm{T}$ are functions of the internal angles $\boldsymbol{\gamma}$ of the neighbouring pentagonal cells.


            \marginnote{
                \begin{center}
                    \includegraphics[width=\marginparwidth]{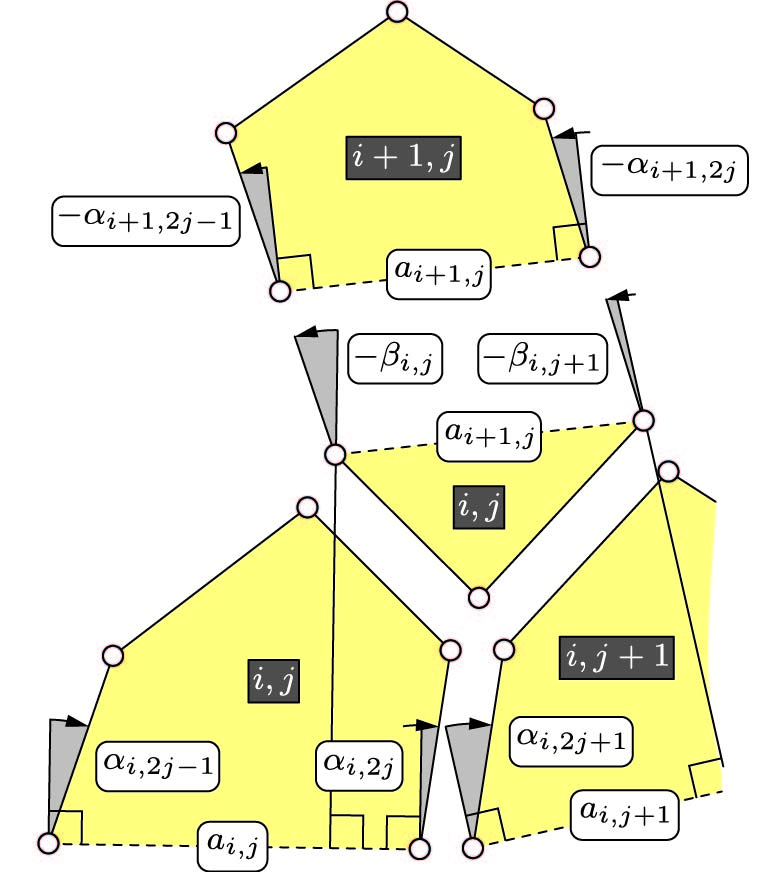}
                    \small(a)\vspace{10mm}
                    \includegraphics[width=\marginparwidth]{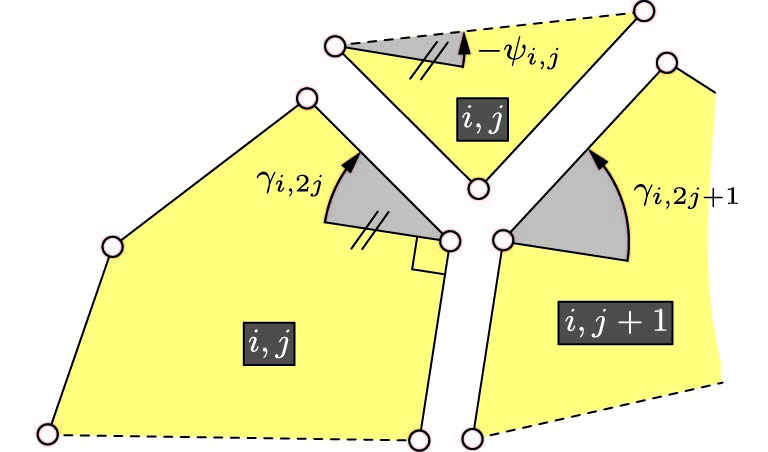}
                    \small(b)
                    \captionof{figure}{(a) Global, local state variables and dependent variable of a hexagonal cell that is cut into a pentagonal and triangular part. (b) Internal angles of triangular and neighbouring pentagonal cells. An exploded view is used to increase the clarity of the presentation.}
                    \label{pic:Figure_9_5}
                \end{center}}[-35mm]

            \subsection{Transformation Matrices}
                The local state variables of the pentagonal $\mathbf{u}_{i,j}^\textrm{P}$ and triangular $\mathbf{u}_{i,j}^\textrm{T}$ cells can be directly assembled from the global state variables $\mathbf{u}$ and $\mathbf{v}$ if they are located in the first cell row ($i=1$). In contrast, local state variables of cells in other rows depend on both, the global state variables $\boldsymbol{\beta}$ that are contained in $\mathbf{u}$ and the local state variables of the previous cell row. As illustrated in Figure~\ref{pic:Figure_9_5}, the state variables $\mathbf{u}^\textrm{P}_{i+1,j}$ of the $j$-th pentagonal cell in the $(i+1)$-th cell row can be expressed in terms of the state variables $\mathbf{u}^\textrm{T}_{i,j}$, $\boldsymbol{\beta}^\textrm{T}_{i,j} = \left[\begin{array}{cc}\beta_{i,j} & \beta_{i,j+1}\end{array}\right]^\top$ and $\mathbf{v}^\textrm{T}_{i,j}$ of the $j$-th triangular cell in the $i$-th cell row such that
                \begin{align}
                    \mathbf{u}^\textrm{P}_{i+1,j}\left(\mathbf{u}^\textrm{T}_{i,j}, \mathbf{v}^\textrm{T}_{i,j},\boldsymbol{\beta}^\textrm{T}_{i,j}\right) =
                    \mathbf{T}^\beta_{i,j} \boldsymbol{\beta}^\textrm{T}_{i,j} + \mathbf{T}^\textrm{lin}_{i,j} \mathbf{u}_{i,j}^\textrm{T} + \mathbf{T}^\textrm{nlin}_{i,j}\left(\mathbf{u}^\textrm{T}_{i,j}, \mathbf{v}^\textrm{T}_{i,j}\right)
                \end{align}
                where the linear matrices are
                \begin{align}
                    \mathbf{T}^\beta_{i,j} =
                    \left[
                    \begin{array}{cc}
                        1 & 0\\
                        0 & 1\\
                        0 & 0
                    \end{array}
                    \right]
                \end{align}
                and
                \begin{align}
                    \mathbf{T}^{\textrm{lin}}_{i,j} =
                    \left[
                    \begin{array}{cccccc}
                        0 & -1 &  0 & 0 & 0 & 0\\
                        0 &  0 & -1 & 0 & 0 & 0\\
                        0 &  0 &  0 & 0 & 0 & 0
                    \end{array}
                    \right].
                \end{align}
                The nonlinear matrix that depends on the internal length $a$ and angle $\psi$ of a triangular cell is
                \begin{align}
                    \mathbf{T}_{i,j}^\textrm{nlin}\left(\mathbf{u}_{i,j}^\textrm{T},\mathbf{v}_{i,j}^\textrm{T}\right) =
                    \left[
                    \begin{array}{c}
                        -\psi_{i,j}\left(\mathbf{u}_{i,j}^\textrm{T},\mathbf{v}_{i,j}^\textrm{T}\right)\\
                        -\psi_{i,j}\left(\mathbf{u}_{i,j}^\textrm{T},\mathbf{v}_{i,j}^\textrm{T}\right)\\
                        a_{i+1,j}\left(\mathbf{u}_{i,j}^\textrm{T},\mathbf{v}_{i,j}^\textrm{T}\right)
                    \end{array}
                    \right].
                \end{align}
                Previous matrices can be used to assemble the transformation matrix $\mathbf{T}^u$ that relates the pentagonal state variables $\mathbf{u}^\textrm{P}$ to the triangular state variables $\mathbf{u}^\textrm{T}$
                \begin{align}
                    \mathbf{T}^u_{i,j} =
                    \frac{\partial\mathbf{u}_{i+1,j}^\textrm{P}} {\partial\mathbf{u}_{i,j}^\textrm{T}} =
                    \mathbf{T}_{i,j}^\textrm{lin} + \frac{\partial\mathbf{T}_{i,j}^\textrm{nlin}}{\partial\mathbf{u}_{i,j}^\textrm{T}}.
                \end{align}
                Similarly, the transformation matrix $\mathbf{T}^v$ relates the pentagonal state variables $\mathbf{u}^\textrm{P}$ to the triangular state variables $\mathbf{v}^\textrm{T}$ so that
                \begin{align}
                    \mathbf{T}^v_{i,j} =
                    \frac{\partial\mathbf{u}_{i+1,j}^\textrm{P}} {\partial\mathbf{v}_{i,j}^\textrm{T}} =
                    \frac{\partial\mathbf{T}_{i,j}^\textrm{nlin}}{\partial\mathbf{v}_{i,j}^\textrm{T}}.
                \end{align}
                Hence it is possible to sequentially compute the local state variables of triangular and pentagonal cells from the local state variables of previous cell rows.


        \section{Corner Geometries}
            \subsection{Global State Variables}
                Compliant pressure actuated cellular structures that are assembled from rectangular cell sides possess pointed cell corners. These corners are subsequently rounded out with circular arcs to increase their aesthetics as well as stiffness and to minimize stress concentrations. This requires, besides the global state variables $\mathbf{u}$ and $\mathbf{v}$ from which the cell corner angles are derived, additional global state variables that describe the cell side thicknesses $\mathbf{t}$ and the target curvature $\kappa^{\textrm{tar}}$ of all cell corner fillets. The ordering of the cell side thicknesses is similar to that of the cell side lengths $\mathbf{v}$ so that
                \begin{align}
                    \mathbf{t} = \left[
                    \begin{array}{cccccc}
                        {\mathbf{t}_{b,1}}^\top & {\mathbf{t}_{c,1}}^\top & \ldots & {\mathbf{t}_{b,nR}}^\top & {\mathbf{t}_{c,nR}}^\top & {\mathbf{t}_a}^\top
                    \end{array}\right]^\top\in\mathbb{R}^{nv}
                \end{align}
                where, for example $\mathbf{t}_{b,i} = \left[\begin{array}{ccc}t_{b,i,1} & \ldots & t_{b,i,nP+2-i}\end{array}\right]^\top$.


            \subsection{Local State Variables}
                The geometry of the $k$-th cell corner that connects $n_s$ cell sides can be expressed in terms of the cell side thicknesses
                \begin{align}
                    \mathbf{t}_k^\textrm{C}\left(\mathbf{t}\right) = \left[
                    \begin{array}{ccc}
                        t_{k,1} & \ldots & t_{k,ns}
                    \end{array}\right]^\top
                \end{align}
                and the local state variables $\mathbf{u}^\textrm{C}$ that are composed of the cell corner angles $\boldsymbol{\theta}$ so that
                \begin{align}
                    \mathbf{u}_k^\textrm{C}\left(\mathbf{u}^\textrm{P},\mathbf{v}^\textrm{P}\right) = \left[
                    \begin{array}{ccc}
                        \theta_{k,1} & \ldots & \theta_{k,ns-1}
                    \end{array}\right]^\top.
                \end{align}
                The latter contains only $n_s-1$ angles since $\theta_{k,ns} = 2\pi-\sum_{i=1}^{n_s-1} \theta_{k,i}$. It can be seen that the superscript ``C" is used to denote local state variables of cell corners. Variables of the $k$-th cell corner that connects $n_s=3$ cell sides are illustrated in Figure~\ref{pic:Figure_9_6}.


            \subsection{Dependent Variables}
                The angles $\boldsymbol{\omega}$ between the cell sides and chords of the circular fillets as well as the chord lengths $\mathbf{x}$ can be computed from the local state variables $\mathbf{u}^\textrm{C}$, $\mathbf{t}^\textrm{C}$ and the unknown cell corner dimensions $\boldsymbol{\xi}$. However, their derivations are straightforward so that they are not explicitly given. A $C^1$ continuous transition between cell corners and sides requires that $\sum_{i=1}^{n_s}\left(\omega_{k,2i}-\omega_{k,2i-1}\right)^2 = 0$. This can not be satisfied for arbitrary cell side thicknesses and cell corner angles. Hence it is necessary to average the kinks between the cell sides and circular arcs so that the fillet curvatures $\boldsymbol{\kappa}$ become
                \begin{align}
                    \kappa_{k,i} = \frac{2}{x_{k,i}} \cos\left(\frac{\theta_{k,i}}{2}\right)
                \end{align}
                where $i=1,\ldots,n_s$.


            \marginnote{
                \begin{center}
                    \includegraphics[width=\marginparwidth]{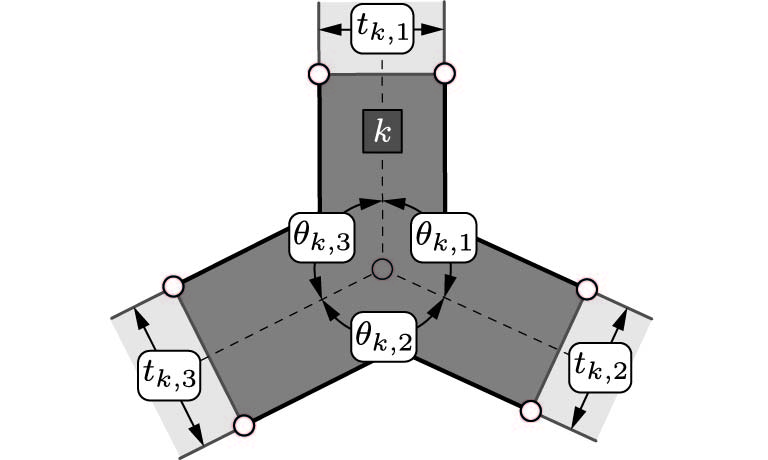}
                    \small(a)\vspace{10mm}
                    \includegraphics[width=\marginparwidth]{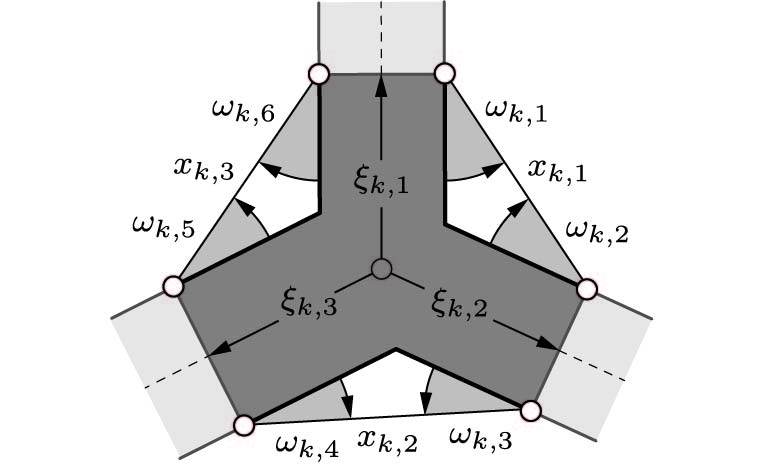}
                    \small(b)\vspace{10mm}
                    \includegraphics[width=\marginparwidth]{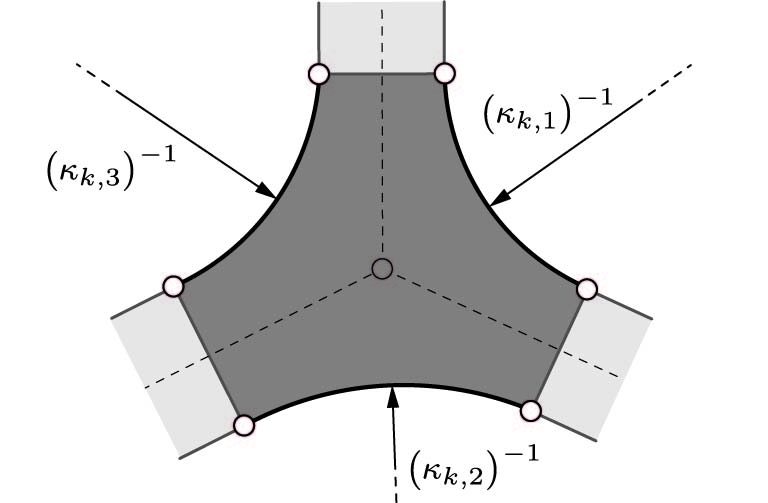}
                    \small(c)
                    \captionof{figure}{Variables of the $k$-th cell corner. (a) Local state variables $\mathbf{u}^\textrm{C}$ and $\mathbf{t}^\textrm{C}$. (b) Internal angles $\boldsymbol{\omega}$, chord lengths $\mathbf{x}$ and cell corner dimensions $\boldsymbol{\xi}$. (d) Fillet curvatures $\boldsymbol{\kappa}$.}
                    \label{pic:Figure_9_6}
                \end{center}}[-62mm]

            \subsection{Optimal Cell Corner Dimensions}
                It is not possible to directly compute the cell corner dimensions $\boldsymbol{\xi}$ from the local state variables of a cell corner. Instead they have to be optimized such that they minimize both, the kinks between the cell sides and circular arcs as well as the deviations of the fillet curvatures from the target value $\kappa^\textrm{tar}$. This is subsequently done in two steps.


                \subsubsection{First Step}
                    Kinks between cell sides and fillets can be minimized in a least square sense by computing
                    \begin{align}
                        \tilde{\boldsymbol{\xi}}_k\left(\mathbf{t}_k^\textrm{C}, \mathbf{u}_k^\textrm{C}, \xi^{\textrm{tar}}_k\right) = \underset{\boldsymbol{\xi}_k}{\underset{\mathcal{C}_k\left(\boldsymbol{\xi}_k,\xi^{\textrm{tar}}_k\right) = 0}{\textrm{argmin}}}\ \mathcal{F}_k\left(\mathbf{t}_k^\textrm{C}, \mathbf{u}_k^\textrm{C}, \boldsymbol{\xi}_k\right)
                    \end{align}
                    where the objective $\mathcal{F}$ is
                    \begin{align}
                        \mathcal{F}_k\left(\mathbf{t}_k^\textrm{C},\mathbf{u}_k^\textrm{C},\boldsymbol{\xi}_k\right) &= \sum_{i=1}^{n_s} \left(\omega_{k,2i} - \omega_{k,2i-1}\right)^2
                    \end{align}
                    and the constraint $\mathcal{C}$ that restricts the solution space for $\boldsymbol{\xi}$ is
                    \begin{align}
                        \mathcal{C}_k\left(\boldsymbol{\xi}_k, \xi^{\textrm{tar}}_k\right) = \xi^{\textrm{tar}}_k - \sum_{i=1}^{n_s} \xi_{k,i}.
                    \end{align}
                    Hence, the Lagrangian of the first optimization step can be written as
                    \begin{align}
                        \underset{\mathbf{u}_k^\textrm{C},\mathbf{t}_k^\textrm{C},\xi^{\textrm{tar}}_k}{\mathcal{L}_k} \left(\boldsymbol{\xi}_k,\lambda_k\right) =
                        \underset{\mathbf{u}_k^\textrm{C},\mathbf{t}_k^\textrm{C}}{\mathcal{F}_k}\left(\boldsymbol{\xi}_k\right) +
                        \lambda_k \underset{\xi^{\textrm{tar}}_k}{\mathcal{C}_k}\left(\boldsymbol{\xi}_k\right).
                    \end{align}
                    The variables $\tilde{\boldsymbol{\upsilon}} = \left[\begin{array}{cc}\tilde{\boldsymbol{\xi}} & \tilde{\lambda}\end{array}\right]^\top$ that satisfy the stationarity condition can be iteratively computed by using a Newton based approach. Variables of the $(i+1)$-th iteration are
                    \begin{align}
                        \boldsymbol{\upsilon}_k^{i+1} = \boldsymbol{\upsilon}_k^i - \left(\mathbf{H}^i_k\right)^{-1} \mathbf{g}^i_k
                    \end{align}
                    where the gradient $\mathbf{g}$ is
                    \begin{align}
                        \underset{\mathbf{u}_k^\textrm{C},\mathbf{t}_k^\textrm{C},\xi^{\textrm{tar}}_k}{\mathbf{g}^i_k}\left(\boldsymbol{\upsilon}_k^i\right) =
                        \left[
                        \begin{array}{cc}
                            \displaystyle \frac{\partial\mathcal{L}_k^i}{\partial\boldsymbol{\xi}_k} &
                            \displaystyle \frac{\partial\mathcal{L}_k^i}{\partial\lambda_k}
                        \end{array}
                        \right]^\top
                    \end{align}
                    and the Hessian $\mathbf{H}$ is
                    \begin{align}
                        &\underset{\mathbf{u}_k^\textrm{C},\mathbf{t}_k^\textrm{C},\xi^{\textrm{tar}}_k}{\mathbf{H}_k^i}\left(\boldsymbol{\upsilon}_k^i\right) =\\\nonumber
                        &\hspace{14mm}\left[
                        \begin{array}{cc}
                            \displaystyle \frac{\partial^2\mathcal{L}_k^i}{\partial{\boldsymbol{\xi}_k}^2} & \displaystyle \frac{\partial^2\mathcal{L}_k^i}{\partial\lambda_k\partial\boldsymbol{\xi}_k}\\
                            \displaystyle \frac{\partial^2\mathcal{L}_k^i}{\partial\boldsymbol{\xi}_k\partial\lambda_k} &
                            \mathbf{0}
                        \end{array}
                        \right] =
                        \left[
                        \begin{array}{cc}
                            \left({\mathbf{H}_k^i}^{-1}\right)_{11} & \left({\mathbf{H}_k^i}^{-1}\right)_{12}\\
                            \left({\mathbf{H}_k^i}^{-1}\right)_{21} & \left({\mathbf{H}_k^i}^{-1}\right)_{22}
                        \end{array}
                        \right]^{-1}.
                    \end{align}


                \subsubsection{Coupling Terms}
                    The first and second optimization step are coupled by the sensitivities $\partial \tilde{\boldsymbol{\xi}}/\partial {\xi^\textrm{tar}}$ and $\partial^2 \tilde{\boldsymbol{\xi}}/\partial {\xi^\textrm{tar}}^2$. Stationarity of $\tilde{\mathbf{g}} = \mathbf{g}\left(\mathbf{t}^\textrm{C},\mathbf{u}^\textrm{C},\xi^{\textrm{tar}},\tilde{\boldsymbol{\upsilon}}\right)$ requires that
                    \begin{align}
                        \tilde{\mathbf{g}}_k\left(\mathbf{t}_k^\textrm{C}, \mathbf{u}_k^\textrm{C}, \xi_k^\textrm{tar}+\Delta \xi_k^\textrm{tar}, \tilde{\boldsymbol{\upsilon}}_k + \Delta \boldsymbol{\upsilon}_k\right) = \mathbf{0}.
                    \end{align}
                    Linearization of the stationarity condition at the optimum leads to
                    \begin{align}
                        \tilde{\mathbf{g}}_k\left(\mathbf{t}_k^\textrm{C},\mathbf{u}_k^\textrm{C},\xi_k^\textrm{tar},\tilde{\boldsymbol{\upsilon}}_k\right) + \frac{\partial\tilde{\mathbf{g}}_k}{\partial\xi_k^\textrm{tar}} \Delta \xi_k^\textrm{tar} + \frac{\partial \tilde{\mathbf{g}}_k}{\partial\boldsymbol{\upsilon}_k} \Delta \boldsymbol{\upsilon}_k = \mathbf{0}
                    \end{align}
                    so that
                    \begin{align}
                        \frac{\partial\tilde{\boldsymbol{\upsilon}}_k}{\partial\xi_k^\textrm{tar}} =
                        -\left(\frac{\partial \tilde{\mathbf{g}}_k}{\partial\boldsymbol{\upsilon}_k}\right)^{-1} \frac{\partial\tilde{\mathbf{g}}_k}{\partial\xi_k^\textrm{tar}}.
                    \end{align}
                    Hence the sensitivities of $\tilde{\boldsymbol{\xi}}$ with respect to $\xi^\textrm{tar}$ are
                    \begin{align}
                        \frac{\partial \tilde{\boldsymbol{\xi}}_k}{\partial \xi_k^\textrm{tar}} = \left({\tilde{\mathbf{H}}_k}^{-1}\right)_{12}
                    \end{align}
                    and
                    \begin{align}
                        \frac{\partial^2\tilde{\boldsymbol{\xi}}_k}{\partial {\xi_k^\textrm{tar}}^2} = - \left({\tilde{\mathbf{H}}_k}^{-1}\right)_{11}
                        \left[
                        \begin{array}{c}
                            \displaystyle
                            \frac{\partial^3\tilde{\mathcal{L}_k}}{\partial{\boldsymbol{\xi}_k}^2 \partial\xi_{k,1}} \left({\tilde{\mathbf{H}}_k}^{-1}\right)_{12}\\
                            \vdots\\
                            \displaystyle
                            \frac{\partial^3\tilde{\mathcal{L}}_k}{\partial{\boldsymbol{\xi}_k}^2 \partial\xi_{k,ns}} \left({\tilde{\mathbf{H}}_k}^{-1}\right)_{12}
                        \end{array}
                        \right]^\top
                        \frac{\partial \tilde{\boldsymbol{\xi}}_k}{\partial \xi_k^\textrm{tar}}.
                    \end{align}


                \subsubsection{Second Step}
                    The optimal value for $\xi^{\textrm{tar}}$ is determined in a second step by computing
                    \begin{align}
                        \tilde{\xi}^\textrm{tar}_k\left(\mathbf{t}_k^\textrm{C}, \mathbf{u}_k^\textrm{C}\right) = \underset{\xi^\textrm{tar}_k}{\textrm{argmin}}\ \mathcal{G}_k\left(\mathbf{t}_k^\textrm{C}, \mathbf{u}_k^\textrm{C}, \tilde{\boldsymbol{\xi}}_k\left(\mathbf{t}_k^\textrm{C}, \mathbf{u}_k^\textrm{C}, \xi^{\textrm{tar}}_k\right)\right)
                    \end{align}
                    where the difference between target and fillet curvatures is taken into account by the objective
                    \begin{align}
                        \mathcal{G}_k\left(\mathbf{u}_k^\mathbf{C}, \mathbf{t}_k^\textrm{C}, \tilde{\boldsymbol{\xi}}_k\left(\mathbf{u}_k^\textrm{C}, \mathbf{t}_k^\textrm{C}, \xi^{\textrm{tar}}_k\right)\right) = \sum_{i=1}^{n_s} \left(\kappa_{k,i}-\kappa ^{\textrm{tar}}_k\right)^2.
                    \end{align}
                    Similar to the first step, a Newton based approach is used to compute the variable $\tilde{\xi}^\textrm{tar}$ that satisfies the stationarity condition. The variable of the $\left(i+1\right)$-th iteration is
                    \begin{align}
                        \xi_k^{\textrm{tar},i+1} = \xi_k^{\textrm{tar},i} - \Delta \xi_k^{\textrm{tar},i}
                    \end{align}
                    where
                    \begin{align}
                        \Delta \xi_k^{\textrm{tar},i} =
                        \begin{cases}
                            \displaystyle
                            \left(\frac{\partial \mathcal{G}_k^i}{\partial \xi_k^\textrm{tar}}\right)^{-1} \mathcal{G}_k^i
                            & \textrm{for} \hspace{5mm} \displaystyle
                            \frac{\partial^2 \mathcal{G}_k^i}{\partial {\xi_k^\textrm{tar}}^2} \leq 0\\
                            \displaystyle
                            \left(\frac{\partial^2 \mathcal{G}_k^i}{\partial {\xi_k^\textrm{tar}}^2}\right)^{-1}
                            \frac{\partial \mathcal{G}_k^i}{\partial \xi_k^\textrm{tar}}
                            & \textrm{for} \hspace{5mm}
                            \displaystyle
                            \frac{\partial^2 \mathcal{G}_k^i}{\partial {\xi_k^\textrm{tar}}^2} > 0.
                        \end{cases}
                    \end{align}
                    The corresponding derivatives are
                    \begin{align}
                        \frac{\partial \mathcal{G}_k^i}{\partial \xi_k^\textrm{tar}} =
                        \frac{\partial \mathcal{G}_k^i}{\partial \tilde{\boldsymbol{\xi}}_k}
                        \frac{\partial \tilde{\boldsymbol{\xi}}_k^i}{\partial \xi_k^\textrm{tar}}
                    \end{align}
                    and
                    \begin{align}
                        \frac{\partial^2 \mathcal{G}_k^i}{\partial {\xi_k^\textrm{tar}}^2} =
                        {\frac{\partial \tilde{\boldsymbol{\xi}}_k^i}{\partial \xi_k^\textrm{tar}}}^\top
                        \frac{\partial^2 \mathcal{G}_k^i}{\partial {\tilde{\boldsymbol{\xi}}_k}^2}
                        \frac{\partial \tilde{\boldsymbol{\xi}}_k^i}{\partial \xi_k^\textrm{tar}} +
                        \frac{\partial \mathcal{G}_k^i}{\partial \tilde{\boldsymbol{\xi}}_k}
                        \frac{\partial^2 \tilde{\boldsymbol{\xi}}_k^i}{\partial {\xi_k^\textrm{tar}}^2}.
                    \end{align}
                    It can be seen that both derivatives are based on the previously derived coupling terms.


                \subsubsection{Sensitivities}
                    The optimization of compliant pressure actuated cellular structures that is outlined in the following chapters requires the sensitivities of $\tilde{\xi}$ with respect to state variables $\mathbf{u}^\textrm{C}$ and $\mathbf{t}^\textrm{C}$. They can be expressed as
                    \begin{align}
                        \frac{d \tilde{\boldsymbol{\xi}}_k}{d \mathbf{u}_k^\textrm{C}} =
                        \frac{\partial \tilde{\boldsymbol{\xi}}_k}{\partial\xi_k^\textrm{tar}}
                        \frac{\partial \tilde{\xi}_k^\textrm{tar}}{\partial \mathbf{u}_k^\textrm{C}} +
                        \frac{\partial \tilde{\boldsymbol{\xi}}_k}{\partial \mathbf{u}_k^\textrm{C}}
                    \end{align}
                    and
                    \begin{align}
                        \frac{d \tilde{\boldsymbol{\xi}}_k}{d \mathbf{t}_k^\textrm{C}} =
                        \frac{\partial \tilde{\boldsymbol{\xi}}_k}{\partial\xi_k^\textrm{tar}}
                        \frac{\partial \tilde{\xi}_k^\textrm{tar}}{\partial \mathbf{t}_k^\textrm{C}} +
                        \frac{\partial \tilde{\boldsymbol{\xi}}_k}{\partial \mathbf{t}_k^\textrm{C}}
                    \end{align}
                    \noindent where
                    \begin{align}
                        \frac{\partial \tilde{\xi}_k^\textrm{tar}}{\partial \mathbf{u}_k^\textrm{C}} &=
                        -\left(\frac{\partial^2 \tilde{\mathcal{G}}_k}{\partial {\xi_k^\textrm{tar}}^2}\right)^{-1}
                        \frac{\partial^2 \tilde{\mathcal{G}}_k}{\partial\xi_k^\textrm{tar}\partial \mathbf{u}_k^\textrm{C}}
                    \end{align}
                    and
                    \begin{align}
                        \frac{\partial \tilde{\xi}_k^\textrm{tar}}{\partial\mathbf{t}_k^\textrm{C}} &=
                        -\left(\frac{\partial^2 \tilde{\mathcal{G}}_k}{\partial {\xi_k^\textrm{tar}}^2}\right)^{-1}
                        \frac{\partial^2 \tilde{\mathcal{G}}_k}{\partial\xi_k^\textrm{tar}\partial\mathbf{t}_k^\textrm{C}}.
                    \end{align}
                    The rounding out of cell corners with circular arcs is surprisingly complex as two coupled objectives need to be minimized. Nonetheless, the computation of optimal cell corner geometries is robust and usually requires only a few iterations.


        \marginnote{
            \begin{center}
                \includegraphics[width=\marginparwidth]{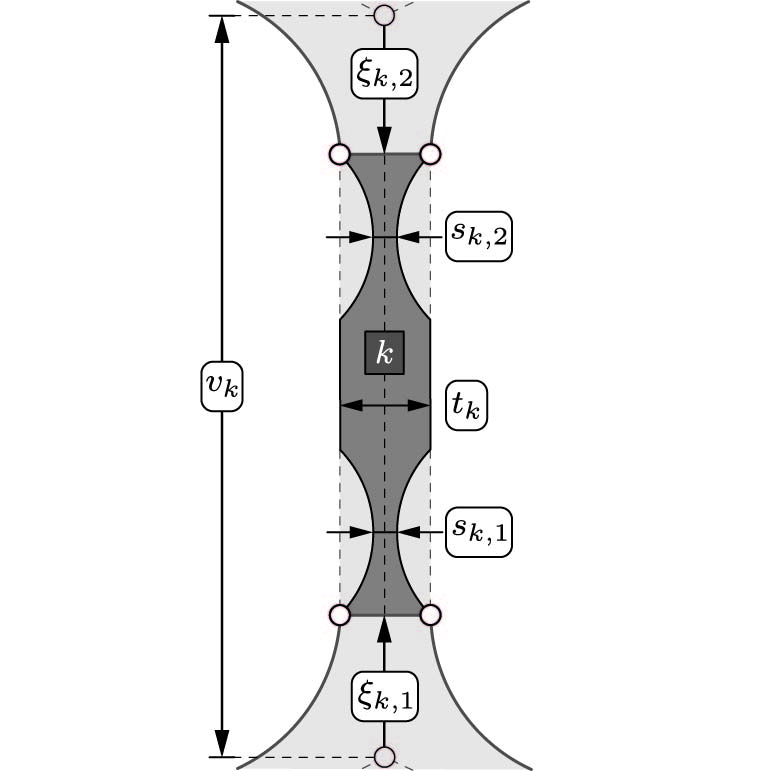}
                \small(a)\vspace{10mm}
                \includegraphics[width=\marginparwidth]{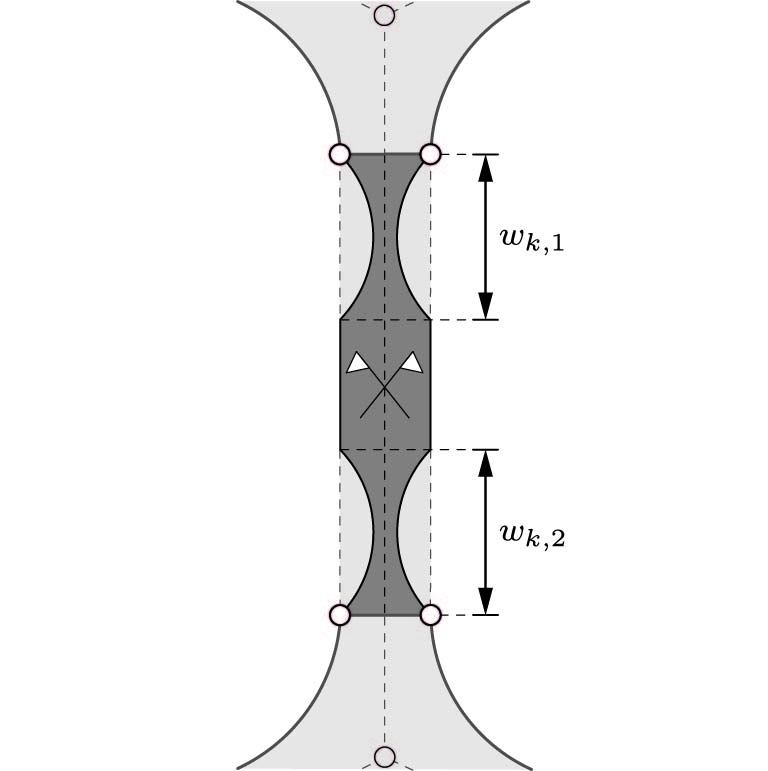}
                \small(b)
                \captionof{figure}{(a) Local state variables $\mathbf{u}^\textrm{S}$, $v^\textrm{S}$, $\mathbf{s}^\textrm{S}$ and $t^\textrm{S}$ of the $k$-th cell side. (b) Internal hinge lengths $\mathbf{w}$.}
                \label{pic:Figure_9_7}
            \end{center}}[-86mm]

        \section{Side Geometries}
            \subsection{Global State Variables}
                The bending deformations of rectangular cell sides are localized by circular cutouts that are symmetrically arranged next to the cell corners as illustrated in Figure~\ref{pic:Figure_9_7}. The description of the cell side geometries thus requires, besides the global state variables $\mathbf{u}$, $\mathbf{v}$, $\mathbf{t}$ and $\kappa^\textrm{tar}$ additional global state variables for the central hinge thicknesses $\mathbf{s}$. The ratio $\mu$ between the width and central hinge thickness is assumed to be constant and identical for all hinges. The ordering of the hinge thicknesses is similar to that of the cell side lengths $\mathbf{v}$ so that
                \begin{align}
                    \mathbf{s} = \left[
                    \begin{array}{cccccc}
                        {\mathbf{s}_{b,1}}^\top & {\mathbf{s}_{c,1}}^\top & \ldots & {\mathbf{s}_{b,nR}}^\top & {\mathbf{s}_{c,nR}}^\top & {\mathbf{s}_a}^\top
                    \end{array}\right]^\top\in\mathbb{R}^{2nv}
                \end{align}
                where, for example
                $\mathbf{s}_{b,i}=\left[
                \begin{array}{cccc}
                    s_{b,i,1} & s_{b,i,2} & \ldots & s_{b,i,2\left(nP+2-i\right)}
                \end{array}\right]^\top$.
                It can be seen that the two hinge thicknesses of a cell side are grouped together.


                \subsection{Local State Variables}
                    The hinge geometries of the $k$-th cell side are fully defined by the cell side thickness
                    \begin{align}
                        t_k^\textrm{S}\left(\mathbf{t}\right) = t_k
                    \end{align}
                    and the central hinge thicknesses
                    \begin{align}
                        \mathbf{s}_k^\textrm{S}\left(\mathbf{s}\right) =
                        \left[
                        \begin{array}{cc}
                            s_{k,1} & s_{k,2}
                        \end{array}
                        \right]^\top.
                    \end{align}
                    The localization of these hinges requires additional variables that include the effective cell side length
                    \begin{align}
                        v_k^\textrm{S}\left(\mathbf{v}\right) = v_k
                    \end{align}
                    and the cell corner dimensions
                    \begin{align}
                        \mathbf{u}_k^\textrm{S}\left(\mathbf{t}^\textrm{C},\kappa^\textrm{tar},\mathbf{u}^\textrm{C}\right) =
                        \left[
                        \begin{array}{cc}
                            \xi_{k,1} & \xi_{k,2}
                        \end{array}
                        \right]^\top
                    \end{align}
                    that depend on the local state variables of the neighbouring cell corners. It can be seen that the superscript ``$\textrm{S}$" is used to denote the local state variables of a cell side.


                \subsection{Dependent Variables}
                    The width $w^\textrm{S}$ of a compliant hinge is assumed to be proportional to the central hinge thickness $s$ so that
                    \begin{align}
                        \mathbf{w}_k^\textrm{S} = \mu \mathbf{s}_k^\textrm{S}.
                    \end{align}
                    The use of a linear relationship is advantageous as it leads to relatively simple expressions. Nonetheless, the consideration of different hinge geometries is straightforward. Furthermore, cell side geometries with a nonconstant thickness or even curved cell sides can be used to minimize the weight of a cellular structure. However, this is not done in this thesis as it unnecessarily complicates the presented approach.


        \section{Summary of Variables}
            \subsection{State Variables}
                In summary it can be said that the centerlines of compliant pressure actuated cellular structures can be fully described by the global state variables $\mathbf{u}$ and $\mathbf{v}$. The additional global state variables $\mathbf{s}$, $\mathbf{t}$, $\kappa^\textrm{tar}$ and $\mu$ are required if thickness variations of cell corners and sides need to be considered. Hence, the design variables $\mathbf{e}^\textrm{G}$ of the geometric model are
                \begin{align}
                    \mathbf{e}^\textrm{G} =
                    \left[
                    \begin{array}{cccccc}
                        \mathbf{u}^\top & \mathbf{v}^\top & \mathbf{t}^\top & \mathbf{s}^\top & \kappa^\textrm{tar} & \mu
                    \end{array}
                    \right]^\top.
                \end{align}
                These global state variables can be sequentially transformed into local state variables of pentagonal and triangular cells as well as cell corners and cell sides. Their dependencies are\\\vspace{2mm}

                \noindent
                \begin{tabular}{lll}
                    \textbf{geometric primitive} & \textbf{local state variables}\\\vspace{-2mm}\\
                    pentagonal cell     &
                    $\mathbf{u}^\textrm{P}\left(\mathbf{u},\mathbf{v}\right)$, $\mathbf{v}^\textrm{P}\left(\mathbf{v}\right)$\\\vspace{-3mm}\\
                    triangular cell     &
                    $\mathbf{u}^\textrm{T}\left(\mathbf{u}^\textrm{P}\right)$, $\mathbf{v}^\textrm{T}\left(\mathbf{v}^\textrm{P}\right)$\\\vspace{-3mm}\\
                    cell corner         & $\mathbf{u}^\textrm{C}\left(\mathbf{u}^\textrm{P},\mathbf{v}^\textrm{P}\right)$, $\mathbf{t}^\textrm{C}\left(\mathbf{t}\right)$, $\kappa^\textrm{tar}$\\\vspace{-3mm}\\
                    cell side           &
                    $\mathbf{u}^\textrm{S}\left(\kappa^\textrm{tar}, \mathbf{u}^\textrm{C}, \mathbf{t}^\textrm{C}\right)$,
                    $v^\textrm{S}\left(\mathbf{v}\right)$,
                    $\mathbf{s}^\textrm{S}\left(\mathbf{s}\right)$,
                    $t^\textrm{S}\left(\mathbf{t}^\textrm{C}\right)$,
                    $\mu$.
                \end{tabular}


            \subsection{Dependent Variables}
                Various geometric properties that are required for the simulation and optimization of compliant pressure actuated cellular structures can be derived from the local state variables of the geometric primitives. They include\\\vspace{2mm}

                \noindent
                \begin{tabular}{ll}
                    \textbf{geometric primitive} & \textbf{dependent variables}\\\vspace{-2mm}\\
                    pentagonal cell & $\boldsymbol{\gamma}\left(\mathbf{u}^\textrm{P},\mathbf{v}^\textrm{P}\right)$,
                    $A^\textrm{P}\left(\mathbf{u}^\textrm{P},\mathbf{v}^\textrm{P}\right)$\\\vspace{-3mm}\\
                    triangular cell &
                    $\psi\left(\mathbf{u}^\textrm{T},\mathbf{v}^\textrm{T}\right)$,
                    $A^\textrm{T}\left(\mathbf{u}^\textrm{T},\mathbf{v}^\textrm{T}\right)$\\\vspace{-3mm}\\
                    cell corner & $\boldsymbol{\xi}\left(\kappa^\textrm{tar},\mathbf{u}^\textrm{C},\mathbf{t}^\textrm{C}\right)$\\\vspace{-3mm}\\
                    cell side & $\mathbf{w}\left(\mu, \mathbf{s}^\textrm{S}\right)$.
                \end{tabular}


            \marginnote{
                \begin{center}
                    \includegraphics[width=\marginparwidth]{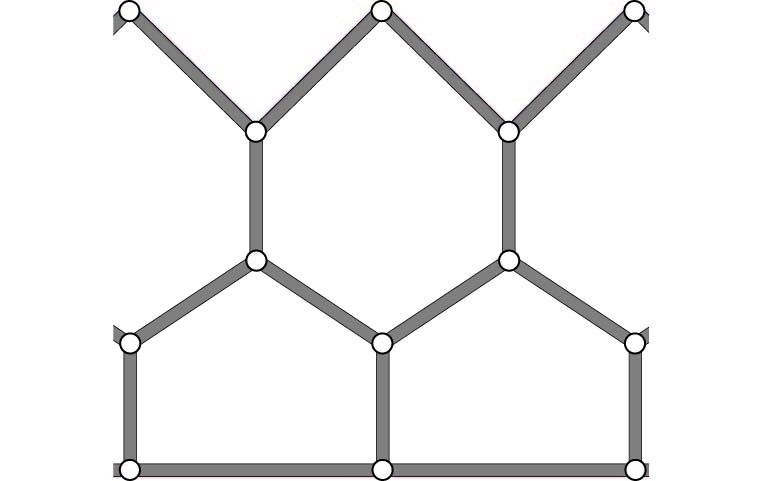}
                    \small(a)\vspace{10mm}
                    \includegraphics[width=\marginparwidth]{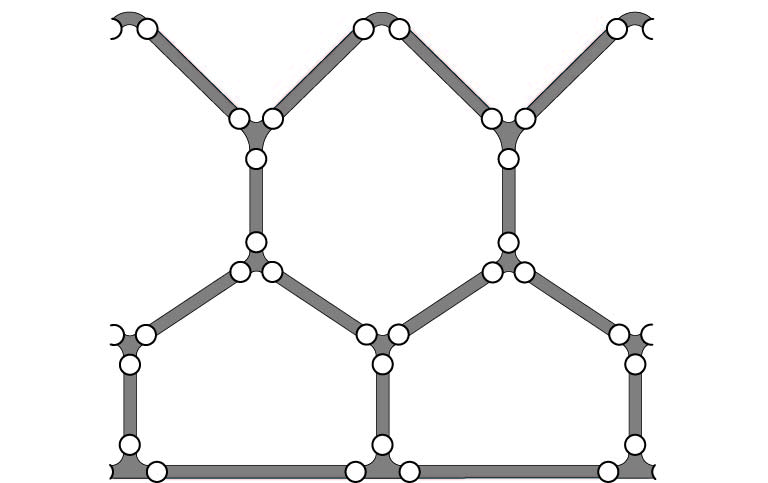}
                    \small(b)
                    \captionof{figure}{Pressure actuated cellular structure with frictionless hinges that are (a) centric and (b) eccentric.}
                    \label{pic:Figure_9_8}
                \end{center}}[-26mm]

            \subsection{Reduced Order Models}
                The geometric model for pressure actuated cellular structures can be considerably simplified if frictionless hinges are used instead of compliant hinges. The corresponding simplifications depend greatly on the hinge locations as illustrated in Figure~\ref{pic:Figure_9_8}. The use of centric hinges reduces the design variables $\mathbf{e}^\textrm{G}$ of the geometric model to
                \begin{align}
                    \mathbf{e}^\textrm{G} =
                    \left[
                    \begin{array}{cc}
                        \mathbf{v}^\top & \mathbf{t}^\top
                    \end{array}
                    \right]^\top
                \end{align}
                since frictionless hinges are capable of arbitrary rotations and cell corners do not exist. The local state variables of a cell side thus reduce to $v^\textrm{S}\left(\mathbf{v}\right)$ and $t^\textrm{S}\left(\mathbf{t}\right)$. Compared to structures with centric hinges, the use of eccentric hinges complicates the geometric model so that
                \begin{align}
                    \mathbf{e}^\textrm{G} =
                    \left[
                    \begin{array}{cccc}
                        \mathbf{u}^\top & \mathbf{v}^\top & \mathbf{t}^\top & \kappa^\textrm{tar}
                    \end{array}
                    \right]^\top.
                \end{align}
                The additional global state variables $\mathbf{u}$ and $\kappa^\textrm{tar}$ are required to describe the cell corner geometries so that their local state variables are identical to those of compliant structures. In contrast, the local state variables of cell sides reduce to $\mathbf{u}^\textrm{S}\left(\kappa^\textrm{tar}, \mathbf{u}^\textrm{C}, \mathbf{t}^\textrm{C}\right)$, $v^\textrm{S}\left(\mathbf{v}\right)$ and $t^\textrm{S}\left(\mathbf{t}^\textrm{C}\right)$ since detailed hinge geometries are not considered.


    \newpage

    \sectionmark{Summary}
    \begin{framed}
        \noindent \textbf{Summary}\\

        \noindent The optimization and manufacturing of compliant pressure actuated cellular structures requires a geometric model that fully describes their cross sectional geometry. It is of importance for the optimization that a minimal set of variables is used that reduces the coupling between different cross sectional regions. The geometric model itself consists of primitives such as cell corners, hinges and sides whose design depends on the manufacturing process. The current model is suited for prototyping and the wire electrical discharge machining of metals. A production of gapless high lift devices that is based on the three-dimensional weaving of fiber reinforced plastics would most likely require a different model.\\

        \noindent \textbf{Conclusion}\\

        \noindent The process dependent geometric model of compliant pressure actuated cellular structures needs to be coupled to a mechanical model for the simulation and optimization. A mechanical model that can be used in conjunction with a wide range of geometric models is subsequently presented.
    \end{framed}

    \newpage
    \thispagestyle{empty}


    \cleardoublepage
    \newgeometry{}
    \thispagestyle{empty}

        \noindent \Large \textbf{Nomenclature} \normalsize
        \vspace{5mm}

        \noindent
        \begin{tabular}{llll}
            \textbf{Superscripts}\hspace{5mm}
            & $\tilde{\square}$ & \hspace{5mm} & optimal value\\\vspace{-2.5mm}\\

            & lin, nlin         &              & linear, nonlinear\\
            & tar               &              & target value\\\vspace{-2.5mm}\\

            & $\beta$, $u$, $v$ &              & derivatives with respect to global state variables \\\vspace{-2mm}\\

            & $C$               &              & cell corner\\
            & $G$               &              & geometric model\\
            & $P$, $T$          &              & pentagonal, triangular cell\\
            & $S$               &              & cell side
        \end{tabular}


        \vspace{5mm}
        \noindent
        \begin{tabular}{llll}
            \textbf{Subscripts}\hspace{9mm}
            & $a$, $b$, $c$     & \hspace{7.5mm} & reference to cell sides
        \end{tabular}


        \vspace{5mm}
        \noindent
        \begin{tabular}{llll}
            \textbf{Numbers}\hspace{11mm}
            & $n_P$, $n_T$ & \hspace{6mm}  & pentagonal, triangular (hexagonal) cells\\
            & $n_R$        &               & cell rows\\
            & $n_s$        &               & cell sides at cell corner\\
            & $n_u$        &               & global state variables\\
            & $n_v$        &               & cell sides
        \end{tabular}


        \vspace{5mm}
        \noindent
        \begin{tabular}{llll}
            \textbf{Greek Letters}\hspace{2.5mm}
            & $\alpha$, $\beta$          &              & global, local state variables\\
            & $\gamma$, $\omega$, $\psi$ &              & internal angles\\
            & $\xi$                      & \hspace{6mm} & local coordinate\\
            & $\theta$                   &              & cell corner angle\\
            & $\kappa$                   &              & fillet curvature\\
            & $\lambda$                  &              & Lagrange multiplier\\
            & $\mu$                      &              & thickness ratio of compliant hinge\\
            & $\upsilon$                 &              & variables for cell corner optimization
        \end{tabular}


        \vspace{5mm}
        \noindent
        \begin{tabular}{llll}
            \textbf{Roman Letters}\hspace{0.5mm}
            & $a$, $b$, $c$     & \hspace{7mm}  & cell side lengths\\
            & $e$               &               & set of design, state variables\\
            & $g$               &               & gradient\\
            & $i$, $j$, $k$     &               & integers\\
            & $s$, $t$          &               & central hinge, cell side thickness\\
            & $u$, $v$          &               & global, local state variables\\
            & $w$               &               & hinge width\\
            & $x$, $y$, $z$     &               & internal lengths\\\vspace{-2.5mm}\\
            & $A$               &               & cross sectional area\\
            & $C$               &               & constraint\\
            & $F$, $G$          &               & objectives\\
            & $H$               &               & Hessian\\
            & $L$               &               & Lagrangian\\
            & $T$               &               & transformation matrix
        \end{tabular}
        \restoregeometry

        \thispagestyle{empty} 
        \cleardoublepage
    \chapter{Mechanical Model}

        \marginnote{
            \begin{center}
                \includegraphics[width=\marginparwidth]{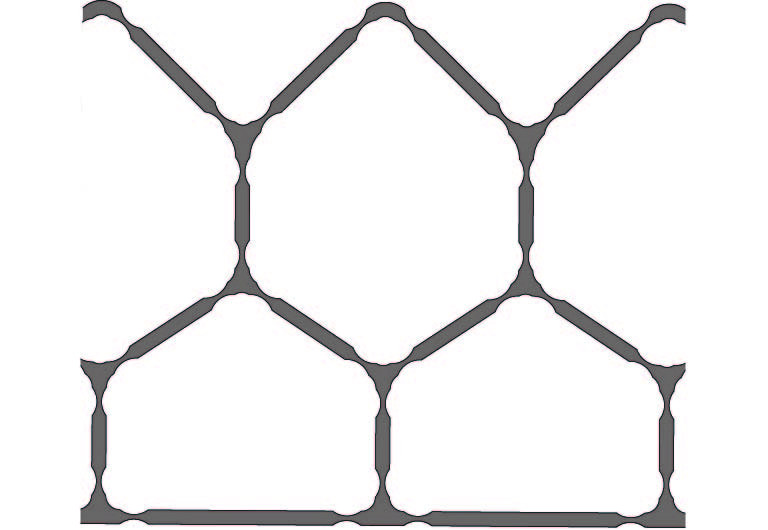}
                \small(a)\vspace{10mm}
                \includegraphics[width=\marginparwidth]{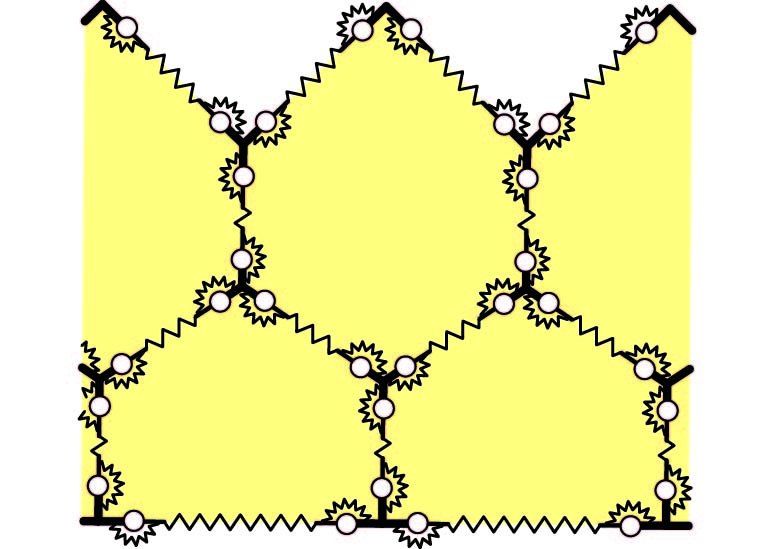}
                \small(b)
                \captionof{figure}{(a) Geometric and (b) mechanical model of a compliant pressure actuated cellular structure. The mechanical model is based on rigid cell corners, frictionless hinges and axial, rotational springs.}
                \label{pic:Figure_10_1}
            \end{center}}[-17mm]

        \noindent The simulation of compliant pressure actuated cellular structures with an arbitrary number of cell rows and geometries is relatively simple and can be done with numerical tools such as the boundary element \cite{Gaul2003-1}, finite difference \cite{Smith1985-1} or finite element \cite{Zienkiewicz2014-1} method. Particularly the latter is commonly used in engineering as it can be applied to a wide range of linear and nonlinear problems with complex geometries and loading conditions \cite{Schweizerhof1984-1}. Depending on the geometry of the structure and the desired accuracy, it is possible to use different kind of finite elements that range from one-dimensional bars and beams to two-dimensional membrane elements. The accuracy of a numerical simulation is therefore only limited by the available computing power and time. Structures with an exceedingly large number of cells can be simulated with multiscale methods \cite{Borst2010-1,Lv2014-1} that trade accuracy against speed. However, these kind of structures are not of practical interest as their manufacturing complexity is prohibitive.\\

        Unfortunately, the optimization of cell geometries for given target shapes, cell pressures and material properties is much harder due to the tight coupling between the geometric and mechanical model. For example, the variation of a single cell side thickness changes the geometries of the neighbouring cell corners. This affects the corresponding hinge eccentricities and therefore all equilibrium configurations of the mechanical model. The latter leads to different maximum hinge and cell side stresses throughout the structure and therefore, in return, alters the optimal geometry of all cell corners and sides. A detailed mechanical model for the simulation and optimization of compliant pressure actuated cellular structures that can be tightly coupled to a wide range of geometric models is subsequently presented. It is important to note that this model builds upon the previously introduced kinematic framework that can be used to describe the centerlines of an undeformed structure.


        \section{Cell Side Deformations}
            \subsection{Global State Variables}

                \marginnote{
                    \begin{center}
                        \includegraphics[width=\marginparwidth]{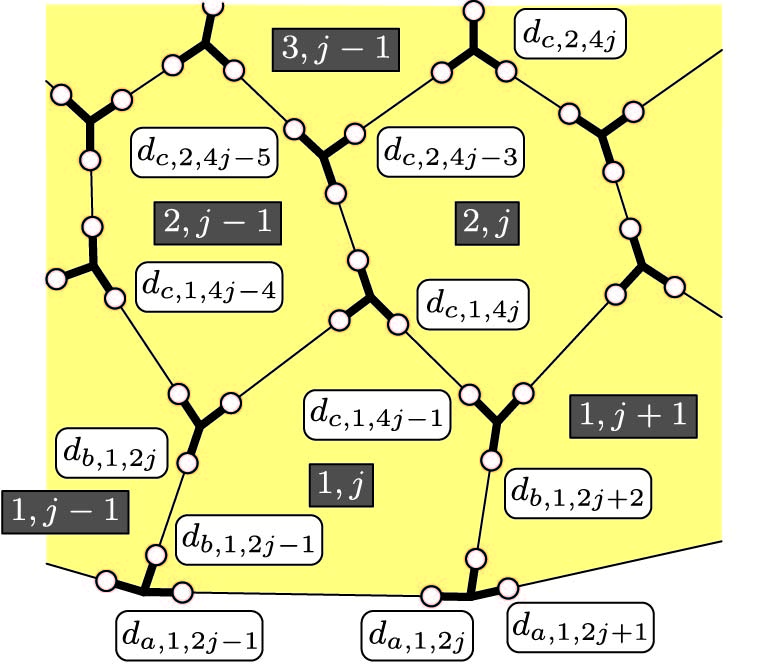}
                        \small(a)\vspace{10mm}
                        \includegraphics[width=\marginparwidth]{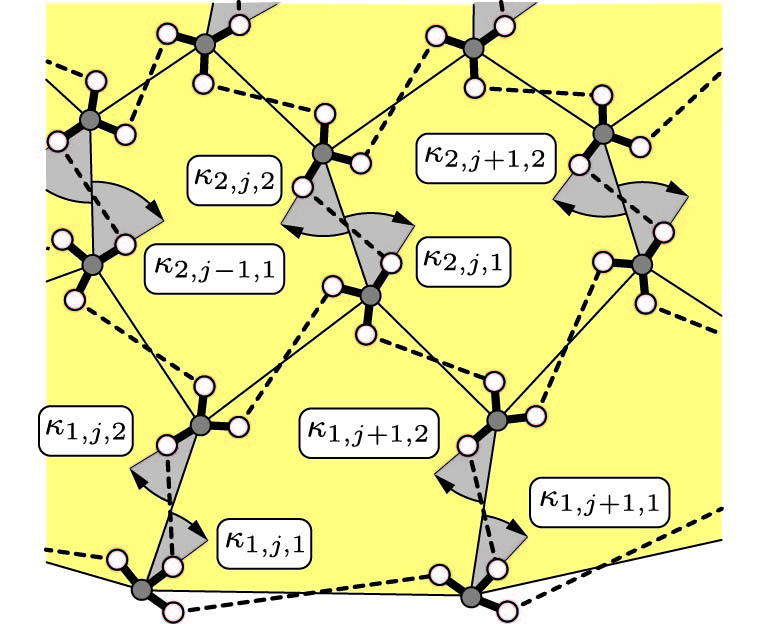}
                        \small(b)
                        \captionof{figure}{Additional global state variables that are needed to describe the deformed configuration of the mechanical model. (a) Hinge eccentricities $\mathbf{d}$ and (b) cell corner rotations $\boldsymbol{\kappa}$.}
                        \label{pic:Figure_10_2}
                    \end{center}}[21mm]

                \noindent The considered geometric and mechanical models of compliant pressure actuated cellular structures are shown in Figure~\ref{pic:Figure_10_1}. It can be seen that the compliant hinges are modeled by rotational springs and frictionless hinges whereas the central cell sides are modeled by axial springs. The hinge eccentricities $\mathbf{d}$ can be derived from the global state variables of the geometric model where it is assumed that
                \begin{align}
                    \mathbf{d}\left(\mathbf{e}^\textrm{G}\right) = \boldsymbol{\xi}\left(\mathbf{e}^\textrm{G}\right) + \frac{1}{2} \mathbf{w}\left(\mathbf{e}^\textrm{G}\right).
                \end{align}
                This assumption is legitimate since cell corners are relatively stiff and compliant hinges relatively small so that their axial elongations can be neglected. The chosen ordering of the hinge eccentricities is shown in Figure~\ref{pic:Figure_10_2}. It can be seen that it is similar to the ordering of the central hinge thicknesses $\mathbf{s}$.\\

                The geometry of an undeformed mechanical model is described by
                \begin{align}
                    \mathbf{e}^\textrm{M}\left(\mathbf{e}^\textrm{G}\right) =
                    \left[
                    \begin{array}{ccc}
                        \mathbf{u}_0^\top & \mathbf{v}_0^\top & \mathbf{d}\left(\mathbf{e}^\textrm{G}\right)^\top
                    \end{array}\right]^\top
                \end{align}
                where the subscript ``0" is used to highlight variables of the reference, undeformed configuration. However, these variables are insufficient to describe the deformed cell sides so that they need to be augmented by the additional global state variables $\boldsymbol{\kappa}$ where
                \begin{align}
                    \boldsymbol{\kappa} = \left[
                    \begin{array}{ccc}
                        {\boldsymbol{\kappa}_1}^\top & \ldots & {\boldsymbol{\kappa}_{nR+1}}^\top
                    \end{array}
                    \right]^\top \in \mathbb{R}^{n\kappa}
                \end{align}
                and the total number of cell corners $n_\kappa$ is
                \begin{align}
                    n_\kappa = 3 n_P + 2 n_T + n_R + 1.
                \end{align}
                The state variables $\boldsymbol{\kappa}$ are defined with respect to the cell sides $\mathbf{b}$. As a consequence, the upper cell corners of a pressure actuated cellular structure need to be described with respect to imaginary cell sides that possess only one hinge so that
                \begin{align}
                    \boldsymbol{\kappa}_i =
                    \begin{cases}
                        \left[
                        \begin{array}{ccccc}
                            \kappa_{i,1,1} & \kappa_{i,1,2} & \ldots & \kappa_{i,nP-i+2,1} & \kappa_{i,nP-i+2,2}
                        \end{array}
                        \right]^\top & i \leq n_R\\
                        \left[
                        \begin{array}{ccc}
                            \kappa_{nR+1,1,1} & \ldots & \kappa_{nR+1,nP-nR+1,1}
                        \end{array}
                        \right]^\top & i > n_R.
                    \end{cases}
                \end{align}
                The deformed geometry of a mechanical model is thus fully described by the global state variables $\mathbf{e}^\textrm{M}$ and
                \begin{align}
                    \mathbf{h}^\textrm{M} =
                    \left[
                    \begin{array}{ccc}
                        \mathbf{u}^\top & \mathbf{v}^\top & \boldsymbol{\kappa}^\top
                    \end{array}\right]^\top.
                \end{align}


            \subsection{Local State Variables}

                \marginnote{
                \begin{center}
                    \includegraphics[width=\marginparwidth]{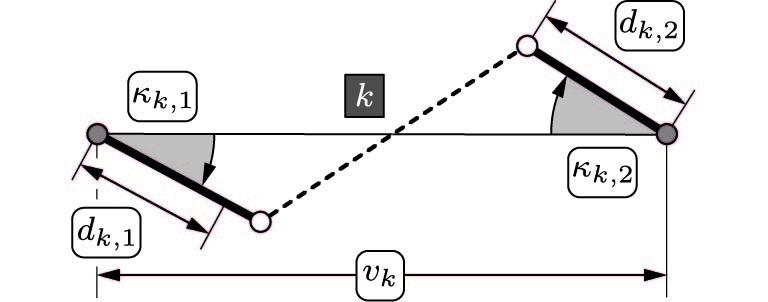}
                    \small(a)\vspace{10mm}
                    \includegraphics[width=\marginparwidth]{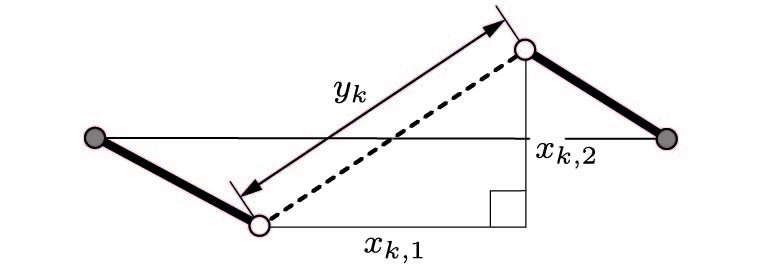}
                    \small(b)\vspace{10mm}
                    \includegraphics[width=\marginparwidth]{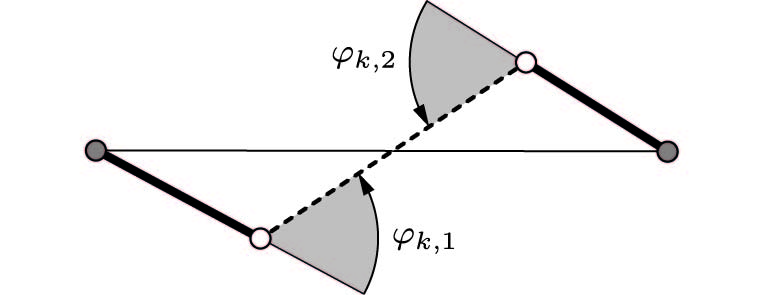}
                    \small(c)\vspace{10mm}
                    \includegraphics[width=\marginparwidth]{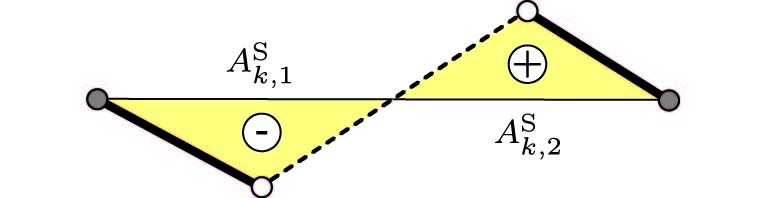}
                    \small(d)
                    \captionof{figure}{Variables of the $k$-th cell side. (a) Local state variables $\mathbf{d}$, $v$ and $\boldsymbol{\kappa}$. (b) Internal lengths $\mathbf{x}$, $y$ and (c) internal bending angles $\boldsymbol{\varphi}$. (d) Effective cell side area $A^\textrm{S}$ that varies the cross sectional area of neighbouring cells. The shown signs highlight the positive and negative terms.}
                    \label{pic:Figure_10_3}
                \end{center}}[-4mm]

                \noindent The deformed geometry of the $k$-th cell side is fully described by the hinge eccentricities
                \begin{align}
                    \mathbf{d}_k^\textrm{S}\left(\mathbf{e}^\textrm{M}\right) =
                    \left[
                    \begin{array}{cc}
                        d_{k,1} & d_{k,2}
                    \end{array}
                    \right]^\top,
                \end{align}
                the effective cell side length
                \begin{align}
                    v_k^\textrm{S}\left(\mathbf{h}^\textrm{M}\right) = v_k
                \end{align}
                and the cell side rotations
                \begin{align}
                    \mathbf{u}_k^\textrm{S}\left(\mathbf{e}^\textrm{M},\mathbf{h}^\textrm{M}\right) =
                    \left[
                    \begin{array}{cc}
                        \kappa_{k,1} & \kappa_{k,2}
                    \end{array}\right]^\top
                \end{align}
                as shown in Figure~\ref{pic:Figure_10_3}. The rotations are special since only those of cell sides $\mathbf{b}$ can be directly expressed in terms of the global state variables $\boldsymbol{\kappa}$. The rotations of the other sides are subsequently derived with the help of the $j$-th pentagonal cell in the $i$-th cell row as illustrated in Figure~\ref{pic:Figure_10_4}. The local state variables
                \begin{align}
                    \boldsymbol{\kappa}^\textrm{P}_{i,j}\left(\boldsymbol{\kappa}\right) = \left[
                    \begin{array}{ccccc}
                        \kappa_{i,2j-1} & \kappa_{i,2j} & \kappa_{i,2j+1} & \kappa_{i,2j+2} & \kappa_{i+1,2j-1}
                    \end{array}\right]^\top
                \end{align}
                of the pentagonal cell sides can be directly obtained from the global state variables $\boldsymbol{\kappa}$. The term $\kappa_{i+1,2j-1}$ is defined with respect to the cell side $b_{i+1,j}$ so that it is necessary to augment the local state variables $\mathbf{u}^\textrm{P}$ and $\mathbf{v}^\textrm{P}$ with the variable
                \begin{align}
                    \beta^\textrm{P}_{i,j}\left(\mathbf{u}\right) = \beta_{i,j}.
                \end{align}
                The internal angles $\boldsymbol{\kappa}_c$ between the rigid cell corners and, for example, the fictitious, straight cell side $c_{i,2j}$ of a pentagonal cell thus become
                \begin{align}
                    \left[
                    \begin{array}{l}
                        \kappa_{c,i,4j-1}\\
                        \kappa_{c,i,4j}
                    \end{array}\right] =
                    \left[
                    \begin{array}{lllllll}
                        \kappa_{i,2j+2}      &\hspace{-2.5mm}-&\hspace{-2.5mm} \Delta\gamma_{i,2j}\\
                        \kappa_{i+1,2j-1}    &\hspace{-2.5mm}-&\hspace{-2.5mm} \Delta\gamma_{i,2j}   &\hspace{-2mm}-&\hspace{-2mm} \Delta\alpha_{i,2j} &\hspace{-2.5mm}+&\hspace{-2.5mm} \Delta\beta_{i,j}
                    \end{array}\right]
                \end{align}
                where, for example $\Delta\alpha = \alpha-\alpha_0$ is the difference of the local pentagonal state variable $\alpha$ between the current (deformed) and the reference (undeformed) configuration. It is assumed that $\Delta \beta=0$ for pentagonal cells that are located in the top, boundary cell row. This is possible since global state variables $\boldsymbol{\beta}$ are not required in the top row so that they can have any constant value, including zero. The corresponding angles for the pentagonal base sides $\mathbf{a}$ that exist for $i=1$ are
                \begin{align}
                    \left[
                    \begin{array}{l}
                        \kappa_{a,i,2j-1}\\
                        \kappa_{a,i,2j}
                    \end{array}\right] =
                    \left[
                    \begin{array}{lll}
                        \kappa_{i,2j-1} &\hspace{-2.5mm}+&\hspace{-2.5mm} \Delta\alpha_{i,2j-1}\\
                        \kappa_{i,2j+1} &\hspace{-2.5mm}+&\hspace{-2.5mm} \Delta\alpha_{i,2j}
                    \end{array}
                    \right].
                \end{align}


            \subsection{Dependent Variables}
                The internal lengths $\mathbf{x}$ of a deformed cell side with eccentric hinges are
                \begin{align}
                    \left[
                    \begin{array}{c}
                        x_{k,1}\\
                        x_{k,2}
                    \end{array}\right] =
                    \left[
                    \begin{array}{r}
                        v_k - \cos\left(\kappa_{k,1}\right) d_{k,1} - \cos\left(\kappa_{k,2}\right) d_{k,2}\\
                        \sin\left(\kappa_{k,1}\right)d_{k,1} + \sin\left(\kappa_{k,2}\right)d_{k,2}
                    \end{array}
                    \right]
                \end{align}
                so that the hinge rotations or bending angles $\boldsymbol{\varphi}$ can be written as
                \begin{align}
                    \boldsymbol{\varphi}_k = \boldsymbol{\kappa}_k + \arcsin\left(\frac{x_{k,2}}{y_k}\right) \left[\begin{array}{cc}1&1\end{array}\right]^\top
                \end{align}
                where the central cell side length is
                \begin{align}
                    y_k = \sqrt{{x_{k,1}}^2 + {x_{k,2}}^2}.
                \end{align}
                The effective cell side area $A^\textrm{S}$ that varies the cross sectional area of a neighbouring cell thus becomes
                \begin{align}
                    A^\textrm{S}_k = \frac{1}{2}\left(\sin\left(\kappa_{k,2}\right) \cos\left(\kappa_{k,2}\right) {d_{k,2}}^2\right. &- \sin\left(\kappa_{k,1}\right) \cos\left(\kappa_{k,1}\right) {d_{k,1}}^2\\\nonumber
                    &\left.+ \left( \sin\left(\kappa_{k,2}\right) d_{k,2} - \sin\left(\kappa_{k,1}\right) d_{k,1} \right) x_{k,1}\right).
                \end{align}


        \blfootnote{\vspace{-5mm}
            \captionof{figure}{Cell side rotations are expressed on the basis of a pentagonal cell. (a) Local pentagonal state variables $\mathbf{u}^\textrm{P}$ are augmented by the global state variable $\beta^\textrm{P}$ that describes the inclination of the cell side $b_{i+1,j}$ and thus serves as a reference for the upper cell corner rotation. Additional local state variables include the (b) hinge eccentricities $\mathbf{d}^\textrm{P}$ and the (c) cell corner rotations $\boldsymbol{\kappa}^\textrm{P}$. (d) Internal variables $\boldsymbol{\kappa}_a$ and $\boldsymbol{\kappa}_c$ describe the angles between the rigid cell corners and the fictitious, straight cell sides $a$ and $\mathbf{c}$.}
            \label{pic:Figure_10_4}}
        \marginnote{
            \begin{center}
                \includegraphics[width=\marginparwidth]{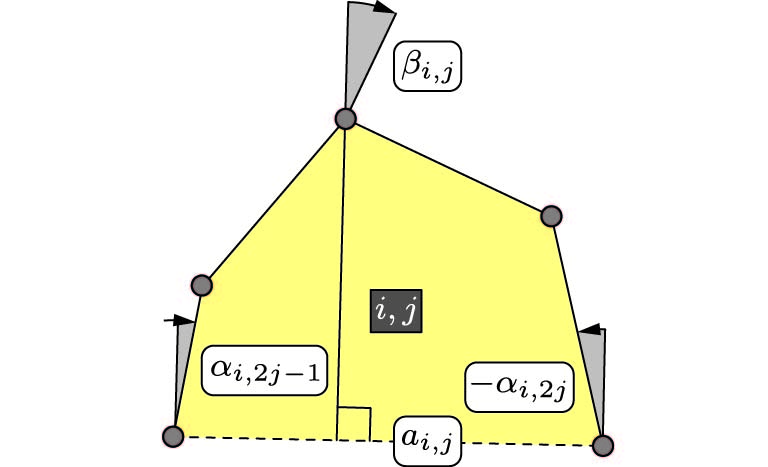}
                \small(a)\vspace{6mm}
                \includegraphics[width=\marginparwidth]{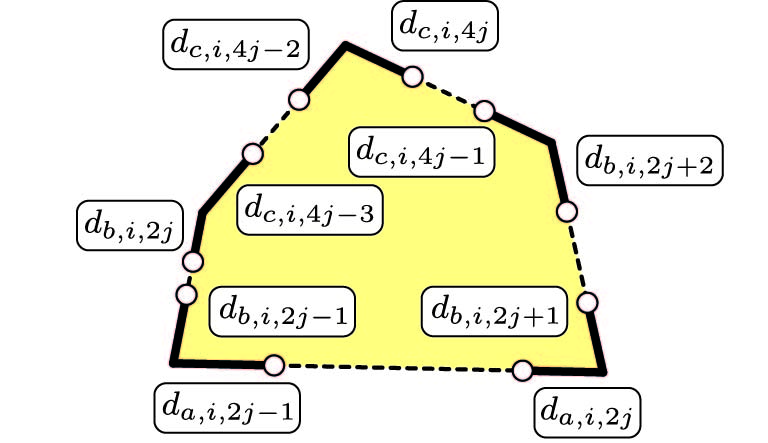}
                \small(b)\vspace{6mm}
                \includegraphics[width=\marginparwidth]{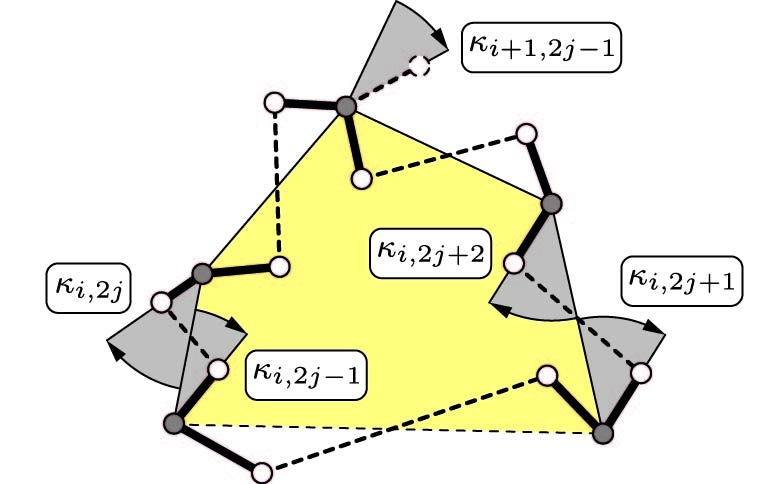}
                \small(c)\vspace{6mm}
                \includegraphics[width=\marginparwidth]{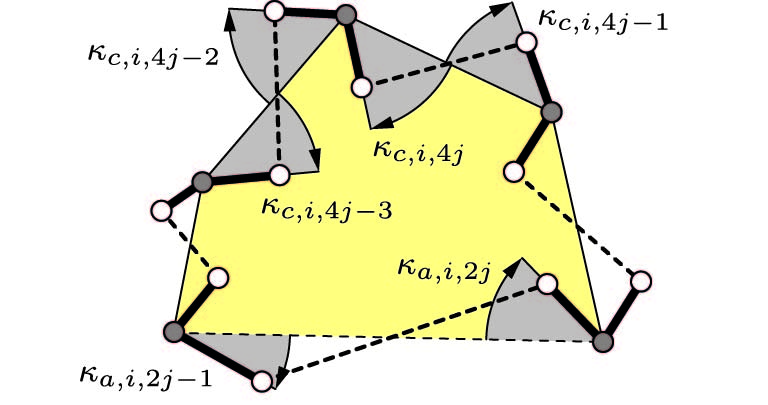}
                \small(d)
            \end{center}}[-96mm]

        \section{Mechanical Properties}
            \subsection{Global State Variables}
                Additional global state variables for the stiffness of the axial $\mathbf{m}$ and rotational $\mathbf{n}$ springs are required to describe the strain energy of the mechanical model. The variables that are illustrated in Figure~\ref{pic:Figure_10_5} can be fully derived from the geometric model so that they are invariant to cell side deformations. The equivalent axial spring stiffness $m_k$ of the $k$-th rectangular cell side is
                \begin{align}
                    m_k\left(\mathbf{e}^\textrm{G}\right) = \frac{E^\textrm{eff} t_k^\textrm{S}}{v_k^\textrm{S} - \sum_{i=1}^2\left(\xi_{k,i}\left(\mathbf{e}^\textrm{G}\right) + w_{k,i}\left(\mathbf{e}^\textrm{G}\right)\right)}
                \end{align}
                where the effective Young's modulus $E^\textrm{eff}$ for the plane strain condition of a material with a Young's modulus $E$ and a Poisson's ratio $\nu$ is
                \begin{align}
                    E^\textrm{eff} = \frac{E}{1-\nu^2}.
                \end{align}
                The stiffness is computed with respect to the cell side length between the compliant hinges as the axial strains of both, hinges and cell corners are neglected. The ordering of the axial spring stiffness is similar to that of the effective cell side lengths $\mathbf{v}$.\\

                \marginnote{
                    \begin{center}
                        \includegraphics[width=\marginparwidth]{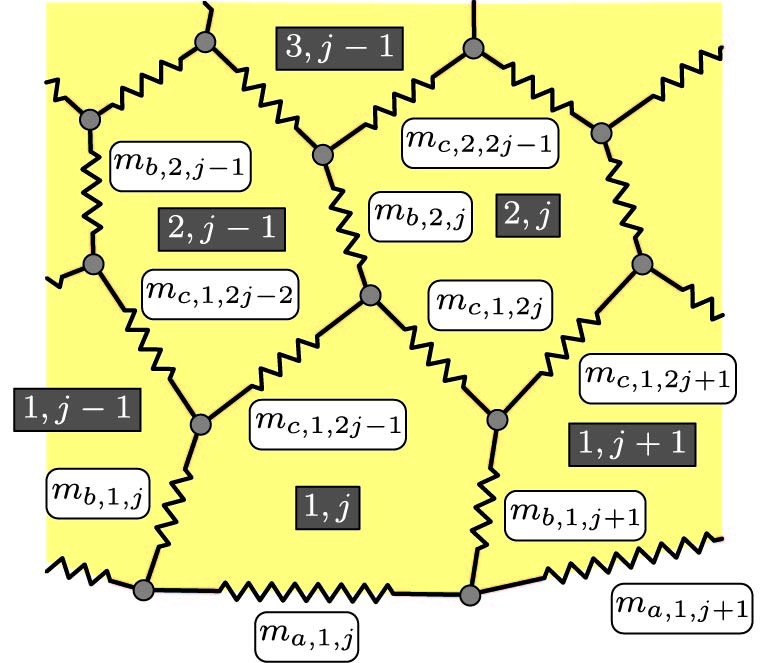}
                        \small(a)\vspace{10mm}
                        \includegraphics[width=\marginparwidth]{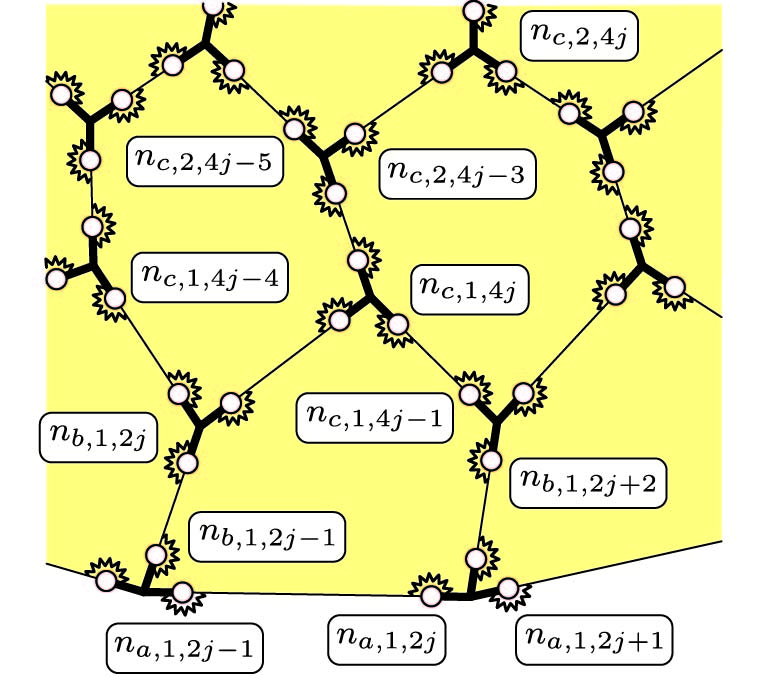}
                        \small(b)
                        \captionof{figure}{Additional global state variables that describe the stiffness of the mechanical model. (a) Axial $\mathbf{m}$ and (b) rotational $\mathbf{n}$ cell side springs.}
                        \label{pic:Figure_10_5}
                        \end{center}}[-50mm]

                As discussed in Chapter~\ref{sec:CompliantStructures}, it is advantageous to derive the rotational spring stiffness from a rectangular beam with a unit length, a thickness $s$ and a correction factor $\chi$. The $i$-th spring stiffness of the $k$-th cell side can be written as
                \begin{align}
                    n_{k,i}\left(\mathbf{e}^\textrm{G}\right) = \frac{E^\textrm{eff} {s_{k,i}^\textrm{S}}^2}{12} \chi_{k,i}\left(\mathbf{e}^\textrm{G}\right)
                \end{align}
                where the correction factor is computed with the help of a finite element model as shown in Figure~\ref{pic:Figure_10_6}. An arctan function\footnote{An analytical expression for the spring stiffness can be computed with the Euler-Bernoulli beam theory. The major terms of the bulky solution are based on an arctan function.} is used to interpolate the simulation results so that
                \begin{align}
                    \chi_{k,i}\left(\mathbf{e}^\textrm{G}\right) \approx \frac{1}{\mu} + \frac{1}{4} \arctan\left(\frac{80-6\mu}{100}\left(\frac{t_k^\textrm{S}}{s_{k,i}^\textrm{S}}-1\right)\right).
                \end{align}

                The stiffness of the axial and rotational springs can be fully derived from the global state variables $\mathbf{e}^\textrm{G}$. Furthermore, the maximum hinge and cell side stresses depend on the central hinge thickness $s$ and the cell side thickness $t$. It is therefore advantageous to augment the previously introduced global state variables $\mathbf{e}^\textrm{M}$ that describe the geometry of the undeformed mechanical model with the stiffness terms $\mathbf{m}$, $\mathbf{n}$ and thicknesses $\mathbf{s}$, $\mathbf{t}$ so that
                \begin{align}
                    \mathbf{e}^\textrm{M}\left(\mathbf{e}^\textrm{G}\right) =
                    \left[
                    \begin{array}{ccccccc}
                        {\mathbf{u}_0}^\top & {\mathbf{v}_0}^\top & \mathbf{d}^\top & \mathbf{m}^\top & \mathbf{n}^\top & \mathbf{s}^\top & \mathbf{t}^\top
                    \end{array}
                    \right]^\top.
                \end{align}


            \subsection{Local State Variables}
                The additional local state variables of the $k$-th cell side due to the axial $m$ and rotational $\mathbf{n}$ spring stiffness are
                \begin{align}
                    m_k^\textrm{S}\left(\mathbf{e}^\textrm{M}\right) = m_k
                \end{align}
                and
                \begin{align}
                    \mathbf{n}_k^\textrm{S}\left(\mathbf{e}^\textrm{M}\right) =
                    \left[
                    \begin{array}{cc}
                        n_{k,1} & n_{k,2}
                    \end{array}
                    \right]^\top.
                \end{align}
                Similarly the terms for the cell side $t$ and hinge $\mathbf{s}$ thicknesses are
                \begin{align}
                    t_k^\textrm{S}\left(\mathbf{e}^\textrm{M}\right) = t_k
                \end{align}
                and
                \begin{align}
                    \mathbf{s}_k^\textrm{S}\left(\mathbf{e}^\textrm{M}\right) =
                    \left[
                    \begin{array}{cc}
                        s_{k,1} & s_{k,2}
                    \end{array}
                    \right]^\top.
                \end{align}

                \marginnote{
                    \begin{center}
                        \includegraphics[width=\marginparwidth]{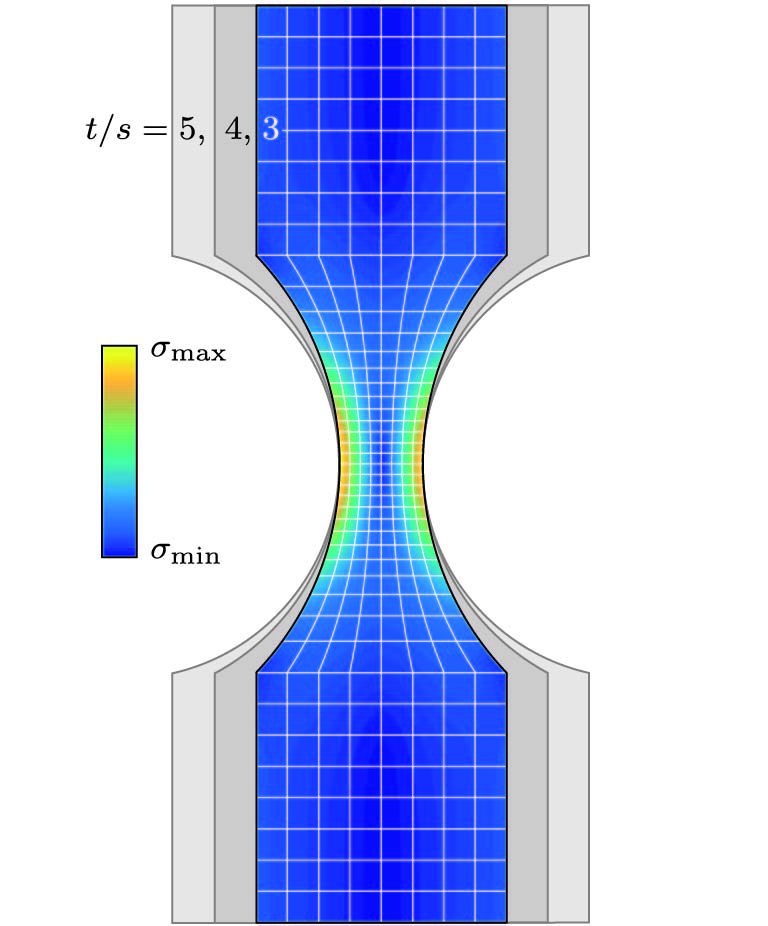}
                        \small(a)\vspace{10mm}
                        \includegraphics[width=\marginparwidth]{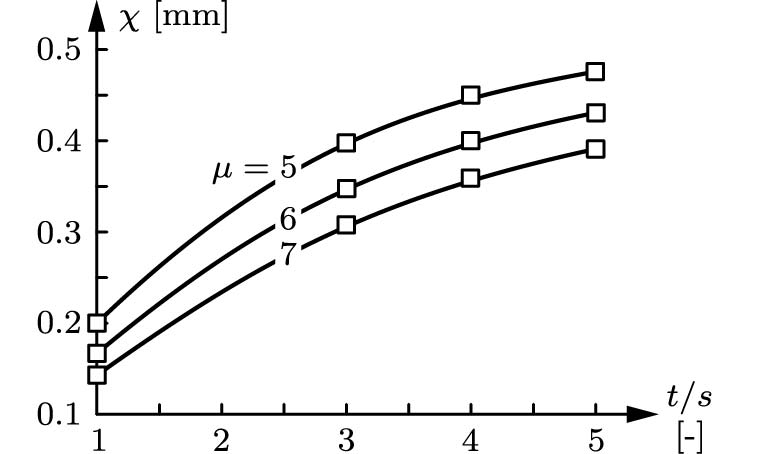}
                        \small(b)
                        \captionof{figure}{(a) Von Mises stresses of a compliant hinge due to end rotations ($t/s=3$, $\mu=5$). (b) Interpolated stiffness parameter $\chi$ for different thickness $t/s$ and aspect ratios $\mu$.}
                        \label{pic:Figure_10_6}
                    \end{center}}[-38mm]


            \subsection{Dependent Variables}
                The stress resultants that act on the $k$-th cell side can be compute from the bending angles $\boldsymbol{\varphi}_k$ and the cell side elongation $\Delta y_k$. The bending moment $M^\textrm{H}_{k,i}$ at the center of the $i$-th hinge is
                \begin{align}
                    M^\textrm{H}_{k,i}\left(\mathbf{e}^\textrm{M},\mathbf{h}^\textrm{M}\right) = n_{k,i}^\textrm{S}\left(\mathbf{e}^\textrm{M}\right) \varphi_{k,i}\left(\mathbf{e}^\textrm{M},\mathbf{h}^\textrm{M}\right)
                \end{align}
                and the axial cell side force $F_k$ can be written as
                \begin{align}
                    F_k\left(\mathbf{e}^\textrm{M},\mathbf{h}^\textrm{M}\right) = m_k^\textrm{S}\left(\mathbf{e}^\textrm{M}\right) \left(y_k\left(\mathbf{e}^\textrm{M},\mathbf{h}^\textrm{M}\right) - y_{0,k}\left(\mathbf{e}^\textrm{M}\right)\right).
                \end{align}
                Unlike the normal force, the bending moment varies along the cell side. Its maximum value $M^\textrm{S}_k$ can be expressed with respect to the stress resultants $\mathbf{M}^\textrm{H}_k$ and the differential pressure $\Delta p_k$ so that
                \begin{align}
                    M_k^\textrm{S} =
                    \begin{cases}
                        \displaystyle
                        \frac{\Delta p_k}{2} {\zeta_k}^2 - M_{k,2}^\textrm{H} &
                        \mbox{if }\ \Delta p_k \neq 0 \ \mbox{and}\
                        0 < \zeta_k < y_k\\
                        \max\left(|M_{k,1}^\textrm{H}|,|M_{k,2}^\textrm{H}|\right) & \mbox{otherwise}
                    \end{cases}
                \end{align}
                where the location $\zeta$ of the maximum bending moment is
                \begin{align}
                    \zeta_k = \frac{y_k}{2} + \frac{M_{k,1}^\textrm{H} + M_{k,2}^\textrm{H}}{\Delta p_k y_k}.
                \end{align}
                Hence, the maximum hinge $\boldsymbol{\sigma}^\textrm{H}$ and cell side $\sigma^\textrm{S}$ stresses are
                \begin{align}
                    \sigma_{k,i}^\textrm{H} = \frac{\rho}{s_{k,i}^\textrm{S}} \left(\frac{6}{s_{k,i}^\textrm{S}}|M_{k,i}^\textrm{H}| + |F_k|\right)
                \end{align}
                and
                \begin{align}
                    \sigma^\textrm{S}_k = \frac{\rho}{t_k^\textrm{S}} \left(\frac{6}{t_k^\textrm{S}}|M_k^\textrm{S}| + |F_k|\right)
                \end{align}
                where the stress reduction factor $\rho$ for the von Mises yield criterion is
                \begin{align}
                    \rho = \sqrt{1-\nu+\nu^2}.
                \end{align}


        \section{Summary of Variables}
            \subsection{State Variables}
                In summary it can be said that the geometry of an undeformed mechanical model can be fully described by the global state variables $\mathbf{u}_0$, $\mathbf{v}_0$ and $\mathbf{d}$. The stiffness of the mechanical model is described by the axial $\mathbf{m}$ and rotational $\mathbf{n}$ springs. The evaluation of maximum stresses requires hinge $\mathbf{s}$ and cell side $\mathbf{t}$ thicknesses. These global state variables are combined into a single vector
                \begin{align}
                    \mathbf{e}^\textrm{M}\left(\mathbf{e}^\textrm{G}\right) =
                    \left[
                    \begin{array}{ccccccc}
                        \mathbf{u}_0^\top & \mathbf{v}_0^\top & \mathbf{d}^\top & \mathbf{m}^\top & \mathbf{n}^\top & \mathbf{s}^\top & \mathbf{t}^\top
                    \end{array}
                    \right]^\top
                \end{align}
                as they can be fully derived from the global state variables of the geometric model. Additional global state variables are required to describe the deformed geometry of the mechanical model.  These variables include the global state variables $\mathbf{u}$ and $\mathbf{v}$ that define the centerline of the deformed configuration as well as $\boldsymbol{\kappa}$ that defines the cell corner rotations
                \begin{align}
                    \mathbf{h}^\textrm{M} =
                    \left[
                    \begin{array}{ccc}
                        \mathbf{u}^\top & \mathbf{v}^\top & \boldsymbol{\kappa}^\top
                    \end{array}
                    \right]^\top.
                \end{align}
                Global state variables can be transformed into local state variables of pentagonal and triangular cells as well as cell sides. Their dependencies are\\\vspace{2mm}

                \noindent
                \begin{tabular}{lll}
                    \textbf{geometric primitive} && \textbf{local state variables}\\\vspace{-2mm}\\
                    pentagonal cell     &&
                    $\mathbf{u}_0^\textrm{P}\left(\mathbf{e}^\textrm{M}\right)$, $\mathbf{v}_0^\textrm{P}\left(\mathbf{e}^\textrm{M}\right)$,
                    $\mathbf{u}^\textrm{P}\left(\mathbf{h}^\textrm{M}\right)$,
                    $\mathbf{v}^\textrm{P}\left(\mathbf{h}^\textrm{M}\right)$\\\vspace{-3mm}\\
                    triangular cell     &&
                    $\mathbf{u}_0^\textrm{T}\left(\mathbf{e}^\textrm{M}\right)$,
                    $\mathbf{u}_0^\textrm{T}\left(\mathbf{e}^\textrm{M}\right)$,
                    $\mathbf{u}^\textrm{T}\left(\mathbf{h}^\textrm{M}\right)$, $\mathbf{v}^\textrm{T}\left(\mathbf{h}^\textrm{M}\right)$\\\vspace{-3mm}\\
                    cell side           &&
                    $v_0^\textrm{S}\left(\mathbf{e}^\textrm{M}\right)$,
                    $\mathbf{u}^\textrm{S}\left(\mathbf{e}^\textrm{M}, \mathbf{h}^\textrm{M}\right)$,
                    $v^\textrm{S}\left(\mathbf{h}^\textrm{M}\right)$,
                    $\mathbf{d}^\textrm{S}\left(\mathbf{e}^\textrm{M}\right)$\\\vspace{-3mm}\\
                    && $m^\textrm{S}\left(\mathbf{e}^\textrm{M}\right)$, $\mathbf{n}^\textrm{S}\left(\mathbf{e}^\textrm{M}\right)$, $\mathbf{s}^\textrm{S}\left(\mathbf{e}^\textrm{M}\right)$, $t^\textrm{S}\left(\mathbf{e}^\textrm{M}\right)$.
                \end{tabular}
                \vspace{3mm}

                \noindent The local state variables of pentagonal and triangular cells are not separately considered in this chapter as they do not differ from the geometric model. Furthermore, cell corners are not a geometric primitive of the mechanical model. Their properties are merged into the local state variables of the cell sides through the hinge eccentricities $\mathbf{d}$ and cell corner rotations $\mathbf{u}$.


            \subsection{Dependent Variables}
                Additional geometric and mechanical properties for the simulation and optimization of pressure actuated cellular structures can be derived from the local state variables of the geometric primitives. They include\\\vspace{2mm}

                \noindent
                \begin{tabular}{ll}
                    \textbf{geometric primitive} & \textbf{dependent variables}\\\vspace{-2mm}\\
                    pentagonal cell &
                    $\boldsymbol{\gamma}_0\left(\mathbf{u}_0^\textrm{P}, \mathbf{v}_0^\textrm{P}\right)$,
                    $\boldsymbol{\gamma}\left(\mathbf{u}^\textrm{P}, \mathbf{v}^\textrm{P}\right)$,
                    $A^\textrm{P}\left(\mathbf{u}^\textrm{P},\mathbf{v}^\textrm{P}\right)$\\\vspace{-3mm}\\
                    triangular cell &
                    $\psi_0\left(\mathbf{u}_0^\textrm{T}, \mathbf{v}_0^\textrm{T}\right)$,
                    $\psi\left(\mathbf{u}^\textrm{T}, \mathbf{v}^\textrm{T}\right)$,
                    $A^\textrm{T}\left(\mathbf{u}^\textrm{T},\mathbf{v}^\textrm{T}\right)$\\\vspace{-3mm}\\
                    cell side &
                    $y_0\left(v_0^\textrm{S}, \mathbf{d}^\textrm{S}\right)$,
                    $\boldsymbol{\varphi}\left(\mathbf{d}^\textrm{S}, \mathbf{u}^\textrm{S}, v^\textrm{S}\right)$,
                    $y\left(\mathbf{d}^\textrm{S}, \mathbf{u}^\textrm{S}, v^\textrm{S}\right)$,\\\vspace{-3mm}\\
                    & $A^\textrm{S}\left(\mathbf{d}^\textrm{S}, \mathbf{u}^\textrm{S}, v^\textrm{S}\right)$, $\sigma^\textrm{S}\left(v_0^\textrm{S}, \mathbf{d}^\textrm{S}, m^\textrm{S}, t^\textrm{S},\mathbf{u}^\textrm{S},v^\textrm{S}\right)$,\\\vspace{-3mm}\\ & $\boldsymbol{\sigma}^\textrm{H}\left(\mathbf{d}^\textrm{S}, \mathbf{n}^\textrm{S},\mathbf{s}^\textrm{S},\mathbf{u}^\textrm{S},v^\textrm{S}\right)$.
                \end{tabular}


            \afterpage{
                \newgeometry{}
                \begin{figure}[htbp]
                    \captionsetup[subfigure]{labelformat=empty}
                    \begin{center}
                        \subfloat[(a)]{
                            \includegraphics[width=57mm]{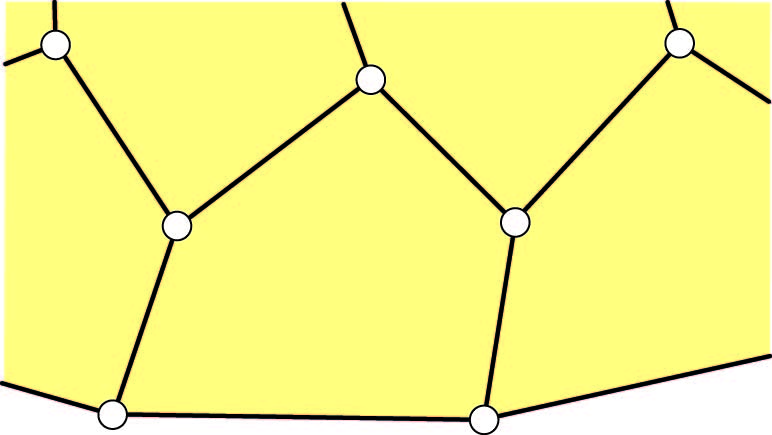}}\hspace{20mm}
                        \subfloat[(e)]{
                            \includegraphics[width=57mm]{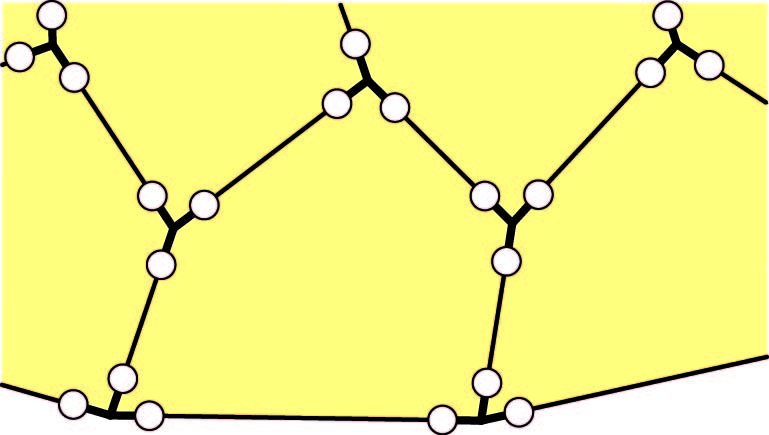}}\vspace{-2mm}

                        \subfloat[(b)]{
                            \includegraphics[width=57mm]{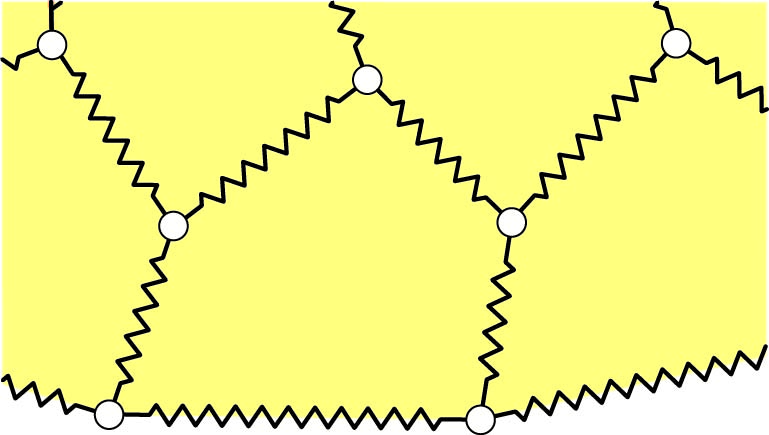}}\hspace{20mm}
                        \subfloat[(f)]{
                            \includegraphics[width=57mm]{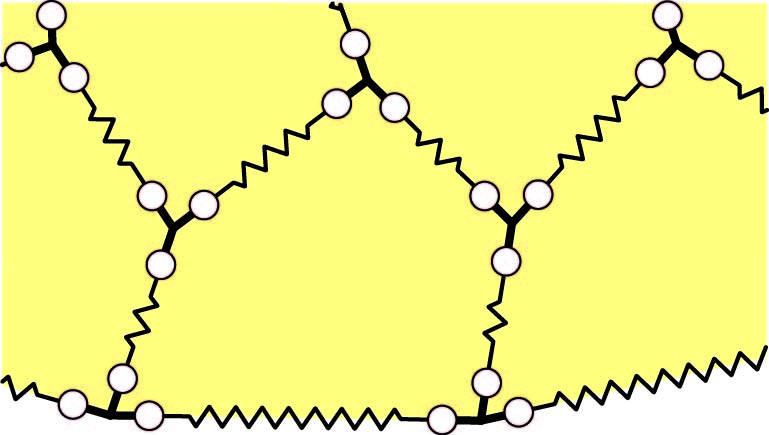}}\vspace{-2mm}

                        \subfloat[(c)]{
                            \includegraphics[width=57mm]{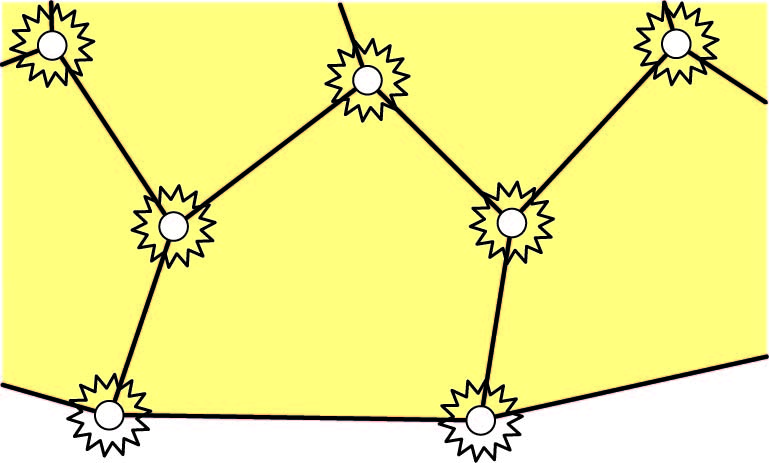}}\hspace{20mm}
                        \subfloat[(g)]{
                            \includegraphics[width=57mm]{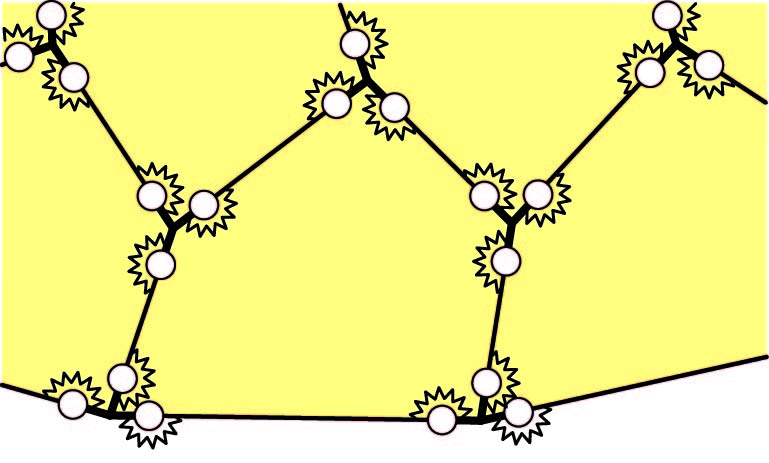}}\vspace{-2mm}

                        \subfloat[(d)]{
                            \includegraphics[width=57mm]{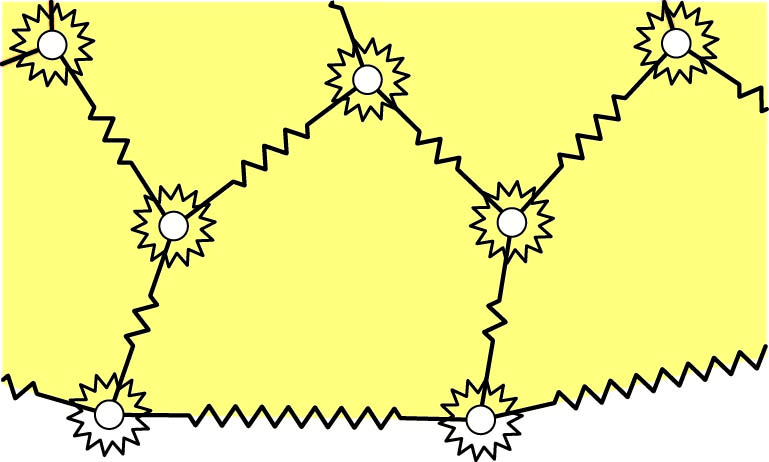}}\hspace{20mm}
                        \subfloat[(h)]{
                            \includegraphics[width=57mm]{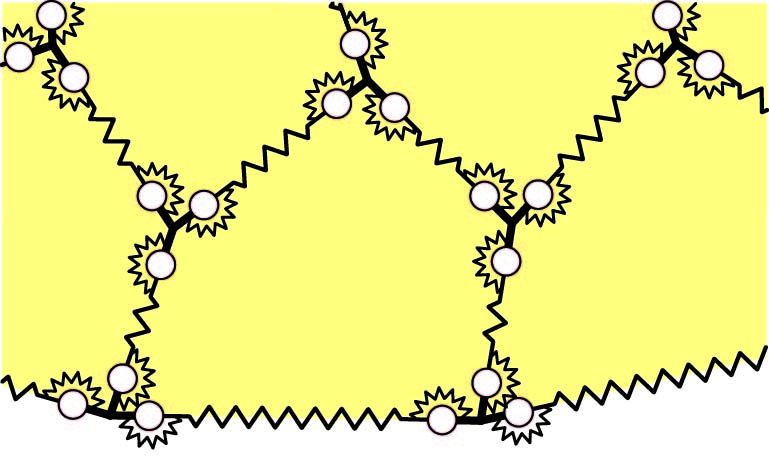}}
                        \caption{The coupling between the geometric and mechanical model can be relaxed if hinge eccentricities or rotational, axial springs are neglected. (a-d) Centric hinges eliminate the global state variables $\mathbf{d}$ and $\boldsymbol{\kappa}$. Additional variables (a-b) $\mathbf{u}_0$, $\mathbf{n}$ and (a,d) $\mathbf{v},\mathbf{m}$ are eliminated for negligible rotational and axial springs, respectively. (e-h) Eccentric hinges increase the coupling between the geometric and mechanical model so that $\mathbf{u}_0$, $\mathbf{d}$ and $\boldsymbol{\kappa}$ can not be eliminated. However, the global state variables (e-f) $\mathbf{n}$ and (e,h) $\mathbf{v},\mathbf{m}$ are eliminated for negligible rotational and axis springs, respectively.}
                        \label{pic:Figure_10_7}
                    \end{center}
                \end{figure}
                \restoregeometry}

            \subsection{Reduced Order Models}
                \label{subsec:ReducedOrderModels}

                \subsubsection{Centric Hinges}
                    The applicability of the mechanical model is by no means limited to compliant pressure actuated cellular structures. As shown in Figure~\ref{pic:Figure_10_7}, the complexity of the mechanical model can be successively reduced so that it can be applied to a wide range of structures. For example, a pressure actuated cellular structure with centric, frictionless hinges and rigid cell sides is fully described by the state variables
                    \begin{alignat}{2}
                        &\mathbf{e}^\textrm{M} &&=
                        \left[
                        \begin{array}{cc}
                            {\mathbf{v}_0}^\top & \mathbf{t}^\top
                        \end{array}
                        \right]^\top\\\nonumber
                        &\mathbf{h}^\textrm{M} &&= \mathbf{u}.
                    \end{alignat}
                    The cell side thicknesses $\mathbf{t}$ are required for the computation of cell side stresses $\boldsymbol{\sigma}^\textrm{S}$. An additional consideration of axial cell side springs leads to
                    \begin{alignat}{2}
                        &\mathbf{e}^\textrm{M} &&=
                        \left[
                        \begin{array}{ccc}
                            {\mathbf{v}_0}^\top & \mathbf{m}^\top & \mathbf{t}^\top
                        \end{array}
                        \right]^\top\\\nonumber
                        &\mathbf{h}^\textrm{M} &&=
                        \left[
                        \begin{array}{cc}
                            \mathbf{u}^\top & \mathbf{v}^\top
                        \end{array}
                        \right]^\top
                    \end{alignat}
                    as cell side elongations need to be taken into account. In contrast, an additional consideration of rotational springs leads to
                    \begin{alignat}{2}
                        &\mathbf{e}^\textrm{M} &&=
                        \left[
                        \begin{array}{ccccc}
                            {\mathbf{u}_0}^\top & {\mathbf{v}_0}^\top & \mathbf{n}^\top & \mathbf{s}^\top & \mathbf{t}^\top
                        \end{array}
                        \right]^\top\\\nonumber
                        &\mathbf{h}^\textrm{M} &&= \mathbf{u}.
                    \end{alignat}

                    \marginnote{
                        \begin{center}
                            \includegraphics[width=\marginparwidth]{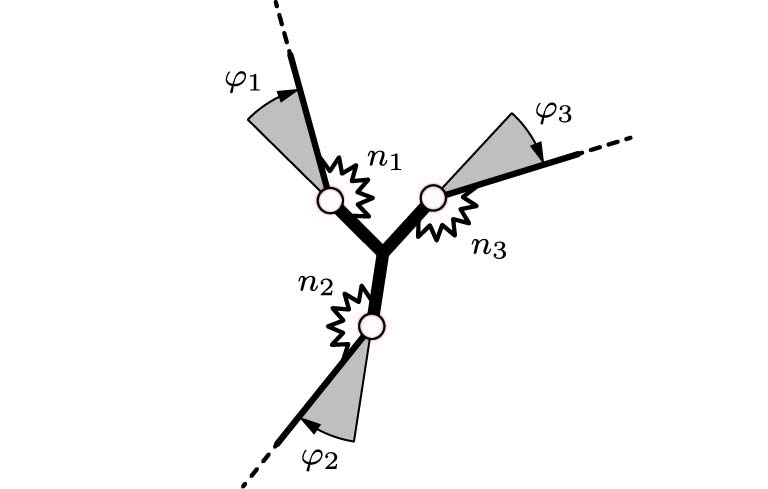}
                            \small(a)\vspace{10mm}
                            \includegraphics[width=\marginparwidth]{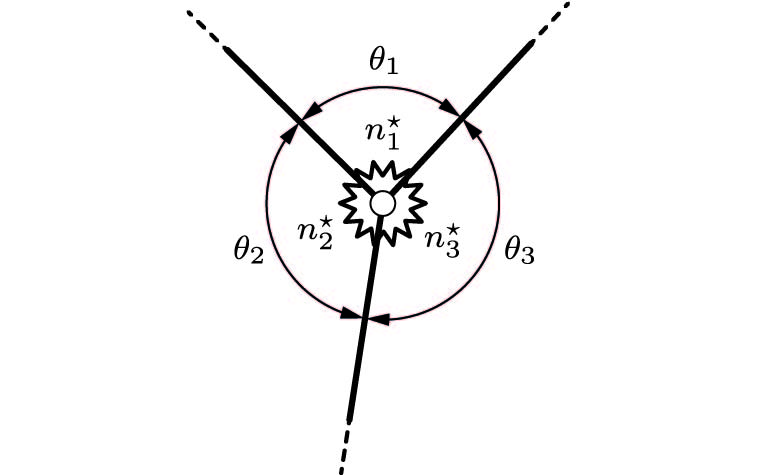}
                            \small(b)
                            \captionof{figure}{Cell corner with (a) eccentric and (b) centric hinges. The latter leads to different cell corner springs.}
                            \label{pic:Figure_10_8}
                        \end{center}}[-83mm]

                    The rotational spring stiffness of a mechanical model with eccentric and centric hinges can be transformed into each other if eccentricities are small. As shown in Figure~\ref{pic:Figure_10_8}, the relationship between cell corner and bending angles is
                    \begin{align}
                        \left[
                        \begin{array}{c}
                            \Delta\theta_1\\
                            \Delta\theta_2\\
                            \Delta\theta_3
                        \end{array}
                        \right] =
                        \left[
                        \begin{array}{rrr}
                            -1 &  0 &  1\\
                             1 & -1 &  0\\
                             0 &  1 & -1
                        \end{array}
                        \right]
                        \left[
                        \begin{array}{c}
                            \varphi_1\\
                            \varphi_2\\
                            \varphi_3
                        \end{array}
                        \right].
                    \end{align}
                    Previous matrix is rank deficient and thus not invertible. This can be overcome by additionally satisfying the equilibrium condition
                    \begin{align}
                        \sum_{i=1}^3 n_i \varphi_i = 0
                    \end{align}
                    so that the relationship between $\boldsymbol{\varphi}$ and $\Delta\boldsymbol{\theta}$ is
                    \begin{align}
                        \left[
                        \begin{array}{c}
                            \varphi_1\\
                            \varphi_2\\
                            \varphi_3
                        \end{array}
                        \right] = \frac{1}{n_1+n_2+n_3}
                        \left[
                        \begin{array}{rrr}
                            -n_3 & n_2 & 0\\
                            -n_3 & -n_1-n_3 & 0\\
                            n_1+n_2 & n_2 & 0
                        \end{array}
                        \right]
                        \left[
                        \begin{array}{c}
                            \Delta \theta_1\\
                            \Delta \theta_2\\
                            \Delta \theta_3
                        \end{array}
                        \right].
                    \end{align}
                    Hence, the relationship between both kind of springs results in
                    \begin{align}
                        \left[
                        \begin{array}{c}
                            n_1^\star\\
                            n_2^\star\\
                            n_3^\star
                        \end{array}
                        \right] =
                        \frac{1}{n_1+n_2+n_3}
                        \left[
                        \begin{array}{c}
                            n_1 n_3\\
                            n_2 n_1\\
                            n_3 n_2
                        \end{array}
                        \right]
                    \end{align}
                    as they need to preserve the total strain energy.


                \subsubsection{Eccentric Hinges}
                    The coupling between the geometric an mechanical model increases considerably if hinge eccentricities are taken into account. The global state variables of a mechanical model with frictionless hinges and rigid cell sides is
                    \begin{alignat}{2}
                        &\mathbf{e}^\textrm{M} &&=
                        \left[
                        \begin{array}{cccc}
                            {\mathbf{u}_0}^\top & {\mathbf{v}_0}^\top & \mathbf{d}^\top & \mathbf{t}^\top
                        \end{array}
                        \right]^\top\\\nonumber
                        &\mathbf{h}^\textrm{M} &&=
                        \left[
                        \begin{array}{cc}
                            \mathbf{u}^\top & \boldsymbol{\kappa}^\top
                        \end{array}
                        \right]^\top.
                    \end{alignat}
                    An additional consideration of axial cell side springs leads to
                    \begin{alignat}{2}
                        &\mathbf{e}^\textrm{M} &&=
                        \left[
                        \begin{array}{ccccc}
                            {\mathbf{u}_0}^\top & {\mathbf{v}_0}^\top & \mathbf{d}^\top & \mathbf{m}^\top & \mathbf{t}^\top
                        \end{array}
                        \right]^\top\\\nonumber
                        &\mathbf{h}^\textrm{M} &&=
                        \left[
                        \begin{array}{ccc}
                            \mathbf{u}^\top & \mathbf{v}^\top & \boldsymbol{\kappa}^\top
                        \end{array}
                        \right]^\top
                    \end{alignat}
                    whereas an additional consideration of rotational springs leads to
                    \begin{alignat}{2}
                        &\mathbf{e}^\textrm{M} &&=
                        \left[
                        \begin{array}{cccccc}
                            {\mathbf{u}_0}^\top & {\mathbf{v}_0}^\top & \mathbf{d}^\top & \mathbf{n}^\top & \mathbf{s}^\top & \mathbf{t}^\top
                        \end{array}
                        \right]^\top\\\nonumber
                        &\mathbf{h}^\textrm{M} &&=
                        \left[
                        \begin{array}{cc}
                            \mathbf{u}^\top & \boldsymbol{\kappa}^\top
                        \end{array}
                        \right]^\top.
                    \end{alignat}


    \newpage

    \sectionmark{Summary}
    \begin{framed}
        \noindent \textbf{Summary}\\

        \noindent The process dependent geometric model needs to be coupled to a mechanical model for the simulation and optimization of pressure actuated cellular structures. The mechanical model that is presented in this thesis is designed such that it fully captures the mechanical essence of a wide range of geometric models while being simple enough to enable the computation of analytical sensitivities. This is achieved by using rigid bodies and rotational, axial springs to represent the cell corners, compliant hinges and cell sides of the geometric model. For the sake of simplicity, it is assumed that cell sides are straight and sufficiently rigid so that their bending deformations can be neglected. This has the added benefit that it reduces hinge deformations and thus increases the fatigue life. Depending on the geometric model, the numerical model can be simplified by neglecting hinge eccentricities and/or rotational, axial springs.\\

        \noindent \textbf{Conclusion}\\

        \noindent The simulation and optimization of pressure actuated cellular structures requires a coupling between the geometric and mechanical model. A strikingly efficient numerical method that processes their analytical sensitivities is subsequently presented.
    \end{framed}

    \newpage
    \thispagestyle{empty}


    \cleardoublepage
    \newgeometry{}
    \thispagestyle{empty}

        \noindent \Large \textbf{Nomenclature} \normalsize
        \vspace{5mm}

        \noindent
        \begin{tabular}{llll}
            \textbf{Superscripts}\hspace{5mm}
            & eff      & \hspace{10mm} & effective value\\\vspace{-2.5mm}\\

            & $G$, $M$ &               & geometric, mechanical model\\
            & $H$      &               & compliant hinge\\
            & $P$, $T$ &               & pentagonal, triangular cell\\
            & $S$      &               & cell side
        \end{tabular}


        \vspace{5mm}
        \noindent
        \begin{tabular}{llll}
            \textbf{Subscripts}\hspace{9mm}
            & $0$           & \hspace{9mm} & reference configuration\\
            & $a$, $b$, $c$ &              & reference to cell sides
        \end{tabular}


        \vspace{5mm}
        \noindent
        \begin{tabular}{llll}
            \textbf{Numbers}\hspace{11mm}
            & $n_\kappa$    & \hspace{7.5mm} & global state variables, cell corners\\
            & $n_P$, $n_T$  &                & pentagonal, triangular (hexagonal) cells\\
            & $n_R$         &                & cell rows
        \end{tabular}


        \vspace{5mm}
        \noindent
        \begin{tabular}{llll}
            \textbf{Greek Letters}\hspace{2.5mm}
            & $\alpha$, $\beta$, $\kappa$   &              & global, local state variables\\
            & $\gamma$, $\psi$              &              & internal angles\\
            & $\zeta$, $\xi$                & \hspace{8mm} & local coordinates\\
            & $\theta$                      &              & cell corner angle\\
            & $\mu$                         &              & thickness ratio of compliant hinge\\
            & $\nu$                         &              & Poisson's ratio\\
            & $\rho$                        &              & stress reduction factor (von Mises)\\
            & $\sigma$                      &              & stress\\
            & $\varphi$                     &              & bending angle of hinge\\
            & $\chi$                        &              & correction factor for hinge geometry
        \end{tabular}


        \vspace{5mm}
        \noindent
        \begin{tabular}{llll}
            \textbf{Roman Letters}\hspace{0.5mm}
            & $a$, $b$, $c$     & \hspace{9.5mm} & side lengths\\
            & $d$               &                & hinge eccentricity\\
            & $e$, $h$          &                & set of design, state variables\\
            & $i$, $j$, $k$     &                & integers\\
            & $m$, $n$          &                & axial, rotational spring stiffness\\
            & $p$               &                & cell row pressure\\
            & $s$, $t$          &                & central hinge, cell side thickness\\
            & $u$, $v$          &                & global, local state variables\\
            & $w$               &                & hinge width\\
            & $x$, $y$          &                & internal lengths\\\vspace{-2.5mm}\\

            & $A$               &                & cross sectional area\\
            & $E$               &                & Young's modulus\\
            & $F$               &                & axial cell side force\\
            & $M$               &                & bending moment
        \end{tabular}
        \restoregeometry

        \thispagestyle{empty} 
        \cleardoublepage
    \chapter{Optimization}
        Analytical solutions for pressure actuated cellular structures can be obtained for special cases. For example, a single variable is sufficient to describe the equilibrium configuration of a cell row under the following conditions. First, identical pentagonal cells that possess a reflection symmetry plane are used. Second, cells are assembled from rigid cell sides and frictionless, centric hinges. The corresponding equilibrium configuration is independent of the cell pressures so that it can be computed by maximizing the cross sectional area of a single cell. The number of variables that are required to describe the equilibrium configuration increases dramatically\footnote{Compliant pressure actuated cellular structures with 536 and 1,616 degrees of freedom are optimized at the end of this chapter.} for arbitrary cell geometries, hinge eccentricities or compliant hinges and cell sides. Furthermore, a coupling between variables occurs in structures with more than one cell row. In general it can be said that it is impossible to find analytical solutions for pressure actuated cellular structures even if they are made from rigid cell sides and frictionless, centric hinges. It is subsequently shown how the previously introduced geometric and mechanical model can be efficiently coupled for the numerical simulation and optimization of a wide range of pressure actuated cellular structures.


        \section{Potential Energy}
            Equilibrium configurations of pressure actuated cellular structures maximize the difference between the pressure and strain energy. It is shown in the following how the potential energy of a structure and its derivatives with respect to the global state variables $\mathbf{e}^\textrm{G}$ of the geometric model can be assembled from the energy terms of the geometric primitives.


            \subsection{Geometric Primitives}
                The pressure potential $\Pi^\textrm{P}$ of the $j$-th pentagonal cell in the $i$-th cell row is the product of the cross sectional area $A^\textrm{P}$ and the cell row pressure $p$ so that
                \begin{align}
                    \Pi_{i,j}^\textrm{P}\left(\mathbf{h}^\textrm{M}\right) = -p_i A^\textrm{P}_{i,j}\left(\mathbf{h}^\textrm{M}\right)
                \end{align}
                if fictitious, straight cell sides are assumed. Similarly, the pressure potential of the corresponding triangular cell is
                \begin{align}
                    \Pi_{i,j}^\textrm{T}\left(\mathbf{h}^\textrm{M}\right) = -p_{i+1} A^\textrm{T}_{i,j}\left(\mathbf{h}^\textrm{M}\right).
                \end{align}
                The potential energy of the $k$-th cell side consists of a pressure and strain part so that
                \begin{align}
                    &\Pi_k^\textrm{S}\left(\mathbf{e}^\textrm{M},\mathbf{h}^\textrm{M}\right) = -\Delta p_k^\textrm{S} A_k^\textrm{S}\left(\mathbf{e}^\textrm{M},\mathbf{h}^\textrm{M}\right)\\\nonumber
                    &\hspace{14mm}+\frac{1}{2} \left( m_k\left(\mathbf{e}^\textrm{M}\right) \Delta {y_k\left(\mathbf{e}^\textrm{M},\mathbf{h}^\textrm{M}\right)}^2 + \sum_{i=1}^2 n_{k,i}\left(\mathbf{e}^\textrm{M}\right) {\varphi_{k,i}\left(\mathbf{e}^\textrm{M},\mathbf{h}^\textrm{M}\right)}^2 \right)
                \end{align}
                where $\Delta p$ is the differential pressure that acts on the cell side and $\Delta y$ is the elongation of the central cell side. Recall that $\Delta \boldsymbol{\varphi} = \boldsymbol{\varphi}$ since undeformed cell sides are straight. It can be seen that only the cell side potentials are a function of the reference (undeformed) configuration.


            \subsection{Structure}
                The potential energy of a cellular structure can be assembled from the contributions of its geometric primitives
                \begin{align}
                    \Pi\left(\mathbf{e}^\textrm{M},\mathbf{h}^\textrm{M}\right) =&
                    \sum_{i=1}^{nR} \left(\ \ \sum_{j=1}^{nP+1-i} \hspace{-1mm} \left(\Pi_{i,j}^\textrm{P} +
                    \delta_i^1 \Pi_{a,j}^\textrm{S} + \Pi_{c,i,2j-1}^\textrm{S} + \Pi_{c,i,2j}^\textrm{S} \right)\right.\\\nonumber
                    &\hspace{7mm} \left.+ \sum_{j=1}^{nP+2-i} \hspace{-2mm} \Pi_{b,i,j}^\textrm{S} + \left(1-\delta_i^{nR}\right) \sum_{j=1}^{nP-i} \Pi_{i,j}^\textrm{T}\right)
                \end{align}
                where $\delta$ is a Kronecker delta.  The gradient $\boldsymbol{\Pi}^u = \partial\Pi/\partial\mathbf{u} \in \mathbb{R}^{nu}$ of the potential energy with respect to the global state variables $\mathbf{u}$ is computed by summing up the contributions of single cell rows. The energy gradient $\boldsymbol{\Pi}_i^u$ incorporates only terms from cells and sides in the $i$-th cell row. It is expressed with respect to the state variables $\mathbf{u}_i$ of a cellular structure that solely consists of cell rows $i \ldots n_R$
                \begin{align}
                    \boldsymbol{\Pi}^u_{i} =&\hspace{4mm}
                    \sum_{j=1}^{nP+1-i} \left(
                    \boldsymbol{\Pi}^{\textrm{P},u}_{i,j}
                    \frac{\partial\mathbf{u}^\textrm{P}_{i,j}}{\partial\mathbf{u}_i} +
                    \boldsymbol{\Pi}^{\textrm{P},v}_{i,j} \frac{\partial \mathbf{v}^\textrm{P}_{i,j}}{\partial \mathbf{u}_i} \right)\\\nonumber
                    & + \hspace{2mm}
                    \sum_{j=1}^{nP-i} \hspace{2mm} \left(
                    \boldsymbol{\Pi}_{i,j}^{\textrm{T},u}
                    \frac{\partial\mathbf{u}^\textrm{T}_{i,j}}{\partial\mathbf{u}_i} +
                    \boldsymbol{\Pi}^{\textrm{T},v}_{i,j}
                    \frac{\partial \mathbf{v}^\textrm{T}_{i,j}}{\partial \mathbf{u}_i}
                    \right) \left(1-\delta_i^{nR}\right)\\\nonumber
                    & +
                    \sum_{j=1}^{nP+1-i} \left(
                    \boldsymbol{\Pi}^{\textrm{S},u}_{a,i,j} \frac{\partial\mathbf{u}^\textrm{S}_{a,i,j}}{\partial\mathbf{u}_i} +
                    \boldsymbol{\Pi}^{\textrm{S},v}_{a,i,j}
                    \frac{\partial v^\textrm{S}_{a,i,j}}{\partial \mathbf{u}_i}
                    \right) \delta_i^1\\\nonumber
                    & +
                    \sum_{j=1}^{nP+2-i}
                    \left(
                    \boldsymbol{\Pi}^{\textrm{S},u}_{b,i,j} \frac{\partial\mathbf{u}^\textrm{S}_{b,i,j}}{\partial\mathbf{u}_i} +
                    \boldsymbol{\Pi}^{\textrm{S},v}_{b,i,j}
                    \frac{\partial v^\textrm{S}_{b,i,j}}{\partial\mathbf{u}_i}
                    \right)\\\nonumber
                    & +
                    \sum_{j=1}^{nP+1-i} \sum_{k=0}^{1}
                    \left(
                    \boldsymbol{\Pi}^{\textrm{S},u}_{c,i,2j-k}
                    \frac{\partial \mathbf{u}^\textrm{S}_{c,i,2j-k}}{\partial \mathbf{u}_i} +
                    \boldsymbol{\Pi}^{\textrm{S},v}_{c,i,2j-k}
                    \frac{\partial v^\textrm{S}_{c,i,2j-k}}{\partial\mathbf{u}_i}\right).
                \end{align}
                Terms such as $\partial\mathbf{u}_{i,j}^\textrm{P}/\partial\mathbf{u}_i$ map state variables of single cells or sides to the state variables $\mathbf{u}_i$. Adding and transforming the gradients of single cell rows from top to bottom leads to
                \begin{align}
                    \boldsymbol{\Pi}^u = \frac{\partial \Pi}{\partial \mathbf{u}} =
                    \left( \left( \boldsymbol{\Pi}^u_{nR} \frac{\partial\mathbf{u}_{nR}}{\partial\mathbf{u}_{nR-1}} + \ldots + \boldsymbol{\Pi}^u_{3} \right) \frac{\partial\mathbf{u}_{3}}{\partial\mathbf{u}_{2}} +
                    \boldsymbol{\Pi}^u_{2} \right) \frac{\partial\mathbf{u}_{2}}{\partial\mathbf{u}_{1}} +
                    \boldsymbol{\Pi}^u_{1}
                \end{align}
                where the terms $\partial\mathbf{u}_{i+1}/\partial\mathbf{u}_i$ are assembled from previously introduced transformation matrices. Gradients such as, for example, $\boldsymbol{\Pi}^v = \partial\Pi/\partial\mathbf{v}$ can be computed in a similar manner.


        \section{Stationarity}
            \subsection{Equilibrium Configurations}
                Cell pressures are assumed to be constant throughout each row. Combinations of cell row pressures are subsequently referred to as pressure sets. A cellular structure is in equilibrium for a pressure set $q$ if its potential energy is stationary i.e.
                \begin{align}
                    \mathbf{f}_q\left(\mathbf{e}^\textrm{M},\mathbf{h}_q^\textrm{M}\right) = \mathbf{0}
                \end{align}
                where the force vector $\mathbf{f}_q$ is
                \begin{align}
                    \mathbf{f}_q\left(\mathbf{e}^\textrm{M},\mathbf{h}_q^\textrm{M}\right) =
                    \left[
                    \begin{array}{ccc}
                        \boldsymbol{\Pi}^u_q\left(\mathbf{e}^\textrm{M},\mathbf{h}_q^\textrm{M}\right) & \boldsymbol{\Pi}^v_q\left(\mathbf{e}^\textrm{M},\mathbf{h}_q^\textrm{M}\right) & \boldsymbol{\Pi}^\kappa_q\left(\mathbf{e}^\textrm{M},\mathbf{h}_q^\textrm{M}\right)
                    \end{array}
                    \right]^\top.
                \end{align}
                This nonlinear set of equations can be solved for the global state variables $\mathbf{h}^\textrm{M}_q$ by using a Newton based approach. State variables of the $(i+1)$-th Newton iteration are
                \begin{align}
                    \mathbf{h}_q^{\textrm{M},i+1} = \mathbf{h}_q^{\textrm{M},i} - {\mathbf{K}_q}^{-1} \mathbf{f}_q
                \end{align}
                where the stiffness matrix $\mathbf{K}_q$ is
                \begin{align}
                    \mathbf{K}_q =
                    \left[
                    \begin{array}{ccc}
                        \boldsymbol{\Pi}^{uu}_q & \boldsymbol{\Pi}^{vu}_q & \boldsymbol{\Pi}^{\kappa u}_q\\
                        \boldsymbol{\Pi}^{uv}_q & \boldsymbol{\Pi}^{vv}_q & \boldsymbol{\Pi}^{\kappa v}_q\\
                        \boldsymbol{\Pi}^{u\kappa}_q & \boldsymbol{\Pi}^{v\kappa}_q & \boldsymbol{\Pi}^{\kappa\kappa}_q
                    \end{array}
                    \right].
                \end{align}


            \subsection{Sensitivities}
                Sensitivities of an equilibrium configuration with respect to the global state variables $\mathbf{e}^\textrm{G}$ of the geometric model are subsequently derived. Infinitesimally small variations of the global state variables $\mathbf{e}^\textrm{M}$ and $\mathbf{h}^\textrm{M}_q$ of the mechanical model need to satisfy
                \begin{align}
                    \mathbf{f}_q\left(\mathbf{e}^\textrm{M}+\Delta\mathbf{e}^\textrm{M},\mathbf{h}_q^\textrm{M}+\Delta\mathbf{h}^\textrm{M}_q\right) = \mathbf{0}.
                \end{align}
                Neglecting higher order terms leads to
                \begin{align}
                    \mathbf{f}_q + \frac{\partial\mathbf{f}_q}{\partial\mathbf{e}^\textrm{M}} \Delta\mathbf{e}^\textrm{M} + \frac{\partial\mathbf{f}_q}{\partial\mathbf{h}^\textrm{M}_q} \Delta\mathbf{h}^\textrm{M}_q = \mathbf{0}
                \end{align}
                so that the gradient $\mathbf{G}_q$ becomes
                \begin{align}
                    \mathbf{G}_q =
                    \frac{\partial\mathbf{h}_q^\textrm{M}}{\partial\mathbf{e}^\textrm{G}} =
                    \frac{\partial\mathbf{h}_q^\textrm{M}}{\partial\mathbf{e}^\textrm{M}} \frac{\partial\mathbf{e^\textrm{M}}}{\partial\mathbf{e}^\textrm{G}} =
                    -{\mathbf{K}_q}^{-1} \frac{\partial \mathbf{f}_q}{\partial \mathbf{e}^\textrm{M}} \frac{\partial\mathbf{e^\textrm{M}}}{\partial\mathbf{e}^\textrm{G}}.
                \end{align}
                Therefore, the sensitivities $\mathbf{S}_q^\textrm{S}$ of the maximum axial cell side stresses $\boldsymbol{\sigma}_q^\textrm{S}$ can be written as
                \begin{align}
                    \mathbf{S}_q^\textrm{S} = \frac{\partial\boldsymbol{\sigma}^\textrm{S}_q}{\partial\mathbf{e}^\textrm{G}} = \left(
                    \frac{\partial\boldsymbol{\sigma}^\textrm{S}_q}{\partial\mathbf{e}^\textrm{M}} + \frac{\partial\boldsymbol{\sigma}_q^\textrm{S}}{\partial\mathbf{h}^\textrm{M}_q} \frac{\partial\mathbf{h}_q^\textrm{M}}{\partial\mathbf{e}^\textrm{M}}\right)
                    \frac{\partial\mathbf{e}^\textrm{M}}{\partial\mathbf{e}^\textrm{G}}
                \end{align}
                and the sensitivities $\mathbf{S}_q^\textrm{H}$ of the maximum hinge stresses $\boldsymbol{\sigma}_q^\textrm{H}$ are
                \begin{align}
                    \mathbf{S}^\textrm{H}_q =
                    \frac{\partial\boldsymbol{\sigma}^\textrm{H}_q}{\partial\mathbf{e}^\textrm{G}} = \left(
                    \frac{\partial\boldsymbol{\sigma}^\textrm{H}_q}{\partial\mathbf{e}^\textrm{M}} + \frac{\partial\boldsymbol{\sigma}_q^\textrm{H}}{\partial\mathbf{h}^\textrm{M}_q} \frac{\partial\mathbf{h}_q^\textrm{M}}{\partial\mathbf{e}^\textrm{M}}\right)
                    \frac{\partial\mathbf{e}^\textrm{M}}{\partial\mathbf{e}^\textrm{G}}.
                \end{align}
                However, only the maximum cell side stresses across all pressure sets are required for the optimization. For example, the $n_R$ independent stress values of $\boldsymbol{\sigma}^\textrm{S}_{q,i}$ of the $i$-th cell side can be condensed into a single value so that $\sigma_i^\textrm{S}$ is
                \begin{align}
                    \sigma_i^\textrm{S} = \max
                    \left[
                    \begin{array}{ccc}
                        |\sigma^\textrm{S}_{1,i}| & \ldots & |\sigma^\textrm{S}_{nR,i}|
                    \end{array}
                    \right]
                \end{align}
                where $i=1,\ldots,n_v$. Based on the previously used maximum stress norm, the corresponding matrices $\mathbf{S}^\textrm{S}$ and $\mathbf{S}^\textrm{H}$ for the sensitivities of the axial cell side and hinge stresses can be assembled in a similar manner. It should be noted that the pressure set that leads to a maximum cell side or hinge stress can vary during an optimization. As a consequence, these gradients are potentially discontinuous.


        \section{Target Shapes}
            \subsection{Residuum}

                \marginnote{
                    \begin{center}
                        \includegraphics[width=\marginparwidth]{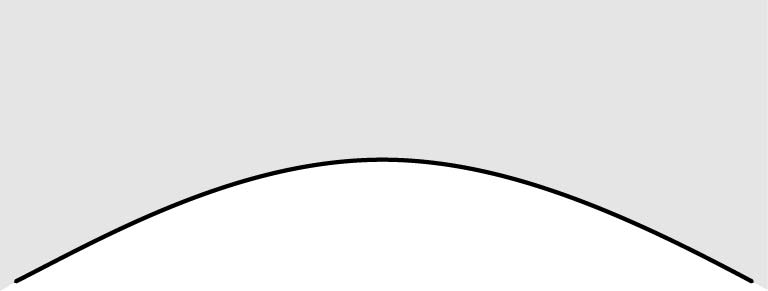}
                        \small(a)\vspace{10mm}
                        \includegraphics[width=\marginparwidth]{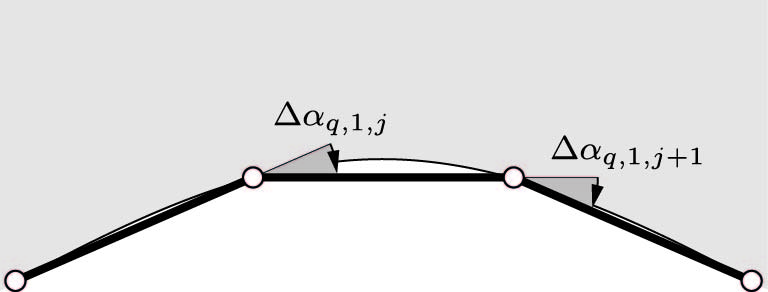}
                        \small(b)\vspace{10mm}
                        \includegraphics[width=\marginparwidth]{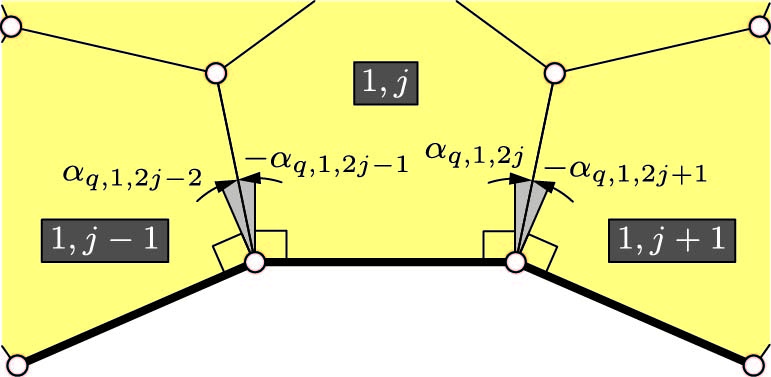}
                        \small(c)
                        \captionof{figure}{Approximation of continuous target shapes with equilibrium configurations of pressure actuated cellular structures. (a) Section of the $q$-th target shape. (b) Approximation of target shape with straight lines of given lengths. (c) The piecewise linear curve defines the required difference between the pentagonal base side angles $\boldsymbol{\alpha}$ for the $q$-th pressure set.}
                        \label{pic:Figure_11_1}
                    \end{center}}[-7mm]

                \noindent The state variables $\mathbf{e}^\textrm{G}$ of the geometric model with $n_R$ cell rows can be optimized such that the outer pentagonal cell corners are, depending on the pressure set $q$, located on $n_R$ different $C^1$ continuous target shapes as illustrated in Figure~\ref{pic:Figure_11_1}. These shapes can be enforced by constraints for the pentagonal base sides angles $\boldsymbol{\alpha}_q \subset \mathbf{u}_q$ that need to satisfy
                \begin{align}
                    \Delta\alpha_{q,1,j} = \alpha_{q,1,2j}-\alpha_{q,1,2j+1}
                \end{align}
                where $i=1$ and $j=1,\ldots,n_P-1$. The deviation between the target shape and an equilibrium configuration at the $j$-th cell corner is given by
                \begin{align}
                    r_{q,1,j} = \Delta \alpha_{q,1,j}-\alpha_{q,1,2j}+\alpha_{q,1,2j+1}
                \end{align}
                and gathered in the residual vector $\mathbf{r}_{q,1}$ for the $q$-th equilibrium configuration
                \begin{align}
                    \mathbf{r}_{q,1} =
                    \left[
                    \begin{array}{ccc}
                        r_{q,1,1} &
                        \hspace{-2mm}\ldots\hspace{-2mm} &
                        r_{q,1,nP-1}
                    \end{array}
                    \right]^\top.
                \end{align}
                In turn, the residual vectors of all equilibrium configurations are gathered in
                \begin{align}
                    \mathbf{r} =
                    \left[
                    \begin{array}{ccc}
                        {\mathbf{r}_1}^\top &
                        \hspace{-2mm}\ldots\hspace{-2mm} &
                        {\mathbf{r}_{nR}}^\top
                    \end{array}
                    \right]^\top.
                \end{align}
                The target angles $\Delta\alpha_{q,1,j}$ are a function of the base lengths $a_{1,j}$ and thus depend on the corresponding axial strains. However, their influence is neglected since these strains are usually small.


            \subsection{Sensitivities}
                Based on the gradients $\mathbf{G}_q$, the matrix $\mathbf{R}$ relates the residual target shape vectors $\mathbf{r}$ to the state variables $\mathbf{e}^\textrm{G}$ of the geometric model for all $n_R$ pressure sets so that
                \begin{align}
                    \mathbf{R} = \frac{\partial \mathbf{r}}{\partial \mathbf{e}^\textrm{G}} =
                    \left[
                    \begin{array}{ccc}
                        {\mathbf{G}_1}^\top \mathbf{B}^\top & \ldots & {\mathbf{G}_{nR}}^\top \mathbf{B}^\top
                    \end{array}
                    \right]^\top
                \end{align}
                where $\mathbf{B}$ is a Boolean matrix.


        \section{Optimization}
            The sensitivity matrix $\mathbf{R}$ of the residual target shape vector $\mathbf{r}$ is not quadratic and thus not invertible. Hence, there exists a null-space $\mathbf{N} = \textrm{null}\left(\mathbf{H}\right)$ with $\dim\left(\mathbf{N}\right) = n_u + 4 n_v-n_R\left(n_P-1\right)$ where changes in state variables $\mathbf{e}^\textrm{G}$ do not affect the residual target shape vector. In other words, it is possible to minimize an arbitrary objective function within the null-space where $\mathbf{r}=\mathbf{0}$. Hence, the underlying optimization problem for compliant pressure actuated cellular structures can be stated as
            \begin{alignat}{2}
                &\textrm{minimize}                 && \mathcal{F}\left(\mathbf{e}^\textrm{G}\right)\\\nonumber
                &\textrm{subject to} \hspace{10mm} && \mathbf{r} = \mathbf{0}\\\nonumber
                &                                  &&
                \boldsymbol{\sigma} = \mathbf{1}\, \sigma_\textrm{max}
            \end{alignat}
            where $\boldsymbol{\sigma}$ is composed of the maximum cell side $\boldsymbol{\sigma}^\textrm{S}$ and hinge $\boldsymbol{\sigma}^\textrm{H}$ stresses, $\mathbf{1}$ is a vector of ones and $\sigma_\textrm{max}$ is the maximum allowed stress. This problem can be solved with Lagrange multipliers and the Newton method so that the Lagrangian
            \begin{align}
                \mathcal{L}\left(\mathbf{e}^\textrm{G},\boldsymbol{\lambda}_r,\boldsymbol{\lambda}_\sigma\right) = \mathcal{F}\left(\mathbf{e}^\textrm{G}\right) + {\boldsymbol{\lambda}_r}^\top \mathbf{r}\left(\mathbf{e}^\textrm{G}\right) + {\boldsymbol{\lambda}_\sigma}^\top \boldsymbol{\sigma}\left(\mathbf{e}^\textrm{G}\right)
            \end{align}
            is stationary if
            \begin{align}
                \frac{\partial\mathcal{L}}{\partial\mathbf{e}^\textrm{G}} = \mathbf{0},\ \
                \frac{\partial\mathcal{L}}{\partial\boldsymbol{\lambda}_r} = \mathbf{0}
                \hspace{5mm}\textrm{and}\hspace{5mm}
                \frac{\partial\mathcal{L}}{\partial\boldsymbol{\lambda}_\sigma} = \mathbf{0}.
            \end{align}

            \marginnote{
                \begin{center}
                    \includegraphics[width=\marginparwidth]{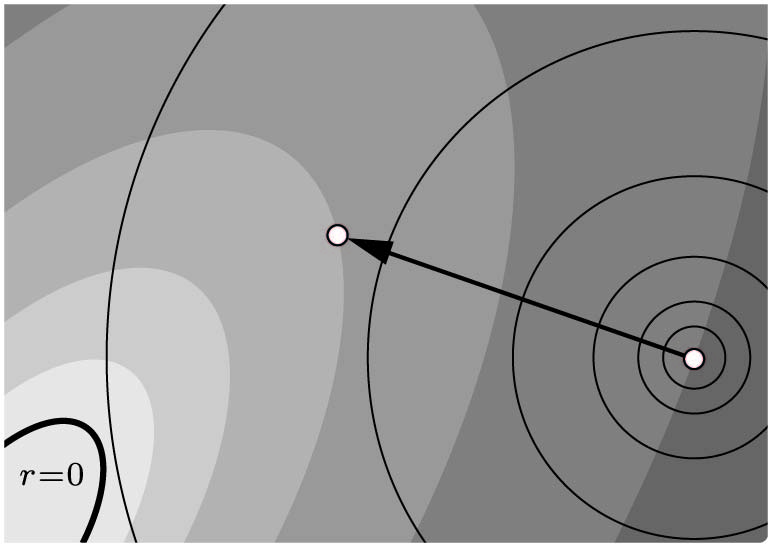}
                    \small(a)\vspace{10mm}
                    \includegraphics[width=\marginparwidth]{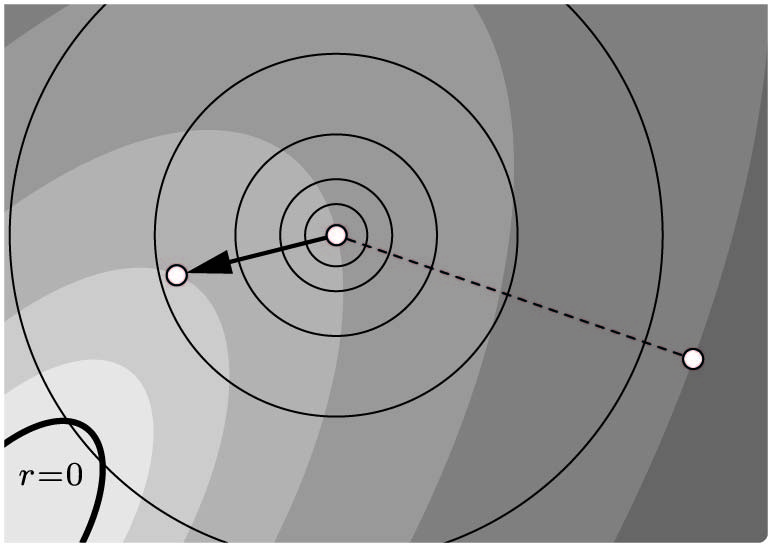}
                    \small(b)
                    \captionof{figure}{Dynamic objective for the optimization of a pressure actuated cellular structure. (a) First and (b) second iteration. The minimum of the objective moves with every iteration whereas the target shapes, constraints are enforced by Lagrange multipliers.}
                    \label{pic:Figure_11_2}
                \end{center}}[-83mm]

            \noindent In the following it is assumed that $\boldsymbol{\mathfrak{e}}^\textrm{G} = \{\mathbf{e}^\textrm{G}\}\backslash\{\mathbf{a}_0\}$ so that only reference cell side lengths other than the pentagonal base sides are varied during the optimization. An objective $\mathcal{F}$ that minimizes the differences between state variables $\boldsymbol{\mathfrak{e}}^\textrm{G}$ and target values $\boldsymbol{\mathfrak{e}}^\textrm{G,tar}$ can be defined as
            \begin{align}
                \mathcal{F}\left(\boldsymbol{\mathfrak{e}}^\textrm{G}\right) = \frac{1}{2}\left(\boldsymbol{\mathfrak{e}}^\textrm{G} - \boldsymbol{\mathfrak{e}}^\textrm{G,tar}\right)^\top
                \left(\boldsymbol{\mathfrak{e}}^\textrm{G} - \boldsymbol{\mathfrak{e}}^\textrm{G,tar}\right).
            \end{align}
            This choice is solely motivated by the objectives relatively simple gradient
            \begin{align}
                \boldsymbol{\mathcal{F}}^\mathfrak{e} = \frac{\partial\mathcal{F}}{\partial\boldsymbol{\mathfrak{e}}^\textrm{G}} = \boldsymbol{\mathfrak{e}^\textrm{G}} - \boldsymbol{\mathfrak{e}^\textrm{G,tar}}
            \end{align}
            and Hessian
            \begin{align}
                \boldsymbol{\mathcal{F}}^\mathfrak{ee} = \frac{\partial\boldsymbol{\mathcal{F}}^\mathfrak{e}}{\partial\boldsymbol{\mathfrak{e}}^\textrm{G}} = \mathbf{I}
            \end{align}
            where $\mathbf{I}$ is an identity matrix of size $n_u+4n_v-n_P$. Therefore, the set of nonlinear equations for $\boldsymbol{\mathfrak{e}}^\textrm{G}$, $\boldsymbol{\lambda}_r$ and $\boldsymbol{\lambda}_\sigma$ that is iteratively solved with the Newton method is
            \begin{align}
                &\left[
                \begin{array}{ccc}
                    \mathbf{I}+\boldsymbol{\mathcal{Z}}^i & {\boldsymbol{\mathcal{R}}^i}^\top & {\boldsymbol{\mathcal{S}}^i}^\top\\
                    \boldsymbol{\mathcal{R}}^i & \mathbf{0} & \mathbf{0}\\
                    \boldsymbol{\mathcal{S}}^i & \mathbf{0} & \mathbf{0}
                \end{array}
                \right]
                \left[
                \begin{array}{l}
                    \Delta\boldsymbol{\mathfrak{e}}^{\textrm{G},i}\\
                    \Delta\boldsymbol{\lambda}_r^i\\
                    \Delta\boldsymbol{\lambda}_\sigma^i
                \end{array}
                \right] = -
                \left[
                \begin{array}{c}
                    \boldsymbol{\mathcal{F}}^{\mathfrak{e},i} + {\boldsymbol{\mathcal{R}}^i}^\top \boldsymbol{\lambda}_r^i + {\boldsymbol{\mathcal{S}}^i}^\top \boldsymbol{\lambda}_\sigma^i\\
                    \mathbf{r}^i\\
                    \boldsymbol{\sigma}^i-\boldsymbol{\sigma}_{\textrm{max}}
                \end{array}
                \right]
            \end{align}
            where
            \begin{align}
                \left[
                \begin{array}{l}
                    \Delta\boldsymbol{\mathfrak{e}}^{\textrm{G},i}\\
                    \Delta\boldsymbol{\lambda}_r^i\\
                    \Delta\boldsymbol{\lambda}_\sigma^i
                \end{array}
                \right] =
                \left[
                    \begin{array}{c}
                        \boldsymbol{\mathfrak{e}}^{\textrm{G},i+1}-\boldsymbol{\mathfrak{e}}^{\textrm{G},i}\\
                        \boldsymbol{\lambda}_r^{i+1} - \boldsymbol{\lambda}_r^i\\
                        \boldsymbol{\lambda}_\sigma^{i+1} - \boldsymbol{\lambda}_\sigma^i
                    \end{array}
                \right].
            \end{align}
            Furthermore, the derivatives of the sensitivity matrices $\boldsymbol{\mathcal{R}} = \partial\mathbf{r}/\partial\boldsymbol{\mathfrak{e}^\textrm{G}}$ and $\boldsymbol{\mathcal{S}} = \partial\boldsymbol{\sigma}/\partial\boldsymbol{\mathfrak{e}}^\textrm{G}$ are required to compute the nonlinear contributions of the constraint equations
            \begin{align}
                \boldsymbol{\mathcal{Z}}_{i,j} = \sum_{k=1}^{nR\left(nP-1\right)} \lambda_{r,k} \frac{\partial\mathcal{R}_{k,i}}{\partial\mathfrak{e}^\textrm{G}_j}
                + \sum_{k=1}^{3nv} \lambda_{\sigma,k} \frac{\partial\mathcal{S}_{k,i}}{\partial\mathfrak{e}^\textrm{G}_j}.
            \end{align}
            This requires third-order derivatives in $\boldsymbol{\mathcal{Z}}$ which are computationally expensive. On the other hand, neglecting these terms can slow down convergence. This problem can be overcome by sacrificing the objective function. If the target values $\boldsymbol{\mathfrak{e}}^\textrm{G,tar}$ are dynamically chosen at each iteration such that
            \begin{align}
                \boldsymbol{\mathfrak{e}}^\textrm{G,i} - \boldsymbol{\mathfrak{e}}^\textrm{G,tar} = \mathbf{0}
            \end{align}
            then the Newton method reduces to
            \begin{align}
                &\left[
                \begin{array}{ccc}
                    \mathbf{I} & {\boldsymbol{\mathcal{R}}^i}^\top & {\boldsymbol{\mathcal{S}}^i}^\top\\
                    \boldsymbol{\mathcal{R}}^i & \mathbf{0} & \mathbf{0}\\
                    \boldsymbol{\mathcal{S}}^i & \mathbf{0} & \mathbf{0}
                \end{array}
                \right]
                \left[
                \begin{array}{c}
                    \boldsymbol{\mathfrak{e}}^{\textrm{G},i+1}-\boldsymbol{\mathfrak{e}}^{\textrm{G},i}\\
                    \boldsymbol{\lambda}_r^{i+1} - \boldsymbol{\lambda}_r^i\\
                    \boldsymbol{\lambda}_\sigma^{i+1} - \boldsymbol{\lambda}_\sigma^i
                \end{array}
                \right] = -
                \left[
                \begin{array}{c}
                    \mathbf{0}\\
                    \mathbf{r}^i\\
                    \boldsymbol{\sigma}^i-\boldsymbol{1}~\sigma_{\textrm{max}}
                \end{array}
                \right]
            \end{align}
            since $\boldsymbol{\mathcal{F}}^\mathfrak{e}= \mathbf{0}$ and therefore $\boldsymbol{\lambda}_r = \boldsymbol{\lambda}_\sigma = \mathbf{0}$. Two iterations during the optimization of a pressure actuated cellular structure with a dynamic objective are visualized in Figure~\ref{pic:Figure_11_2}.


        \marginnote{
            \begin{center}
                \includegraphics[width=\marginparwidth]{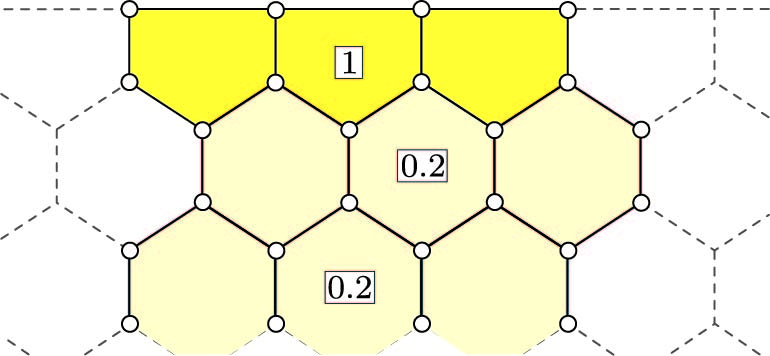}
                \small(a)\vspace{10mm}
                \includegraphics[width=\marginparwidth]{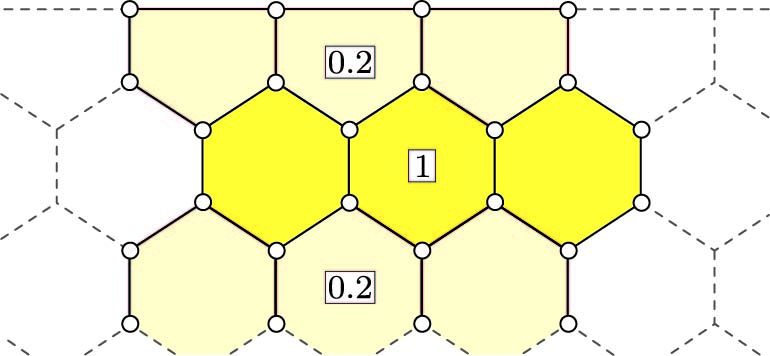}
                \small(b)\vspace{10mm}
                \includegraphics[width=\marginparwidth]{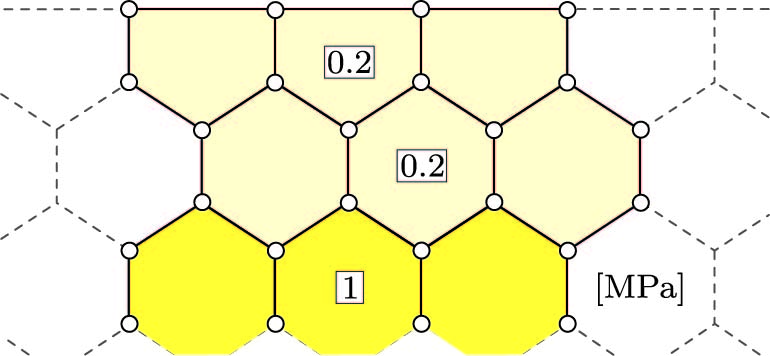}
                \small(c)
                \captionof{figure}{(a) First, (b) second and (c) third pressure sets of example structures.}
                \label{pic:Figure_11_3}
            \end{center}}[-53mm]

        \section{Examples}
            It was shown in Chapter~\ref{subsec:ReducedOrderModels} that eight different mechanical models can be used in conjunction with a geometric model of varying complexity. For example, the use of frictionless instead of compliant hinges eliminates rigid cell corners and hinge eccentricities of the geometric and mechanical model, respectively. It is not helpful to demonstrate the optimization of pressure actuated cellular structures for all possible simplifications and combinations thereof. Two kind of example structures with extreme properties are subsequently used instead.\\

            The first kind of structures possess centric, frictionless hinges and rigid cell sides. Their geometric and mechanical models are solely coupled via the cell side lengths and thicknesses. In contrast, the second kind of structures possess eccentric, compliant hinges and elastic cell sides so that their geometric and mechanical models are fully coupled. Example structures are optimized for two and three target shapes that range from circular arcs with moderate and extreme curvatures to the cruise and takeoff, landing configuration of a gapless leading edge. All of these structures have in common that they use cell row pressures of either $p=0.2$ or 1~MPa as illustrated in Figure~\ref{pic:Figure_11_3}.\\

            \marginnote{
                \begin{center}
                    \includegraphics[width=\marginparwidth]{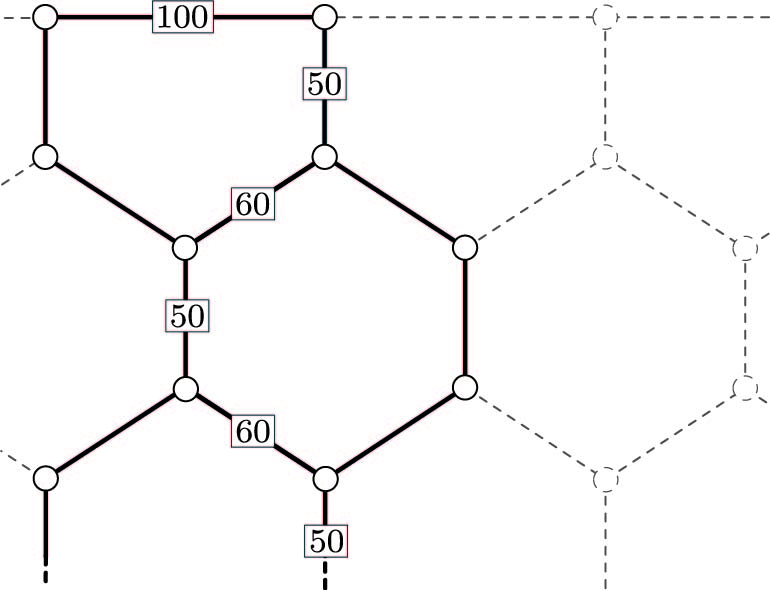}
                    \small(a)\vspace{6mm}
                    \includegraphics[width=\marginparwidth]{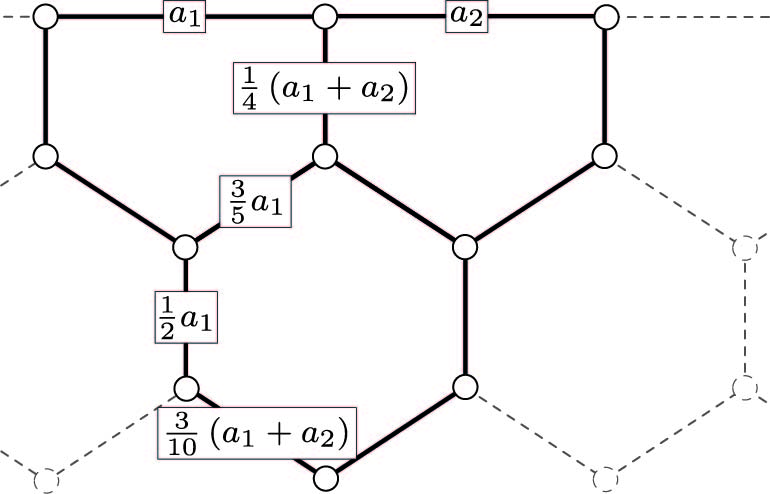}
                    \small(b)\vspace{6mm}
                    \includegraphics[width=\marginparwidth]{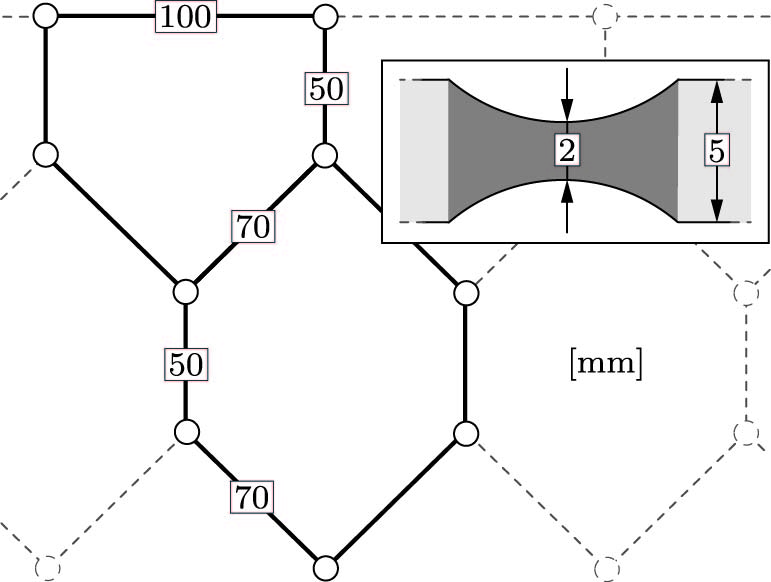}
                    \small(c)
                    \captionof{figure}{Initial cell side lengths and hinge thicknesses of example structures. (a-b) Structures with frictionless hinges and (a) constant, (b) varying pentagonal base side lengths. (c) Compliant pressure actuated cellular structures.}
                    \label{pic:Figure_11_4}
                \end{center}}[-20mm]

            The shape changing potential of pressure actuated cellular structures is reduced at both ends by the decreasing number of cells in each additional cell row. It is thus assumed that, depending on the total number of cell rows, the shape changes between the first and the last $n_R$ pentagonal base sides are zero. Furthermore, for the sake of simplicity, it is assumed that the maximum step length during the optimization is constant so that the total number of iterations could be reduced with a line search algorithm.


            \subsection{Rigid Body Structures}
                Shape changes of pressure actuated cellular structures with centric, frictionless hinges and rigid cell sides depend solely on the cell geometries and the pressure ratios between cell rows. Their shape changing potential is therefore an upper bound of what can be achieved by structures with compliant hinges and sides. Four different example structures are subsequently used to demonstrate the limits of pressure actuated cellular structures.


                \afterpage{
                    \newgeometry{}
                    \begin{figure}[htbp]
                        \begin{center}
                            \subfloat[]{
                                \includegraphics[width=0.98\textwidth]{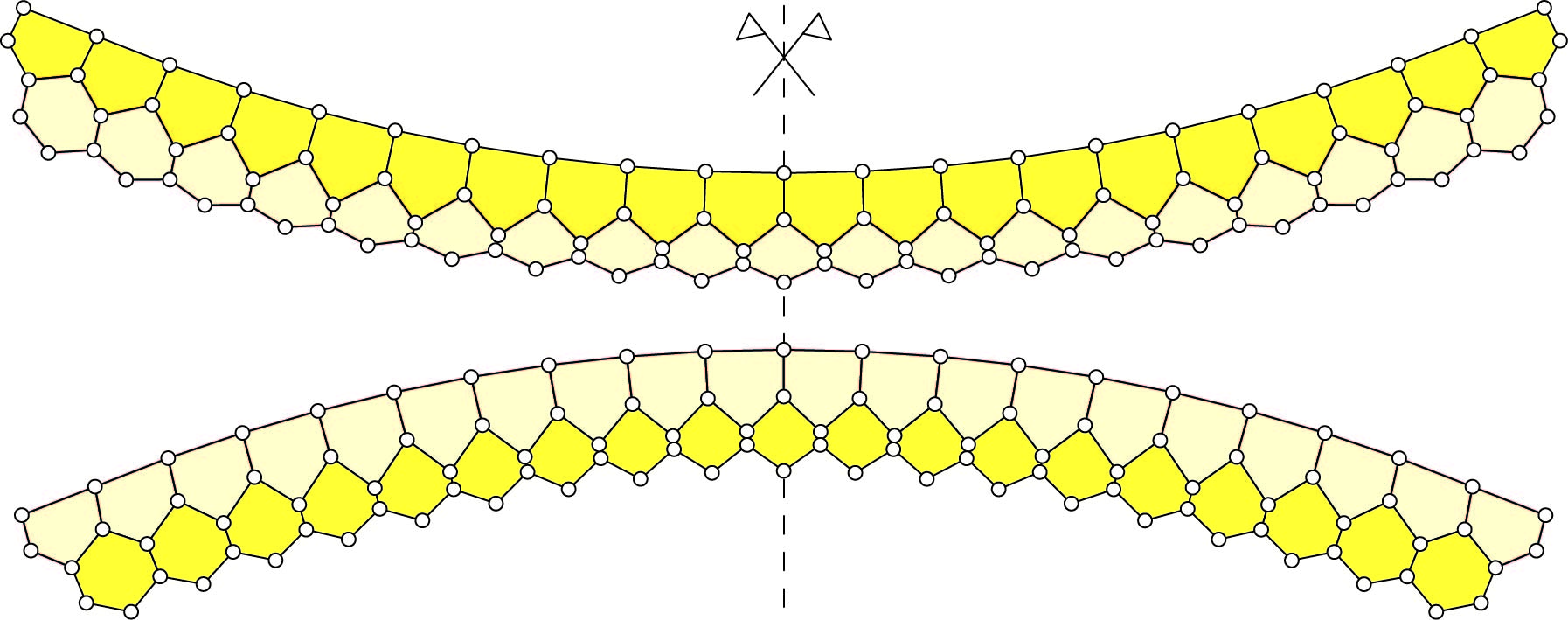}}\vspace{-2mm}

                            \subfloat[]{
                                \includegraphics[width=0.98\textwidth]{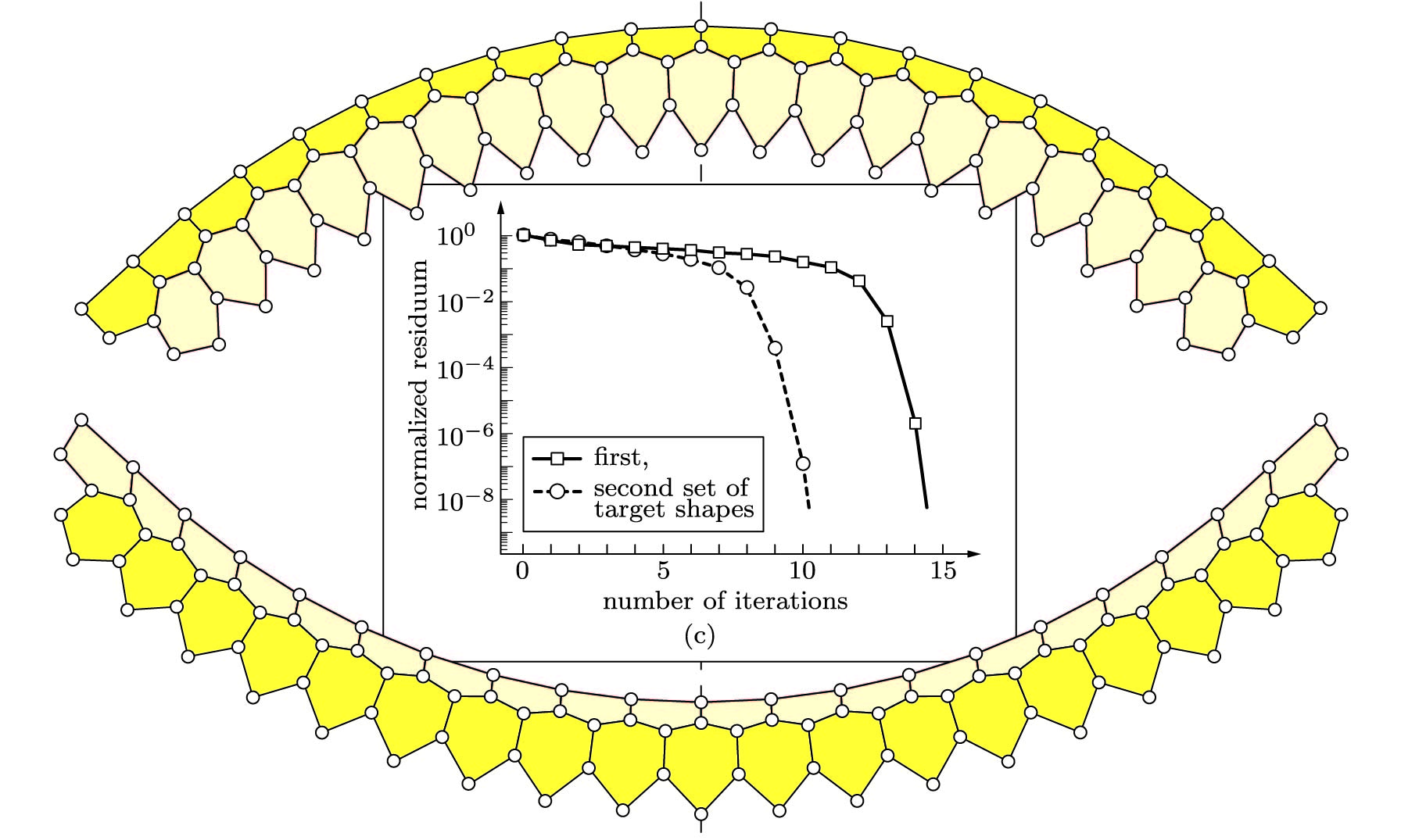}}
                            \caption{Pressure actuated cellular structures that consist of two cell rows and twenty base pentagons. Structures are assembled from centric, frictionless hinges and rigid cell sides. The cell side lengths are optimized such that the equilibrium shapes resemble convex, concave circular arcs with an identical absolute curvature. The maximum shape changes that can be achieved depend on the sign of the curvature for a given pressure set. Equilibrium shapes for the (a) first $\left(\Delta\alpha=\pm 2.5^\circ\right)$ and (b) second $\left(\Delta\alpha=\mp 5^\circ\right)$ set of target shapes. (c) Convergence of the residual target shape vectors.}
                            \label{pic:Figure_11_5}
                        \end{center}
                    \end{figure}
                    \restoregeometry}

                \subsubsection{Extreme Curvatures}
                    The achievable shape changes of pressure actuated cellular structures depend on the target shapes and the associated pressure sets. This is demonstrated with the help of a structure that consists of two cell rows and twenty base pentagons. Its 119 cell side lengths\footnote{The twenty pentagonal base side lengths are not varied during the optimization so that the total number of variables reduces from 139 to 119.} are optimized such that the equilibrium shapes resemble a convex and concave circular arc with an identical absolute curvature. Despite the constant curvature of the target shapes, edge effects prohibit the solution of this problem by investigating only a few cells.\\

                    As previously discussed, the computation of third order derivatives is avoided by using a dynamic objective that is updated after each iteration. As a consequence, the optimization results depend on the maximum step length during the optimization that is limited to $\Delta v_\textrm{max}=3$~mm and the initial cell side lengths that are summarized in Figure~\ref{pic:Figure_11_4}~(a). The optimization results for the target shapes
                    \begin{align}\nonumber
                        \Delta\alpha_{q,i} =
                        \begin{cases}
                            \hspace{2.7mm}2.5^\circ & \textrm{for } q=1\\
                            -2.5^\circ & \textrm{otherwise}
                        \end{cases}
                    \end{align}
                    and
                    \begin{align}\nonumber
                        \Delta\alpha_{q,i} =
                        \begin{cases}
                            -5^\circ & \textrm{for } q=1\\
                            \hspace{2.7mm}5^\circ & \textrm{otherwise}.
                        \end{cases}
                    \end{align}
                    that depend on the pressure set $q$ are shown in Figure~\ref{pic:Figure_11_5}. The non-zero shape changes between the pentagonal base sides are in both cases limited to $i=2,\ldots,18$. It can be seen that the achievable absolute curvature is halved if the structure needs to deform into a convex arc for the first pressure set $\left(q=1\right)$. In both cases, the structures shape changing capability is limited by extreme pentagonal or hexagonal cell geometries.\\

                    The proposed algorithm requires at most fifteen iterations for the optimization of the example structure if a fixed maximum step length is used. Furthermore, the runtime is in each case less than twelve seconds on a single Intel i5-4250U core with 1.3~GHz and Matlab 2014b\footnote{MathWorks claims that an average speedup of 2.1 can be achieved by simply upgrading from 2015a to 2018a.}. This value can be seen as an upper bound as the symmetry and the small bandwidth of the stiffness matrix are not fully exploited. In general it can be said that practically relevant structures can be conveniently optimized even on aged hardware without the need for extremely refined software.


                \afterpage{
                    \newgeometry{}
                    \begin{figure}[htbp]
                        \begin{center}
                            \subfloat[]{
                                \includegraphics[height=115mm]{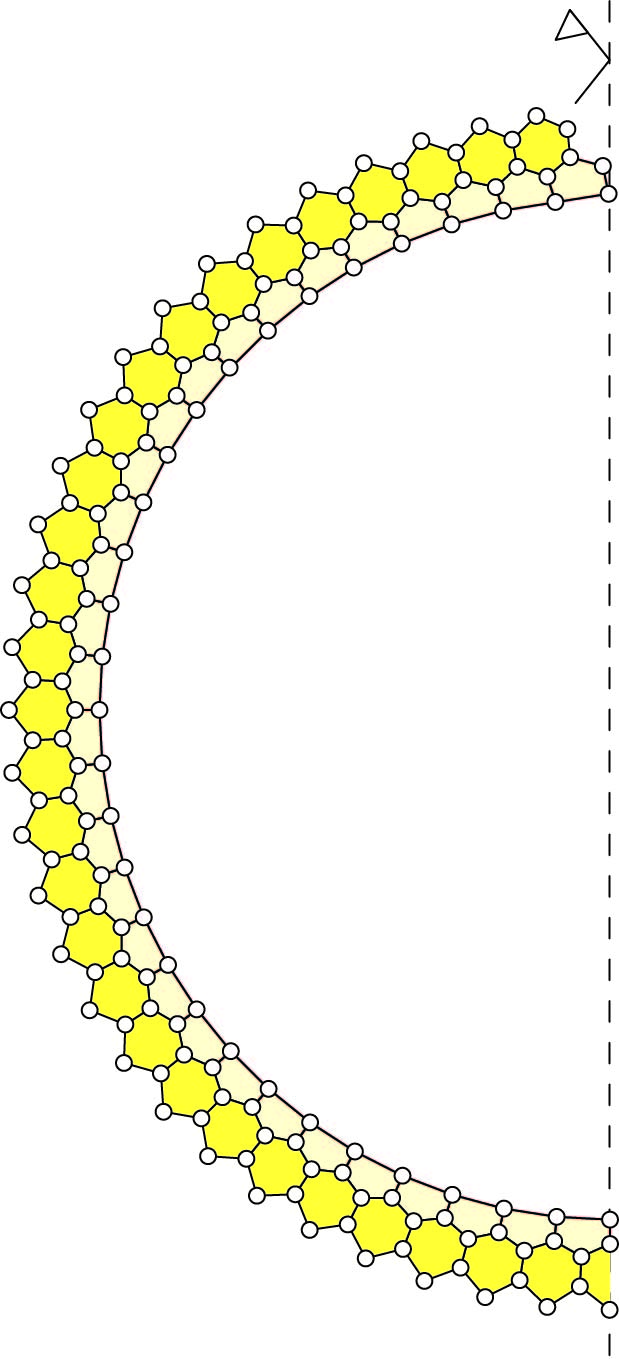}}\hfill
                            \subfloat[]{
                                \includegraphics[height=115mm]{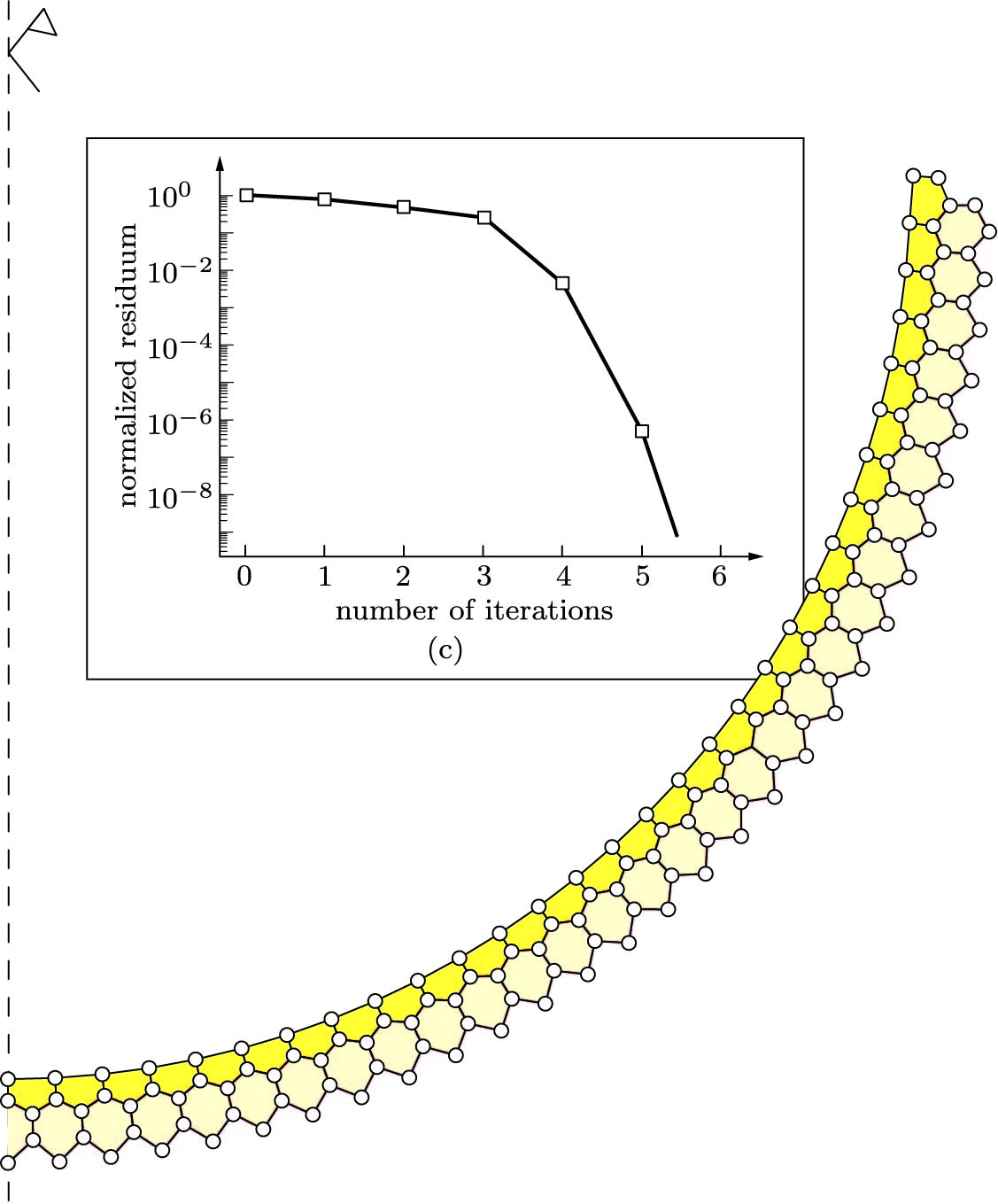}}
                            \caption{Pressure actuated cellular structure that consists of two cell rows and sixty base pentagons. The structure is assembled from centric, frictionless hinges and rigid cell sides. Cell side lengths are optimized such that the equilibrium shapes resemble a (a) full and (b) half circle with a convex curvature for the first and second pressure set, respectively. (c) Convergence of the residual target shape vector.}
                            \label{pic:Figure_11_6}
                        \end{center}
                    \end{figure}
                    \restoregeometry}

                \subsubsection{Moderate Curvatures}
                    Pressure actuated cellular structures that fully utilize their shape changing potential suffer from extreme cell geometries. For example, the pentagonal cells of the previous example possess almost rectangular geometries for $\Delta\alpha_2=5^\circ$. These kind of cell geometries are disadvantageous as they lead to increased cell side forces and manufacturing sensitivities. The following example shows that balanced cell geometries can be obtained if pressure actuated cellular structures are optimized for moderate shape changes and appropriate cell row pressures. This is demonstrated with the help of a structure that consists of two cell rows and sixty base pentagons as illustrated in Figure~\ref{pic:Figure_11_6}. The 359 cell side lengths are optimized such that the equilibrium shapes resemble a half and a full circle with a convex curvature so that
                    \begin{align}\nonumber
                        \Delta\alpha_{q,i} =
                        \begin{cases}
                            3^\circ & \textrm{for } q=1\\
                            6^\circ & \textrm{otherwise}.
                        \end{cases}
                    \end{align}
                    Similar to the previous example, the non-zero shape changes between pentagonal base sides are limited to $i=2,\ldots,58$. Furthermore, the initial cell side lengths and the maximum step length during the optimization are unchanged.\\

                    The difference between the pentagonal base side angles of both equilibrium shapes is $\Delta\alpha_2-\Delta\alpha_1=3^\circ$ and thus considerably smaller than the $10^\circ$ of the previous example. It can be seen that the moderate shape changing requirements in combination with an appropriate choice of cell row pressures leads to well balanced pentagonal and hexagonal cell geometries. This is also reflected by the required number of iterations that reduces from fifteen to six.


                \begin{figure}[htbp]
                    \begin{center}
                        \subfloat[]{
                            \includegraphics[height=184mm]{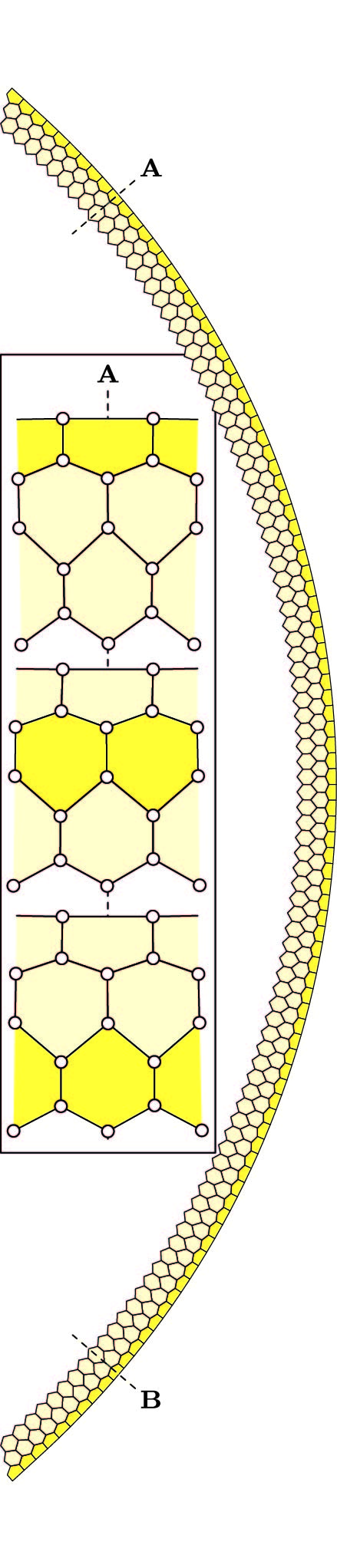}}
                        \subfloat[]{
                            \includegraphics[height=184mm]{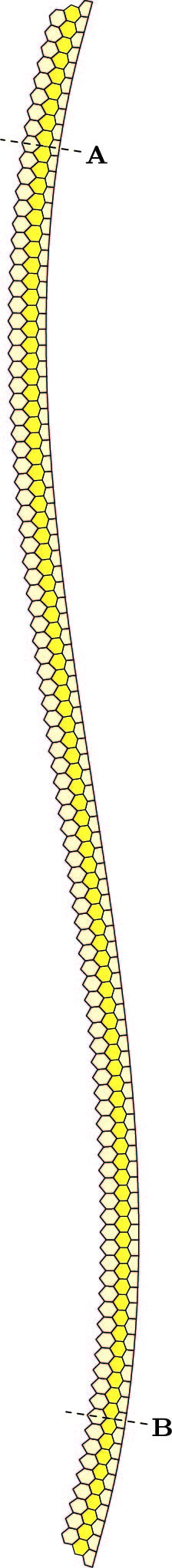}}
                        \subfloat[]{
                            \includegraphics[height=184mm]{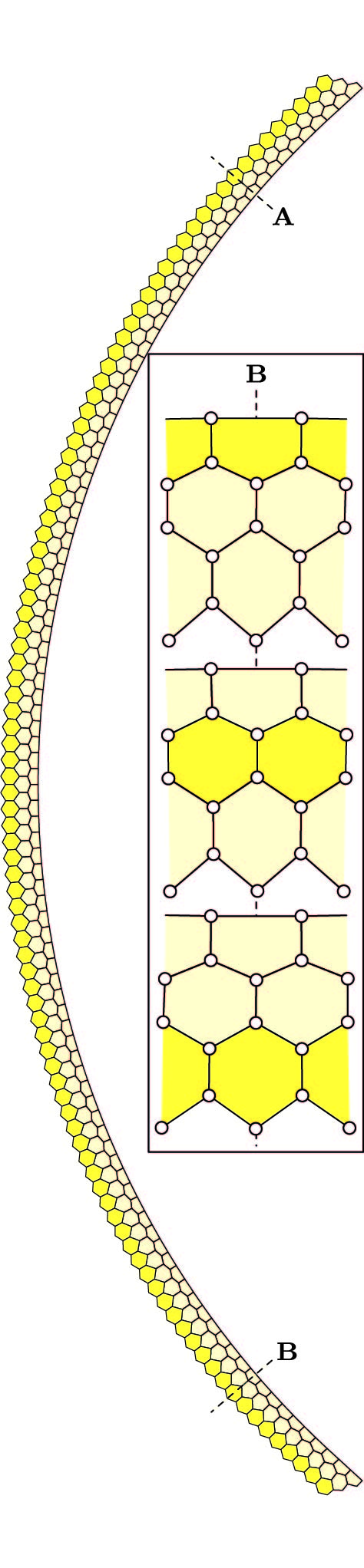}}
                    \end{center}

                    \marginnote{
                        \begin{center}
                            \vspace{5mm}
                            \includegraphics[width=\marginparwidth]{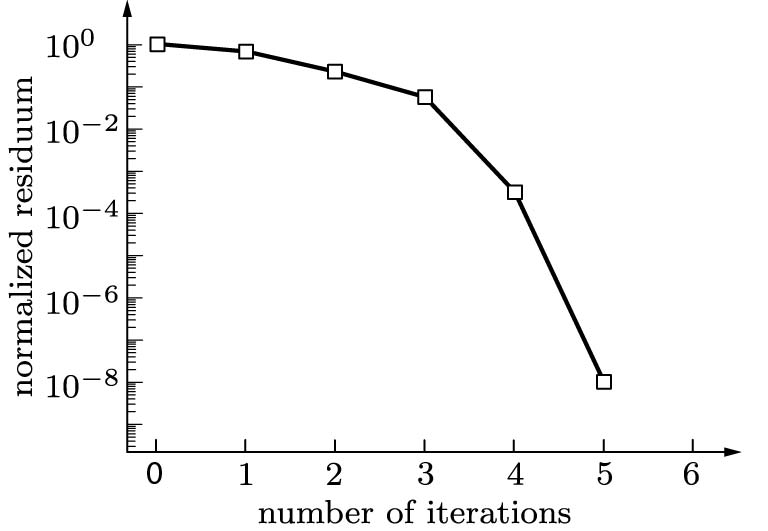}
                            \small(d)
                            \captionof{figure}{Pressure actuated cellular structure that consists of three cell rows and one-hundred base pentagons. The structure is assembled from centric, frictionless hinges and rigid cell sides. Cell side lengths are optimized such that the equilibrium shapes resemble a circular arc with a (a) concave and (c) convex curvature as well as an (b) Euler spiral. Detailed views of two opposite regions are shown. (d) Convergence of the residual target shape vector.}
                            \label{pic:Figure_11_7}
                        \end{center}}[-207mm]
                \end{figure}

                \subsubsection{Euler Spiral}
                    The presented framework for the optimization of pressure actuated cellular structures is by no means limited to two cell rows and target shapes with constant curvatures. This is demonstrated in the following with the help of a structure that consists of three cell rows and one-hundred base pentagons as illustrated in Figure~\ref{pic:Figure_11_7}. The 894 cell side lengths are optimized such that the equilibrium shapes resemble a convex and a concave circular arc with an identical absolute curvature as well as an Euler spiral
                    \begin{align}\nonumber
                        \Delta\alpha_{q,i} =
                        \begin{cases}
                            \hspace{12.2mm}-1^\circ & \textrm{for } q=1\\
                            \displaystyle \frac{1}{47} \left(i-50\right)^\circ & \textrm{for } q=2\\
                            \hspace{14.8mm}1^\circ & \textrm{otherwise}.
                        \end{cases}
                    \end{align}
                    The potential shape changes at both ends are further reduced by the third cell row so that $i=3,\ldots,97$. Hence, the angles between the first and the last three pentagonal base sides are constant. The initial cell side lengths and the maximum step length during the optimization are identical to the first two examples.\\

                    The consideration of three target shapes comes at the cost of a reduced shape changing capability. For example, the maximum difference between the pentagonal base side angles of all equilibrium shapes is $\Delta\alpha_3-\Delta\alpha_1=2^\circ$ and thus smaller than in the previous examples. It can be seen that, unlike the previous examples, the structure does not possess a symmetry plane for any of the equilibrium shapes. Furthermore, it requires only five iterations to find an optimal solution that satisfies the target shapes. Hence, the convergence properties are not significantly affected by the varying curvature of the second target shape and the increased coupling between the state variables that is caused by the additional cell row.


                \subsubsection{Gapless Leading Edge}
\begin{figure}[htbp]
                        \begin{center}
                            \subfloat[]{
                                \includegraphics[width=\textwidth]{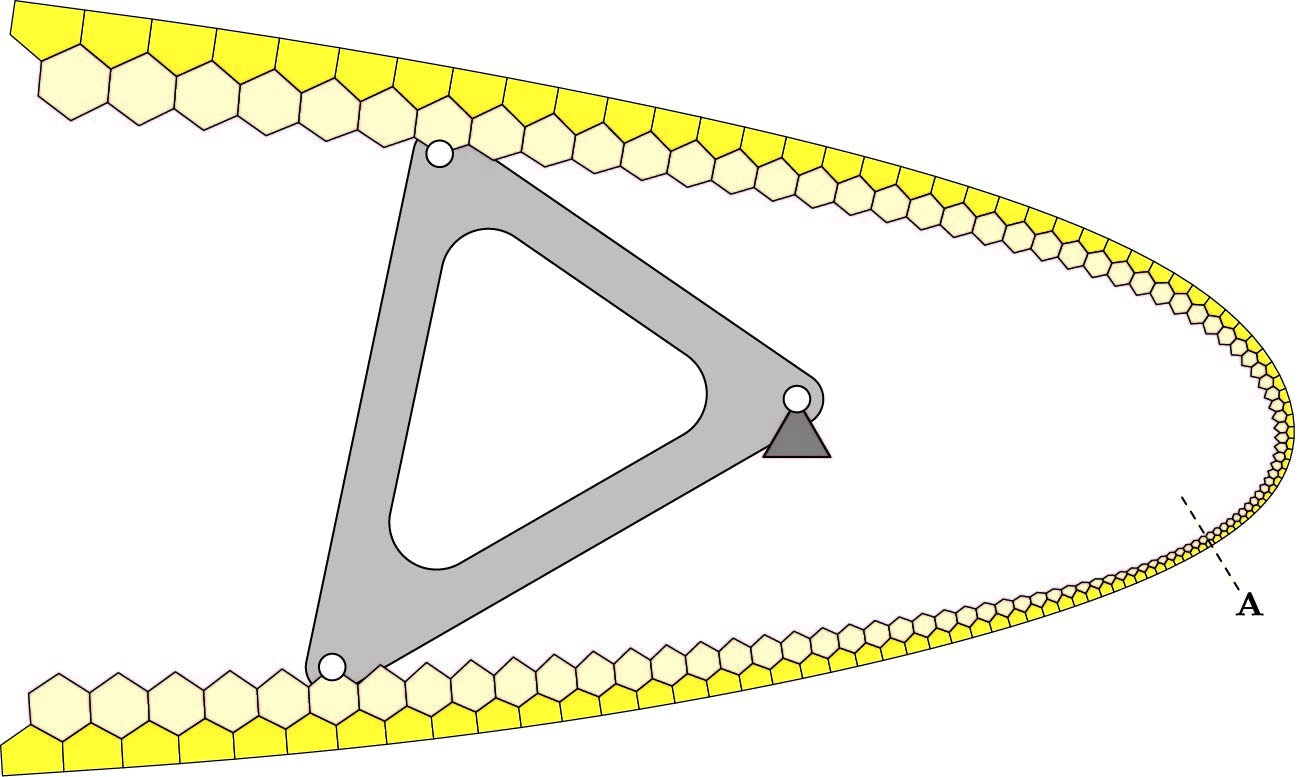}}\vspace{4mm}

                            \subfloat[]{
                                \includegraphics[width=\textwidth]{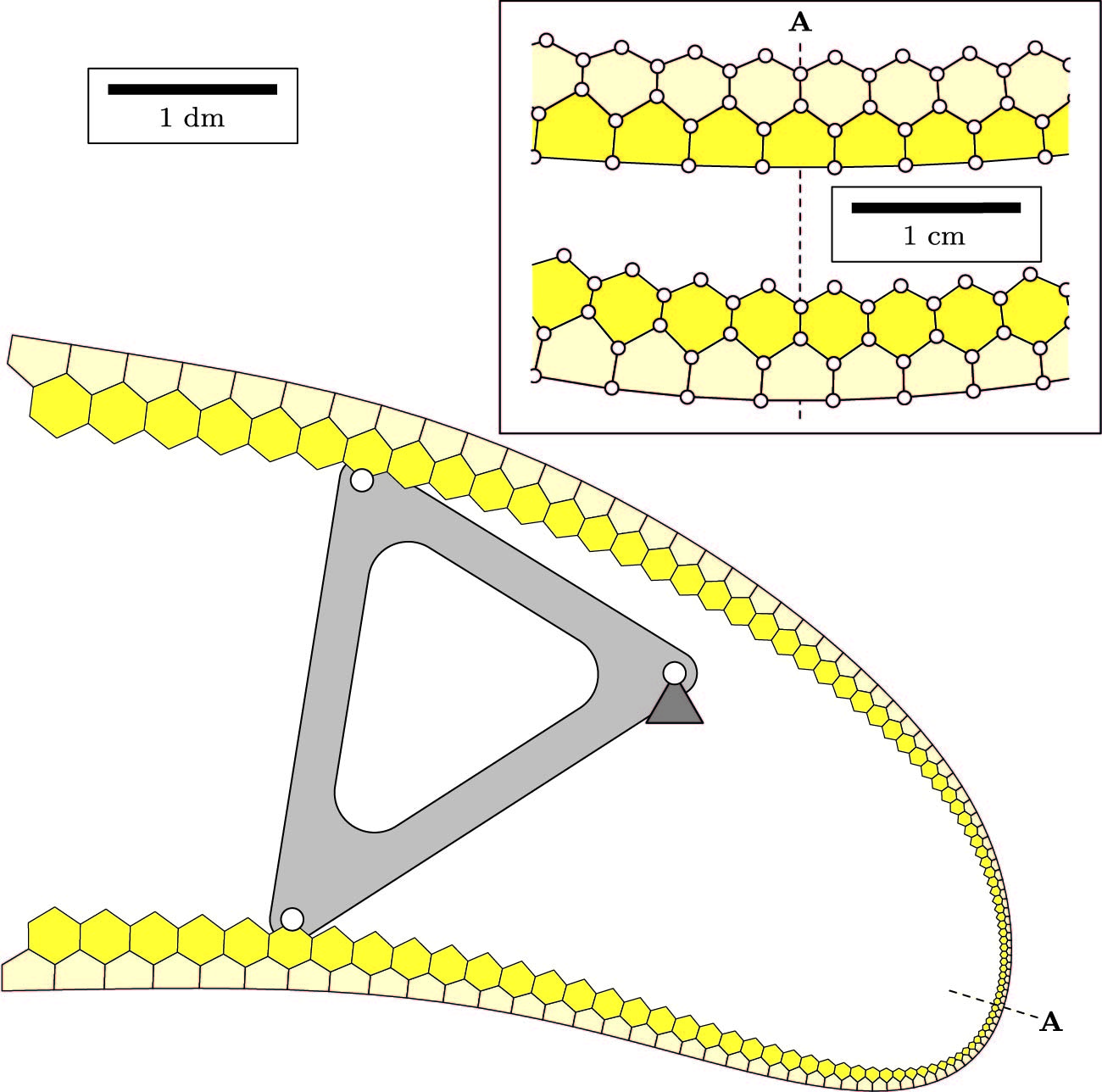}}
                        \end{center}

                        \marginnote{
                            \begin{center}
                                \subfloat[]{
                                    \includegraphics[width=\marginparwidth]{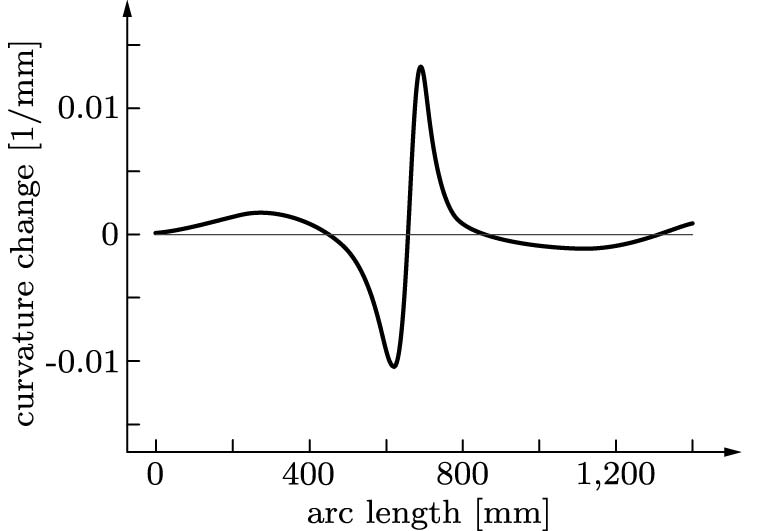}}\\
                                \subfloat[]{
                                    \includegraphics[width=\marginparwidth]{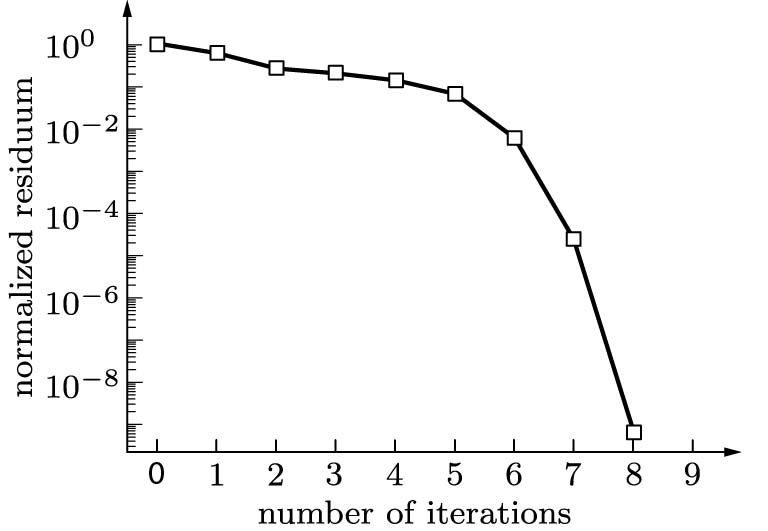}}
                                \captionof{figure}{Pressure actuated cellular structure that consists of two cell rows and one-hundred base pentagons. The structure is assembled from centric, frictionless hinges and rigid cell sides. Cell side lengths are optimized such that the equilibrium shapes resemble the (a) cruise and (b) high lift configuration of a gapless leading edge for the first and second pressure set, respectively. Detailed view of region with smallest cell sizes is shown. The overall stiffness can be increased by an internal, passive mechanism that connects the upper and lower surface. (c) Curvature change of target shapes as a function of the arc length in counterclockwise direction. (d) Convergence of the residual target shape vector.}
                                \label{pic:Figure_11_8}
                            \end{center}}[-205mm]
                    \end{figure}

                    The shape changing potential of pressure actuated cellular structures with constant pentagonal base side lengths can not be fully utilized for arbitrary target shapes. This is disadvantageous as it leads to smaller average cell sizes that depend on the maximum, local shape changing requirements. A structure that consists of two cell rows and one-hundred base pentagons is subsequently used to demonstrate the benefits of varying base side lengths. The 599 cell sides are optimized such that the equilibrium shapes resemble the cruise and high lift configuration of a gapless leading edge as illustrated in Figure~\ref{pic:Figure_11_8}. As for the previous examples, shape changes at both ends are assumed to be zero so that $i=2,\ldots,98$.\\

                    The geometry of the leading edge is based on parametric least square interpolations of discrete points with octic polynomials. Details about the raw data and the interpolations can be found in the appendix. Both target shapes possess identical arc lengths so that axial strains during shape changes are avoided. The curvature change along the arc-length are shown in Figure~\ref{pic:Figure_11_8}~(c). It can be seen that the shape changing requirements vary considerably so that the use of constant pentagonal base side lengths is not optimal. However, the computation of side lengths that lead to uniform shape changing requirements is far from trivial and an interesting problem by itself. The base side length $a_j$ of the $j$-th pentagonal cell is subsequently expressed with the help of a quadratic function
                    \begin{align}
                        a_j = b_1 j^2 + b_2 j + b_3
                    \end{align}
                    where $\mathbf{b}$ are unknown coefficients. The arc length $L$ of the leading edge needs to be identical to the sum of the $n_P$ pentagonal base side lengths so that
                    \begin{align}
                        L = \sum_{j=1}^{n_P} a_j = n_P \left(\frac{n_P+1}{2} \left(\frac{2n_P+1}{3} b_1 + b_2\right) + b_3\right).
                    \end{align}
                    Hence, the third coefficient becomes
                    \begin{align}
                        b_3 = \frac{L}{n_P} - \frac{n_P+1}{2} \left(\frac{2n_P+1}{3} b_1 + b_2\right)
                    \end{align}
                    and the remaining two coefficients are optimized such that the maximum change between base sides is minimal. An extension of this approach to polynomials with higher degrees is straightforward. The initial cell side lengths of the remaining sides are computed according to Figure~\ref{pic:Figure_11_4}~(b).\\

                     The convergence of the optimization is not significantly affected by the relatively complex target shapes and the varying cell sizes as it requires only eight iterations to find an optimal solution. The stiffness of the structure can be enhanced by passive or active mechanisms that internally connect the upper and lower surface and thus provide additional supports. Constant distances between certain cell corners can be enforced with the help of constraints so that these mechanisms can be easily retrofitted. This is of great interest as the large variation of curvature changes leads to relatively thin regions that decrease the overall bending stiffness. Their effect on the leading edge stiffness can be significantly reduced by a better choice of pentagonal base side lengths and nonuniform cell row pressures that are chosen according to the curvature changes of certain regions.


            \subsection{Compliant Structures}
                The first two examples of this chapter are subsequently revisited to demonstrate the effects of compliant hinges and elastic cell sides on the shape changing potential of pressure actuated cellular structures. The equilibrium shapes of compliant structures depend additionally on the material properties and the geometry of the reference configuration. The initial cell side lengths and hinge, cell side thicknesses of both examples are summarized in Figure~\ref{pic:Figure_11_4}~(c). It is assumed that the hinge ratio is $\mu=4$ and that the target curvature of cell corners is $\kappa^\textrm{tar}=0.5$~mm$^{-1}$. The maximum step lengths during the optimization are limited to $\Delta v_\textrm{max}=3$~mm and $\Delta s_\textrm{max}=\Delta t_\textrm{max}=0.3$~mm.\\

                Depending on the target shapes, the reference state variables $\boldsymbol{\alpha}_0$ are chosen such that the bending deformations of the pentagonal base side hinges are minimal. The reference state variables $\boldsymbol{\beta}_0$ are assumed to be zero. Furthermore, reference state angles are constrained during the optimization so that they remain unchanged. This is motivated by the different sensitivities of the equilibrium shapes with respect to cell side lengths, thicknesses and reference state angles. However, their values can be optimized in a subsequent step by minimizing the total bending energy of all equilibrium shapes for given cell row pressures, cell side lengths and thicknesses\footnote{This step can affect the structures equilibrium shapes so that it needs to be complemented by another optimization of the cell side lengths and hinge, cell side thicknesses.}.\\

                It is subsequently assumed that the example structures are made from polyphenylensulfon\footnote{PPSU is a moldable plastic that is commonly used in rapid prototyping.} so that

                \vspace{1mm}
                \noindent
                \begin{center}
                    \begin{tabular}{lll}
                        \textbf{material}\\
                        \hspace{4mm} $E=2,340$~MPa                && Young's modulus\\
                        \hspace{4mm} $\nu=0.4$                    && Poisson's ratio\\
                        \hspace{4mm} $\sigma_\textrm{max}=50$~MPa && yield strength.
                    \end{tabular}
                \end{center}
                \vspace{1mm}

                \noindent It is usually not possible to fully utilize the yield strength of the material as it depends on the manufacturing process, environmental conditions, load history and size effects. However, the influence of these factors is difficult to quantify and thus neglected. As a consequence, the presented results are nonconservative and potential shape changes are likely to be smaller.


                \afterpage{
                    \newgeometry{}
                    \begin{figure}[htbp]
                        \begin{center}
                            \subfloat[]{
                                \includegraphics[width=\textwidth]{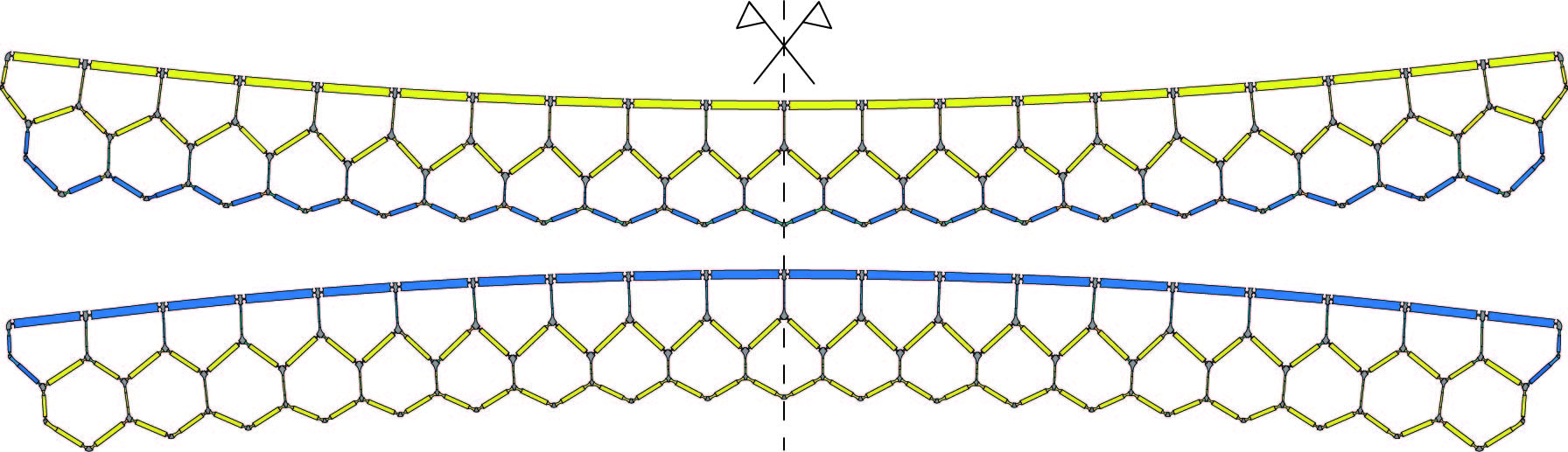}}\vspace{-2mm}

                            \subfloat[]{
                                \includegraphics[width=\textwidth]{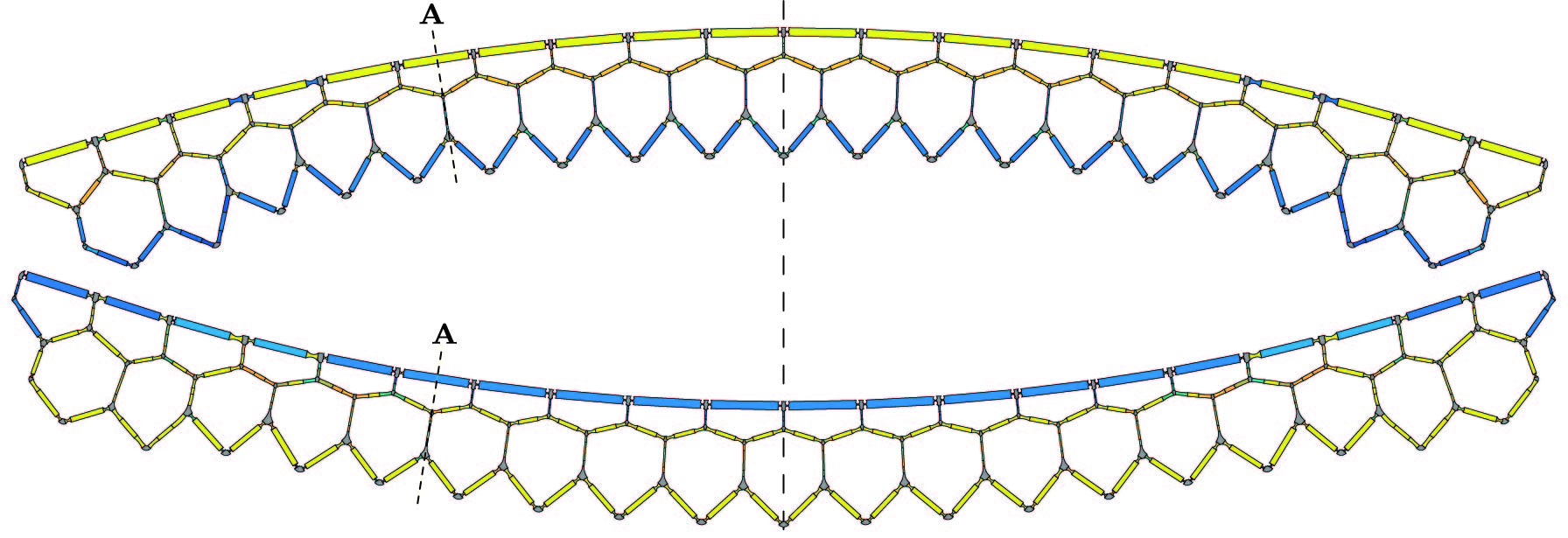}}\vspace{-2mm}

                            \subfloat[]{
                                \includegraphics[height=45mm]{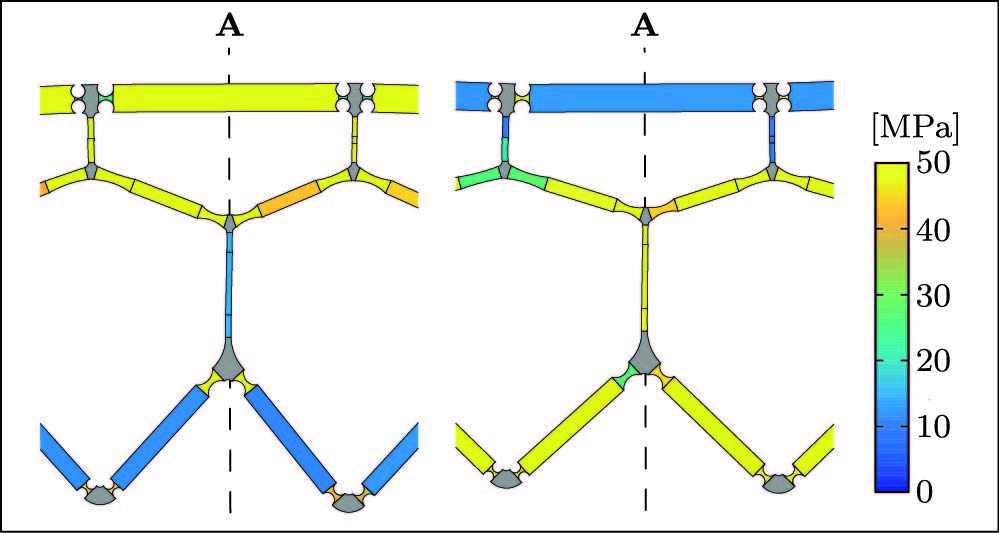}}\hfill
                            \subfloat[]{
                                \includegraphics[height=45mm]{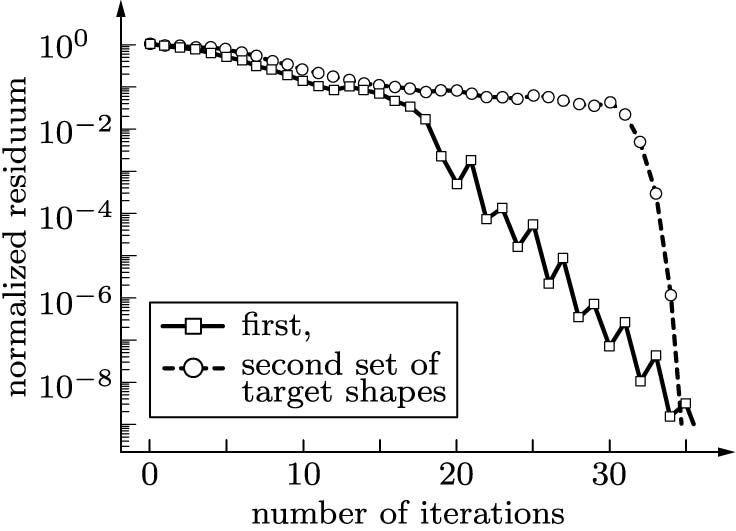}}
                            \caption{Compliant pressure actuated cellular structures that consist of two cell rows and twenty base pentagons. Cell side lengths and hinge, side thicknesses are optimized such that the equilibrium shapes resemble convex, concave circular arcs with an identical absolute curvature. The maximum shape changes that can be achieved depend on the sign of the curvature for a given pressure set. Equilibrium shapes and von Mises stresses for the (a) first $\left(\Delta\alpha=\pm0.75^\circ\right)$ and (b) second $\left(\Delta\alpha=\mp2.00^\circ\right)$ set of target shapes. (c) Detailed view of the structure that is optimized for the second set of target shapes. (d) Convergence of the residual target shape and stress vectors.}
                            \label{pic:Figure_11_9}
                        \end{center}
                    \end{figure}
                    \restoregeometry}

                \subsubsection{Extreme Curvatures}
                    It was previously shown that the shape changing potential of pressure actuated cellular structures with centric, frictionless hinges is limited by extreme cell geometries. This effect is magnified in compliant structures as each cell side needs to accommodate two cell corners and hinges. This limits the minimum cell side lengths and thus decreases the range of possible cell geometries. The shape changing potential is further reduced by the bending stiffness of compliant hinges and the elasticity of cell sides. Similar to the first example, this is demonstrated with the help of a structure that consists of two cell rows and twenty base pentagons.\\

                    The 536 optimization variables that include 119 cell side lengths, 139 cell side thicknesses and 278 hinge thicknesses\footnote{The pentagonal base side lengths are assumed to be invariant whereas their hinge and cell side thicknesses are varied.} are optimized such that the equilibrium shapes resemble a convex and concave circular arc with an identical absolute curvature. The corresponding target shapes are
                    \begin{align}\nonumber
                        \Delta\alpha_{q,i} =
                        \begin{cases}
                            \hspace{2.7mm}0.75^\circ & \textrm{for } q=1\\
                            -0.75^\circ & \textrm{otherwise}
                        \end{cases}
                    \end{align}
                    and
                    \begin{align}\nonumber
                        \Delta\alpha_{q,i} =
                        \begin{cases}
                            -2^\circ & \textrm{for } q=1\\
                            \hspace{2.7mm}2^\circ & \textrm{otherwise}
                        \end{cases}
                    \end{align}
                    as shown in Figure~\ref{pic:Figure_11_9}. As before, the non-zero shape changes between the pentagonal base sides are limited to $i=2,\ldots,18$. The reference configuration is chosen such that the bending deformations of the base side hinges are minimal so that $\Delta\alpha_{0,i}=0$.\\

                    It can be seen that the structures shape changing capabilities are severely limited by the minimum cell side lengths that are needed to accommodate the rigid cell corners and compliant hinges. Furthermore, all cell sides and hinges are fully stressed for at least one of the two pressure sets. The corresponding stress constraints lead to a wide range of hinge geometries. For example, hinges of cell sides that are not exposed to differential pressures possess a nearly uniform thickness. In contrast, pentagonal base side hinges possess a large thickness variation that is driven by large bending angles and small axial cell side forces. This is problematic as the effect of transverse forces on the maximum hinge stresses is not considered. Hence it would be best to use an advanced hinge model instead that additionally takes shear deformations into account. Nonetheless, the proposed algorithm requires at most 35 iterations for the optimization despite the considerably increased nonlinearity of the underlying problem.


                \subsubsection{Moderate Curvatures}
                    The finite size of cell corners and hinges limits the minimum cell side lengths and thus the shape changing potential of compliant pressure actuated cellular structures. However, this is usually not a problem as moderate shape changing requirements lead to balanced cell geometries. Similar to the second example, this is demonstrated with the help of a structure that consists of two cell rows and sixty base pentagons as illustrated in Figure~\ref{pic:Figure_11_10}.\\

                    The 1,616 optimization variables that include 359 cell side lengths, 419 cell side thicknesses and 838 hinge thicknesses are optimized for a half and full circle with a convex curvature so that
                    \begin{align}\nonumber
                        \Delta\alpha_{q,i} =
                        \begin{cases}
                            3^\circ & \textrm{for } q=1\\
                            6^\circ & \textrm{otherwise}
                        \end{cases}
                    \end{align}
                    where $i=2,\ldots,58$. The reference configuration is chosen such that the bending deformations of the base side hinges are minimal. This is the case for $\Delta\alpha_{0,i}=4.5^\circ$ so that the shape of the undeformed structure is between a half and a full circle. It can be seen that the cell corners and hinges are easily integrated into the balanced cell geometries. Furthermore, if compared with the second example, the influence of compliant hinges and sides on the cell geometries is surprisingly small. As before, the optimization requires about 35 iterations.

                    \afterpage{
                        \newgeometry{}
                        \begin{figure}[htbp]
                            \begin{center}
                                \subfloat[]{
                                    \includegraphics[height=102mm]{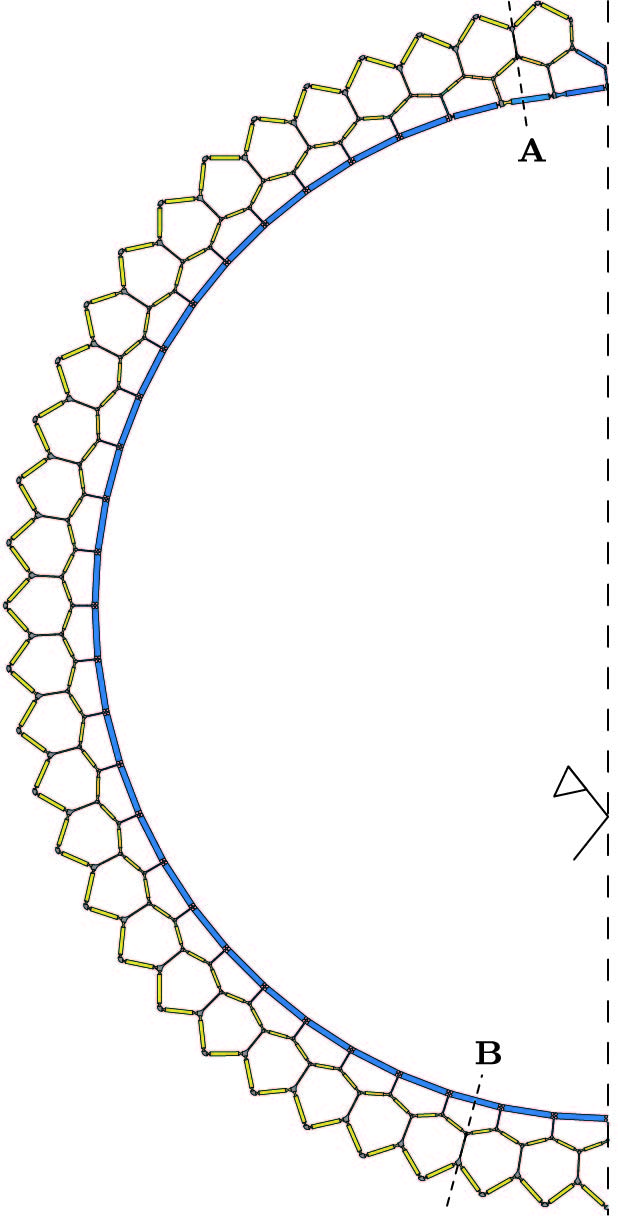}}\hfill
                                \subfloat[]{
                                    \includegraphics[height=102mm]{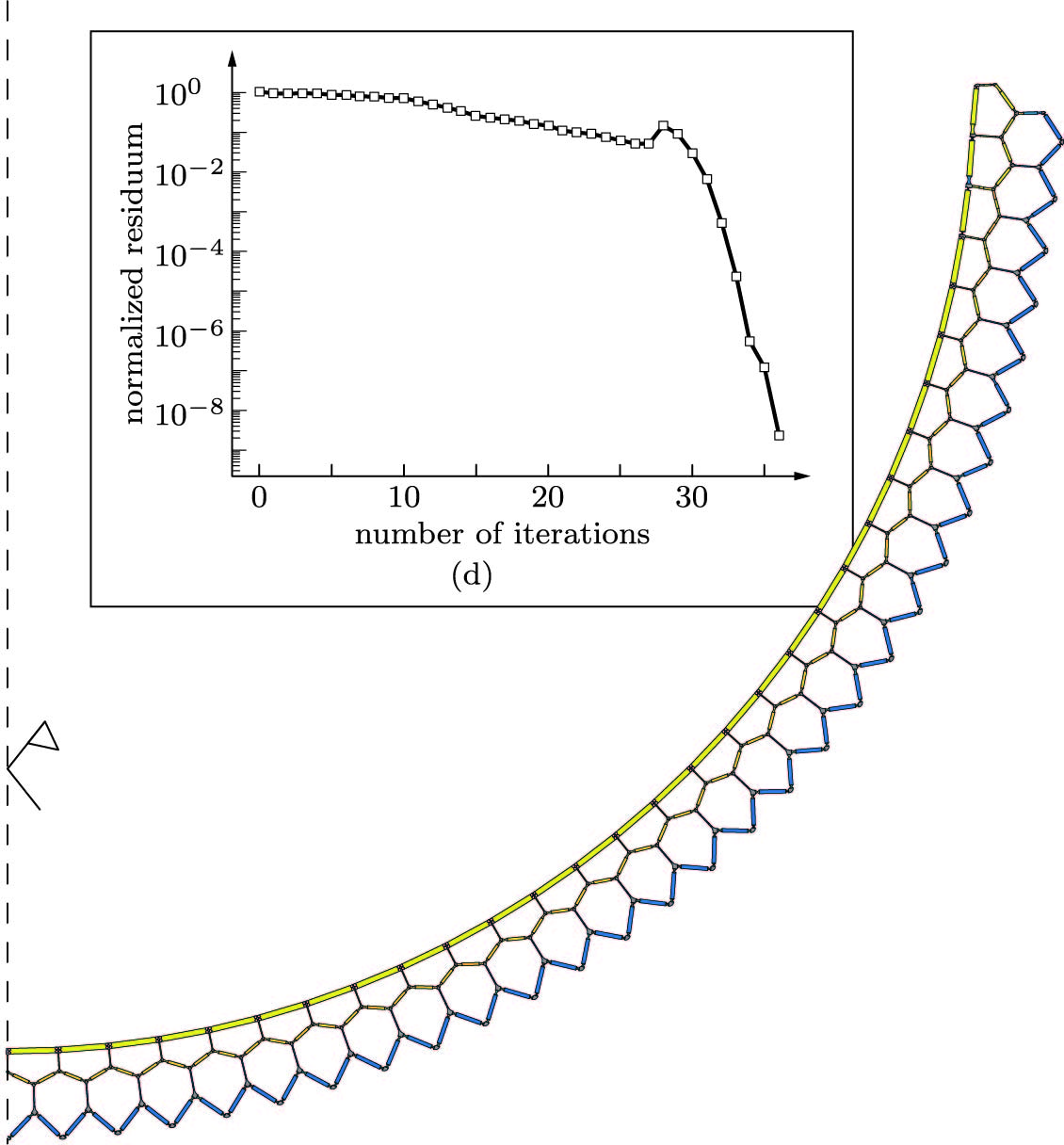}}\vspace{-2mm}

                                \subfloat[]{
                                    \includegraphics[width=\textwidth]{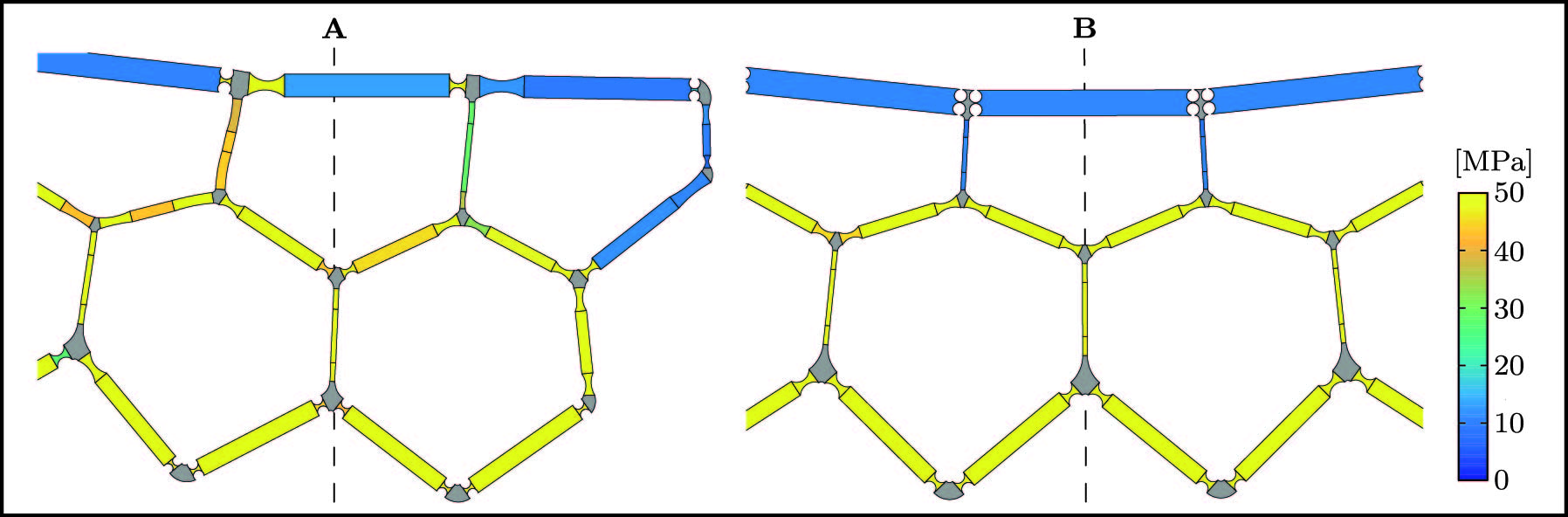}}
                                \caption{Compliant pressure actuated cellular structure that consists of two cell rows and sixty base pentagons. Cell side lengths and hinge, side thicknesses are optimized such that the equilibrium shapes resemble a (a) full and (b) half circle with a convex curvature for the first and second pressure set, respectively. (c) Detailed views of the equilibrium shape for the first pressure set. (d) Convergence of the residual target shape and stress vector.}
                                \label{pic:Figure_11_10}
                            \end{center}
                        \end{figure}
                        \restoregeometry}


    \newpage

    \sectionmark{Summary}
    \begin{framed}
        \noindent \textbf{Summary}\\

        \noindent The geometric and mechanical model can be efficiently coupled for the simulation and optimization of pressure actuated cellular structures. The redundancy of optimization variables leads to a null space in which an arbitrary objective function can be minimized without affecting the equilibrium shapes of the pentagonal base sides. However, this requires third-order derivatives that are expensive to compute. In summary it can be said that the optimization of practically relevant structures can be reliably done within a relatively short time frame even on aged hardware.\\

        \noindent \textbf{Conclusion}\\

        \noindent The simulation and optimization of pressure actuated cellular structures is by no means a bottleneck for the future development of gapless high lift devices. However, an industrial realization of this novel concept requires geometric models that are tailored to certain manufacturing processes and mechanical models that take shear deformations of hinges into account. Some of the required improvements are discussed in the following chapter.
    \end{framed}

    \newpage
    \thispagestyle{empty}


    \cleardoublepage
    \newgeometry{}
    \thispagestyle{empty}

        \noindent \Large \textbf{Nomenclature} \normalsize
        \vspace{5mm}

        \noindent
        \begin{tabular}{llll}
            \textbf{Superscripts}\hspace{5mm}
            & tar       & \hspace{5mm} & target value\\\vspace{-2.5mm}\\

            & $\kappa$, $\mathfrak{e}$, $u$, $v$ & & derivatives with respect to global state variables\\ \vspace{-2mm}\\

            & $G$, $M$  &              & geometric, mechanical model\\
            & $H$       &              & compliant hinge\\
            & $P$, $T$  &              & pentagonal, triangular cell\\
            & $S$       &              & cell side
        \end{tabular}


        \vspace{5mm}
        \noindent
        \begin{tabular}{llll}
            \textbf{Subscripts}\hspace{9mm}
            & max                          & \hspace{1mm} & maximum\\
            & $0$                          &              & reference configuration\\
            & $\sigma$, $a$, $b$, $c$, $r$ &              & reference to stresses, cell sides, residuum\\
            & $q$                          &              & pressure set
        \end{tabular}


        \vspace{5mm}
        \noindent
        \begin{tabular}{llll}
            \textbf{Numbers}\hspace{11mm}
            & $n_P$ & \hspace{14mm} & pentagonal cells\\
            & $n_R$ &               & cell rows\\
            & $n_u$ &               & global state variables\\
            & $n_v$ &               & cell sides
        \end{tabular}


        \vspace{5mm}
        \noindent
        \begin{tabular}{llll}
            \textbf{Greek Letters}\hspace{2.5mm}
            & $\alpha$, $\beta$ & \hspace{12mm} & global state variables\\
            & $\kappa$          &               & cell corner curvature\\
            & $\lambda$         &               & Lagrange multiplier\\
            & $\mu$             &               & thickness ratio of compliant hinge\\
            & $\nu$             &               & Poisson's ratio\\
            & $\sigma$          &               & stress\\
            & $\varphi$         &               & bending angle of hinge\\\vspace{-2.5mm}\\
            & $\Pi$             &               & potential energy
        \end{tabular}


        \vspace{5mm}
        \noindent
        \begin{tabular}{llll}
            \textbf{Roman Letters}\hspace{0.5mm}
            & $a$                 & \hspace{10mm} & base side length\\
            & $b$                 &               & interpolation parameter\\
            & $e$, $h$            &               & set of design, state variables\\
            & $f$                 &               & force vector\\
            & $i$, $j$, $k$       &               & integers\\
            & $m$, $n$            &               & axial, rotational spring stiffness\\
            & $p$                 &               & cell row pressure\\
            & $q$                 &               & pressure set\\
            & $r$                 &               & residuum\\
            & $s$, $t$            &               & central hinge, cell side thickness\\
            & $u$, $v$            &               & global state variables\\
            & $y$                 &               & internal length
        \end{tabular}

    \newgeometry{}
    \thispagestyle{empty}

        \begin{tabular}{llll}
            \textbf{Roman Letters}\hspace{0.5mm}
            & $A$                 &               & cross sectional area\\
            & $B$                 &               & Boolean matrix\\
            & $E$                 &               & Young's modulus\\
            & $F$                 &               & objective\\
            & $G$, $R$, $S$       &               & gradients\\
            & $H$, $K$            & \hspace{10mm} & Hessian, stiffness matrix\\
            & $L$                 &               & arc length \textbf{AND} Lagrangian\\
            & $N$                 &               & null space\\
            & $Z$                 &               & constraint derivative
        \end{tabular}
        \restoregeometry

        \thispagestyle{empty} 
        \cleardoublepage
    \chapter{Conclusions}
        \section{Summary}

        \marginnote{
        \begin{center}
            \includegraphics[width=\marginparwidth]{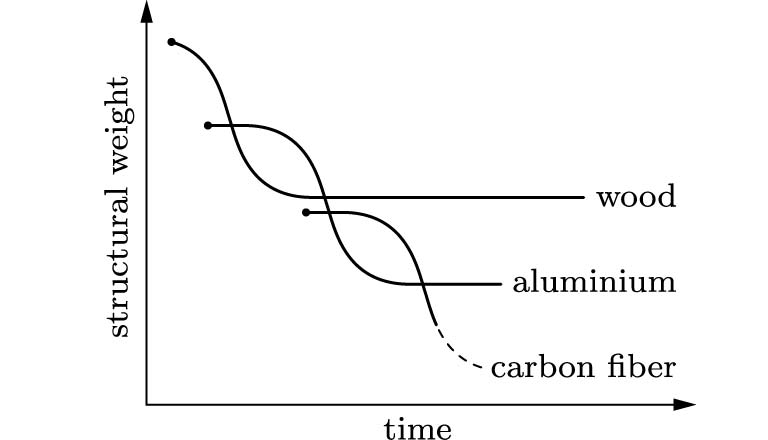}
            \small(a)\vspace{10mm}
            \includegraphics[width=\marginparwidth]{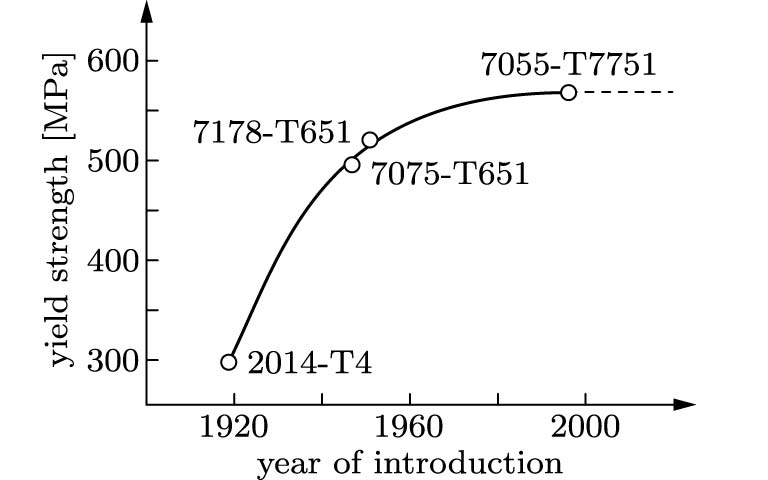}
            \small(b)
            \captionof{figure}{(a) Material driven innovation cycles of airplane structures. (b) Yield strength of aluminium alloys since the 1920s (data from \cite{Campbell2006-1}).}
            \label{pic:Figure_12_1}
        \end{center}}[-2mm]

        \noindent Airplanes consist of a fuselage, wings, high lift devices, control surfaces and engines. These components are, as far as possible, independently developed and manufactured. They are incrementally improved within the limits of existing technologies until the required efforts outweigh the potential gains. Significant progress at this point can only be achieved by innovations that need to occur in one or more fields. Some of the most important innovations for the aerospace industry were made in material science. For example, great weight savings were achieved by the transition from wood to aluminium and the subsequent transition from aluminium to carbon fiber reinforced plastics. The latter occurred after the yield strength of newly developed aluminium alloys leveled off in the second half of the twentieth century as illustrated in Figure~\ref{pic:Figure_12_1}.\\

        A myriad of similar innovation cycles within many fields led to highly optimized structures. It is therefore increasingly difficult to significantly improve the performance of airplanes by solely optimizing their components. On the other hand, the steadily increasing passenger numbers, oil prices and population densities around airports put a huge pressure on the airplane industry. Another aspect that adds to this pressure is the need for military airplanes with ever improving stealth properties. These opposing trends an only be countered by a shift towards an integrated development of airplane components. A complete fusion between wings and high lift devices seems to be particularly promising as they are a main source of noise emissions and radar reflections. Furthermore, they hinder the use of new technologies that improve the laminar flow around wings and thus the fuel consumption of airplanes.\\

        Existing approaches towards gapless high lift devices lead to structures that are complex and heavy. It is ironic that this is mostly due to their separate consideration of actuators, mechanisms and skins. In contrast, plants that can undergo rapid shape changes are highly integrated. It is virtually impossible to distinguish between their components. Nastic plants consist of a large number of three-dimensional cells with tailored geometries and material properties. Their overall shape and stiffness depends on the cell pressures that are varied through osmosis. The pressure induced tension forces are advantageous since they enable the use of relatively thin and flexible cell walls that, in return, minimize the deformation energy during shape changes. Another remarkable property of nastic plants is that they function without a central control system although they have to vary the pressures of a large number of cells.\\

        Despite the attractive properties of nastic plants, a translation of their working principles into gapless high lift devices seemed to be elusive. It is believed by the author that some of these obstacles were overcome by the work in this thesis. It was shown that cell geometries of pressure actuated cellular structures can be simplified if their surfaces remain developable during shape changes\footnote{The taper ratio of most wings is close to one so that their surfaces can be considered to be developable.}. This enables the use of prismatic cells with tailored pentagonal or hexagonal cross sections. A further simplification was achieved by the stacking of equally pressurized cell rows. A major problem that remained was the optimization of cell geometries for given target shapes, cell row pressures and material properties. This led to the development of a kinematic framework for compliant pressure actuated cellular structures that efficiently links their geometric and mechanical models. The optimization itself is based on a second order approach that minimizes an arbitrary objective function and enforces the required target shapes with the help of Lagrange multipliers. An object-oriented design of the algorithm was implement in Matlab together with a postprocessor that enables rapid prototyping. The outstanding properties of this approach were demonstrated by several example structures with two and three cell rows. Different material properties, cell row pressures and target shapes were considered.\\

        It was shown that compliant pressure actuated cellular structures can be made from a wide range of materials if cell sizes and pressures are chosen accordingly. Particularly advantageous is the use of fiber reinforced composites that can be tailored to efficiently carry the cell side forces. Cytoskeletons can be used to support the cell sides that are exposed to differential pressures. This reduces their thicknesses and thus the overall weight. The pressurization of prismatic cells requires hermetic seals at both ends that can be realized as minimal surfaces that are locally reinforced by cytoskeletons. These end-caps possess an isotropic stress state in the undeformed configuration so that they are relatively thin and flexible. Despite their many advantages, the integration of cytoskeletons and end-caps into compliant pressure actuated cellular structures is difficult. The use of hermetic, elastic tubes that are inserted into each cell is therefore often advantageous.\\

        Pressure actuated cellular structures with developable surfaces can be manufactured by various processes such as the wire electrical discharge machining of metals or the extrusion of plastics. Particularly promising is the three-dimensional weaving of high strength fibers as it enables relatively lightweight structures that can undergo large shape changes. In summary it can be said that the tight integration, shape changing capabilities, stiffness and simplicity of pressure actuated cellular structures is unique. It seems likely that, in one way or another, they will find their way into future airplanes.


        \section{Future Work}
            The presented framework for the simulation and optimization of compliant pressure actuated cellular structures is remarkably efficient. With regards to the total number of cells and cell rows, there seems to be no limit for the simulation and optimization of practically relevant structures. Nonetheless, several improvements can be made that range from the geometric and mechanical model to the software development and manufacturing of compliant pressure actuated cellular structures. A brief overview of the main areas that need to be addressed for an industrial use of this technology is given in the following.


            \subsection{Geometric Model}

                \marginnote{
                \begin{center}
                    \includegraphics[width=\marginparwidth]{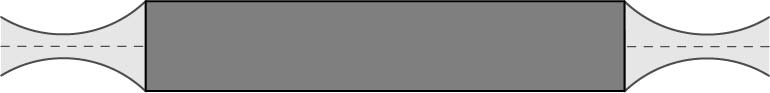}
                    \small(a)\vspace{15mm}
                    \includegraphics[width=\marginparwidth]{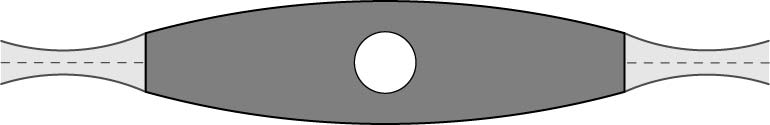}
                    \small(b)\vspace{15mm}
                    \includegraphics[width=\marginparwidth]{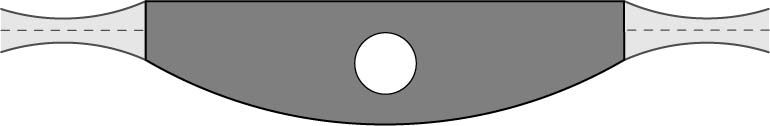}
                    \small(c)\vspace{15mm}
                    \includegraphics[width=\marginparwidth]{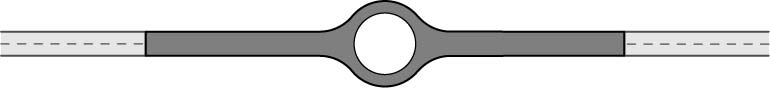}
                    \small(d)
                    \captionof{figure}{Different cell side designs for compliant pressure actuated cellular structures. (a) Current cell side with a constant thickness. (b-d) Improved cell side for segmented structures with a central opening for a tension rod. (b-c) Cell sides with a varying thickness that are exposed to differential pressures. (b) Internal and (c) boundary side that provides a smooth surface. (d) Side that is only subjected to an axial force.}
                    \label{pic:Figure_12_2}
                \end{center}}[-90mm]

                \noindent Geometric models that are designed for different manufacturing processes can be incorporated into the presented framework. The current model with rectangular cell sides is particularly suited for the wire electrical discharge machining of metals. The addition of a central cell side opening, as shown in Figure~\ref{pic:Figure_12_2}, enables prototype structures that consist of a large number of cross sectional segments that are connected via tension rods. Particularly interesting is the development of a geometric model for woven structures. This needs to be done such that the weaving parameters are directly optimized.\\

                The equal pressurization of cells within a single row is optimal as long as curvature changes along the arc length are of the same sign. In all other cases it is best to separate cell rows into regions with different pressures. This needs to be done in conjunction with an algorithm that splits the parametric target shapes into pentagonal base sides of varying lengths such that the required shape changes between cells are uniform.


            \subsection{Mechanical Model}

                \marginnote{
                \begin{center}
                    \includegraphics[width=\marginparwidth]{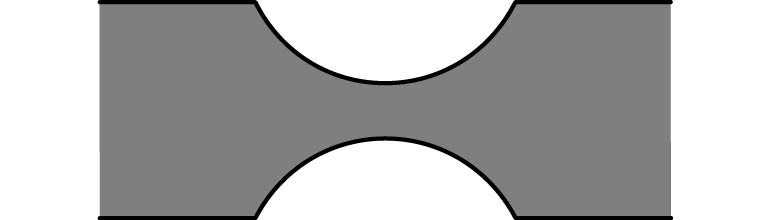}
                    \small(a)\vspace{15mm}
                    \includegraphics[width=\marginparwidth]{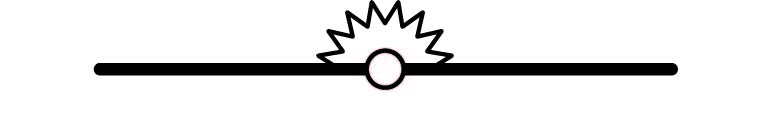}
                    \small(b)\vspace{15mm}
                    \includegraphics[width=\marginparwidth]{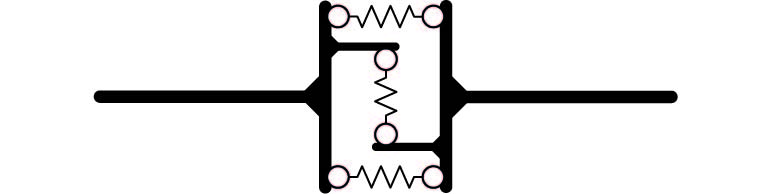}
                    \small(c)
                    \captionof{figure}{(a) Compliant hinge of the geometric model. (b) Current and (c) advanced mechanical representation that takes shear and axial deformations into account.}
                    \label{pic:Figure_12_3}
                \end{center}}[-6mm]

                \noindent Simulation and optimization results are accurate as long as the geometric model is properly represented. The current model has two shortcomings in this regard. First, as shown in Figure~\ref{pic:Figure_12_2}~(d), cell sides that are not exposed to differential pressures possess hinges with a nearly constant thickness. Their bending deformations are not localized by the hinge geometry so that current hinge locations are probably inaccurate. Second, as shown in Figure~\ref{pic:Figure_12_3}, compliant hinges are represented by rotational springs. Their shear stresses are neglected so that some hinges can become unrealistically thin. A remedy is the use of an advanced hinge representation that takes shear and axial deformations into account.


            \subsection{Simulation and Optimization}
                The required number of iterations for the optimization of pressure actuated cellular structures with centric, frictionless hinges and rigid cell sides is remarkably small and essentially independent of the number of cells and cell rows. This does not significantly change for compliant structures that possess an increased nonlinearity. Nonetheless, it might be possible to further reduce the number of iterations with the help of line search algorithms or momentum methods particularly in the presence of large nonlinearities.\\

                The exact consideration of an objective leads to optimization results that are independent of the initial state variables and step length. This requires third order derivatives that are difficult to implement for models with eccentric hinges, and rotational, axial springs. It is therefore possible to reduce the sensitivity of the optimization results with respect to the initial state variables and step length by using a staged optimization approach with an increasing model complexity.


            \subsection{Software Development}
                The mechanical aspect of pressure actuated cellular structures is well understood and an integration of previously discussed improvements into the existing framework is, at least from a theoretical point of view, straightforward. However, the corresponding development of a modular, highly versatile and efficient software is not trivial\footnote{The currently used software is a result of five major revisions. Although highly optimized, it encompasses nearly 10k lines of object oriented Matlab code.}. Such a software should be able to seamlessly switch between different geometric and mechanical models and to interface with different solution methods as well as pre- and post processors. Furthermore, it needs to minimize the runtime by exploiting the one-dimensional nature of pressure actuated cellular structures that leads to sparse, symmetric matrices with a small bandwidth. This enables the use of domain decomposition methods that distribute the workload between several processors. However, the amount of time and effort that is required for the development of such a software can be hardly overestimated.\\

                An important step towards the development of a modular code is to split the software into four parts:
                \begin{itemize}
                    \item The geometric and mechanical module are based on a kinematic framework that describes the centerline of cell sides. It is thus advantageous to create a \textbf{\textit{Kinematic Module}} that deals with common geometric properties.

                    \item Depending on the manufacturing process, a large number of different \textbf{\textit{Geometric Modules}} can be created. These modules need to additionally describe cell pressures, material properties and the stiffness of axial, rotational and transverse springs so that they become interchangeable.

                    \item The current hinge representation leads to eight different \textbf{\textit{Mechanical Modules}}. A consideration of hinge representations with axial and transverse springs increases this number to 32. Interchangeability of modules requires a sole focus on energy terms that are directly derived from cell pressures and stiffness terms.

                    \item A separate \textbf{\textit{Numerical Module}} processes the sensitivities of the geometric and mechanical modules for the simulation and optimization of compliant pressure actuated cellular structures. It needs to be implemented such that it can easily switch between different simulation and optimization approaches.
            \end{itemize}


            \subsection{Manufacturing}
                The next software generation enables an accurate simulation and optimization of compliant pressure actuated cellular structures. A fast and inexpensive realization of prototypes is therefore of utmost importance for a successful push towards gapless high lift devices. This needs to be tackled in two ways.\\

                First, pneumatic and hydraulic pouches with varying sizes and pressure ratings need to be developed. These pouches can be used in prototype structures that are assembled from a large number of segments or in production structures that are woven from high strength fibers. Previous work in this regard by Gram\"uller et al~\cite{Gramueller2014-1,Gramueller2015-1} is too sketchy and insufficient. Second, existing three-dimensional weaving techniques for cellular structures need to be translated into a geometric model. An approach that directly optimizes the weaving parameters increases the accuracy and decreases the turnaround time. 
        \renewcommand{\chaptermark}[1]{\markboth{}{#1}}

    \cleardoublepage
    \chapter*{Appendix}
    \chaptermark{Appendix}
    \addcontentsline{toc}{chapter}{Appendix}
        \renewcommand\thefigure{A\arabic{figure}}
        \setcounter{figure}{0}

        \renewcommand\thetable{A\arabic{table}}
        \setcounter{table}{0}

        This appendix serves two purposes. First, it provides data that enables the verification of the load carrying capacity and shape changing capabilities of a pressure actuated cellular structure via simple hand calculations. Second, detailed information about the considered geometries of the passenger seat and the gapless high lift devices is provided. This data can be used to verify the presented results or serve as a basis for further investigations.


        \section*{Cantilever}
        \addcontentsline{toc}{section}{Cantilever}
            The used node and cell side numbering of the tip loaded cantilever that was introduced in Chapter~7 is illustrated in Figure~\ref{pic:Figure_A_1}. The nodal coordinates of the equilibrium configurations for cell row pressures of either 0.4 or 2.0~MPa are summarized in Table~\ref{tab:Table_A_1} and the corresponding cell side forces are summarized in Table~\ref{tab:Table_A_2}.
            \vspace{5mm}

            \begin{center}
                \includegraphics[width=\textwidth]{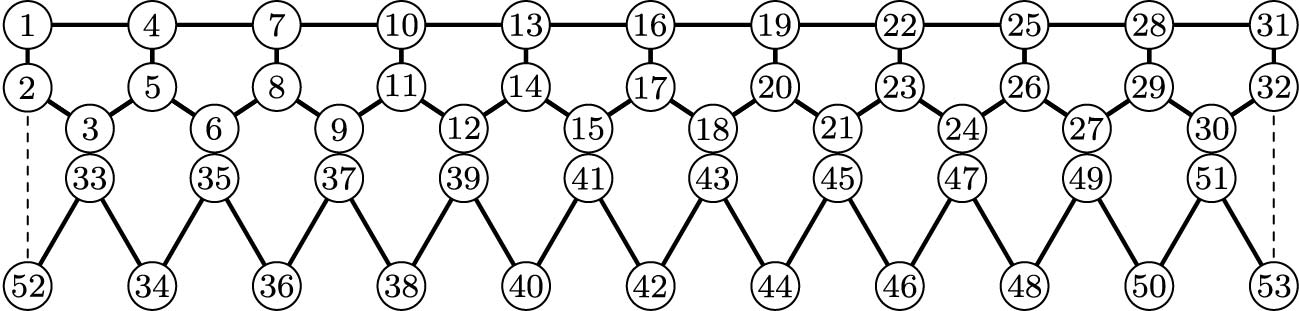}
                \small(a)\vspace{5mm}
                \includegraphics[width=\textwidth]{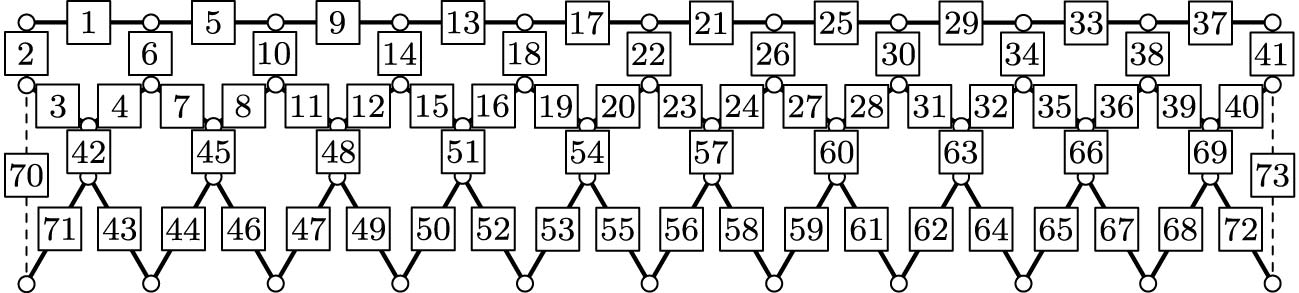}
                \small(b)
                \captionof{figure}{(a) Node and (b) cell side numbering of cantilever.}
                \label{pic:Figure_A_1}
            \end{center}

            \newpage

            \begin{table}[H]
                \begin{center}
                    \footnotesize
                    \begin{tabular}{rrrrr}
                       \multirow{2}{*}{$N$} & $x_1$ & $y_1$ & $x_2$ & $y_2$\\
                       & [mm] & [mm] & [mm] & [mm]\\\hline
                         1 &          0 &          0 &          0 &          0\\
                         2 &          0 &   −50.0000 &          0 &   −50.0000\\
                         3 &    55.2935 &   −73.2946 &    44.9419 &   −89.7522\\
                         4 &    99.9717 &     2.3803 &    99.7155 &    −7.5388\\
                         5 &   109.1154 &   −46.7765 &    95.4444 &   −57.3562\\
                         6 &   166.4943 &   −64.3172 &   137.7908 &   −99.8625\\
                         7 &   199.0197 &    16.1460 &   198.5391 &   −22.8334\\
                         8 &   216.2532 &   −30.7902 &   191.0712 &   −72.2727\\
                         9 &   275.5689 &   −39.8271 &   231.8457 &  −116.2892\\
                        10 &   294.9835 &    44.2696 &   296.4087 &   −43.3651\\
                        11 &   319.7106 &     0.8118 &   286.9084 &   −92.4543\\
                        12 &   379.7027 &     1.7879 &   327.0753 &  −137.0259\\
                        13 &   385.0069 &    87.8104 &   393.6093 &   −66.8610\\
                        14 &   416.6285 &    49.0796 &   383.1809 &  −115.7615\\
                        15 &   475.4671 &    60.8281 &   423.6836 &  −160.0282\\
                        16 &   465.8645 &   146.6494 &   490.6004 &   −91.2076\\
                        17 &   503.5866 &   113.8310 &   480.2891 &  −140.1329\\
                        18 &   559.1272 &   136.5303 &   522.0143 &  −183.2492\\
                        19 &   534.1871 &   219.6702 &   587.8749 &  −114.3957\\
                        20 &   576.9816 &   193.8123 &   578.6864 &  −163.5443\\
                        21 &   626.9387 &   227.0433 &   622.4210 &  −204.6209\\
                        22 &   586.7487 &   304.7423 &   685.8386 &  −134.4733\\
                        23 &   633.3788 &   286.6968 &   678.7550 &  −183.9691\\
                        24 &   675.5251 &   329.4017 &   725.1370 &  −222.0311\\
                        25 &   620.8486 &   398.7486 &   784.7016 &  −149.5105\\
                        26 &   669.9165 &   389.1391 &   780.6846 &  −199.3490\\
                        27 &   702.3229 &   439.6351 &   830.1427 &  −233.3182\\
                        28 &   634.7420 &   497.7787 &   884.3763 &  −157.5712\\
                        29 &   684.7360 &   496.9998 &   884.3378 &  −207.5713\\
                        30 &   706.0260 &   553.0957 &   937.0112 &  −236.3032\\
                        31 &   628.0109 &   597.5519 &   984.3722 &  −156.6585\\
                        32 &   677.3280 &   605.7876 &   989.0510 &  −206.4391\\
                        33 &    57.2529 &  −113.2467 &    43.3326 &  −129.7198\\
                        34 &   121.7017 &  −189.7083 &    78.2896 &  −223.4109\\
                        35 &   172.8900 &  −103.8027 &   132.7541 &  −139.5442\\
                        36 &   248.0438 &  −169.7718 &   163.4523 &  −234.7157\\
                        37 &   287.6632 &   −77.9550 &   224.6643 &  −155.6393\\
                        38 &   372.6743 &  −130.6155 &   253.0009 &  −251.5405\\
                        39 &   397.8790 &   −33.8440 &   318.6826 &  −176.1355\\
                        40 &   490.8541 &   −70.6632 &   346.6743 &  −272.1379\\
                        41 &   499.6303 &    28.9510 &   414.9356 &  −199.0599\\
                        42 &   597.8391 &    10.1082 &   444.5336 &  −294.5793\\
                        43 &   588.7928 &   109.6983 &   513.7245 &  −222.3808\\
                        44 &   688.7908 &   110.3399 &   546.7814 &  −316.7590\\
                        45 &   661.2429 &   206.4707 &   615.3785 &  −243.9961\\
                        46 &   759.0693 &   227.2075 &   653.6184 &  −336.3958\\
                        47 &   713.2453 &   316.0904 &   720.1230 &  −261.7156\\
                        48 &   804.7591 &   356.4048 &   765.1069 &  −351.0266\\
                        49 &   741.9552 &   434.2231 &   827.9739 &  −273.2594\\
                        50 &   823.3175 &   492.3625 &   881.1414 &  −357.9542\\
                        51 &   745.9406 &   555.7095 &   938.8169 &  −276.2624\\
                        52 &          0 &  −195.2352 &          0 &  −219.8436\\
                        53 &   814.3314 &   628.6665 &  1002.8343 &  −353.0854
                    \end{tabular}
                    \caption{Nodal coordinates of the cantilever for a tip load of 5~kN/m and cell row pressures of either 0.4 or 2.0~MPa.}
                    \label{tab:Table_A_1}
                \end{center}
            \end{table}
            \normalsize

            \begin{table}[htbp]
                \begin{center}
                    \footnotesize
                    \begin{tabular}{rrrrr|rrrrr}
                        \multirow{2}{*}{$S$} & \multirow{2}{*}{$N_1$} & \multirow{2}{*}{$N_2$} & $F_1$ & $F_2$ & \multirow{2}{*}{$S$} & \multirow{2}{*}{$N_1$} & \multirow{2}{*}{$N_2$} & $F_1$ & $F_2$\\
                        & & & [kN/m]& [kN/m] & & & & [kN/m]& [kN/m]\\\hline
                         1&  4&  1&   5.6967& 101.1813&     38& 28& 29&  33.7915& 204.0862\\
                         2&  1&  2&  25.1299&  97.0876&     39& 29& 30& 192.5062& 133.5500\\
                         3&  2&  3& 178.6629&  92.2729&     40& 30& 32& 193.2165& 131.7476\\
                         4&  3&  5& 173.6615&  90.7198&     41& 31& 32&  17.8448&  98.2794\\
                         5&  7&  4&   1.5200&  95.6071&     42&  3& 33& 233.6901&  33.7887\\
                         6&  4&  5&  40.5616& 192.2466&     43& 33& 34&  71.2196&  10.0724\\
                         7&  5&  6& 178.1628&  96.3843&     44& 34& 35&  71.2196&  10.0724\\
                         8&  6&  8& 174.8691&  95.1198&     45&  6& 35& 238.5784&  35.8049\\
                         9& 10&  7&  −4.1023&  89.7046&     46& 35& 36&  72.7398&  10.5494\\
                        10&  7&  8&  40.0964& 195.0859&     47& 36& 37&  72.7398&  10.5494\\
                        11&  8&  9& 180.4924& 100.7557&     48&  9& 37& 241.5641&  37.3859\\
                        12&  9& 11& 177.6238&  99.5173&     49& 37& 38&  73.7565&  10.9766\\
                        13& 13& 10&  −9.9525&  83.6211&     50& 38& 39&  73.7565&  10.9766\\
                        14& 10& 11&  39.1389& 197.4369&     51& 12& 39& 243.6188&  38.8057\\
                        15& 11& 12& 183.3486& 105.1662&     52& 39& 40&  74.5794&  11.3855\\
                        16& 12& 14& 180.7461& 103.9376&     53& 40& 41&  74.5794&  11.3855\\
                        17& 16& 13& −15.4578&  77.3521&     54& 15& 41& 245.2704&  40.1158\\
                        18& 13& 14&  37.9728& 199.3913&     55& 41& 42&  75.2853&  11.7804\\
                        19& 14& 15& 186.0857& 109.6317&     56& 42& 43&  75.2853&  11.7804\\
                        20& 15& 17& 183.7634& 108.4100&     57& 18& 43& 246.6331&  41.3322\\
                        21& 19& 16& −20.2787&  70.8695&     58& 43& 44&  75.8726&  12.1639\\
                        22& 16& 17&  36.7531& 200.9847&     59& 44& 45&  75.8726&  12.1639\\
                        23& 17& 18& 188.4157& 114.1835&     60& 21& 45& 247.6905&  42.4668\\
                        24& 18& 20& 186.4534& 112.9662&     61& 45& 46&  76.3101&  12.5382\\
                        25& 22& 19& −24.1811&  64.1432&     62& 46& 47&  76.3101&  12.5382\\
                        26& 19& 20&  35.6046& 202.2369&     63& 24& 47& 248.3767&  43.5349\\
                        27& 20& 21& 190.2044& 118.8527&     64& 47& 48&  76.5516&  12.9102\\
                        28& 21& 23& 188.7079& 117.6375&     65& 48& 49&  76.5516&  12.9102\\
                        29& 25& 22& −27.0161&  57.1543&     66& 27& 49& 248.5886&  44.5968\\
                        30& 22& 23&  34.6538& 203.1596&     67& 49& 50&  76.5321&  13.3235\\
                        31& 23& 24& 191.4190& 123.6629&     68& 50& 51&  76.5321&  13.3235\\
                        32& 24& 26& 190.5010& 122.4428&     69& 30& 51& 248.2835&  45.8038\\
                        33& 28& 25& −28.7356&  49.9398&     70&  2& 52&        0&        0\\
                        34& 25& 26&  34.0214& 203.7648&     71& 52& 33&  69.8304&   9.6162\\
                        35& 26& 27& 192.1228& 128.6126&     72& 51& 53&  76.3720&  13.7124\\
                        36& 27& 29& 191.8973& 127.3305&     73& 53& 32&        0&        0\\
                        37& 31& 28& −29.4297&  42.7883
                    \end{tabular}
                    \caption{Cell side forces of the cantilever for a tip load of 5~kN/m and cell row pressures of either 0.4 or 2.0~MPa.}
                    \label{tab:Table_A_2}
                \end{center}
            \end{table}
            \normalsize


        \marginnote{
        \begin{center}
            \includegraphics[width=\marginparwidth]{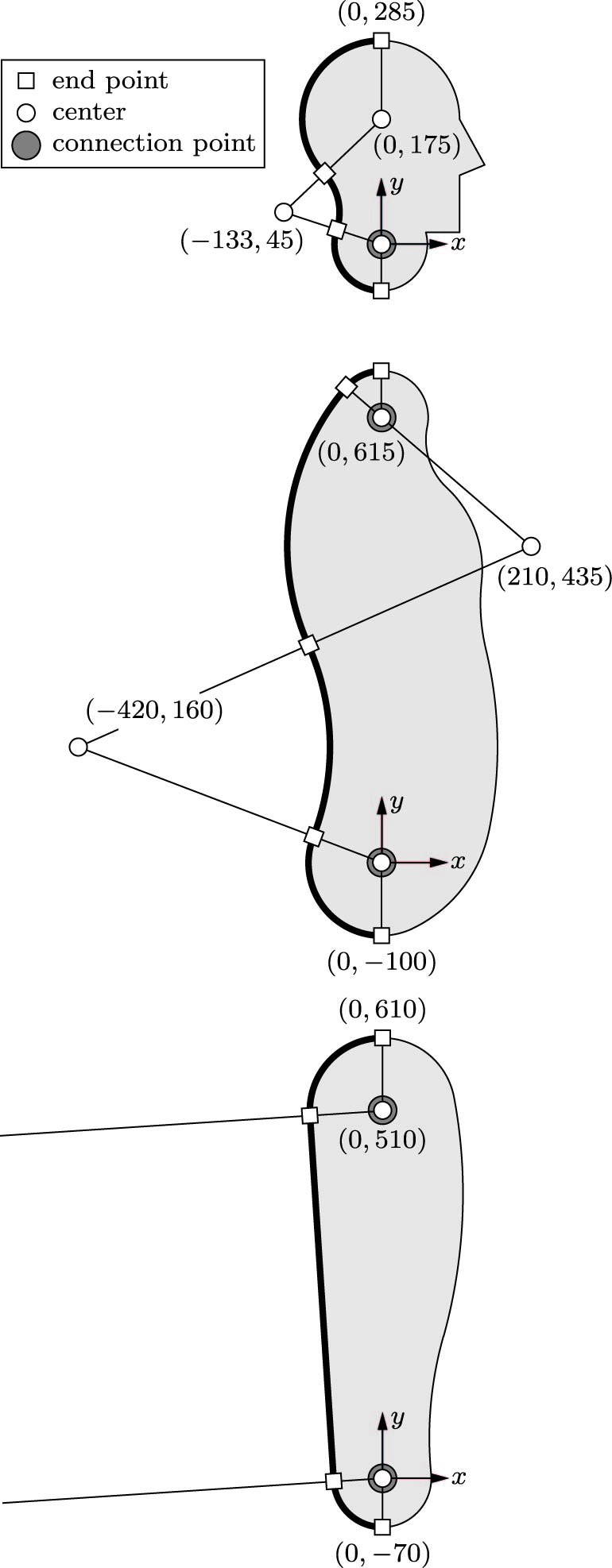}
            \captionof{figure}{Coordinates of center, connection and end points of a two meter tall dummy. Center and end points define sequences of circular arcs that represent the boundaries of the head, torso and upper legs that are in contact with a seat.}
            \label{pic:Figure_A_2}
        \end{center}}[-157mm]

        \section*{Applications}
        \addcontentsline{toc}{section}{Geometry}
            Pressure actuated cellular structures seem to be primarily of interest to the aerospace industry. However, there might be other applications that benefit from lightweight, shape changing structures. For example, many cars and trucks possess pneumatic systems so that their passenger seats could be made, at least partially, from pressure actuated cellular structures. The target shapes for such a seat can be derived from a dummy with a rigid head, torso and extremities that are connected via hinges. The coordinates of the connection points and a description of the boundaries that are in contact with the seat is given in Figure~\ref{pic:Figure_A_2} for a two meter tall dummy. The boundaries are described by $C^1$ continuous sequences of circular arcs\footnote{The approximation of the human body with circular arcs is by no means motivated by a potential discretization with shape changing modules.} so that they are fully defined by the coordinates of their center and end points. The corresponding data for persons with different body sizes can be obtained by scaling. The considered seating postures for a forward, central and backward position are illustrated in Figure~\ref{pic:Figure_A_3}.\\

            \marginnote{
            \begin{center}
                \includegraphics[width=\marginparwidth]{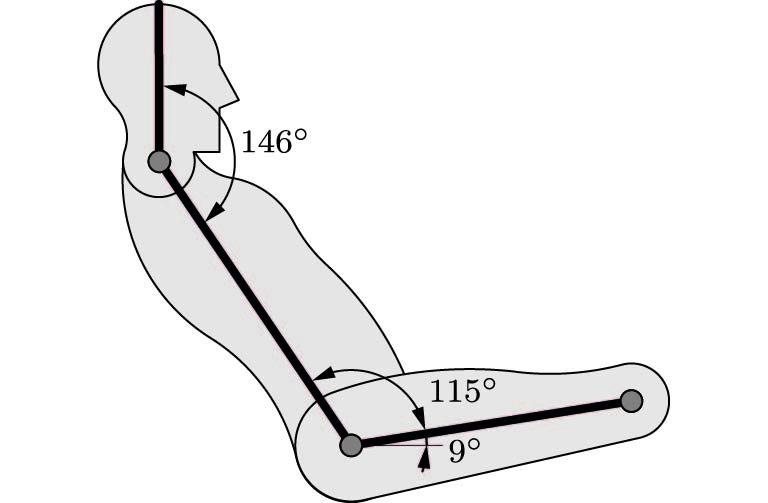}
                \small(a)\vspace{5mm}
                \includegraphics[width=\marginparwidth]{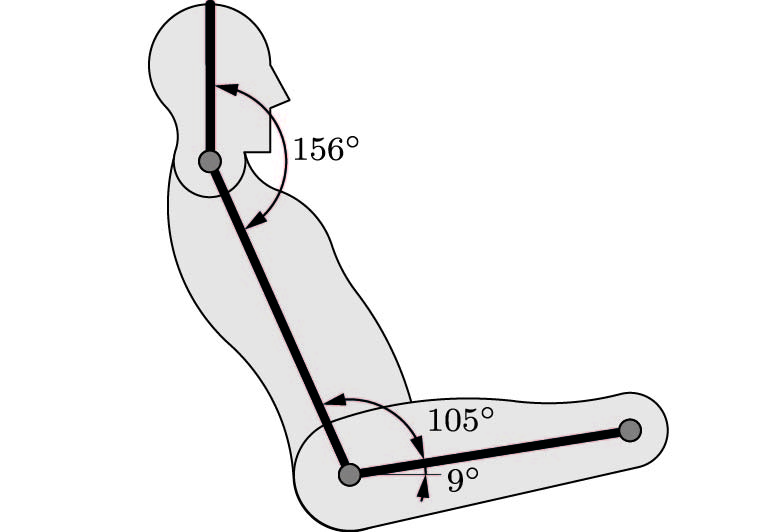}
                \small(b)\vspace{5mm}
                \includegraphics[width=\marginparwidth]{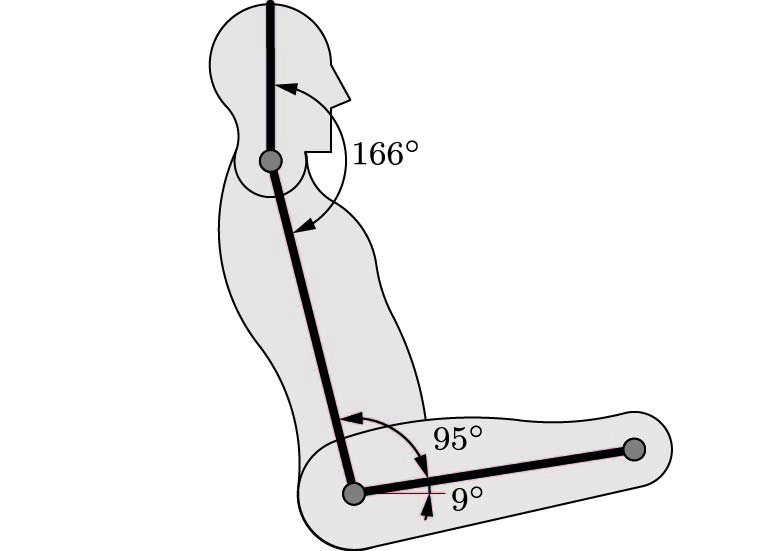}
                \small(c)
                \captionof{figure}{Seating postures of the dummy for the (a) backward, (b) central and (c) forward position.}
                \label{pic:Figure_A_3}
            \end{center}}[-42mm]

            \begin{table}[htbp]
                \begin{center}
                    \footnotesize
                    \begin{tabular}{ccc|ccc|ccc}
                        \multirow{2}{*}{$N$} & $x$ & $y$ & \multirow{2}{*}{$N$} & $x$ & $y$ &
                        \multirow{2}{*}{$N$} & $x$ & $y$\\
                        & [mm] & [mm] & & [mm] & [mm] & & [mm] & [mm]\\\hline
                         1 &       0 &      0 & 23 &  834.26 & 113.18 & 45 & 657.62 & 203.53\\
                         2 &   27.11 &   8.06 & 24 &  896.83 & 111.64 & 46 & 630.37 & 209.40\\
                         3 &   78.96 &  13.69 & 25 &  953.57 & 108.56 & 47 & 603.05 & 215.26\\
                         4 &  110.26 &  19.35 & 26 & 1003.82 & 104.50 & 48 & 575.52 & 221.11\\
                         5 &  141.37 &  25.02 & 27 & 1048.91 &  99.87 & 49 & 547.28 & 226.93\\
                         6 &  172.50 &  30.60 & 28 & 1090.23 &  94.89 & 50 & 518.30 & 232.72\\
                         7 &  203.38 &  36.37 & 29 & 1090.23 & 109.43 & 51 & 488.79 & 238.47\\
                         8 &  233.95 &  42.08 & 30 & 1060.82 & 115.19 & 52 & 458.76 & 244.20\\
                         9 &  264.58 &  47.77 & 31 & 1031.76 & 120.97 & 53 & 428.04 & 249.89\\
                        10 &  295.48 &  53.45 & 32 & 1003.11 & 126.76 & 54 & 396.53 & 255.54\\
                        11 &  326.65 &  59.12 & 33 &  974.87 & 132.58 & 55 & 364.29 & 261.14\\
                        12 &  358.04 &  64.78 & 34 &  947.01 & 138.42 & 56 & 331.30 & 266.71\\
                        13 &  389.73 &  70.41 & 35 &  919.54 & 144.28 & 57 & 297.46 & 272.22\\
                        14 &  421.98 &  76.02 & 36 &  892.51 & 150.15 & 58 & 262.70 & 277.68\\
                        15 &  455.23 &  81.57 & 37 &  865.90 & 156.05 & 59 & 226.95 & 283.06\\
                        16 &  490.02 &  87.02 & 38 &  839.69 & 161.96 & 60 & 190.22 & 288.39\\
                        17 &  526.84 &  92.33 & 39 &  813.87 & 167.89 & 61 & 152.45 & 293.64\\
                        18 &  566.29 &  97.46 & 40 &  788.21 & 173.83 & 62 & 113.55 & 298.81\\
                        19 &  609.27 & 102.29 & 41 &  762.90 & 179.78 & 63 &  73.44 & 303.88\\
                        20 &  656.78 & 106.67 & 42 &  736.98 & 185.70 & 64 &  32.04 & 308.85\\
                        21 &  710.06 & 110.32 & 43 &  715.17 & 190.44 & 65 &      0 & 312.48\\
                        22 &  770.05 & 112.69 & 44 &  682.42 & 197.91
                    \end{tabular}
                    \caption{Trailing edge coordinates for the cruise configuration.}
                    \label{tab:Table_A_3}
                \end{center}
            \end{table}
            \normalsize

            However, existing passenger seats are inexpensive, lightweight and sufficiently comfortable so that it is unlikely that the integration of pressure actuated cellular structures would provide any noteworthy benefit. In contrast, shape changing airfoils would not only benefit from a reduced weight but also from a substantial improvement of aerodynamic and stealth properties. Optimal airfoil geometries for both high and low speed configurations were difficult to come by as most work focuses on gapless leading edges \cite{Kintscher2016-1}. The cruise geometry of a trailing edge as used in this thesis is given by a number of discrete points whose coordinates are summarized in Table~\ref{tab:Table_A_3}. The corresponding coordinates of a leading edge for a high and low speed configuration are summarized in Table~\ref{tab:Table_A_4}.
            \begin{table}[htbp]
                \begin{center}
                    \footnotesize
                    \begin{tabular}{ccccc|ccccc}
                        \multirow{2}{*}{$N$} &  $x_1$ &   $y_1$ &  $x_2$ &   $y_2$ &
                        \multirow{2}{*}{$N$} &  $x_1$ &   $y_1$ &  $x_2$ &   $y_2$\\
                        & [mm] & [mm] & [mm] & [mm] & & [mm] & [mm] & [mm] & [mm]\\\hline
                         1&      0& -184.00&      0 & -184.00 & 34 & 643.23 &   13.32 & 595.67 & -125.05\\
                         2&  63.35& -179.02&  62.93 & -181.89 & 35 & 641.25 &   19.76 & 594.08 & -117.33\\
                         3& 107.56& -174.33& 106.64 & -180.41 & 36 & 638.65 &   26.20 & 591.97 & -109.27\\
                         4& 149.70& -169.46& 148.06 & -180.17 & 37 & 635.39 &   32.63 & 589.34 & -100.86\\
                         5& 190.38& -164.46& 187.74 & -181.36 & 38 & 631.42 &   39.06 & 586.12 &  -92.08\\
                         6& 228.29& -159.25& 224.42 & -183.58 & 39 & 626.61 &   45.49 & 582.22 &  -82.90\\
                         7& 263.60& -153.86& 258.29 & -186.65 & 40 & 620.86 &   51.91 & 577.52 &  -73.25\\
                         8& 298.19& -148.42& 291.15 & -191.03 & 41 & 614.03 &   58.32 & 571.93 &  -63.12\\
                         9& 330.92& -142.87& 321.95 & -196.17 & 42 & 606.04 &   64.71 & 565.37 &  -52.49\\
                        10& 361.15& -137.18& 350.10 & -201.50 & 43 & 596.81 &   71.09 & 557.73 &  -41.37\\
                        11& 389.36& -131.39& 376.10 & -206.96 & 44 & 586.25 &   77.45 & 548.95 &  -29.77\\
                        12& 415.82& -125.51& 400.24 & -212.39 & 45 & 574.25 &   83.78 & 538.90 &  -17.70\\
                        13& 440.50& -119.56& 422.51 & -217.54 & 46 & 560.69 &   90.08 & 527.47 &   -5.19\\
                        14& 463.47& -113.54& 443.04 & -222.23 & 47 & 545.63 &   96.34 & 514.66 &    7.66\\
                        15& 484.77& -107.45& 461.89 & -226.27 & 48 & 529.22 &  102.58 & 500.59 &   20.69\\
                        16& 504.39& -101.32& 479.08 & -229.45 & 49 & 511.65 &  108.78 & 485.40 &   33.76\\
                        17& 522.36&  -95.13& 494.68 & -231.63 & 50 & 493.00 &  114.95 & 469.12 &   46.79\\
                        18& 538.85&  -88.90& 508.86 & -232.79 & 51 & 473.21 &  121.08 & 451.68 &   59.71\\
                        19& 554.07&  -82.64& 521.85 & -232.98 & 52 & 452.18 &  127.17 & 432.97 &   72.50\\
                        20& 568.20&  -76.35& 533.82 & -232.25 & 53 & 429.77 &  133.21 & 412.86 &   85.11\\
                        21& 581.34&  -70.04& 544.85 & -230.60 & 54 & 405.88 &  139.20 & 391.20 &   97.49\\
                        22& 593.52&  -63.71& 555.00 & -228.02 & 55 & 380.43 &  145.12 & 367.90 &  109.56\\
                        23& 604.61&  -57.36& 564.18 & -224.38 & 56 & 353.41 &  150.97 & 342.92 &  121.23\\
                        24& 614.36&  -50.99& 572.21 & -219.47 & 57 & 324.91 &  156.75 & 316.31 &  132.43\\
                        25& 622.55&  -44.60& 578.90 & -213.12 & 58 & 294.97 &  162.45 & 288.11 &  143.06\\
                        26& 629.16&  -38.19& 584.28 & -205.36 & 59 & 263.56 &  168.08 & 258.26 &  153.08\\
                        27& 634.35&  -31.77& 588.50 & -196.35 & 60 & 230.61 &  173.62 & 226.65 &  162.46\\
                        28& 638.34&  -25.34& 591.72 & -186.30 & 61 & 196.10 &  179.06 & 193.30 &  171.15\\
                        29& 641.30&  -18.90& 594.11 & -175.41 & 62 & 160.06 &  184.40 & 158.19 &  179.13\\
                        30& 643.39&  -12.46& 595.79 & -163.82 & 63 & 122.43 &  189.63 & 121.29 &  186.42\\
                        31& 644.64&   -6.01& 596.80 & -151.63 & 64 &  83.18 &  194.74 &  82.56 &  193.01\\
                        32& 645.03&       0& 597.11 & -139.74 & 65 &  42.43 &  199.74 &  42.16 &  198.97\\
                        33& 644.53&    6.87& 596.71 & -132.40 & 66 &      0 &  204.36 &      0 &  204.36
                    \end{tabular}
                    \caption{Leading edge coordinates for the high and low speed configuration.}
                    \label{tab:Table_A_4}
                \end{center}
            \end{table}
            \normalsize

            Parametric least square interpolations with octic polynomials where the first and the last two coordinates are weighted\footnote{The first and the last coordinate are weighted by a factor of ten whereas the second and penultimate coordinate are weighted by a factor of five.} to improve the continuity between the wing and the leading edge leads to
            \begin{align*}
                x\left(\xi\right) &= a_1 \xi^8 + a_2 \xi^7 + a_3 \xi^6 + a_4 \xi^5 + a_5 \xi^4 + a_6 \xi^3 + a_7 \xi^2 + a_8 \xi + a_9\\
                y\left(\xi\right) &= b_1 \xi^8 + b_2 \xi^7 + b_3 \xi^6 + b_4 \xi^5 + b_5 \xi^4 + b_{y,6} \xi^3 + b_7 \xi^2 + b_8 \xi + b_9
            \end{align*}
            where
            \begin{align*}
                \mathbf{a}_1 =
                \left[
                \begin{array}{r}
                    -110895.50\\
                    470701.20\\
                    -816461.56\\
                    744751.96\\
                    -384866.56\\
                    114494.55\\
                    -21951.76\\
                    4227.14\\
                    0.40
                \end{array}
                \right]^\top,\hspace{3mm}
                \mathbf{b}_1 =
                \left[
                \begin{array}{r}
                    -4513.08\\
                    18500.59\\
                    -30982.30\\
                    27008.63\\
                    -12833.30\\
                    2979.79\\
                    -91.03\\
                    319.06\\
                    -183.99
                \end{array}
                \right]^\top
            \end{align*}
            and
            \begin{align*}
                \mathbf{a}_2 =
                \left[
                \begin{array}{r}
                    -100003.75\\
                    427199.76\\
                    -745882.30\\
                    685297.61\\
                    -358234.95\\
                    109427.92\\
                    -22005.47\\
                    4200.64\\
                    0,40
                \end{array}
                \right]^\top,\hspace{3mm}
                \mathbf{b}_2 =
                \left[
                \begin{array}{r}
                    113773.88\\
                    -450565.97\\
                    708034.57\\
                    -553978.53\\
                    216539.15\\
                    -33260.99\\
                    -304.26\\
                    150.68\\
                    -184.01
                \end{array}
                \right]^\top
            \end{align*}
            for $\xi\in\left[0,1\right]$. 
        \newgeometry{}

    \cleardoublepage
    \bibliographystyle{plain}

        \markboth{}{}
    \newpage\null
    \pagenumbering{gobble}
    \includepdf{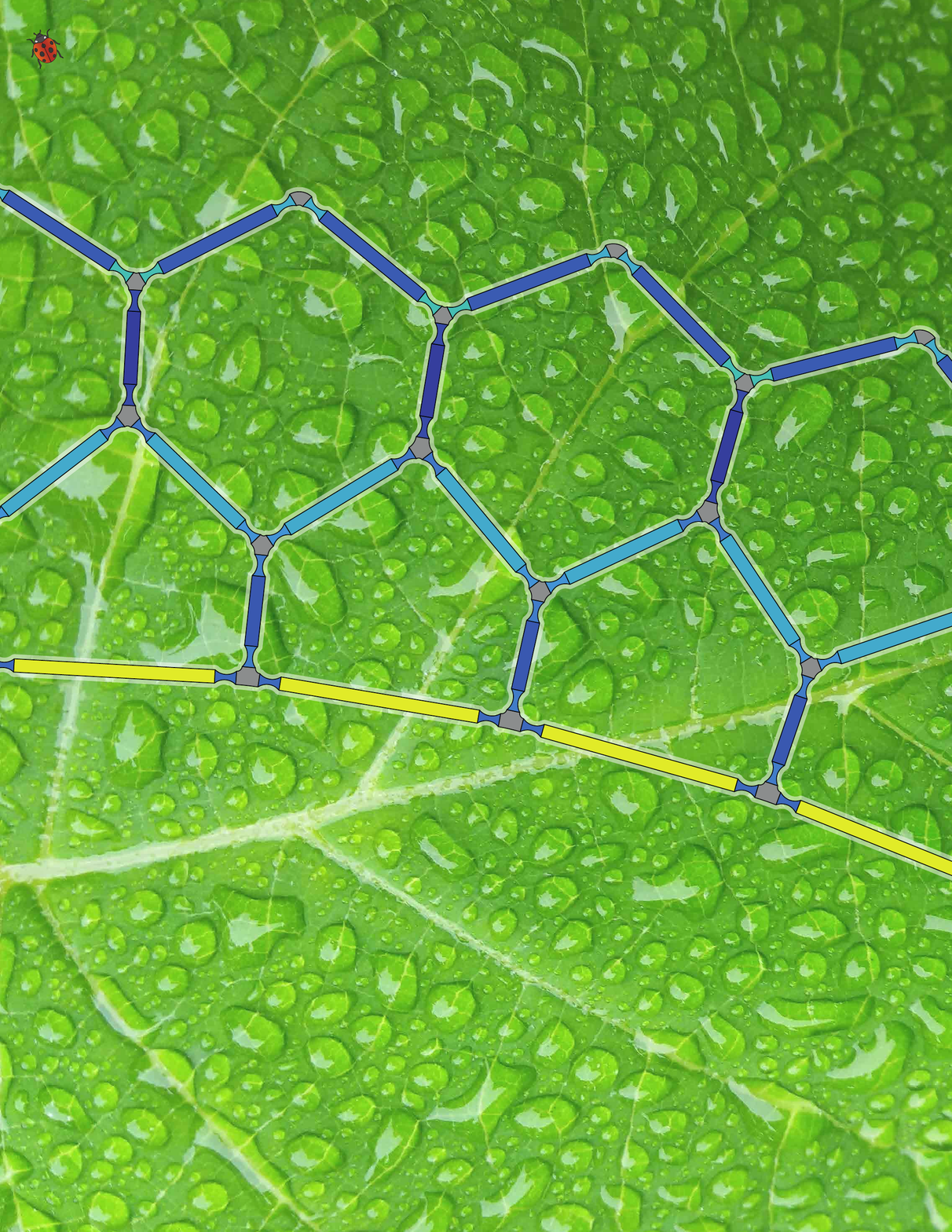} 
\end{document}